\RequirePackage{fix-cm}
\documentclass[twocolumn]{svjour3}          
\smartqed  
\usepackage{appendix}
\usepackage{amsmath}
\usepackage{graphicx}
\usepackage[switch]{lineno}
\usepackage{array}
\usepackage{longtable}
\usepackage[numbers]{natbib}
\usepackage{hyphenat}
\setcitestyle{square}
\setcitestyle{citesep={,}}

\newcommand*\patchAmsMathEnvironmentForLineno[1]{%
\expandafter\let\csname old#1\expandafter\endcsname\csname #1\endcsname
\expandafter\let\csname oldend#1\expandafter\endcsname\csname end#1\endcsname
\renewenvironment{#1}%
{\linenomath\csname old#1\endcsname}%
{\csname oldend#1\endcsname\endlinenomath}}%
\newcommand*\patchBothAmsMathEnvironmentsForLineno[1]{%
\patchAmsMathEnvironmentForLineno{#1}%
\patchAmsMathEnvironmentForLineno{#1*}}%
\AtBeginDocument{%
\patchBothAmsMathEnvironmentsForLineno{equation}%
\patchBothAmsMathEnvironmentsForLineno{align}%
\patchBothAmsMathEnvironmentsForLineno{flalign}%
\patchBothAmsMathEnvironmentsForLineno{alignat}%
\patchBothAmsMathEnvironmentsForLineno{gather}%
\patchBothAmsMathEnvironmentsForLineno{multline}%
}

\usepackage{todonotes}
\usepackage{comment}
\usepackage{graphicx}

\title{Computational Study of Transition-Metal Substitutions in Rutile TiO$_2$ (110) for Photoelectrocatalytic Ammonia Synthesis}

\author{Benjamin M. Comer$^{1 \dagger}$, Max H. Lenk$^{2 \dagger}$, Aradhya P. Rajanala$^{3}$, Emma L. Flynn$^{4}$, Andrew J. Medford$^{1}$*}
\date{\today}

\institute{
$^{1}$ School of Chemical and Biomolecular Engineering, Georgia Institute of Technology\\
$^{2}$ School of Materials Science and Engineering, Georgia Institute of Technology\\
$^{3}$ School of Physics, Georgia Institute of Technology\\
$^{4}$ School of Computer Science, Georgia Institute of Technology\\
$\dagger$ These authors contributed equally to this work. \\
* Correspondence \email{andrew.medford@chbe.gatech.edu}\\
  311 Ferst Drive NW, Atlanta, Georgia 30318 \\
  Tel.:+1 (404) 385-5531\\
}
\authorrunning{B. Comer et. al.}
\titlerunning{Transition-Metal Substitutions in Rutile TiO$_2$ (110)}

\begin{document}

\maketitle
\begin{abstract}

Synthesis of ammonia through photo- and electrocatalysis is a rapidly growing field. Titania-based catalysts are widely reported for photocatalytic ammonia synthesis and have also been suggested as electrocatalysts. The addition of transition-metal dopants is one strategy for improving the performance of titania-based catalysts. In this work, we screen \textit{d}\hyp{}block transition-metal dopants for surface site stability and evaluate trends in their performance as the active site for the reduction of nitrogen to ammonia on TiO$_2$. We find a linear relationship between the \textit{d}\hyp{}band center and formation energy of the dopant site, while the binding energies of N$_2$, N$_2$H, and NH$_2$ all are strongly correlated with the cohesive energies of the dopant metals. The activity of the metal-doped systems shows a volcano type relationship with the NH$_2$ and N$_2$H energies as descriptors. Some metals such as Co, Mo, and V are predicted to slightly improve photo- and electrocatalytic performance, but most metals inhibit the ammonia synthesis reaction. The results provide insight into the role of transition-metal dopants for promoting ammonia synthesis, and the trends are based on unexpected electronic structure factors that may have broader implications for single-atom catalysis and doped oxides.
\end{abstract}
\section{Introduction}
The fixation of atmospheric nitrogen has long been one of the prime challenges in chemistry and chemical engineering \cite{ritter_18, Schloegl_2003}. The Haber-Bosch process has been the route of choice for performing nitrogen fixation for the past century, permitting much of the population growth over that period \cite{Smil_1999}. However, this process has significant drawbacks, including high CO$_2$ emissions and centralized production due to large capital requirements \cite{Comer_2019}. The Haber-Bosch process's considerable contribution to CO$_2$ emissions has been an increasingly pressing concern for the global community, as it accounts for 340 million tonnes of CO$_2$---fully 2\% of the carbon emissions worldwide \cite{gross_12, Schiffer_2017}. For this reason, supplanting the Haber-Bosch process would represent a significant contribution to global efforts to curb climate change. Another drawback lies in the centralization of the Haber-Bosch process, which leads to high transportation costs and contributes to global economic inequality \cite{Comer_2019}.  Due to the high pressures and pure feedstocks required, Haber-Bosch has significant economies of scale. Constructing a new plant requires significant capital and natural resources, as well as a critical mass of demand. Industrialized nations meet these criteria through a constant availability of capital and industrialized agricultural with a steady demand for fertilizer \cite{McArthur_2017}. However, these barriers have prevented developing regions, such as Sub-Saharan Africa, from developing Haber-Bosch plants. The lack of local production results in high costs of fertilizers due to transportation, corresponding to reduced crop yields, which lowers demand and further increases prices \cite{IFDC_2012}. This causes the fertilizer to be most expensive in regions where it is most needed.

Due to the various drawbacks of the Haber-Bosch process, researchers have sought alternative means of producing fixed nitrogen \cite{Comer_2019, McPherson_2019, WANG20181055, Kyriakou_2017, de_Bruijn_2015, Michalsky_2015}. 
Two strategies that have received significant recent interest are electrocatalysis \cite{McPherson_2019} and photocatalysis \cite{Medford_2017}. However, making either of these technologies viable presents a significant challenge. Electrochemical nitrogen fixation requires generating electricity and transporting electrons to the catalyst surface to perform the reaction \cite{Kyriakou_2017}. The need for both solar and electrocatalytic cells may limit the viability of electrochemical processes in the developing world \cite{Comer_2019}. An alternative route is photochemical nitrogen fixation, where the catalyst is placed in direct contact with sunlight, air, and water vapor  to produce fixed nitrogen. The photochemical route has the potential to operate with low capital investment and simpler infrastructure, making it promising for low-resource environments.

The scientific community has known of photochemical nitrogen fixation for decades, but inconsistent results and low rates have discouraged further study \cite{Medford_2017}. N. Dhar \cite{Dhar_1941} was the first to investigate photocatalytic nitrogen, and the first well\hyp{}controlled experiments were performed decades later when Schrauzer and Guth independently re-discovered the process \cite{Schrauzer_1977}. Schrauzer and Guth were able to establish the production of NH$_3$ in sterilized desert sands under illumination, \cite{Schrauzer_1977} including confirmation via isotopic labeling \cite{Schrauzer_1983}. Numerous independent experiments have been performed over the years and reported photochemical production of NH$_3$ over titania materials \cite{Bickley_1979,Augugliaro_1982,Soria_1991,Li_2018,Yuan_2013,Hirakawa_2017}, though legitimate skepticism has remained due to issues with contamination \cite{edwards1992opinion, Davies1995, davies1993reply}, interference \cite{Gao_2018,Cui2018}, and inconsistent results \cite{Medford_2017}. However, recent experiments utilizing ambient pressure X\hyp{}ray Photoelectron Spectroscopy (AP\hyp{}XPS) have observed reduced nitrogen compounds under only under illumination \cite{Comer_2018b}. These experiments provide direct experimental evidence that photo\hyp{}induced nitrogen reduction occurs on titania surfaces, though the presence of carbon-based impurities was found to be a critical enabler of the process. 

Despite improved understanding of photocatalytic nitrogen fixation, the rates of reaction on titania-based catalysts remain relatively low ($\mu mol$ scale \cite{Hirakawa_2017}). The competition of NH$_3$ production with H$_2$ evolution is a key issue for electrocatalytic or photocatalytic nitrogen fixation and has been dubbed the ``selectivity challenge'' \cite{Singh_2017}. For this reason, high Faradic efficiency is also often sought in the electrochemical literature \cite{McPherson_2019}. The driver of this is the opportunity cost of using electricity for catalysis over other possible uses. For this reason, many electrochemical studies focus on low-overpotentials where the selectivity is generally highest, but the overall reaction rate is relatively low.

In contrast, photo-excited electrons are difficult to harvest for alternate uses, and hence the overall solar-to-ammonia efficiency is the key metric for assessing photocatalytic ammonia production. Rates on the order of 0.02\% solar-to-ammonia efficiency have been reported for pure titania catalysts \cite{Hirakawa_2017}. These rates are orders of magnitude lower than the $\sim$20\% solar-to-hydrogen efficiency achieved by state-of-the-art catalytic systems for solar hydrogen production \cite{Nakamura_2015, Jia_2016}. However, it has been posited that a comparatively low solar efficiency of $\sim$0.1-1\% may be sufficient to enable solar fertilizer technology \cite{Comer_2019, Medford_2017}. With sufficiently low capital cost, this system could see use in areas that are far from fertilizer plants due to the lowering of transportation costs. Thus, strategies for enhancing the rate of photocatalytic ammonia production on titania catalysts may be a viable route for designing nitrogen fixation catalysts.

One route to increasing reaction rates is the inclusion of transition\hyp{}metal dopants in TiO$_2$ \cite{Zaleska_2008}.  Transition\hyp{}metal dopants can increase rates via at least two distinct mechanisms: increasing the amount of photo-generated electrons that reach the surface by improving absorption and charge separation, or by altering the kinetics of the surface reaction.
Transition-metal doping has been previously explored to improve the performance of TiO$_2$ photocatalysts \cite{Schneider_2014, Li_2007, Dozzi_2013}. In particular, early work on photocatalytic nitrogen fixation tested a variety of metal dopants. These studies found that several noble metals \cite{Ranjit_1996} and iron in particular increase yields \cite{Schrauzer_1977, Schrauzer_1983, Augugliaro_1982, Soria_1991, Ranjit_1996, Ranjit_1997}. More recently, Hirakawa et al. found that depositing noble metals (Ru, Pt, Pd) onto an already prepared rutile (110) surface led to a decrease in reaction rates \cite{Hirakawa_2017}. This decrease in rates, along with detailed experimental and theoretical studies on the role of iron dopants  \cite{Soria_1991, Comer_2018} suggests that the primary mechanism of these previously-reported dopants is enhanced charge separation. However, transition-metal dopants are also known to affect the surface properties in a range of other materials and reactions \cite{Khan_2018, Gu_2014, Ammal_2016, Gu_2017, Comer_2018, Garc_a_Mota_2011, Yao_2017}. In particular, the field of ``single-atom catalysis'' has revealed that isolated transition-metal sites supported on oxide materials can exhibit remarkable catalytic properties \cite{Liu_2016, Qiao_2011, O_Connor_2018}. However, relatively little effort has been focused on understanding how isolated transition-metal atom dopants affect the surface reactivity of oxides for the conversion of nitrogen to ammonia \cite{Tao_2019, Liu_2019, Zhao_2019, Cheng_2019, Li_2017}.

In this work, we focus on the potential of isolated transition-metal atoms substituted into the rutile (110) titania surface as a potential route to improve the surface kinetics of the nitrogen reduction reaction. We screen The \textit{d}-block transition-metals substituted onto the (110) surface in two different configurations corresponding to formal oxidation states of 2+ and 4+. 
We analyze the trends present across the periodic table for the dopant formation energy, N$_2$, N$_2$H, and NH$_2$ adsorption energy. We also map out the thermodynamics of all N$_2$ reduction pathways and use this to assess the most favorable reaction mechanism. Finally, we assess the expected improvement in reaction rates from forming metal dopant sites on the surface. In this analysis, we consider both electrocatalytic and photocatalytic N$_2$ reduction. The results illustrate that there are clear correlations in the formation and adsorption energies with the \textit{d}-band center and cohesive energies, respectively. We also find scaling relations between the surface species. These scaling relations result in an optimum in the rate-limiting potential for nitrogen reduction as a function of N$_2$H and NH$_2$ adsorption energies. 

\section{Results and Discussion}

Rutile TiO$_2$ (110) is chosen as a model surface based on the experimentally observed correlation between rutile content and reaction rates for photocatalytic nitrogen fixation \cite{Schrauzer_1977}. Additionally, there is a rich literature on the surface science of rutile (110) \cite{Diebold_2003, Yates_1991, Lu1994, Walle2009}, and recent surface-science experiments and DFT calculations indicate that carbon substitution defects on the rutile (110) surface are active for photocatalytic nitrogen reduction \cite{Comer_2018b}. From this model surface, slabs containing metal dopants at the surface in the 2+ and 4+ oxidation states are generated for each dopant metal.
In total, all d-block transition-metals, except Mn and Cr (23 total), are screened for their surface formation energy and activity for nitrogen reduction.  The binding energies of N$_2$H and NH$_2$ have been identified as descriptors for activity in the literature as they are typically involved in the rate-limiting steps \cite{Hoskuldsson_2017, Montoya_2015, Comer_2018}. Thus, these energies have been calculated to assess the activity of generated surfaces.  Full details of the calculation methodology are in the Methods section (Sec. \ref{sec:methods}).


\begin{figure}
    \centering
    \includegraphics[width=0.5\linewidth]{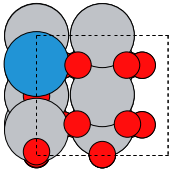}

    \centering
    \includegraphics[width=0.5\linewidth]{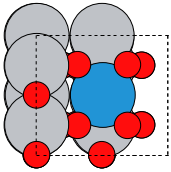}
    \caption{An example of the screened 2+ (top) and 4+ (bottom) slabs. For 2+ sites the substituent metal has replaced a 6 fold Ti atom (seen in blue) and a bridging oxygen vacancy has been formed to allow the metal to enter the 2+ oxidation state.  For 4+ sites the substituent metal has replaced a 5 fold Ti atom (seen in blue) resulting in a 4+ formal oxidation state.}
    \label{fig:2ex_slab}
\end{figure}



\subsection{Trends In Active Site Formation Energies}

The stability of substituted metal surface sites is examined with respect to the position of their \textit{d}-band center. In Fig. \ref{fig:d_band}a, the formation energy of the studied active sites has been plotted against the location of the \textit{d}-band center of the corresponding transition-metal. The pure metallic form is the reference for the formation energy of each metal substituted site. The \textit{d}-band centers are also calculated from the metallic bulk state rather than the single atom. The plot indicates there is a strong correlation between the \textit{d}-band center and formation energy of the metal substitution (R$^2$ = 0.89). We can rationalize the observed correlation in the context of the \textit{d}-band model of chemical bonding \cite{Hammer_1995, Nilsson_2008} summarized in Equation \ref{eq:d_band} below:
\begin{figure*}
    \centering
    \includegraphics[width=0.9\linewidth]{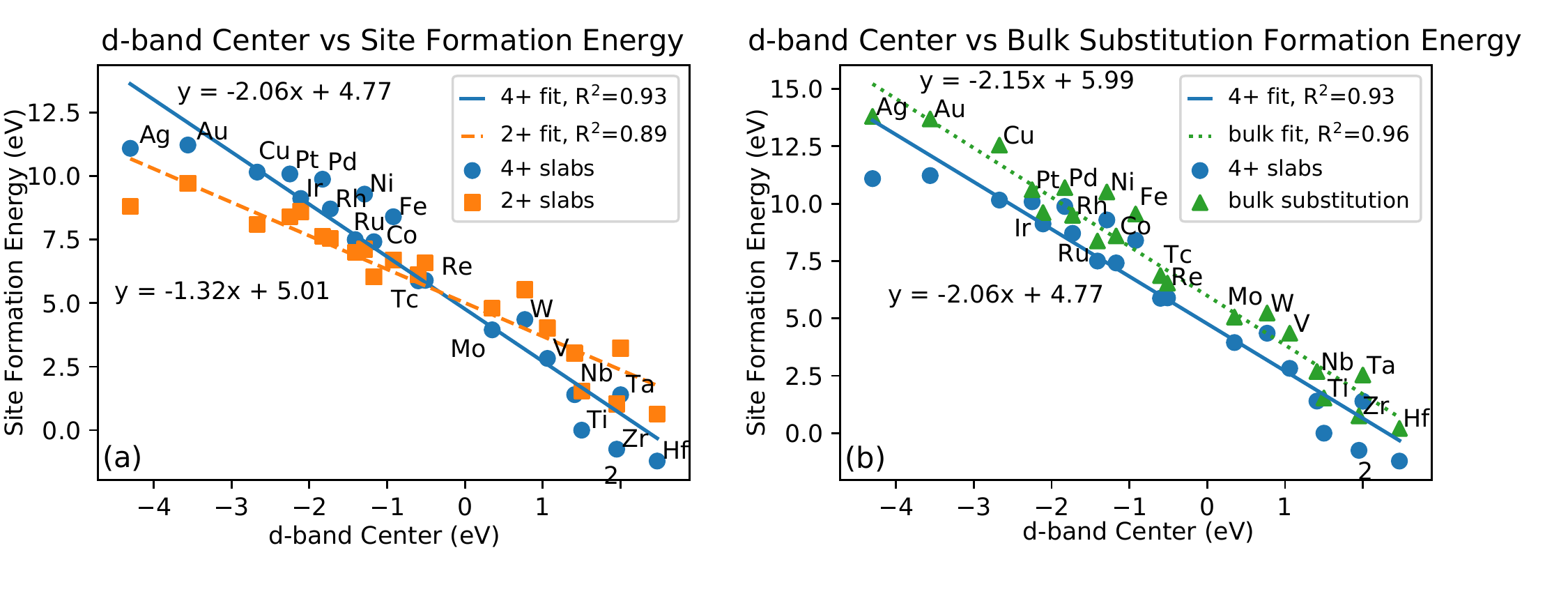}
    \caption{(a) The formation energy of 4+ surface sites (blue) and 2+ surface sites (orange) with respect to their bulk metallic state vs. the metallic \textit{d}-band center (b) The formation energy of 4+ surface sites (blue) and bulk substitutions (green) with respect to their bulk metallic state vs. the metallic \textit{d}-band center. \textit{d}-band centers were obtained from Ref. \citenum{Ruban_1997}. Only metals whose \textit{d}-band center was previously reported in Ref. \citenum{Ruban_1997} are included.}
    \label{fig:d_band}
\end{figure*}

\begin{equation}
    \Delta E_d = \int^{E_F} E(\rho'(E) - \rho(E))dE
    \label{eq:d_band}
\end{equation}
where $\Delta E_d$ is the binding energy associated with interaction with the \textit{d}-band, $E_F$ is the Fermi level energy, $\rho(E)$ is the \textit{d}-band density of states before adsorption, and $\rho'(E)$ is the density of states after adsorption. The interaction between adsorbates and the metal's \textit{s}-states are assumed to be approximately constant for all metals, such that variations in binding energies are controlled primarily through bonding interactions with the \textit{d}-band. Interaction with the \textit{d}-band causes the orbitals of the adsorbate to separate into bonding and anti-bonding orbitals. As the \textit{d}-band center approaches the Fermi-level, the anti-bonding orbitals increasingly fill, leading to a weaker bond.

Our system involves a metal atom interacting with an oxide surface rather than an adsorbate binding to the metal surface. We hypothesize that the \textit{d}-orbitals of the integrated metal atom interact with the \textit{p}-orbitals of oxygen atoms in the surface similar to the way a metal surface interacts with adsorbing oxygen atoms. This explanation is consistent with the observation that the interaction weakens from left to right on the periodic table, as predicted in the literature \cite{Hammer_2000}. This trend implies that the metals most able to integrate into a surface are those with the most favorable interaction with oxygen. A similar relationship has been reported previously for doped rutile oxides \cite{Xu_2015} and oxide-supported single-atom catalysts \cite{O_Connor_2018}. Other reports suggest that the electronegativity of the substituted metal is the relevant descriptor predicting stability \cite{Garc_a_Mota_2011}. The electronegativity is also correlated with the formation energy (R$^2$=0.78 and 0.58 for 2+ and 4+ respectively, see Fig. S3), but not as strong as the correlation with the \textit{d}-band center of the metal (R$^2$=0.89 and 0.93, see Fig. \ref{fig:d_band}). The fact that both of these quantities correlate with the formation energy is not surprising, as a lower energy \textit{d}-center indicates a more favorable addition of electrons, which is similar to the concept of electronegativity. 
The main exceptions to this trend are Ti, Zr, Hf, and Ag. The first three can be rationalized easily since all three lie in the same column of the periodic table, which is the same as the host metal, Ti. The improved stability of substituent metals within the group lines up with the chemical intuition since these elements have the same number of valence \textit{d} electrons. This chemical similarity affords approximately 1.5eV of improved stability relative to the trend. The final outlier, Ag, is more difficult to explain. However, the \textit{d}-band center of Ag is itself an outlier for its position on the periodic table. This deviation may indicate that more complex bonding interactions are involved that are not easily described by the \textit{d}-band model.


These results have implications for the relative stability of single-atom sites over surface metal clusters or bulk substitutions in TiO$_2$, and will relate to the feasibility of synthesizing metal-doped surfaces experimentally. Some elements (Y, Sc, Zr, Hf) favor integration into the surface structure rather than the formation of surface metal clusters (see Table S1 and S2). Conversely, noble metals such as Rh and Pt do not integrate into the surface favorably and will tend to form surface nano-clusters. This result agrees with TEM measurements in the experimental literature, indicating that clusters of metals such as platinum, silver, gold, nickel, rhodium form on a TiO$_2$ surface \cite{Iliev_2006, Dung_Dang_2010, Shinde_2013, Yu_2019} and rutile's reputation as a support \cite{Bagheri_2014}. A metal's ability to form surface sites is also dependent on the relative stability of bulk substitution, since a dopant that is more stable in the bulk than the surface will tend to segregate into the bulk rather than forming surface sites. Fig. \ref{fig:d_band}b shows that the 4+ surface sites are more stable than the bulk substitutions for all metals studied. The relative stability of surface sites relative to bulk integration suggests that bulk synthesis techniques such as co-precipitation should lead to a concentration of surface sites that exceeds the concentration of bulk sites for all metals considered.
The correlation between the bulk formation energies of dopant metals and their corresponding 4+ surface sites (Fig. \ref{fig:d_band}b) is also striking, indicating that bulk and surface integration are controlled by similar electronic structure interactions. 

Fig. \ref{fig:d_band}a also indicates that the oxidation state of the surface site that forms is dependent on the energy of the substituent metal's \textit{d}-band center. Elements with more negative \textit{d}-band centers tend to favor forming 2+ surface sites, whereas more positive \textit{d}-band centers favor 4+ sites, with the cross-over point being approximately 0.8eV below the fermi-level. This trend makes intuitive sense, as a more negative \textit{d}-band center implies that the addition of electrons is more favorable, making the more negative oxidation state more stable. For most metals studied the 2+ site is either more stable or nearly as stable, suggesting that the 2+ substitutions are generally more favorable. An alternative interpretation is that the inclusion of metal dopants favors the formation of surface oxygen vacancies, since the 2+ site involves an oxygen vacancy. The reactivity of oxygen vacancies is typically greater than the pristine surface, so promoting oxygen vacancy formation may be yet another indirect mechanism through which metal dopants affect catalytic activity.



\subsection{Trends In nitrogen adsorption and cohesive energies}
\label{sec:reactivity}

The adsorption of the inert N$_2$ molecule is required for nitrogen fixation, and the first hydrogenation to N$_2$H is known to be the potential-limiting step on pure TiO$_2$ \cite{Comer_2018}. In addition, the NH$_2$ $\rightarrow$ NH$_3$ reaction has been identified as potential limiting on some materials \cite{Hoskuldsson_2017}. This suggests that the trends in N$_2$, N$_2$H, and NH$_2$ binding will provide an indication of a metal's ability to promote nitrogen reduction. The N$_2$ and N$_2$H energies are calculated for both 2+ and 4+ slabs to screen the surface's ability to reduce N$_2$. The N$_2$H binding energy is $>$1.5eV for all 4+ sites (see Table S2), therefore the subsequent analysis focuses exclusively on 2+ sites.

\begin{figure}
    \centering
    \includegraphics[width=0.4\textwidth]{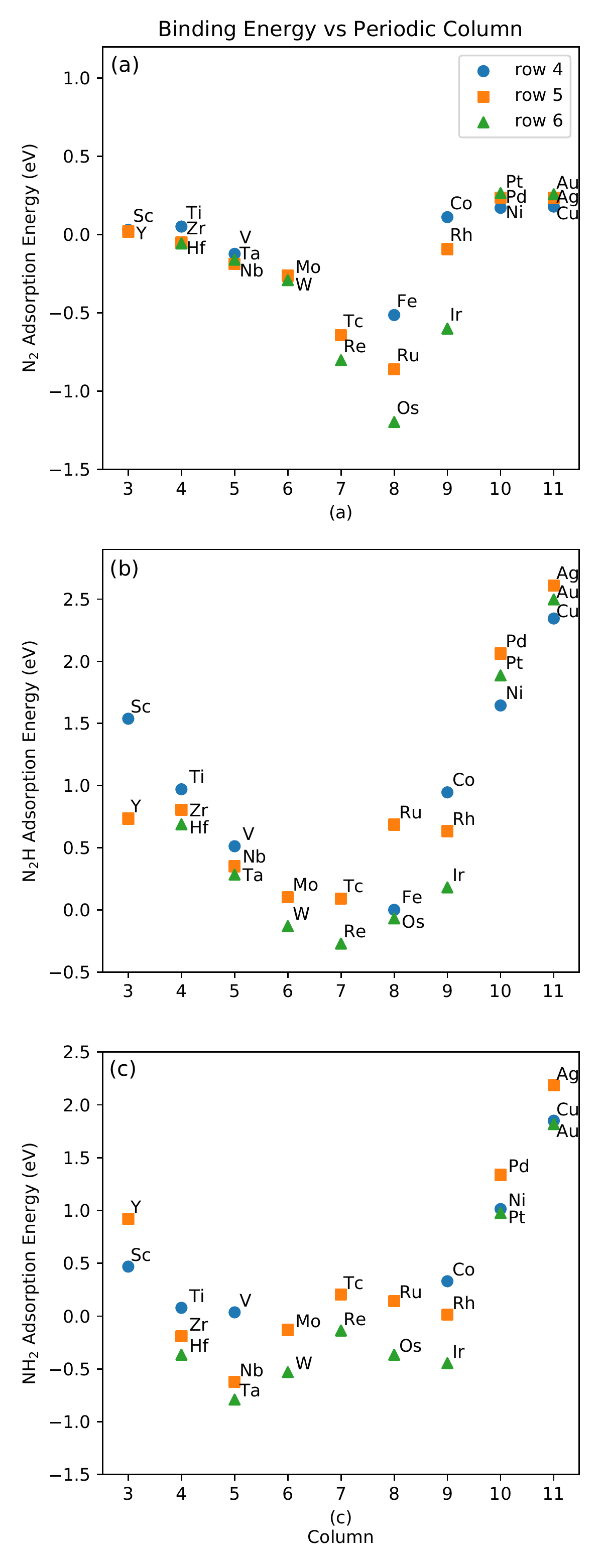}
    \caption{The binding energies of (a) N$_2$, (b) N$_2$H and (c) NH$_2$ plotted against the periodic column for 2+ metal substituent sites}
    \label{fig:column_trends}
\end{figure}

The results for N$_2$, N$_2$H, and NH$_2$ adsorption on 2+ sites as a function of periodic table group are shown in Fig. \ref{fig:column_trends}. The results differ from the typical linear correlation that we expect from the \textit{d}-band model \cite{Nilsson_2008}, and instead, show relatively quadratic behavior with a maximum near the middle of the \textit{d}-block at Os and Re  for N$_2$ and N$_2$H respectively. Similar results are found for NH$_2$ adsorption (Fig. \ref{fig:column_trends}c), though the magnitude of the adsorption energy varies, and there is small upward trend near the middle of the \textit{d}-block.

\begin{figure}
    \centering
    \includegraphics[width=0.4\textwidth]{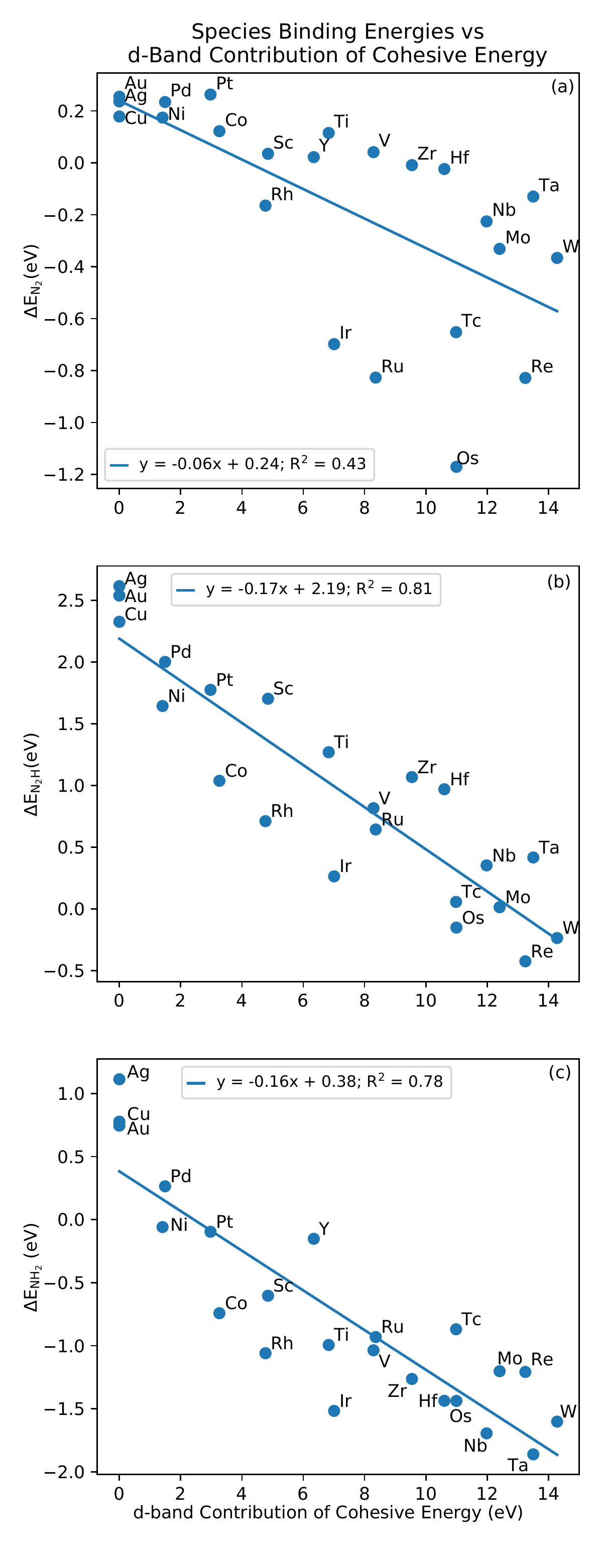}
    \caption{the \textit{d}-band contribution to cohesive energies vs. the binding energies of (a) N$_2$, (b) N$_2$H, (c) NH$_2$ for 2+ metal substituent sites. The \textit{d}-band cohesive energy contributions obtained from Turchanin and Agraval \cite{Turchanin_2008}}
    \label{fig:cohesive}
\end{figure}

While the N$_2$, N$_2$H, and NH$_2$ binding observed deviates from the near-linear correlation expected from the \textit{d}-band model, we find a linear correlation between the binding energies and the \textit{d}-band contributions of the cohesive energies of the corresponding bulk systems (Fig. \ref{fig:cohesive}). Cohesive Energy is defined as the change in energy associated with isolated, neutrally charged atoms being brought together to form a bulk material \cite{Laughlin_2014}. A metal's cohesive energy is made up of a \textit{d} contribution and a \textit{s} contribution (Eq. \ref{eq:cohesive}). Cohesive energies have generally been a measure of the ``bulk-nobleness'' of a metal \cite{Hammer_1995}, with higher cohesive energies correlating to a more noble character. The metals with the highest ``bulk-nobleness'' are in the center of the \textit{d}-block and resist corrosion due to the difficulty of breaking their strong metal-metal bonds. In our case, the inverse is true: the stronger the metal-metal bonds of the bulk material, the stronger the interaction between the metal and a given nitrogen species.
\begin{equation}
    E_{coh} = \epsilon_d + \epsilon_s
    \label{eq:cohesive}
\end{equation}
where $E_{coh}$ is the total cohesive energy, $\epsilon_d$ is the \textit{d} contribution and $\epsilon_s$ is the \textit{s} contribution.

The correlation between \textit{d}-band contribution to cohesive energy and binding is the strongest for N$_2$H and NH$_2$ (Fig. \ref{fig:cohesive}b-c). These two species show a relatively strong quadratic dependence (Fig. \ref{fig:column_trends}b-c) suggesting that the bonding of nitrogen species to these substituent metals is similar to that of forming metal-metal bonds of the original bulk material. Thus, we hypothesize that the physics of nitrogen bonding to these substituent sites is similar to the bonding between single metal atoms and a bulk metal.
A similar quadratic trend is seen for N$_2$ adsorption in Fig. \ref{fig:cohesive}a, though there are several outliers near the middle of the \textit{d}-block (Tc, Ru, Re, Os, Ir) that bind N$_2$ substantially stronger than predicted by the cohesive energy descriptor. The origin of this anomalously-high reactivity toward N$_2$ is not clear, though we note that the bonding mechanism changes between physisorption for early/late metals and chemisorption for more reactive metals, indicating that the quadratic trend may still hold for chemisorption.

The trends observed for site formation energy (Fig. \ref{fig:d_band}) and nitrogen compound adsorption energy (Fig. \ref{fig:cohesive}) differ qualitatively from trends observed in bulk metals. For single transition-metal dopant atoms, the \textit{d}-band center controls formation energy, while the cohesive energy controls adsorption energy. In bulk metals the inverse is true: the d-band center controls a material's ability to bind gas-phase species, whereas the cohesive energy controls how stable the material is \cite{Hammer_1995}. This suggests that the origins of scaling relations for single-atom catalysts or dopant sites may differ from the case of bulk metals. However, we note that the trend does not seem to hold in the case of 4+ sites (see Table S2), and prior work suggests that adsorption energy oxygen is correlated to the \textit{d}-band center \cite{Hammer_2000}, so the trend is not general. The implication of different factors controlling the scaling relations of different adsorbates is that these adsorbates will also not scale with each other. This suggests that single metal atoms or dopants may be able to ``break'' the scaling relations between adsorbates and reach more active regions of the catalytic phase space \cite{Gani_2018, Darby_2018}.

\subsection{Trends in Catalytic Activity for Nitrogen Reduction}
\label{sec:cat_trends}
The photocatalytic activity of doped TiO$_2$ surfaces can be assessed by computing the maximum thermodynamic barrier with electrons at the conduction band edge potential \cite{Comer_2018}, while the electrocatalytic activity of doped TiO$_2$ surfaces can be assessed by computing the thermodynamic limiting potential \cite{Norskov_2004,Garc_a_Mota_2011}. The computational hydrogen electrode (CHE) provides a route to computing the thermochemical potential of electrons at the TiO$_2$ surface (Sec. \ref{sec:PEC_methods}), and the resulting analysis provides only a thermodynamic picture of the reaction pathway. This analysis establishes a lower bound on the kinetics and correlates well with experimental trends in the literature \cite{Seh_2017}.

Computing the maximum barrier or limiting potential requires the free energies of each state along a given reaction pathway. The full thermodynamics of the N$_2$ reduction reaction pathways on all 2+ sites were calculated, allowing the generation of free energy diagrams for all possible reaction pathways (Fig. S4-S109, Table S1). H\"oskuldsson et al. \cite{Hoskuldsson_2017} also previously found strong scaling relations between the binding of nitrogen compounds and the N$_2$H binding energy for rutile metal oxides. The binding energies of all species are fit to linear scaling relations with N$_2$H and NH$_2$ as descriptors to assess the scaling relations for this system (Fig. S1). The scaling relations have a root mean squared error on the order of 0.2 eV, consistent with general scaling relations for other reactions \cite{Wang_2011}. The N$_2$H and NH$_2$ were also used to fit scaling relationships for all electrochemical steps, yielding similar accuracy to scaling relations for individual species (Fig. S2). These scaling relations directly predicting reaction energies are used for subsequent analyses.



\begin{figure}

    \centering
    \includegraphics[width=0.45\textwidth]{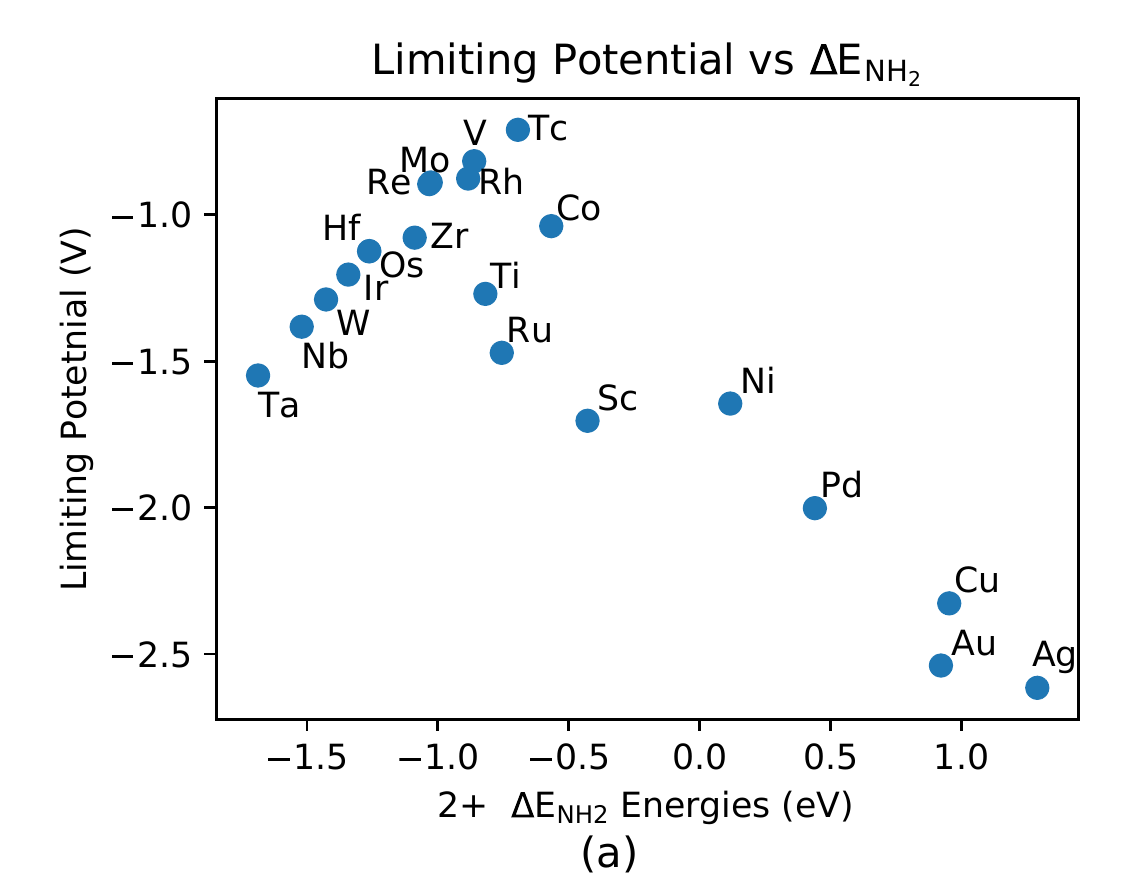}
    \includegraphics[width=0.6\textwidth]{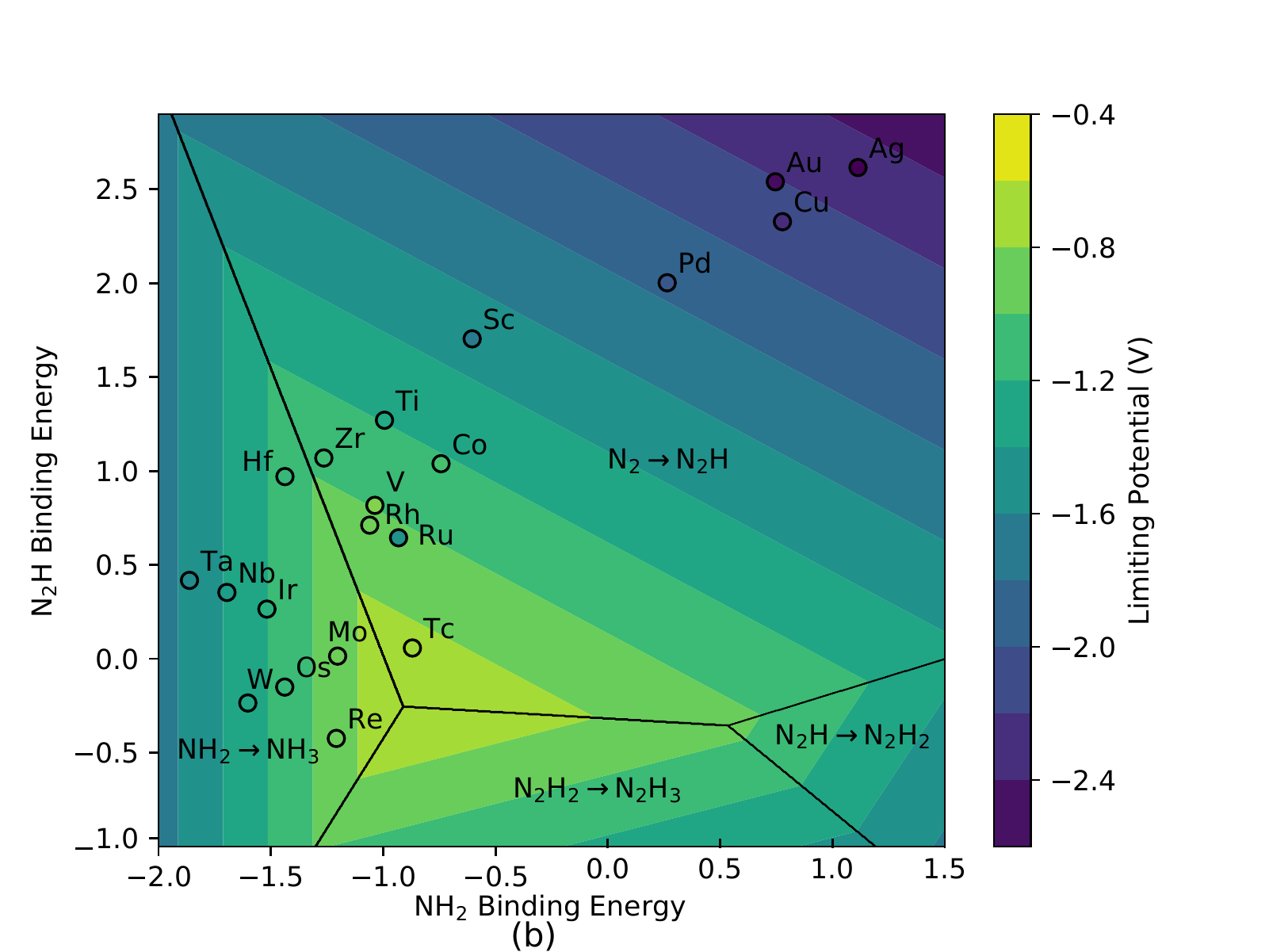}
    \caption{(a)The limting potential vs the NH$_2$ binding energy. The data for this plot can be seen in Table S3 (b) A two dimensional volcano plot for electrochemical limiting potential using N$_2$H and NH$_2$ as descriptors for the studied systems. Points are colored in with the limiting potential calculated from DFT. Any surface for which a full path was not available has been excluded.}
    \label{fig:2d_plot}
\end{figure}
\begin{figure}
    \centering
    \includegraphics[width=0.5\textwidth]{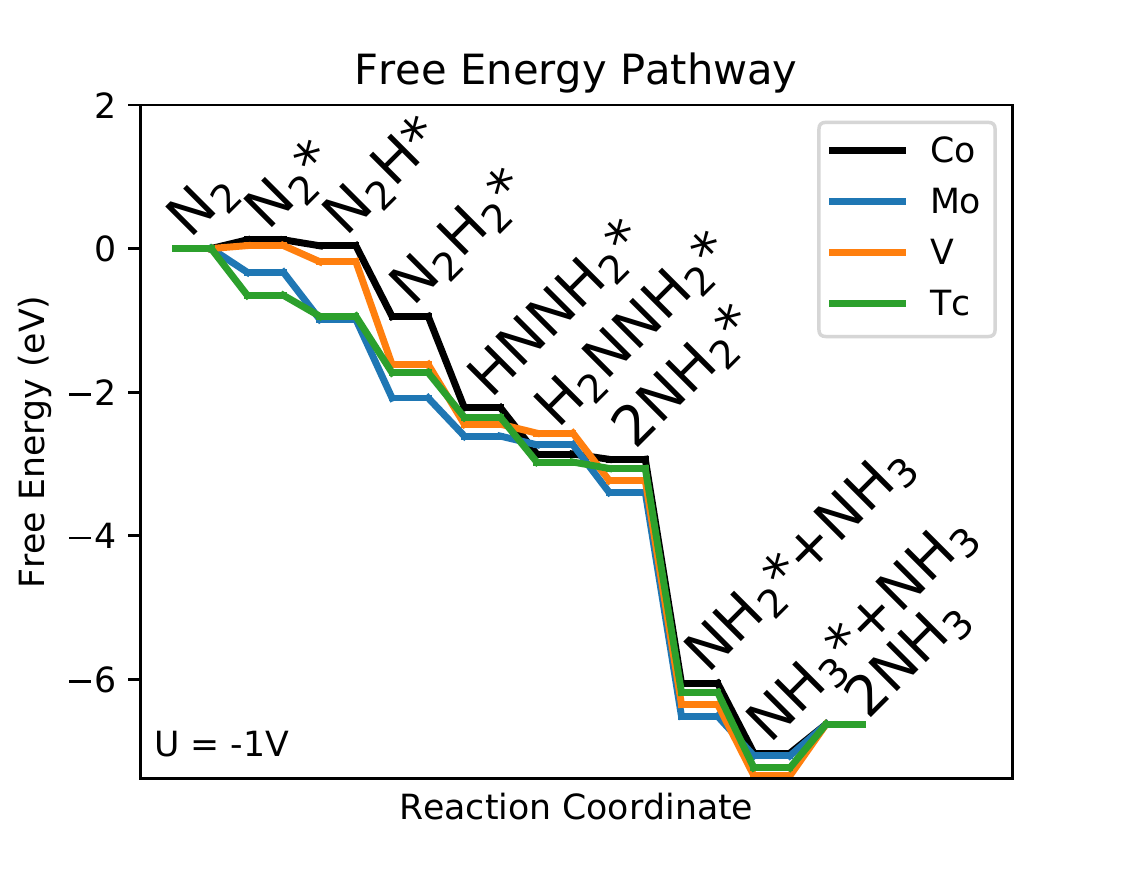}
    \caption{A free energy diagram at -1V relative to RHE for the most promising elements: Co, Mo, V, and Tc.}
    \label{fig:FED}
\end{figure}

The electrochemical limiting potential is calculated for all surfaces to assess their ability to reduce N$_2$ under applied bias. The results are plotted against the NH$_2$ binding energy in Fig. \ref{fig:2d_plot}a. This plot reveals a clear volcano relationship between the NH$_2$ binding energy and the limiting potential. Free energy diagrams have been generated for selected elements and can be seen in Fig. \ref{fig:FED}. In contrast to prior work by H\"oskuldsson et. al.\cite{Hoskuldsson_2017} and Montoya et. al. \cite{Montoya_2015}, we find that the NH$_2$ binding energy is a slightly more reliable descriptor than N$_2$H binding; however these quantities are linked by scaling relations, indicating that either descriptor will provide consistent trends. In this case, the limiting step shifts from NH$_2$ desorption on the left to N$_2$ hydrogenation on the right, with most dopants being limited by NH$_2$ desorption. This means that NH$_2$ adsorption energy directly controls the reactive side of the volcano, and explains why it is an accurate descriptor in this case. 

To better understand the relationship between the descriptors, limiting potential, and limiting steps, the fits of the scaling relations (see Fig. S2) were used to generate a two dimensional volcano plot (Fig. \ref{fig:2d_plot}b). As with the scaling relations, the root mean squared error of the predicted limiting potential in Fig. \ref{fig:2d_plot}b is roughly 0.2V. The results confirm the findings from Fig. \ref{fig:2d_plot}a, but provide additional insight into the limiting steps. The deviation from the NH$_2$/N$_2$H scaling for reactive dopants (NH$_2$ binding $<$-0.75 eV) leads to a cluster of dopants that behave similarly for different reasons. For example, the limiting potential and NH$_2$ adsorption for Mo and V are nearly identical, but the mechanism shifts between the two with V being limited by N$_2$H formation and Mo being limited by NH$_2$ desorption. The results also show that the optimal limiting potential is still relatively large ($\sim$-0.4 V), and that Tc is near-optimal.

Overall, the results suggest that several dopants are capable of improving the performance over pure TiO$_2$. The elements that show significant improvement are Tc, Co, Mo, V, Rh, and Re. Tc sits at the top of the volcano. The high activity of Tc presents a serious problem for experimental testing or practical application since Tc is a scarce, synthetically produced element. Further complicating matters, most isotopes of Tc are radioactive. In addition, the active sites may be challenging to synthesize due to their relatively high formation energy (Fig. \ref{fig:d_band}). On the other hand, the relatively low limiting potential of Co, Mo, V, Rh, and Re are promising results. Co, Mo, and V are relatively inexpensive and abundant, whereas Re and Rh are relatively scarce \cite{Vesborg_2012}. These elements are promising dopants for improving catalytic rates on TiO$_2$. However, synthesis may be a challenge since the formation energy of the surface sites is generally positive relative to the bulk metals (Fig. \ref{fig:d_band}). Nonetheless, the results indicate that Co is the most promising dopant for reducing the thermodynamic limiting potential of ammonia synthesis on TiO$_2$.

A second consideration when assessing the electrocatalytic activity of a surface is the largest thermochemical barrier. Thermochemical steps do not involve electron transfers, so they are not considered when computing the limiting potential. However, they may still present a substantial barrier that will affect the overall rate. The largest thermochemical barriers for each surface can be seen in Table S4. The three steps with significant thermochemical barriers are N$_2$ adsorption, NH$_2$-NH$_2$ scission, and NH$_3$ desorption. For the case of the most promising dopant, Co, 
the adsorption of N$_2$ is endergonic by 0.12 eV, which may lead to low N$_2$* coverages under competitive adsorption with H$_2$O. Moreover, there is a substantial thermochemical barrier of 0.43 eV for NH$_3$ desorption. Desorption of NH$_3$ is the thermochemical limiting step for most dopants, suggesting that NH$_3$ may exist at high coverages or even poison the surface. However, solvation effects have been neglected, and the free energy is computed at a chemical potential of NH$_3$ equivalent to 1 bar, suggesting that NH$_3$ desorption may not be limiting in aqueous solutions with low NH$_3$ concentrations. Some of the noble dopants (Pd, Ag, Au, and Cu) also show substantial thermochemical barriers of 0.5-1.5 eV for NH$_2$-NH$_2$ scission, indicating that rates for these metals will be low even at the limiting potential. A more detailed kinetic analysis of both electrochemical and thermochemical activation energies is required to predict the electrocatalytic rate for any dopant, but this thermochemical analysis provides lower bound for the kinetic barrier.

\begin{figure}
    \centering
    \includegraphics[width=0.45\textwidth]{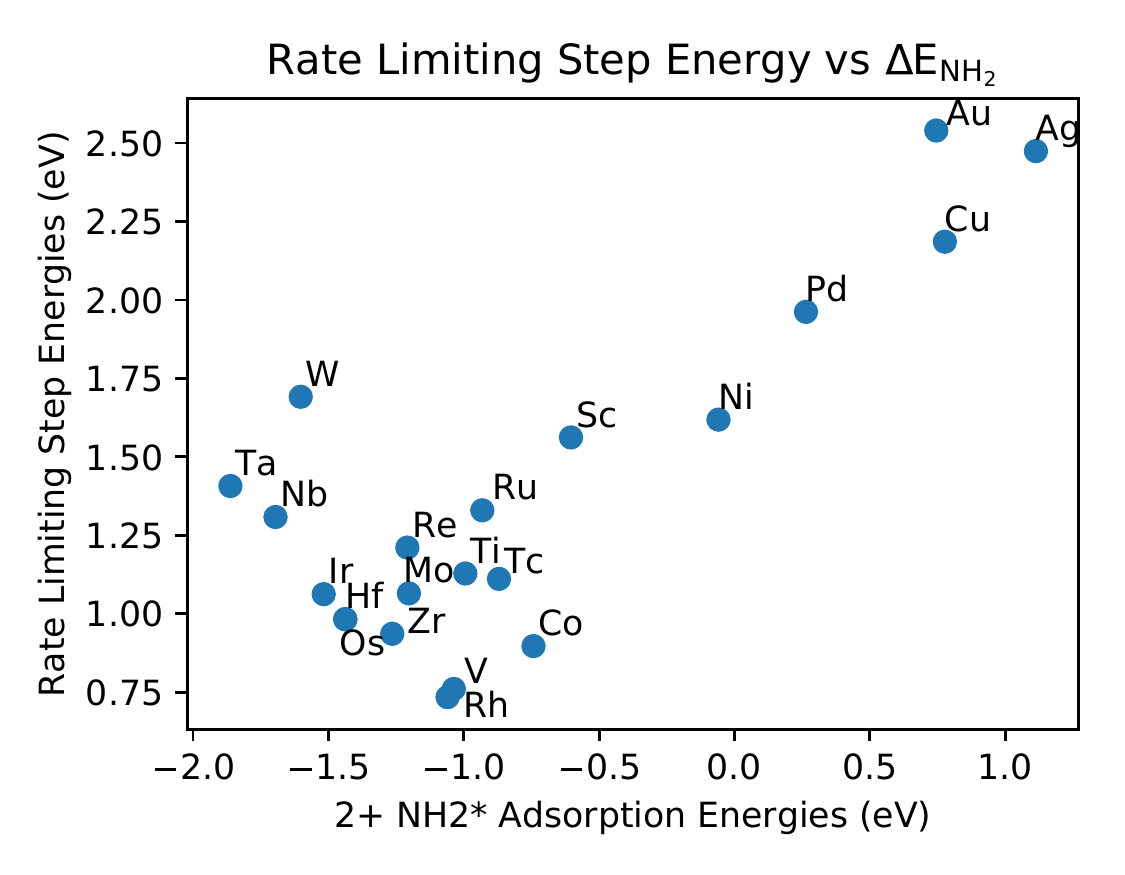}

    \caption{The highest barrier observed vs the NH$_2$ binding energy with the potential set to band edge of rutile. The data for this plot can be seen in Table S5. Any surface for which a full path was not available has been excluded.}
    \label{fig:NH2_thermo}
    
\end{figure}

Finally, we assess the ability of dopant metals to improve photocatalytic nitrogen reduction. This is calculated based on the largest thermodynamic barrier at a reductive potential equal to the conduction band edge of TiO$_2$ (approximately -0.15 V vs. RHE \cite{Nozik_1996}). This approach assumes that the conduction band edge of TiO$_2$ is not significantly affected by the presence of the dopant, and neglects improvements in other bulk photochemical properties such as charge separation or carrier lifetime. Nonetheless, it provides a good starting point for assessing the impact of dopant metals on the surface catalytic properties. 
The highest thermodynamic barrier for the best reaction pathway is plotted vs. the NH$_2$* binding energy in Fig. \ref{fig:NH2_thermo}. The results are qualitatively similar to the electrochemical limiting potentials in Fig. \ref{fig:2d_plot}, but there are some deviations that occur for two reasons. The first is that the photochemical analysis includes both thermochemical and electrochemical steps. The desorption of NH$_3$* is a thermochemical step that becomes rate-limiting for reactive surfaces. For less reactive surfaces the electrochemical step of N$_2$ hydrogenation is rate-limiting, which becomes slightly more favorable under the applied bias, effectively shifting the right side of the volcano downward. The second reason for deviation is that multi-electron transfers are less sensitive to small potentials, so dopants such as Tc which have relatively unstable N$_2$H$_{\mathrm{x}>2}$ states are not as favorable under photocatalytic conditions. Overall, the results suggest that the minimum thermodynamic barriers of 0.93 eV, 0.90 eV, 0.89 eV, 0.73 eV, for Zr, Co, Mo, and  Rh, respectively. This represents a substantial improvement over the 1.21 eV limiting potential for pure Ti, indicating that these metals may act as surface promoters for photocatalytic nitrogen reduction if kinetic barriers are low.

Experimental observations can provide further insight into the computational predictions. Several prior reports have investigated transition-metal dopants for enhancing photocatalytic ammonia production on TiO$_2$ \cite{Schrauzer_1977, Ranjit_1996, Hirakawa_2017}. Interestingly Schrauzer et al. reports increases in ammonia yield in the presence of Co and Mo dopants \cite{Schrauzer_1977}, though this report comes from the early literature and rigorous controls \cite{Greenlee_2018} or isotopic labeling studies \cite{Andersen_2019} were not included. Moreover, the same report revealed enhanced rates for Fe and Ni, so the confirmation of the prediction regarding the former two should be treated with caution. The rate enhancement for noble metals such as Ru, Rh, and Pd is conflicting even in the early literature, with Schrauzer and Guth reporting no enhancement \cite{Schrauzer_1977}. However, Ranjit et al. reported enhancement for all noble metals with the most significant improvement for Ru \cite{Ranjit_1996}. In all of these systems, the metal dopants were incorporated via co-precipitation, and catalysts were polycrystalline TiO$_2$, indicating that the metals may also enhance yields via charge separation, mediation of crystallization, or other mechanisms\cite{Medford_2017}. Hirakawa et al. added Ru, Pt, and Pd to pre-synthesized TiO$_2$ particles and reported no significant improvement in the reaction rates \cite{Hirakawa_2017}. These experiments are more consistent with the computational model system used in this study since only surface properties are affected, and the results are consistent with the prediction that these noble metals will not affect the rate. However, further systematic and well-controlled experiments that characterize the state of the metal incorporation in the TiO$_2$ surface are required to validate the predicted trends.



\section{Conclusions}
The stability of metal dopant surface sites and their effects on the reaction thermodynamics of N$_2$ reduction on rutile (110) are studied using DFT. We find that the formation energy of these doped surface states is strongly related to the location of the \textit{d}-band center of the substituted metal, with a trend consistent with the \textit{d}-band model. We also find a correlation between the cohesive energy of metals and their N$_2$H and NH$_2$ binding energy on the surface, suggesting that the bonding of nitrogen species is similar to that of bulk metals. Finally, we investigate the effects of dopant sites on the full reaction pathways for 2+ sites on all studied metals. We find a clear volcano relationship between NH$_2$ binding and both the electrochemical limiting potential and the highest thermodynamic barrier for photocatalytic reactions. The formation of Co and Mo 2+ sites is proposed to yield a slight improvement of reaction rates in both electrocatalysis and photocatalysis. Other metals commonly used in catalysis, such as Ir, and Pd are predicted to have a limited or detrimental effect on the surface catalytic properties of TiO$_2$ for nitrogen reduction. This suggests that the role of metal dopants in photocatalytic ammonia synthesis by TiO$_2$ is likely related to modifications of bulk properties in most cases. However, the existence of clear trends in the formation energy and reactivity of single metal atom dopants toward nitrogen intermediates suggests that computational design of metal-doped oxide materials is a promising strategy for other oxide systems and/or other nitrogen conversion reactions.


\section{Methods}
\label{sec:methods}

\subsection{Density Functional Theory Calculations}
All first principles calculations are performed in the Quantum Espresso software package \cite{QE-2009}.
The TiO$_2$ slabs and atomistic images are created using the Atomic Simulation Environment (ASE) package \cite{ase-paper}.  Spin polarization is used in all simulations to ensure the lowest energy spin state is obtained for each site. The BEEF-vdW functional \cite{Wellendorff_2012} is used with plane wave cutoff of 400 eV and a Monkhorst-Pack k-point grid spacing of 4$\times$4$\times$1 \cite{Monkhorst_1976}. The effciency versions of the standard solid state psuedopotentials \cite{SSSP_pseudos} (SSSP) are used for all calculations because of their high reported accuracy \cite{Lejaeghereaad3000}. The convergence threshold is set at $10^{-6}$ eV and Fermi-Dirac smearing of 0.1 eV is used. All structures are converged to a maximum force smaller than 0.05 eV/A using the BFGS line search algorithm. Adsorption energies are obtained from the DFT calculations by subtracting the energy of a clean slab and the energy of the free gas molecule from the energy of the gas adsorbed to the slab:
\begin{equation}
E_{adsorption} = E_{slab+adsorbate} - E_{slab} - E_{adsorbate}
\end{equation}

\subsection{Thermochemistry}

To calculate the adsorption energy at standard temperature and pressure, the thermochemistry package from ASE is used \cite{ase-paper}. The contributions of zero point energy (ZPE) have been included for all systems. Free gasses are approximated in the ideal gas limit, and adsorbed gasses in the harmonic limit \cite{Reuter_2005}. A frequency cutoff of 33 cm$^{-1}$ for low frequency modes was selected. The vibrational mode for all metals were assumed to be approximately the same as those of the respective species adsorbed to the oxygen vacant TiO$_2$ (110) surface, thus the same thermodynamic correction values are applied to all species of the same type.

\subsection{Photochemistry}
\label{sec:PEC_methods}
The photochemistry has been treated using the methods outlined by Hellman et. al. \cite{Hellman2017}. Within this framework, the effects of excited states are neglected allowing the treatment of excited electrons and holes using the computational hydrogen electrode model (CHE). In this formalism, the reference electrode is set by setting the free energy of hydrogen evolution reaction (HER) to zero:
\begin{equation}
    H^+ + e^- \rightarrow H_2;\; \Delta G = 0\:at\: U = 0
\end{equation}

The potentials of electrons and holes are set at the value of the band edges of the rutile \cite{Nozik_1996}.

\subsection{Model Surface Generation}
 The surfaces are constructed with 4 TiO$_2$ tri-layers (bottom 2 layers constrained to bulk positions), with a 1$\times$2 supercell repeat. The pristine slab totals 48 atoms, with 4 Ti and 8 O per tri-layer. 6 \AA  of vacuum are added on both top and bottom of the slab and a dipole corrections is applied in the $z$ direction\cite{Dipole_paper}. The lattice parameters of the unit cells are fixed at the calculated value for pure rutile TiO$_2$. 2+ sites were created by replacing the six-fold titanium site with the substituent metal and removing a single bridging oxygen (see Fig. \ref{fig:2ex_slab}) 4+ sites were generated by replacing the five-fold titanium site with the substituent metal. The adsorption site was selected to directly over the substituent metals. 

\section{Acknowledgements}
We would like to thank Fuhzu Liu and Gabriel Gusm\~ao for their comments and suggestions for improving this manuscript. This material is based upon work supported by the U. S. Department of Energy, Office of Science, Office of Basic Energy Sciences Computational Chemical Sciences program under Award No.DE-SC0019410. 

\bibliography{main}

\begin{thebibliography}{10}

\bibitem{Ammal_2016}
S.~C. Ammal and A.~Heyden.
\newblock Water-gas shift activity of atomically dispersed cationic platinum
  versus metallic platinum clusters on titania supports.
\newblock {\em {ACS} Catalysis}, 7(1):301--309, dec 2016.

\bibitem{Andersen_2019}
S.~Z. Andersen, V.~{\v{C}}oli{\'{c}}, S.~Yang, J.~A. Schwalbe, A.~C. Nielander,
  J.~M. McEnaney, K.~Enemark-Rasmussen, J.~G. Baker, A.~R. Singh, B.~A. Rohr,
  M.~J. Statt, S.~J. Blair, S.~Mezzavilla, J.~Kibsgaard, P.~C.~K. Vesborg,
  M.~Cargnello, S.~F. Bent, T.~F. Jaramillo, I.~E.~L. Stephens, J.~K.
  N{\o}rskov, and I.~Chorkendorff.
\newblock A rigorous electrochemical ammonia synthesis protocol with
  quantitative isotope measurements.
\newblock {\em Nature}, 570(7762):504--508, may 2019.

\bibitem{Augugliaro_1982}
V.~Augugliaro, A.~Lauricella, L.~Rizzuti, M.~Schiavello, and A.~Sclafani.
\newblock {Conversion of solar energy to chemical energy by photoassisted
  processes{\textemdash}I. Preliminary results on ammonia production over doped
  titanium dioxide catalysts in a fluidized bed reactor}.
\newblock {\em Int. J. Hydrogen Energy}, 7(11):845--849, 1982.

\bibitem{Bagheri_2014}
S.~Bagheri, N.~M. Julkapli, and S.~B.~A. Hamid.
\newblock Titanium dioxide as a catalyst support in heterogeneous catalysis.
\newblock {\em The Scientific World Journal}, 2014:1--21, 2014.

\bibitem{Dipole_paper}
L.~Bengtsson.
\newblock Dipole correction for surface supercell calculations.
\newblock {\em Phys. Rev. B}, 59:12301--12304, May 1999.

\bibitem{Bickley_1979}
R.~I. Bickley and V.~Vishwanathan.
\newblock {Photocatalytically induced fixation of molecular nitrogen by near
  {UV} radiation}.
\newblock {\em Nature}, 280(5720):306--308, jul 1979.

\bibitem{Cheng_2019}
S.~Cheng, Y.-J. Gao, Y.-L. Yan, X.~Gao, S.-H. Zhang, G.-L. Zhuang, S.-W. Deng,
  Z.-Z. Wei, X.~Zhong, and J.-G. Wang.
\newblock Oxygen vacancy enhancing mechanism of nitrogen reduction reaction
  property in {Ru}/{TiO}$_2$.
\newblock {\em Journal of Energy Chemistry}, 39:144--151, dec 2019.

\bibitem{Comer_2019}
B.~M. Comer, P.~Fuentes, C.~O. Dimkpa, Y.-H. Liu, C.~A. Fernandez, P.~Arora,
  M.~Realff, U.~Singh, M.~C. Hatzell, and A.~J. Medford.
\newblock Prospects and challenges for solar fertilizers.
\newblock {\em Joule}, 3(7):1578--1605, jul 2019.

\bibitem{Comer_2018b}
B.~M. Comer, Y.-H. Liu, M.~B. Dixit, K.~B. Hatzell, Y.~Ye, E.~J. Crumlin, M.~C.
  Hatzell, and A.~J. Medford.
\newblock The role of adventitious carbon in photo-catalytic nitrogen fixation
  by titania.
\newblock {\em Journal of the American Chemical Society}, 140(45):15157--15160,
  2018.

\bibitem{Comer_2018}
B.~M. Comer and A.~J. Medford.
\newblock Analysis of photocatalytic nitrogen fixation on rutile
  {TiO}$_2$(110).
\newblock {\em {ACS} Sustainable Chemistry {\&} Engineering}, 6(4):4648--4660,
  feb 2018.

\bibitem{Cui2018}
X.~Cui, C.~Tang, and Q.~Zhang.
\newblock {A Review of Electrocatalytic Reduction of Dinitrogen to Ammonia
  under Ambient Conditions}.
\newblock {\em Advanced Energy Materials}, 8(22):1--25, 2018.

\bibitem{Dung_Dang_2010}
T.~M.~D. Dang, T.~M.~H. Nguyen, and H.~P. Nguyen.
\newblock The preparation of nano-gold catalyst supported on iron doped
  titanium oxide.
\newblock {\em Advances in Natural Sciences: Nanoscience and Nanotechnology},
  1(2):025011, jun 2010.

\bibitem{Darby_2018}
M.~T. Darby, M.~Stamatakis, A.~Michaelides, and E.~C.~H. Sykes.
\newblock Lonely atoms with special gifts: Breaking linear scaling
  relationships in heterogeneous catalysis with single-atom alloys.
\newblock {\em The Journal of Physical Chemistry Letters}, 9(18):5636--5646,
  sep 2018.

\bibitem{Davies1995}
J.~A. Davies, D.~L. Boucher, and J.~G. Edwards.
\newblock The question of artificial photosynthesis of ammonia on heterogeneous
  catalysts.
\newblock In {\em Advances in Photochemistry}, pages 235--310. John Wiley {\&}
  Sons, Inc., 1995.

\bibitem{davies1993reply}
J.~A. Davies and J.~G. Edwards.
\newblock {Reply: Standards of Demonstration for the Heterogeneous
  Photoreactions of N$_2$ with {H}$_2${O}}.
\newblock {\em Angew. Chem. Int. Ed.}, 32(4):552--553, 1993.

\bibitem{de_Bruijn_2015}
F.~J. de~Bruijn.
\newblock The quest for biological nitrogen fixation in cereals: A perspective
  and prospective.
\newblock In {\em Biological Nitrogen Fixation}, pages 1087--1101. John Wiley
  {\&} Sons, Inc, jul 2015.

\bibitem{Dhar_1941}
N.~Dhar, E.~Seshacharyulu, and N.~Biswas.
\newblock {New aspects of nitrogen fixation and loss in soils}.
\newblock {\em Proceedings of the National institute of sciences of India},
  7:115--131, 1941.

\bibitem{Diebold_2003}
U.~Diebold.
\newblock The surface science of titanium dioxide.
\newblock {\em Surface Science Reports}, 48(5-8):53--229, jan 2003.

\bibitem{Dozzi_2013}
M.~V. Dozzi and E.~Selli.
\newblock Doping {TiO}$_2$ with p-block elements: Effects on photocatalytic
  activity.
\newblock {\em Journal of Photochemistry and Photobiology C: Photochemistry
  Reviews}, 14:13--28, mar 2013.

\bibitem{edwards1992opinion}
J.~G. Edwards, J.~A. Davies, D.~L. Boucher, and A.~Mennad.
\newblock {An Opinion on the Heterogeneous Photoreactions of N$_2$ with
  {H}$_2${O}}.
\newblock {\em Angew. Chem. Int. Ed.}, 31(4):480--482, 1992.

\bibitem{Gani_2018}
T.~Z.~H. Gani and H.~J. Kulik.
\newblock Understanding and breaking scaling relations in single-site
  catalysis: Methane to methanol conversion by {Fe$^{IV}$=O}.
\newblock {\em {ACS} Catalysis}, 8(2):975--986, jan 2018.

\bibitem{Gao_2018}
X.~Gao, Y.~Wen, D.~Qu, L.~An, S.~Luan, W.~Jiang, X.~Zong, X.~Liu, and Z.~Sun.
\newblock Interference effect of alcohol on nessler's reagent in photocatalytic
  nitrogen fixation.
\newblock {\em {ACS} Sustainable Chemistry {\&} Engineering}, 6(4):5342--5348,
  mar 2018.

\bibitem{Garc_a_Mota_2011}
M.~Garc{\'{\i}}a-Mota, A.~Vojvodic, H.~Metiu, I.~C. Man, H.-Y. Su,
  J.~Rossmeisl, and J.~K. N{\o}rskov.
\newblock Tailoring the activity for oxygen evolution electrocatalysis on
  rutile {TiO}$_2$(110) by transition-metal substitution.
\newblock {\em {ChemCatChem}}, 3(10):1607--1611, aug 2011.

\bibitem{SSSP_pseudos}
I.~E. C. N. M. N.~M. Gianluca~Prandini, Antimo~Marrazzo.
\newblock A standard solid state pseudopotentials (sssp) library optimized for
  accuracy and efficiency (version 1.0, data download), 2018.

\bibitem{QE-2009}
P.~Giannozzi, S.~Baroni, N.~Bonini, M.~Calandra, R.~Car, C.~Cavazzoni,
  D.~Ceresoli, G.~L. Chiarotti, M.~Cococcioni, I.~Dabo, A.~{Dal Corso},
  S.~de~Gironcoli, S.~Fabris, G.~Fratesi, R.~Gebauer, U.~Gerstmann,
  C.~Gougoussis, A.~Kokalj, M.~Lazzeri, L.~Martin-Samos, N.~Marzari, F.~Mauri,
  R.~Mazzarello, S.~Paolini, A.~Pasquarello, L.~Paulatto, C.~Sbraccia,
  S.~Scandolo, G.~Sclauzero, A.~P. Seitsonen, A.~Smogunov, P.~Umari, and R.~M.
  Wentzcovitch.
\newblock Quantum espresso: a modular and open-source software project for
  quantum simulations of materials.
\newblock {\em Journal of Physics: Condensed Matter}, 21(39):395502 (19pp),
  2009.

\bibitem{Greenlee_2018}
L.~F. Greenlee, J.~N. Renner, and S.~L. Foster.
\newblock The use of controls for consistent and accurate measurements of
  electrocatalytic ammonia synthesis from dinitrogen.
\newblock {\em {ACS} Catalysis}, 8(9):7820--7827, jul 2018.

\bibitem{gross_12}
M.~Gross.
\newblock We need to talk about nitrogen.
\newblock {\em Current Biology}, 22, 2012.

\bibitem{Gu_2017}
X.-K. Gu, C.-Q. Huang, and W.-X. Li.
\newblock First-principles study of single transition metal atoms on {ZnO} for
  the water gas shift reaction.
\newblock {\em Catalysis Science {\&} Technology}, 7(19):4294--4301, 2017.

\bibitem{Gu_2014}
X.-K. Gu, B.~Qiao, C.-Q. Huang, W.-C. Ding, K.~Sun, E.~Zhan, T.~Zhang, J.~Liu,
  and W.-X. Li.
\newblock Supported single pt1/au1 atoms for methanol steam reforming.
\newblock {\em {ACS} Catalysis}, 4(11):3886--3890, oct 2014.

\bibitem{Hammer_2000}
B.~Hammer and J.~N{\o}rskov.
\newblock Theoretical surface science and catalysis{\textemdash}calculations
  and concepts.
\newblock In {\em Advances in Catalysis}, pages 71--129. Elsevier, 2000.

\bibitem{Hammer_1995}
B.~Hammer and J.~K. Norskov.
\newblock Why gold is the noblest of all the metals.
\newblock {\em Nature}, 376(6537):238--240, jul 1995.

\bibitem{Hellman2017}
A.~Hellman and B.~Wang.
\newblock {First-Principles View on Photoelectrochemistry: Water-Splitting as
  Case Study}.
\newblock {\em Inorganics}, 5(2):37, 2017.

\bibitem{Hirakawa_2017}
H.~Hirakawa, M.~Hashimoto, Y.~Shiraishi, and T.~Hirai.
\newblock Photocatalytic conversion of nitrogen to ammonia with water on
  surface oxygen vacancies of titanium dioxide.
\newblock {\em Journal of the American Chemical Society}, 139(31):10929--10936,
  jul 2017.

\bibitem{Hoskuldsson_2017}
{\'{A}}.~B. H\"oskuldsson, Y.~Abghoui, A.~B. Gunnarsd{\'{o}}ttir, and
  E.~Sk{\'{u}}lason.
\newblock Computational screening of rutile oxides for electrochemical ammonia
  formation.
\newblock {\em {ACS} Sustainable Chemistry {\&} Engineering},
  5(11):10327--10333, sep 2017.

\bibitem{Iliev_2006}
V.~Iliev, D.~Tomova, L.~Bilyarska, A.~Eliyas, and L.~Petrov.
\newblock Photocatalytic properties of {TiO}$_2$ modified with platinum and
  silver nanoparticles in the degradation of oxalic acid in aqueous solution.
\newblock {\em Applied Catalysis B: Environmental}, 63(3-4):266--271, mar 2006.

\bibitem{Jia_2016}
J.~Jia, L.~C. Seitz, J.~D. Benck, Y.~Huo, Y.~Chen, J.~W.~D. Ng, T.~Bilir, J.~S.
  Harris, and T.~F. Jaramillo.
\newblock Solar water splitting by photovoltaic-electrolysis with a
  solar-to-hydrogen efficiency over 30{\%}.
\newblock {\em Nature Communications}, 7(1), oct 2016.

\bibitem{Khan_2018}
M.~E. Khan, M.~M. Khan, and M.~H. Cho.
\newblock Recent progress of metal{\textendash}graphene nanostructures in
  photocatalysis.
\newblock {\em Nanoscale}, 10(20):9427--9440, 2018.

\bibitem{Kyriakou_2017}
V.~Kyriakou, I.~Garagounis, E.~Vasileiou, A.~Vourros, and M.~Stoukides.
\newblock Progress in the electrochemical synthesis of ammonia.
\newblock {\em Catalysis Today}, 286:2--13, may 2017.

\bibitem{ase-paper}
A.~H. Larsen, J.~J. Mortensen, J.~Blomqvist, I.~E. Castelli, R.~Christensen,
  M.~Dułak, J.~Friis, M.~N. Groves, B.~Hammer, C.~Hargus, E.~D. Hermes, P.~C.
  Jennings, P.~B. Jensen, J.~Kermode, J.~R. Kitchin, E.~L. Kolsbjerg, J.~Kubal,
  K.~Kaasbjerg, S.~Lysgaard, J.~B. Maronsson, T.~Maxson, T.~Olsen, L.~Pastewka,
  A.~Peterson, C.~Rostgaard, J.~Schiøtz, O.~Schütt, M.~Strange, K.~S.
  Thygesen, T.~Vegge, L.~Vilhelmsen, M.~Walter, Z.~Zeng, and K.~W. Jacobsen.
\newblock The atomic simulation environment—a python library for working with
  atoms.
\newblock {\em Journal of Physics: Condensed Matter}, 29(27):273002, 2017.

\bibitem{Lejaeghereaad3000}
K.~Lejaeghere, G.~Bihlmayer, T.~Bj{\"o}rkman, P.~Blaha, S.~Bl{\"u}gel, V.~Blum,
  D.~Caliste, I.~E. Castelli, S.~J. Clark, A.~Dal~Corso, S.~de~Gironcoli,
  T.~Deutsch, J.~K. Dewhurst, I.~Di~Marco, C.~Draxl, M.~Du{\l}ak, O.~Eriksson,
  J.~A. Flores-Livas, K.~F. Garrity, L.~Genovese, P.~Giannozzi, M.~Giantomassi,
  S.~Goedecker, X.~Gonze, O.~Gr{\r a}n{\"a}s, E.~K.~U. Gross, A.~Gulans,
  F.~Gygi, D.~R. Hamann, P.~J. Hasnip, N.~A.~W. Holzwarth, D.~Iu{\c s}an, D.~B.
  Jochym, F.~Jollet, D.~Jones, G.~Kresse, K.~Koepernik, E.~K{\"u}{\c
  c}{\"u}kbenli, Y.~O. Kvashnin, I.~L.~M. Locht, S.~Lubeck, M.~Marsman,
  N.~Marzari, U.~Nitzsche, L.~Nordstr{\"o}m, T.~Ozaki, L.~Paulatto, C.~J.
  Pickard, W.~Poelmans, M.~I.~J. Probert, K.~Refson, M.~Richter, G.-M.
  Rignanese, S.~Saha, M.~Scheffler, M.~Schlipf, K.~Schwarz, S.~Sharma,
  F.~Tavazza, P.~Thunstr{\"o}m, A.~Tkatchenko, M.~Torrent, D.~Vanderbilt, M.~J.
  van Setten, V.~Van~Speybroeck, J.~M. Wills, J.~R. Yates, G.-X. Zhang, and
  S.~Cottenier.
\newblock Reproducibility in density functional theory calculations of solids.
\newblock {\em Science}, 351(6280), 2016.

\bibitem{Li_2018}
C.~Li, T.~Wang, Z.-J. Zhao, W.~Yang, J.-F. Li, A.~Li, Z.~Yang, G.~A. Ozin, and
  J.~Gong.
\newblock Promoted fixation of molecular nitrogen with surface oxygen vacancies
  on plasmon-enhanced {TiO}$_2$ photoelectrodes.
\newblock {\em Angewandte Chemie International Edition}, 57(19):5278--5282, mar
  2018.

\bibitem{Li_2017}
S.-J. Li, D.~Bao, M.-M. Shi, B.-R. Wulan, J.-M. Yan, and Q.~Jiang.
\newblock Amorphizing of au nanoparticles by {CeOx}-{RGO} hybrid support
  towards highly efficient electrocatalyst for n$_2$reduction under ambient
  conditions.
\newblock {\em Advanced Materials}, 29(33):1700001, jul 2017.

\bibitem{Li_2007}
Y.~Li, S.~Peng, F.~Jiang, G.~Lu, and S.~Li.
\newblock Effect of doping {TiO}$_2$ with alkaline-earth metal ions on its
  photocatalytic activity.
\newblock {\em Journal of the Serbian Chemical Society}, 72(4):393--402, 2007.

\bibitem{Liu_2016}
J.~Liu.
\newblock Catalysis by supported single metal atoms.
\newblock {\em {ACS} Catalysis}, 7(1):34--59, nov 2016.

\bibitem{Liu_2019}
S.~Liu, Y.~Wang, S.~Wang, M.~You, S.~Hong, T.-S. Wu, Y.-L. Soo, Z.~Zhao,
  G.~Jiang, J.~Qiu, B.~Wang, and Z.~Sun.
\newblock Photocatalytic fixation of nitrogen to ammonia by single {Ru} atom
  decorated {TiO}$_2$ nanosheets.
\newblock {\em {ACS} Sustainable Chemistry {\&} Engineering}, 7(7):6813--6820,
  feb 2019.

\bibitem{Lu1994}
G.~Lu, a.~Linsebigler, and J.~T. Yates.
\newblock {Ti3+ Defect Sites on {TiO}$_2$(110): Production and Chemical
  Detection of Active Sites}.
\newblock {\em Journal of Physical Chemistry}, 98(45):11733--11738, 1994.

\bibitem{McArthur_2017}
J.~W. McArthur and G.~C. McCord.
\newblock Fertilizing growth: Agricultural inputs and their effects in economic
  development.
\newblock {\em Journal of Development Economics}, 127:133--152, jul 2017.

\bibitem{McPherson_2019}
I.~J. McPherson, T.~Sudmeier, J.~Fellowes, and S.~C.~E. Tsang.
\newblock Materials for electrochemical ammonia synthesis.
\newblock {\em Dalton Transactions}, 48(5):1562--1568, 2019.

\bibitem{Medford_2017}
A.~J. Medford and M.~C. Hatzell.
\newblock Photon-driven nitrogen fixation: Current progress, thermodynamic
  considerations, and future outlook.
\newblock {\em {ACS} Catalysis}, pages 2624--2643, mar 2017.

\bibitem{Michalsky_2015}
R.~Michalsky, A.~M. Avram, B.~A. Peterson, P.~H. Pfromm, and A.~A. Peterson.
\newblock Chemical looping of metal nitride catalysts: low-pressure ammonia
  synthesis for energy storage.
\newblock {\em Chem. Sci.}, 6:3965--3974, 2015.

\bibitem{Laughlin_2014}
U.~Mizutani, M.~Inukai, H.~Sato, and E.~Zijlstra.
\newblock 2 - electron theory of complex metallic alloys.
\newblock In D.~E. Laughlin and K.~Hono, editors, {\em Physical Metallurgy
  (Fifth Edition)}, pages 103 -- 202. Elsevier, Oxford, fifth edition edition,
  2014.

\bibitem{Monkhorst_1976}
H.~J. Monkhorst and J.~D. Pack.
\newblock Special points for brillouin-zone integrations.
\newblock {\em Physical Review B}, 13(12):5188--5192, jun 1976.

\bibitem{Montoya_2015}
J.~H. Montoya, C.~Tsai, A.~Vojvodic, and J.~K. N{\o}rskov.
\newblock {The challenge of electrochemical ammonia synthesis: A new
  perspective on the role of nitrogen scaling relations}.
\newblock {\em ChemSusChem}, 8(13):2180--2186, 2015.

\bibitem{Nakamura_2015}
A.~Nakamura, Y.~Ota, K.~Koike, Y.~Hidaka, K.~Nishioka, M.~Sugiyama, and
  K.~Fujii.
\newblock A 24.4\% solar to hydrogen energy conversion efficiency by combining
  concentrator photovoltaic modules and electrochemical cells.
\newblock {\em Applied Physics Express}, 8(10):107101, 2015.

\bibitem{Nilsson_2008}
A.~Nilsson and L.~G. Pettersson.
\newblock Adsorbate electronic structure and bonding on metal surfaces.
\newblock In {\em Chemical Bonding at Surfaces and Interfaces}, pages 57--142.
  Elsevier, 2008.

\bibitem{Norskov_2004}
J.~K. N{\o}rskov, J.~Rossmeisl, A.~Logadottir, L.~Lindqvist, J.~R. Kitchin,
  T.~Bligaard, and H.~J{\'{o}}nsson.
\newblock Origin of the overpotential for oxygen reduction at a fuel-cell
  cathode.
\newblock {\em J. Phys. Chem. B}, 108(46):17886--17892, nov 2004.

\bibitem{Nozik_1996}
A.~J. Nozik and R.~Memming.
\newblock Physical chemistry of semiconductor-liquid interfaces.
\newblock {\em The Journal of Physical Chemistry}, 100(31):13061--13078, jan
  1996.

\bibitem{O_Connor_2018}
N.~J. O'Connor, A.~S.~M. Jonayat, M.~J. Janik, and T.~P. Senftle.
\newblock Interaction trends between single metal atoms and oxide supports
  identified with density functional theory and statistical learning.
\newblock {\em Nature Catalysis}, 1(7):531--539, jul 2018.

\bibitem{IFDC_2012}
{P.A. Fuentes, B. Bumb, and M. Johnson}.
\newblock Improving fertilizer markets in west africa: The fertilizer supply
  chain in senegal.
\newblock Technical report, International Fertilizer Development Center and
  International Food Policy Research Institute, Muscle Shoals, Alabama,
  December 2012.

\bibitem{Qiao_2011}
B.~Qiao, A.~Wang, X.~Yang, L.~F. Allard, Z.~Jiang, Y.~Cui, J.~Liu, J.~Li, and
  T.~Zhang.
\newblock Single-atom catalysis of {CO} oxidation using pt1/{FeOx}.
\newblock {\em Nature Chemistry}, 3(8):634--641, jul 2011.

\bibitem{Ranjit_1996}
K.~Ranjit, T.~Varadarajan, and B.~Viswanathan.
\newblock Photocatalytic reduction of dinitrogen to ammonia over
  noble-metal-loaded {TiO}$_2$.
\newblock {\em Journal of Photochemistry and Photobiology A: Chemistry},
  96(1-3):181--185, may 1996.

\bibitem{Ranjit_1997}
K.~Ranjit and B.~Viswanathan.
\newblock Photocatalytic reduction of nitrite and nitrate ions to ammonia on
  m/{TiO}$_2$ catalysts.
\newblock {\em Journal of Photochemistry and Photobiology A: Chemistry},
  108(1):73--78, jul 1997.

\bibitem{Reuter_2005}
K.~Reuter, C.~Stampf, and M.~Scheffler.
\newblock {AB} initio atomistic thermodynamics and statistical mechanics of
  surface properties and functions.
\newblock In {\em Handbook of Materials Modeling}, pages 149--194. Springer
  Nature, 2005.

\bibitem{ritter_18}
S.~K. Ritter.
\newblock The haber-bosch reaction: An early chemical impact on sustainability.
\newblock {\em C{\&}EN}, 86, 2018.

\bibitem{Ruban_1997}
A.~Ruban, B.~Hammer, P.~Stoltze, H.~Skriver, and J.~N{\o}rskov.
\newblock Surface electronic structure and reactivity of transition and noble
  metals1communication presented at the first francqui colloquium, brussels,
  19{\textendash}20 february 1996.1.
\newblock {\em Journal of Molecular Catalysis A: Chemical}, 115(3):421--429,
  feb 1997.

\bibitem{Schiffer_2017}
Z.~J. Schiffer and K.~Manthiram.
\newblock Electrification and decarbonization of the chemical industry.
\newblock {\em Joule}, 1(1):10--14, Sep 2017.

\bibitem{Schloegl_2003}
R.~Schl\"ogl.
\newblock Catalytic synthesis of ammonia - a ``never-ending story''?
\newblock {\em Angewandte Chemie International Edition}, 42(18):2004--2008, may
  2003.

\bibitem{Schneider_2014}
J.~Schneider, M.~Matsuoka, M.~Takeuchi, J.~Zhang, Y.~Horiuchi, M.~Anpo, and
  D.~W. Bahnemann.
\newblock Understanding {TiO}$_2$ photocatalysis: Mechanisms and materials.
\newblock {\em Chemical Reviews}, 114(19):9919--9986, 2014.
\newblock PMID: 25234429.

\bibitem{Schrauzer_1977}
G.~Schrauzer and T.~Guth.
\newblock {Photocatalytic reactions. 1. Photolysis of water and photoreduction
  of nitrogen on titanium dioxide}.
\newblock {\em J. Am. Chem. Soc.}, 99(22):7189--7193, 1977.

\bibitem{Schrauzer_1983}
G.~N. Schrauzer, N.~Strampach, L.~N. Hui, M.~R. Palmer, and J.~Salehi.
\newblock {Nitrogen photoreduction on desert sands under sterile conditions}.
\newblock {\em Proc. Natl. Acad. Sci. U.S.A.}, 80(12):3873--3876, 1983.

\bibitem{Seh_2017}
Z.~W. Seh, J.~Kibsgaard, C.~F. Dickens, I.~Chorkendorff, J.~K. N{\o}rskov, and
  T.~F. Jaramillo.
\newblock Combining theory and experiment in electrocatalysis: Insights into
  materials design.
\newblock {\em Science}, 355(6321), 2017.

\bibitem{Shinde_2013}
V.~M. Shinde and G.~Madras.
\newblock {CO} methanation toward the production of synthetic natural gas over
  highly active {Ni}/{TiO}2catalyst.
\newblock {\em {AIChE} Journal}, 60(3):1027--1035, nov 2013.

\bibitem{Singh_2017}
A.~R. Singh, B.~A. Rohr, J.~A. Schwalbe, M.~Cargnello, K.~Chan, T.~F.
  Jaramillo, I.~Chorkendorff, and J.~K. N{\o}rskov.
\newblock Electrochemical ammonia synthesis-the selectivity challenge.
\newblock {\em ACS Catal.}, 7(1):706--709, jan 2017.

\bibitem{Smil_1999}
V.~Smil.
\newblock Detonator of the population explosion.
\newblock {\em Nature}, 400(6743):415--415, jul 1999.

\bibitem{Soria_1991}
J.~Soria, J.~C. Conesa, V.~Augugliaro, L.~Palmisano, M.~Schiavello, and
  A.~Sclafani.
\newblock {Dinitrogen photoreduction to ammonia over titanium dioxide powders
  doped with ferric ions}.
\newblock {\em J. Phys. Chem.}, 95(1):274--282, jan 1991.

\bibitem{Tao_2019}
H.~Tao, C.~Choi, L.-X. Ding, Z.~Jiang, Z.~Han, M.~Jia, Q.~Fan, Y.~Gao, H.~Wang,
  A.~W. Robertson, S.~Hong, Y.~Jung, S.~Liu, and Z.~Sun.
\newblock Nitrogen fixation by ru single-atom electrocatalytic reduction.
\newblock {\em Chem}, 5(1):204--214, jan 2019.

\bibitem{Turchanin_2008}
M.~A. Turchanin and P.~G. Agraval.
\newblock Cohesive energy, properties, and formation energy of transition metal
  alloys.
\newblock {\em Powder Metallurgy and Metal Ceramics}, 47(1-2):26--39, jan 2008.

\bibitem{Vesborg_2012}
P.~C.~K. Vesborg and T.~F. Jaramillo.
\newblock Addressing the terawatt challenge: scalability in the supply of
  chemical elements for renewable energy.
\newblock {\em {RSC} Advances}, 2(21):7933, 2012.

\bibitem{Walle2009}
L.~E. Walle, A.~Borg, P.~Uvdal, and A.~Sandell.
\newblock {Experimental evidence for mixed dissociative and molecular
  adsorption of water on a rutile {TiO}$_2$ (110) surface without oxygen
  vacancies}.
\newblock {\em Physical Review B}, 80(23):235436, 2009.

\bibitem{WANG20181055}
L.~Wang, M.~Xia, H.~Wang, K.~Huang, C.~Qian, C.~T. Maravelias, and G.~A. Ozin.
\newblock Greening ammonia toward the solar ammonia refinery.
\newblock {\em Joule}, 2(6):1055 -- 1074, 2018.

\bibitem{Wang_2011}
S.~Wang, V.~Petzold, V.~Tripkovic, J.~Kleis, J.~G. Howalt, E.~Sk{\'{u}}lason,
  E.~M. Fern{\'{a}}ndez, B.~Hvolb{\ae}k, G.~Jones, A.~Toftelund, H.~Falsig,
  M.~Björketun, F.~Studt, F.~Abild-Pedersen, J.~Rossmeisl, J.~K. N{\o}rskov,
  and T.~Bligaard.
\newblock Universal transition state scaling relations for (de)hydrogenation
  over transition metals.
\newblock {\em Physical Chemistry Chemical Physics}, 13(46):20760, 2011.

\bibitem{Wellendorff_2012}
J.~Wellendorff, K.~T. Lundgaard, A.~M\o{}gelh\o{}j, V.~Petzold, D.~D. Landis,
  J.~K. N\o{}rskov, T.~Bligaard, and K.~W. Jacobsen.
\newblock Density functionals for surface science: Exchange-correlation model
  development with bayesian error estimation.
\newblock {\em Physical Review B}, 85(23):235149--235149, 2012.

\bibitem{Xu_2015}
Z.~Xu and J.~R. Kitchin.
\newblock Relationships between the surface electronic and chemical properties
  of doped 4d and 5d late transition metal dioxides.
\newblock {\em The Journal of Chemical Physics}, 142(10):104703, mar 2015.

\bibitem{Yao_2017}
Z.~Yao and K.~Reuter.
\newblock First-principles computational screening of dopants to improve the
  deacon process over {RuO}$_2$.
\newblock {\em {ChemCatChem}}, 10(2):465--469, dec 2017.

\bibitem{Yates_1991}
J.~T. Yates, A.~Szab{\'{o}}, and M.~A. Henderson.
\newblock The influence of surface defect sites on chemisorption and catalysis.
\newblock In {\em Structure-Activity and Selectivity Relationships in
  Heterogeneous Catalysis, Proceedings of the {ACS} Symposium on
  Structure-Activity Relationships in Heterogeneous Catalysis}, pages 273--290.
  Elsevier, 1991.

\bibitem{Yu_2019}
J.~Yu, J.~Yu, Z.~Shi, Q.~Guo, X.~Xiao, H.~Mao, and D.~Mao.
\newblock The effects of the nature of {TiO}2 supports on the catalytic
  performance of {Rh}{\textendash}{Mn}/{TiO}2 catalysts in the synthesis of
  {C}2 oxygenates from syngas.
\newblock {\em Catalysis Science {\&} Technology}, 9(14):3675--3685, 2019.

\bibitem{Yuan_2013}
S.-J. Yuan, J.-J. Chen, Z.-Q. Lin, W.-W. Li, G.-P. Sheng, and H.-Q. Yu.
\newblock {Nitrate formation from atmospheric nitrogen and oxygen
  photocatalysed by nano-sized titanium dioxide}.
\newblock {\em Nat. Commun.}, 4, 2013.

\bibitem{Zaleska_2008}
A.~Zaleska.
\newblock Doped-{TiO}$_2$: A review.
\newblock {\em Recent Patents on Engineering}, 2(3):157--164, nov 2008.

\bibitem{Zhao_2019}
Z.~Zhao, S.~Hong, C.~Yan, C.~Choi, Y.~Jung, Y.~Liu, S.~Liu, X.~Li, J.~Qiu, and
  Z.~Sun.
\newblock Efficient visible-light driven n$_2$ fixation over two-dimensional
  sb/{TiO}$_2$ composites.
\newblock {\em Chemical Communications}, 55(50):7171--7174, 2019.

\end{thebibliography}
\appendix
\end{document}


\maketitle\begin{table}
\setlength\tabcolsep{2pt}
\begin{center}
\begin{tabular}{| c | c | c | c | c | c | c | c | c | c | c | c | c | c |}
\hline
Element & H$_2$NNH$_2$ & HNNH & N & N$_2$ & N$_2$H & N$_2$H$_2$ & N$_2$H$_3$ & NH & NH$_2$ & NH$_3$ & Formation Energy\\
\hline

Y & 1.01 & 1.69 & 2.72 & 0.02 &  & 2.0 & 1.34 & 2.58 & -0.15 & -0.77 & -1.38 \\
Rh & 1.16 & 1.3 & 1.86 & -0.16 & 0.71 & 0.85 & 0.44 & 1.34 & -1.06 & -0.87 & 6.01 \\
Pt &  & 2.16 & 2.77 & 0.26 & 1.78 & 2.06 & 1.44 & 1.69 & -0.1 & -0.09 & 6.86 \\
Ir & 0.82 & 0.74 & 1.03 & -0.7 & 0.26 & 0.19 & -0.01 & 0.54 & -1.52 & -1.2 & 7.07 \\
Ta & 1.1 & 0.31 & -0.99 & -0.13 & 0.42 & -0.32 & -0.22 & -0.95 & -1.86 & -0.85 & 1.69 \\
Zr & 1.23 & 1.39 & 1.71 & -0.01 & 1.07 & 0.74 & 0.35 & 0.16 & -1.26 & -0.88 & -0.51 \\
Re & 0.95 & 0.67 & -1.5 & -0.83 & -0.42 & -0.15 & 0.32 & -0.18 & -1.21 & -0.96 & 5.06 \\
Hf & 1.21 & 1.32 & 1.59 & -0.02 & 0.97 & 0.6 & 0.2 & 0.02 & -1.44 & -0.95 & -0.92 \\
Pd & 1.53 & 2.12 & 3.55 & 0.23 & 2.0 & 2.25 & 1.67 & 2.5 & 0.26 & -0.22 & 6.08 \\
Ti & 1.42 & 1.64 & 1.73 & 0.11 & 1.27 & 0.85 & 0.61 & 0.37 & -0.99 & -0.6 & 0.0 \\
Ag & 1.44 & 2.33 & 5.3 & 0.24 & 2.62 & 2.65 & 2.04 & 3.83 & 1.11 & -0.18 & 7.28 \\
Sc & 1.06 & 1.76 & 3.51 & 0.03 & 1.7 & 1.59 & 1.0 & 2.18 & -0.6 & -0.76 & -1.71 \\
Ru & 0.82 & 0.76 & 0.48 & -0.83 & 0.64 & 0.17 & 0.57 & 0.86 & -0.93 & -1.13 & 5.45 \\
Co & 1.14 & 1.53 & 2.34 & 0.12 & 1.04 & 1.05 & 0.78 & 1.82 & -0.74 & -0.72 & 4.49 \\
Nb & 1.23 & 0.43 & -1.04 & -0.23 & 0.35 & -0.36 & -0.04 & -0.84 & -1.7 & -0.86 & 1.5 \\
Mo & 1.27 & 1.11 & -1.2 & -0.33 & 0.01 & -0.08 & 0.39 & -0.24 & -1.2 & -0.75 & 3.26 \\
Ni & 1.75 & 1.94 & 3.35 & 0.17 & 1.64 & 1.9 & 1.09 &  & -0.06 & -0.43 & 5.58 \\
Tc & 1.03 & 0.95 & -0.87 & -0.65 & 0.06 & 0.27 & 0.65 & 0.52 & -0.87 & -0.92 & 4.58 \\
Os & 0.61 & 0.39 & -0.7 & -1.17 & -0.15 & -0.36 & 0.06 & 0.06 & -1.44 & -1.29 & 6.31 \\
Cu & 1.33 & 2.07 & 4.51 & 0.18 & 2.33 & 2.4 & 1.65 & 3.41 & 0.78 & -0.45 & 6.55 \\
V & 1.43 & 1.5 & -0.33 & 0.04 & 0.82 & 0.38 & 0.55 & 0.17 & -1.04 & -1.03 & 2.48 \\
Au & 1.68 & 2.36 & 4.05 & 0.25 & 2.54 & 2.82 & 2.15 & 2.95 & 0.75 & -0.08 & 8.18 \\
W & 1.25 & 0.84 & -1.8 & -0.37 & -0.24 & -0.73 & 0.01 & -1.06 & -1.6 & -0.8 & 3.99 \\
\hline
\end{tabular}
\end{center}
\caption{The calculated relative energies of all 2+ surface species on all metal substituents at standard state. All energies are referenced with respect to N$_2$ gas and H$_2$ gas at 300K and 1 bar of pressure. Blank spaces represent calculations that could not be converged}
\label{table:energies}
\end{table}

\begin{table}
\begin{center}
\begin{tabular}{| c | c | c | c |}
\hline
Element & N$_2$ & N$_2$H & Formation Energy \\
\hline
Y & 0.04 & 2.6 & 1.1 \\
Rh & -0.14 & 2.22 & 8.71 \\
Pt & -0.34 & 1.77 & 10.08 \\
Ir & -0.39 & 2.1 & 9.12 \\
Ta & 0.1 & 2.37 & 1.39 \\
Zr & 0.07 & 2.46 & -0.75 \\
Fe & -0.12 &  & 8.4 \\
Hf & 0.06 & 2.43 & -1.22 \\
Re & 0.12 &  & 5.9 \\
Pd & -0.01 & 1.7 & 9.88 \\
Ti & 0.16 & 2.54 & -0.0 \\
Ag & 0.24 &  & 11.09 \\
Sc & 0.08 & 2.34 & 0.63 \\
Ru & -0.0 &  & 7.5 \\
Co & 0.2 &  & 7.42 \\
Nb & 0.11 & 2.45 & 1.4 \\
Mo & 0.14 &  & 3.95 \\
Ni & 0.2 & 1.75 & 9.29 \\
Tc & 0.12 &  & 5.88 \\
Os & -0.22 & 1.96 & 7.69 \\
Cu & 0.22 &  & 10.16 \\
V & 0.19 & 2.81 & 2.82 \\
Au & 0.27 & 2.23 & 11.22 \\
W & 0.13 & 2.37 & 4.36 \\
\hline
\end{tabular}
\end{center}
\label{table:4+_energies}
\caption{The calculated relative energies of all 4+ surface species on all metal substituents at standard state. All energies are referenced with respect to N$_2$ gas and H$_2$ gas at 300K and 1 bar of pressure. Blank spaces represent calculations that could not be converged}
\end{table}

\begin{table}
\begin{center}
\begin{tabular}{| c | c |c |}
\hline
Element & Limiting Potential & Limiting Step \\
\hline
Sc & -1.7 & N$_2$ $\rightarrow$ N$_2$H*\\
Ti & -1.27 & N$_2$ $\rightarrow$ N$_2$H*\\
V & -0.82 & N$_2$ $\rightarrow$ N$_2$H*\\
Co & -1.04 & N$_2$ $\rightarrow$ N$_2$H*\\
Ni & -1.64 & N$_2$ $\rightarrow$ N$_2$H*\\
Cu & -2.33 & N$_2$ $\rightarrow$ N$_2$H*\\
Zr & -1.08 & N$_2$* $\rightarrow$ N$_2$H*\\
Nb & -1.38 & NH$_2$*+NH$_3$ $\rightarrow$ 2NH$_3$\\
Mo & -0.89 & NH$_2$*+NH$_3$ $\rightarrow$ 2NH$_3$\\
Tc & -0.71 & N$_2$* $\rightarrow$ N$_2$H*\\
Ru & -1.47 & N$_2$* $\rightarrow$ N$_2$H*\\
Rh & -0.88 & N$_2$* $\rightarrow$ N$_2$H*\\
Pd & -2.0 & N$_2$ $\rightarrow$ N$_2$H*\\
Ag & -2.62 & N$_2$ $\rightarrow$ N$_2$H*\\
Hf & -1.12 & NH$_2$*+NH$_3$ $\rightarrow$ 2NH$_3$\\
Ta & -1.55 & NH$_2$*+NH$_3$ $\rightarrow$ 2NH$_3$\\
W & -1.29 & NH$_2$*+NH$_3$ $\rightarrow$ 2NH$_3$\\
Re & -0.9 & NH$_2$*+NH$_3$ $\rightarrow$ 2NH$_3$\\
Os & -1.12 & NH$_2$*+NH$_3$ $\rightarrow$ 2NH$_3$\\
Ir & -1.2 & NH$_2$*+NH$_3$ $\rightarrow$ 2NH$_3$\\
Au & -2.54 & N$_2$ $\rightarrow$ N$_2$H*\\
\hline
\end{tabular}
\end{center}
\caption{The limiting potentials and limiting steps for each dopant metal on 2+ surfaces}\label{table:pot_limiting_steps}\end{table}\begin{table}
\begin{center}
\begin{tabular}{| c | c |c |}
\hline
Element & Largest Thermodynamic Step & Limiting Step \\
\hline
Sc & 0.44 & NH$_3$*+NH$_3$ $\rightarrow$ 2NH$_3$\\
Ti & 0.68 & NH$_3$*+NH$_3$ $\rightarrow$ 2NH$_3$\\
V & 0.72 & NH$_3$*+NH$_3$ $\rightarrow$ 2NH$_3$\\
Co & 0.43 & NH$_3$*+NH$_3$ $\rightarrow$ 2NH$_3$\\
Ni & 0.17 & N$_2$ $\rightarrow$ N$_2$*\\
Cu & 1.25 & H$_2$NNH$_2$* $\rightarrow$ 2NH$_2$*\\
Zr & 0.95 & NH$_3$*+NH$_3$ $\rightarrow$ 2NH$_3$\\
Nb & 1.38 & NH$_3$*+NH$_3$ $\rightarrow$ 2NH$_3$\\
Mo & 0.89 & NH$_3$*+NH$_3$ $\rightarrow$ 2NH$_3$\\
Tc & 0.61 & NH$_3$*+NH$_3$ $\rightarrow$ 2NH$_3$\\
Ru & 0.82 & NH$_3$*+NH$_3$ $\rightarrow$ 2NH$_3$\\
Rh & 0.75 & NH$_3$*+NH$_3$ $\rightarrow$ 2NH$_3$\\
Pd & 0.54 & H$_2$NNH$_2$* $\rightarrow$ 2NH$_2$*\\
Ag & 1.48 & H$_2$NNH$_2$* $\rightarrow$ 2NH$_2$*\\
Hf & 1.12 & NH$_3$*+NH$_3$ $\rightarrow$ 2NH$_3$\\
Ta & 1.55 & NH$_3$*+NH$_3$ $\rightarrow$ 2NH$_3$\\
W & 1.29 & NH$_3$*+NH$_3$ $\rightarrow$ 2NH$_3$\\
Re & 0.9 & NH$_3$*+NH$_3$ $\rightarrow$ 2NH$_3$\\
Os & 1.12 & NH$_3$*+NH$_3$ $\rightarrow$ 2NH$_3$\\
Ir & 1.2 & NH$_3$*+NH$_3$ $\rightarrow$ 2NH$_3$\\
Au & 0.88 & H$_2$NNH$_2$* $\rightarrow$ 2NH$_2$*\\
\hline
\end{tabular}
\end{center}
\caption{The largest barrier for thermochemical steps and corresponding steps for each dopant metal on 2+ surfaces}\label{table:thermo_limiting_steps}\end{table}\begin{table}
\begin{center}
\begin{tabular}{| c | c |c |}
\hline
Element & Rate Limiting Step & Limiting Step \\
\hline
Sc & 1.56 & N$_2$ $\rightarrow$ N$_2$H*\\
Ti & 1.13 & N$_2$ $\rightarrow$ N$_2$H*\\
V & 0.76 & N$_2$H$_2$* $\rightarrow$ H$_2$NNH$_2$*\\
Co & 0.9 & N$_2$ $\rightarrow$ N$_2$H*\\
Ni & 1.62 & N$_2$ $\rightarrow$ N$_2$H$_2$*\\
Cu & 2.19 & N$_2$ $\rightarrow$ N$_2$H*\\
Zr & 0.94 & N$_2$* $\rightarrow$ N$_2$H*\\
Nb & 1.31 & N$_2$H$_2$* $\rightarrow$ H$_2$NNH$_2$*\\
Mo & 1.07 & N$_2$H$_2$* $\rightarrow$ H$_2$NNH$_2$*\\
Tc & 1.11 & N$_2$* $\rightarrow$ H$_2$NNH$_2$*\\
Ru & 1.33 & N$_2$* $\rightarrow$ N$_2$H*\\
Rh & 0.73 & N$_2$* $\rightarrow$ N$_2$H*\\
Pd & 1.96 & N$_2$ $\rightarrow$ N$_2$H$_2$*\\
Ag & 2.47 & N$_2$ $\rightarrow$ N$_2$H*\\
Hf & 0.98 & NH$_2$*+NH$_3$ $\rightarrow$ 2NH$_3$\\
Ta & 1.41 & NH$_2$*+NH$_3$ $\rightarrow$ 2NH$_3$\\
W & 1.69 & N$_2$H$_2$* $\rightarrow$ H$_2$NNH$_2$*\\
Re & 1.21 & N$_2$* $\rightarrow$ H$_2$NNH$_2$*\\
Os & 0.98 & NH$_2$*+NH$_3$ $\rightarrow$ 2NH$_3$\\
Ir & 1.06 & NH$_2$*+NH$_3$ $\rightarrow$ 2NH$_3$\\
Au & 2.54 & N$_2$ $\rightarrow$ N$_2$H$_2$*\\
\hline
\end{tabular}
\end{center}
\caption{The largest thermodynamic barrier and corresponding steps for each dopant metal on 2+ surfaces when set at the band edge of rutile, -0.142V}\label{table:rate_limiting_steps}\end{table}\begin{figure}
\centering
\includegraphics[width=0.8\linewidth]{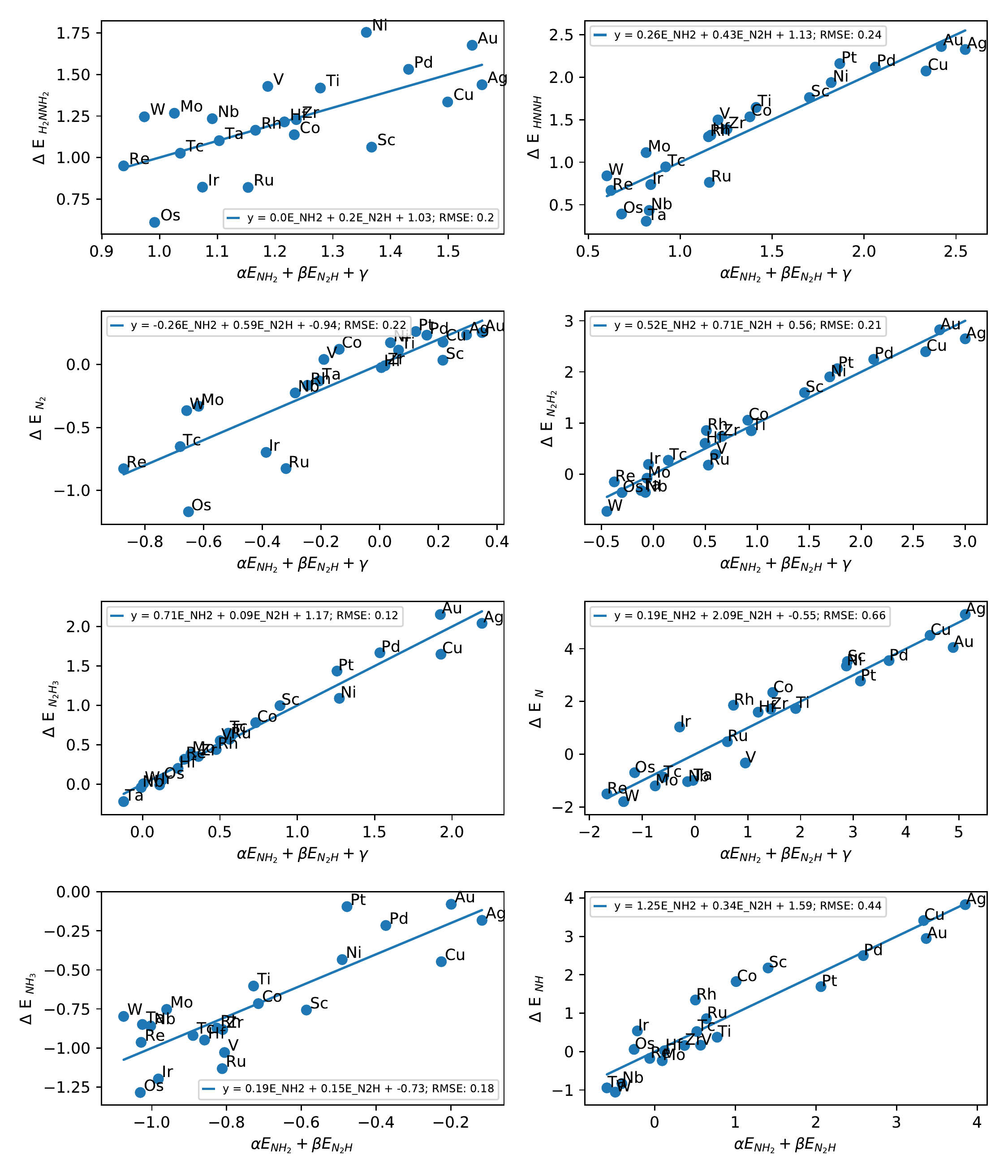}
\caption{The calculated scaling relations between the binding energies of various species and the binding energies of N$_2$H and NH$_2$ on 2+ dopant sites}
\label{fig:scaling_species}
\end{figure}

\begin{figure}
\centering
\includegraphics[width=0.8\linewidth]{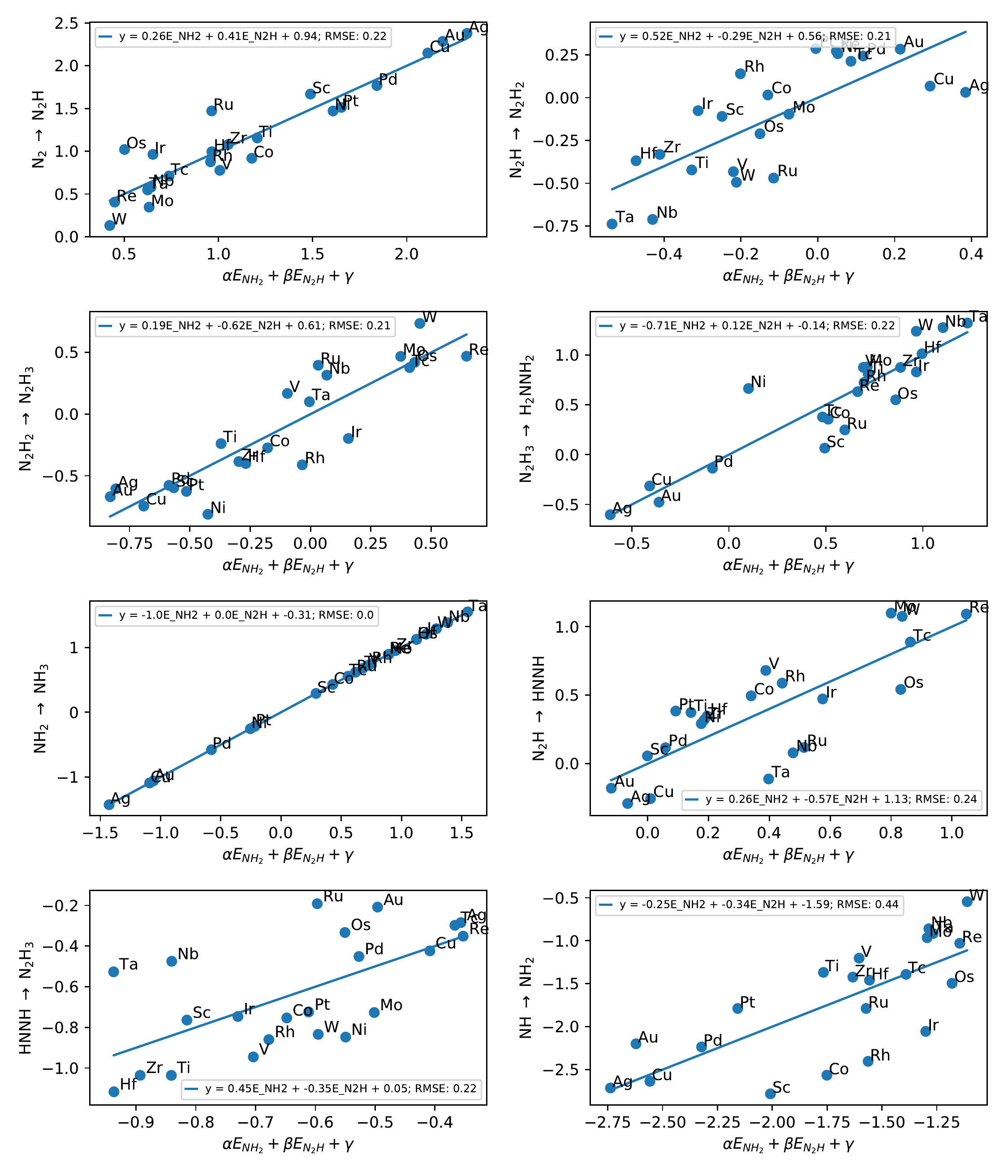}
\caption{The calculated scaling relations between the reaction energies energies of all electrochemical reations and the binding energies of N$_2$H and NH$_2$ on 2+ dopant sites}
\label{fig:scaling_reactions}
\end{figure}

\begin{figure}
\centering
\includegraphics[width=0.8\linewidth]{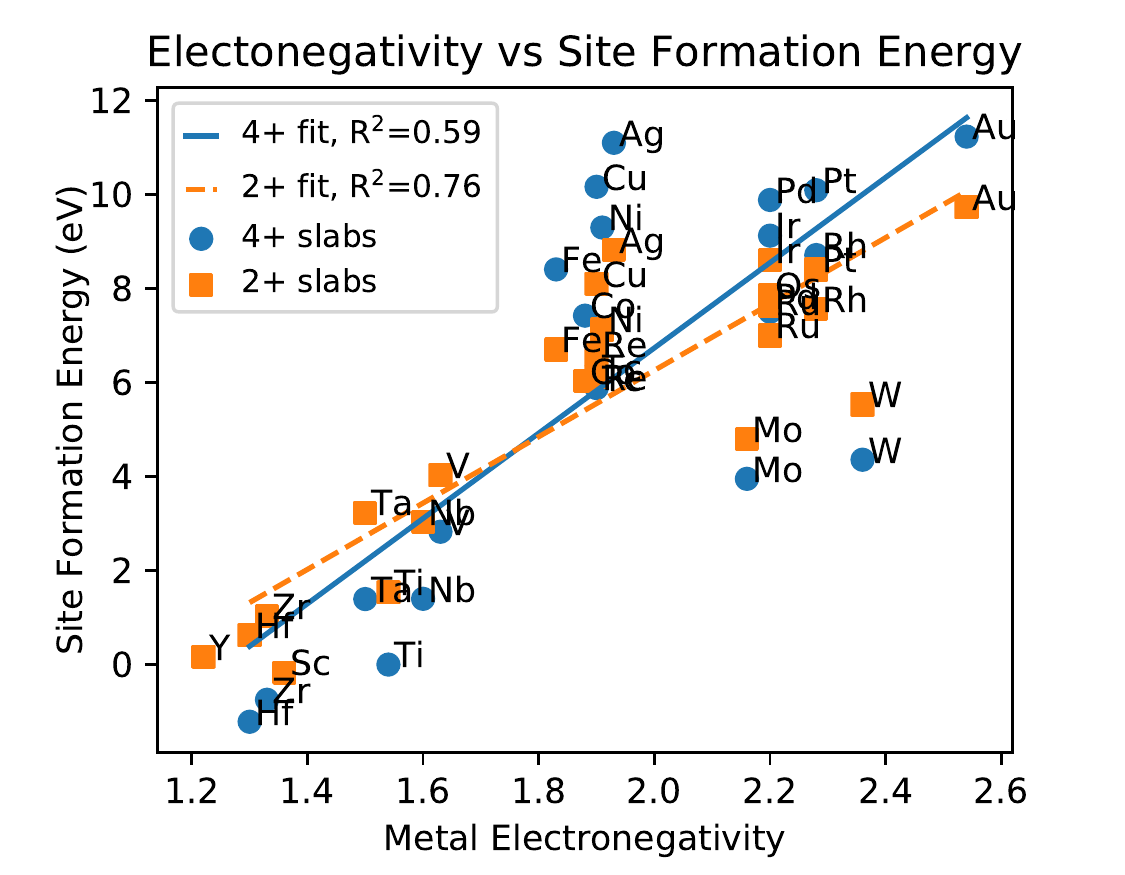}
\caption{Electronegativity vs formation energy of 2+ dopant site}
\label{fig:electronegativity}
\end{figure}

\twocolumn
\newpage
\begin{figure}
\includegraphics[width=1\linewidth]{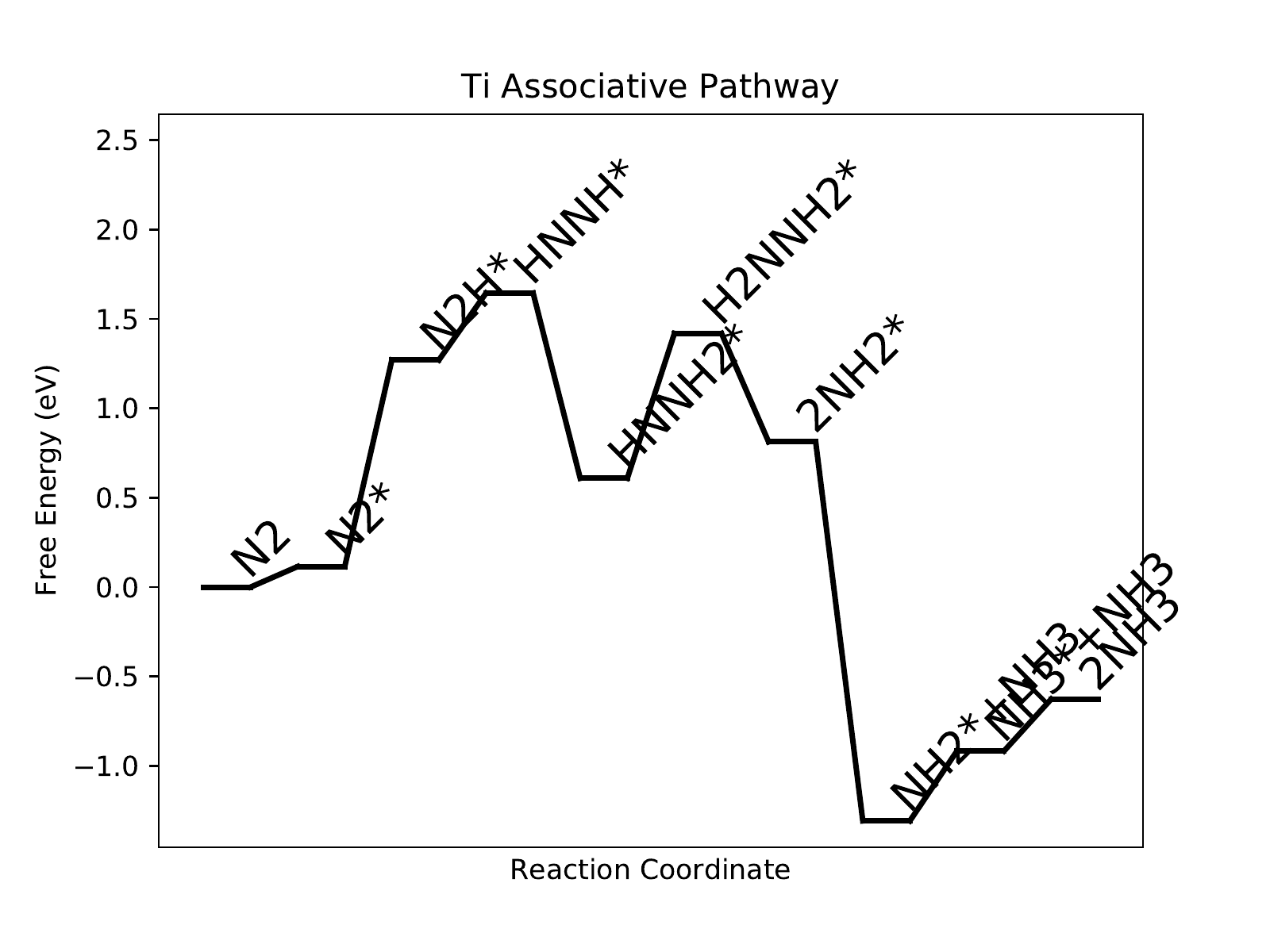}
\label{fig:Ti_associative}
\caption{Free energy diagram for Ti}
\end{figure}

\begin{figure}
\includegraphics[width=1\linewidth]{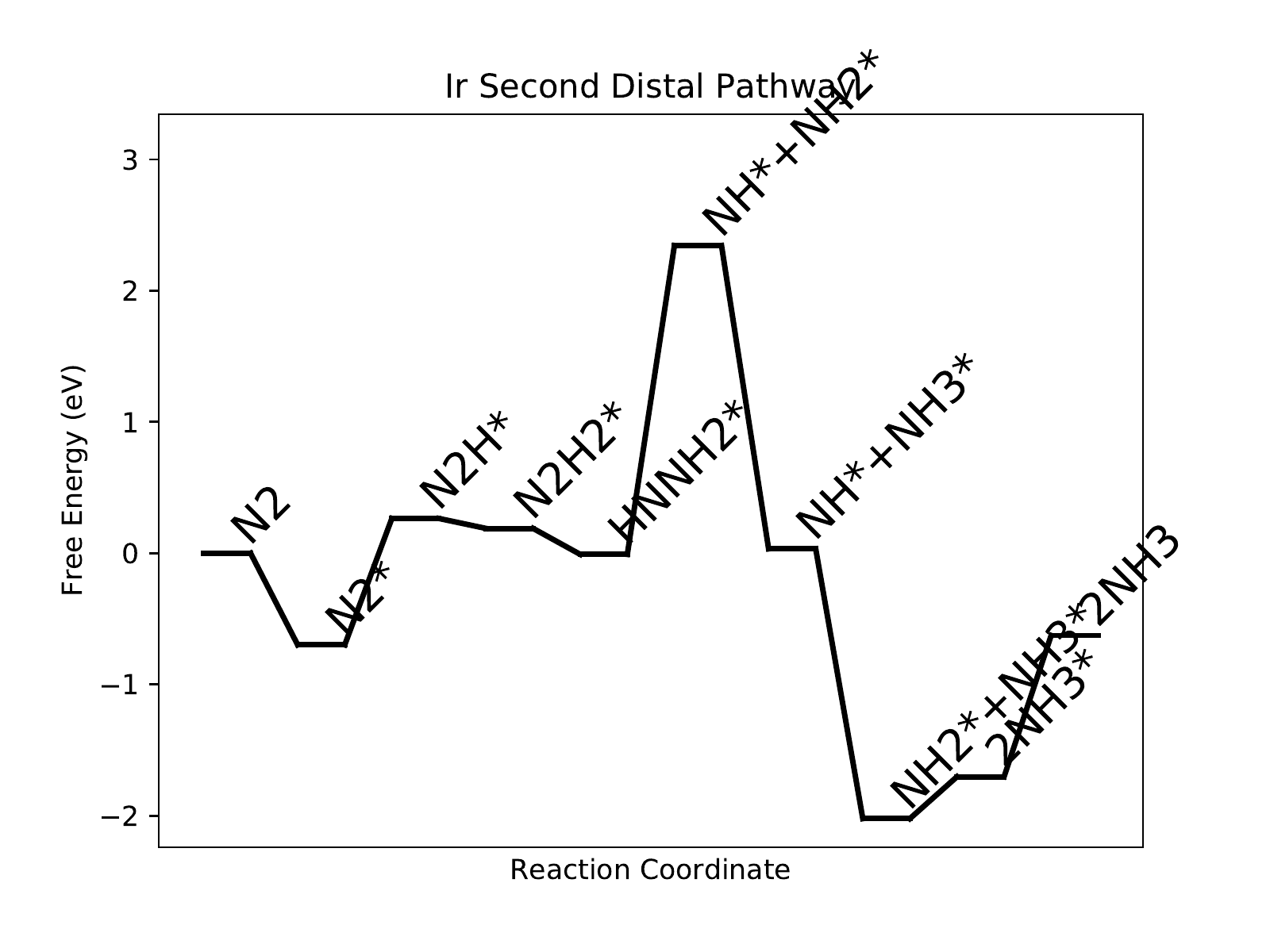}
\label{fig:Ir_distal_2}
\caption{Free energy diagram for Ir}
\end{figure}

\newpage
\begin{figure}
\includegraphics[width=1\linewidth]{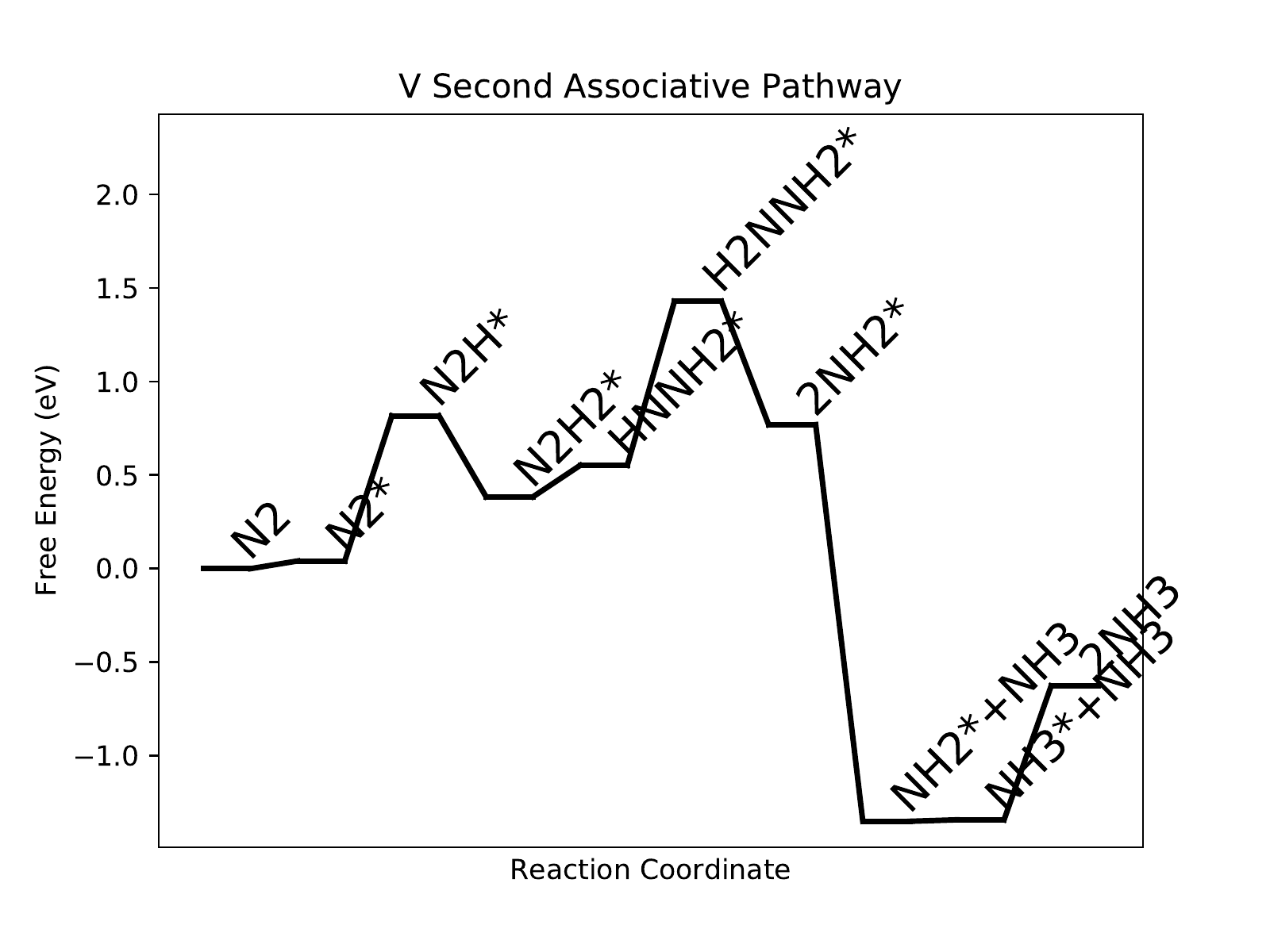}
\label{fig:V_associative_2}
\caption{Free energy diagram for V}
\end{figure}

\begin{figure}
\includegraphics[width=1\linewidth]{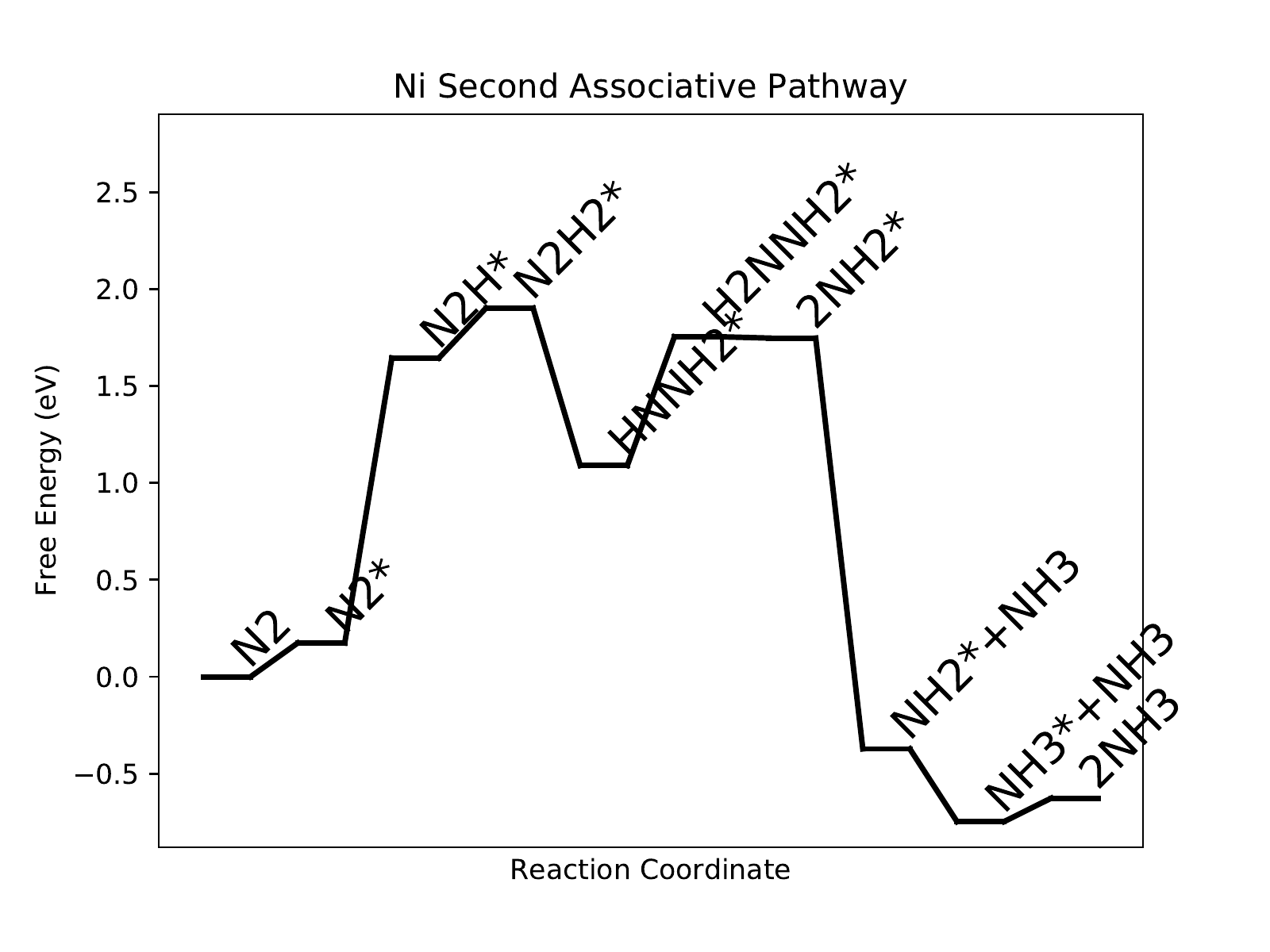}
\label{fig:Ni_associative_2}
\caption{Free energy diagram for Ni}
\end{figure}

\newpage
\begin{figure}
\includegraphics[width=1\linewidth]{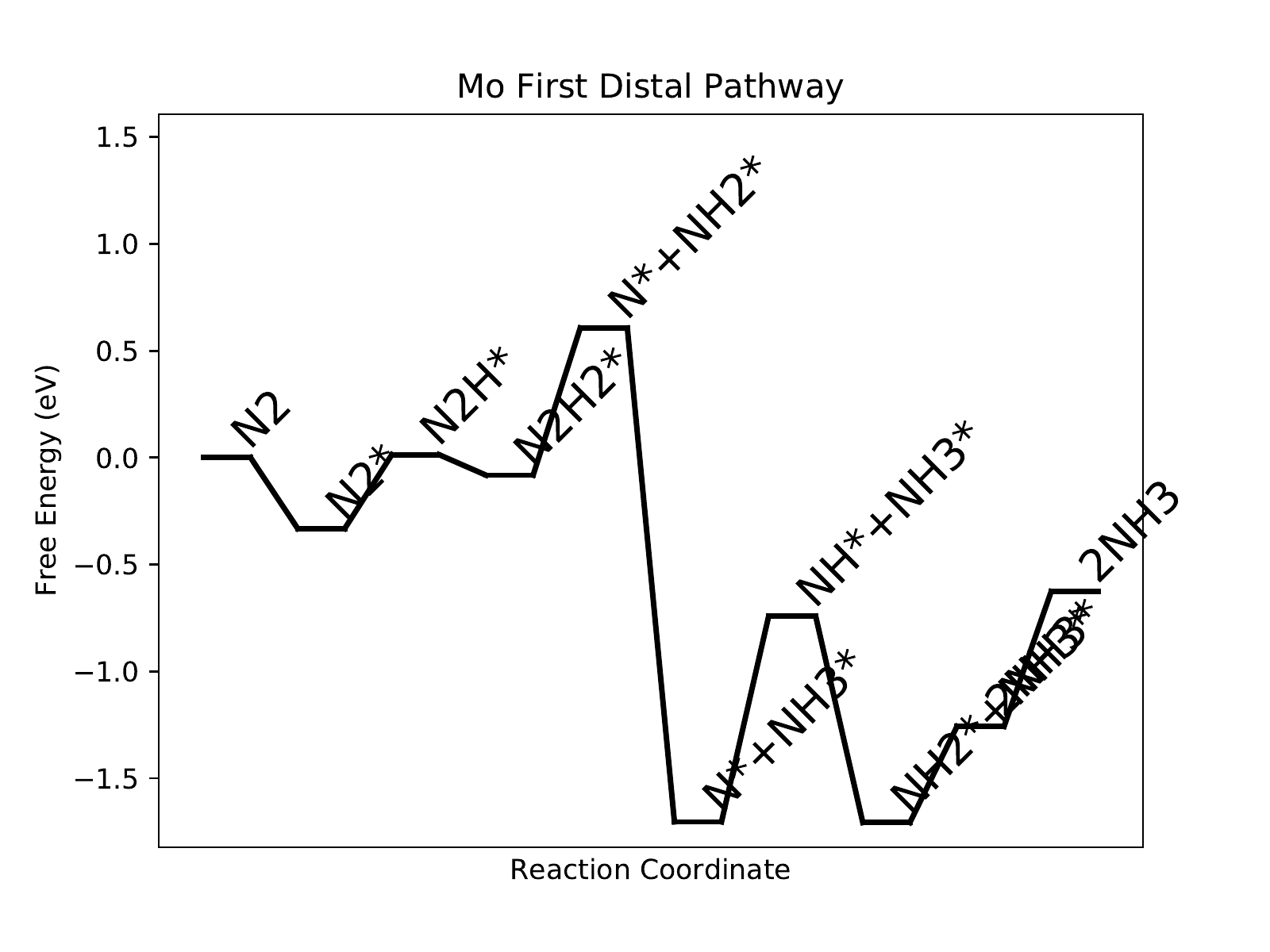}
\label{fig:Mo_distal_1}
\caption{Free energy diagram for Mo}
\end{figure}

\begin{figure}
\includegraphics[width=1\linewidth]{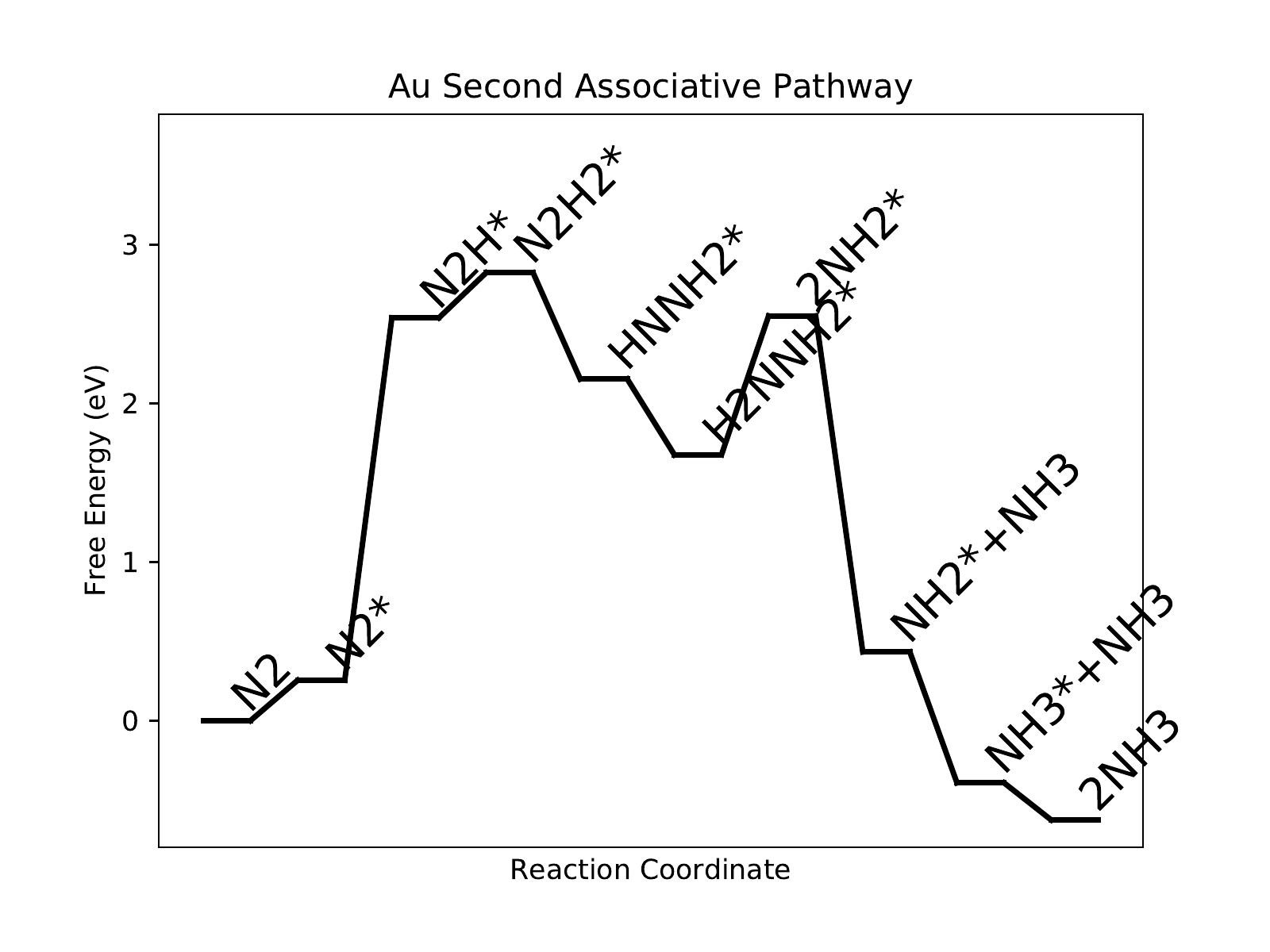}
\label{fig:Au_associative_2}
\caption{Free energy diagram for Au}
\end{figure}

\newpage
\begin{figure}
\includegraphics[width=1\linewidth]{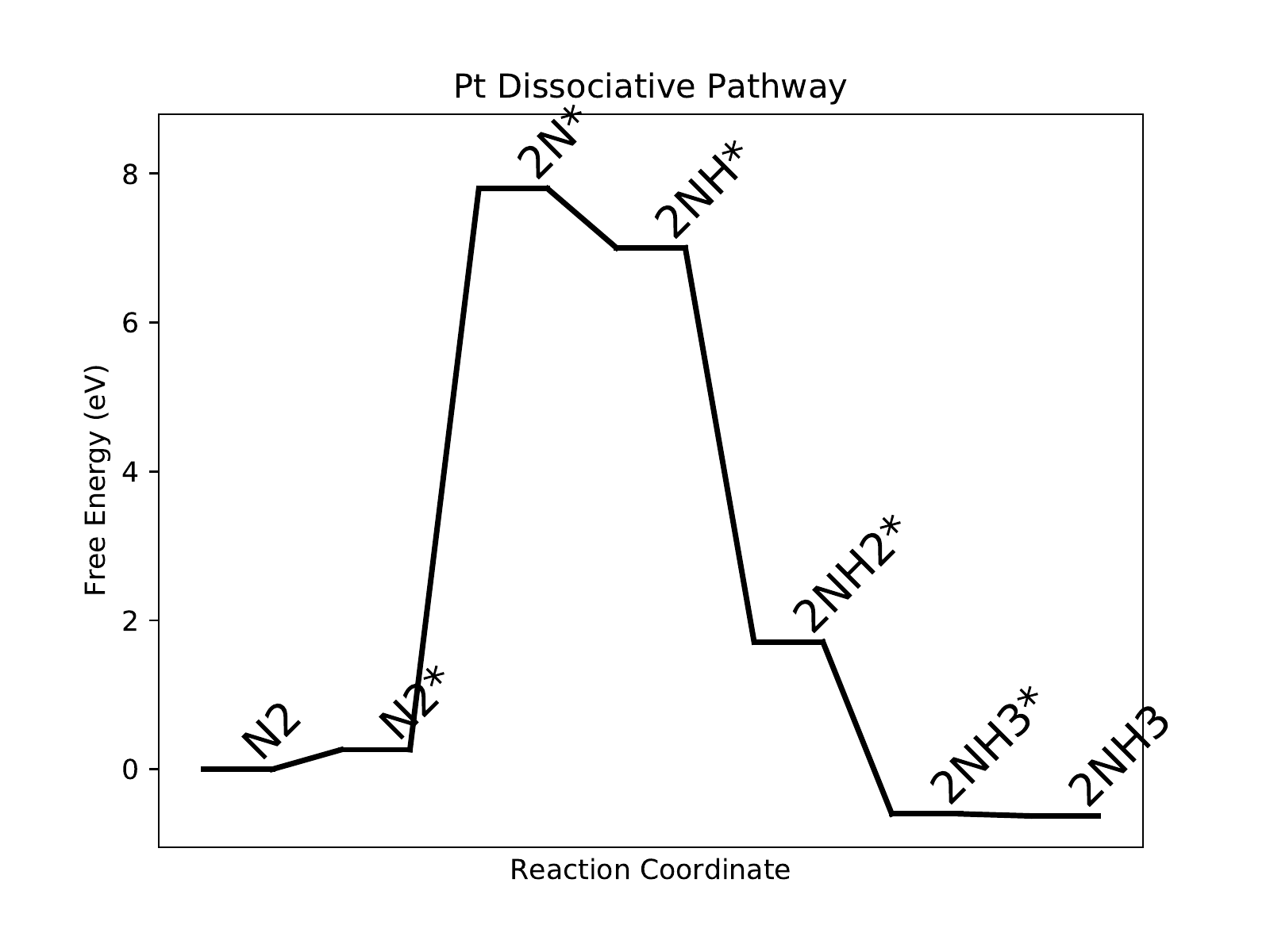}
\label{fig:Pt_dissociative}
\caption{Free energy diagram for Pt}
\end{figure}

\begin{figure}
\includegraphics[width=1\linewidth]{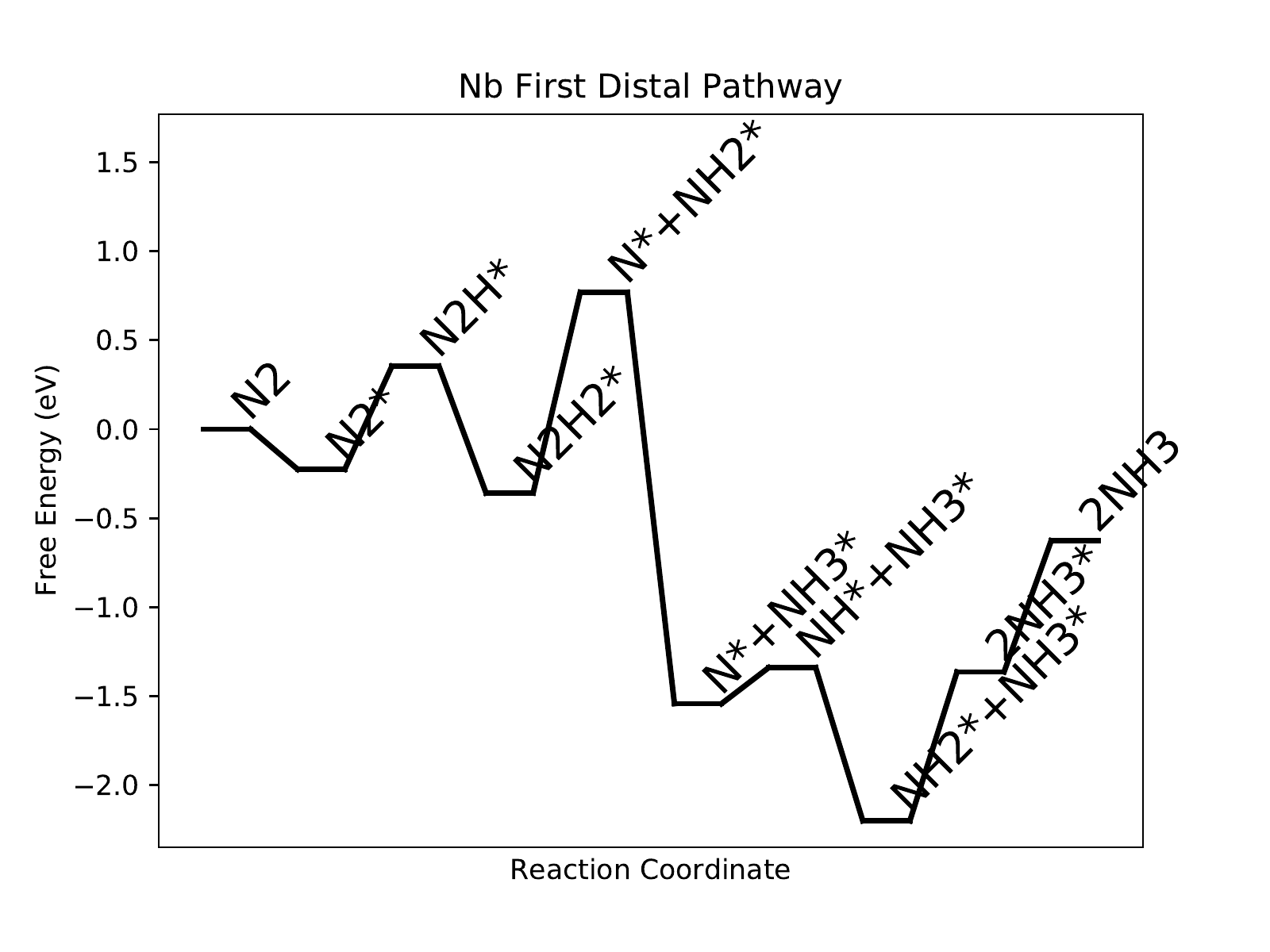}
\label{fig:Nb_distal_1}
\caption{Free energy diagram for Nb}
\end{figure}

\newpage
\begin{figure}
\includegraphics[width=1\linewidth]{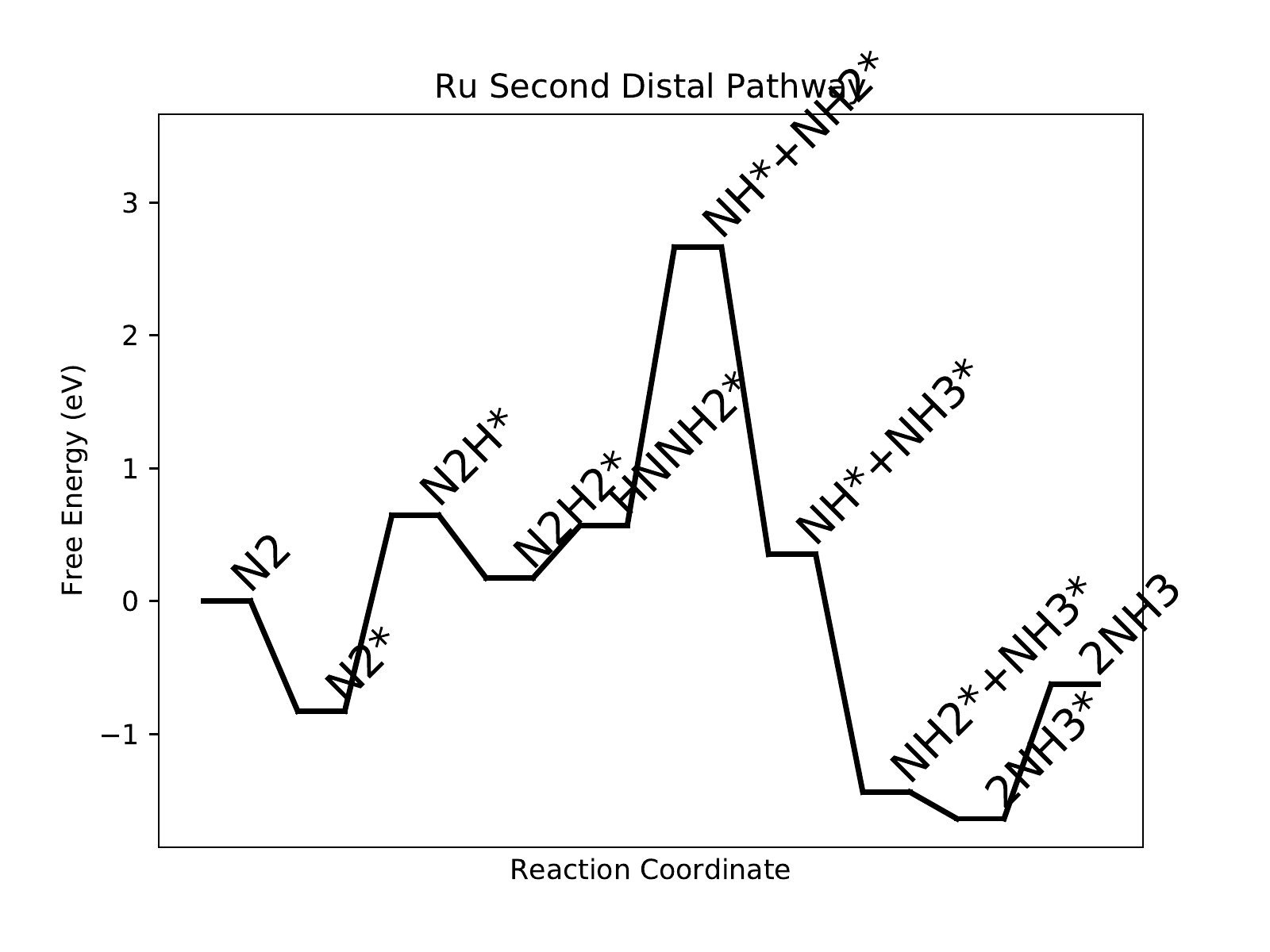}
\label{fig:Ru_distal_2}
\caption{Free energy diagram for Ru}
\end{figure}

\begin{figure}
\includegraphics[width=1\linewidth]{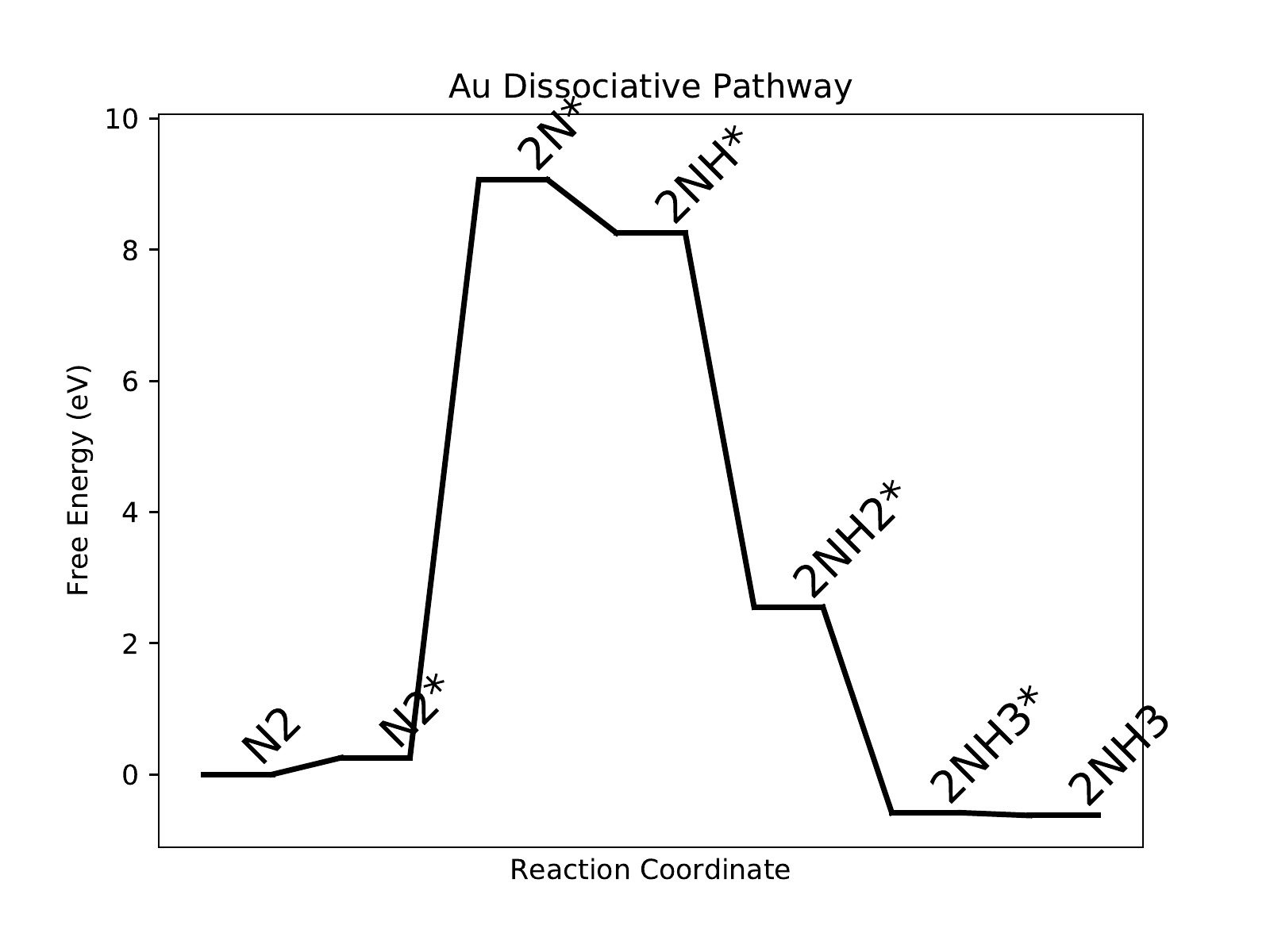}
\label{fig:Au_dissociative}
\caption{Free energy diagram for Au}
\end{figure}

\newpage
\begin{figure}
\includegraphics[width=1\linewidth]{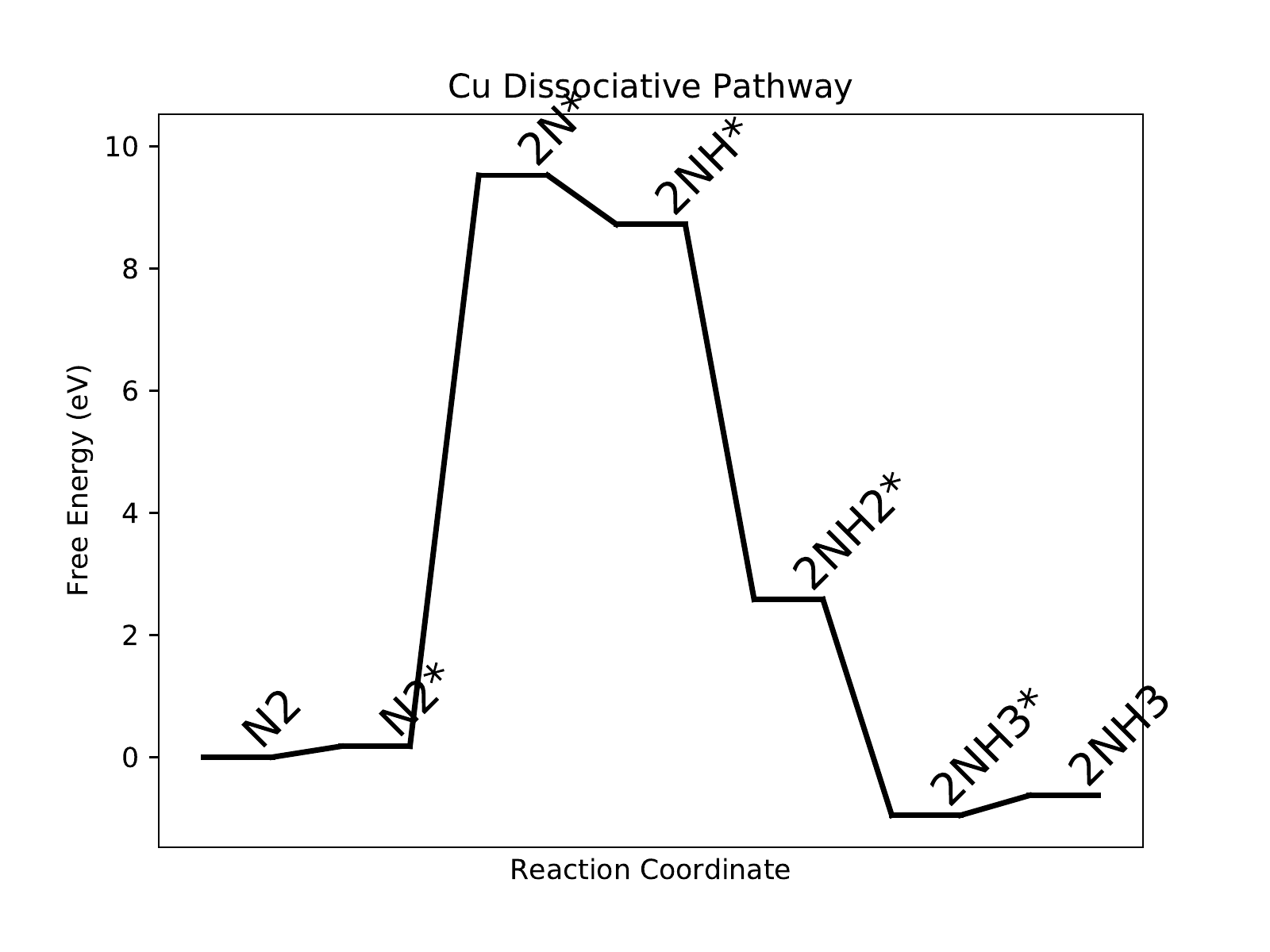}
\label{fig:Cu_dissociative}
\caption{Free energy diagram for Cu}
\end{figure}

\begin{figure}
\includegraphics[width=1\linewidth]{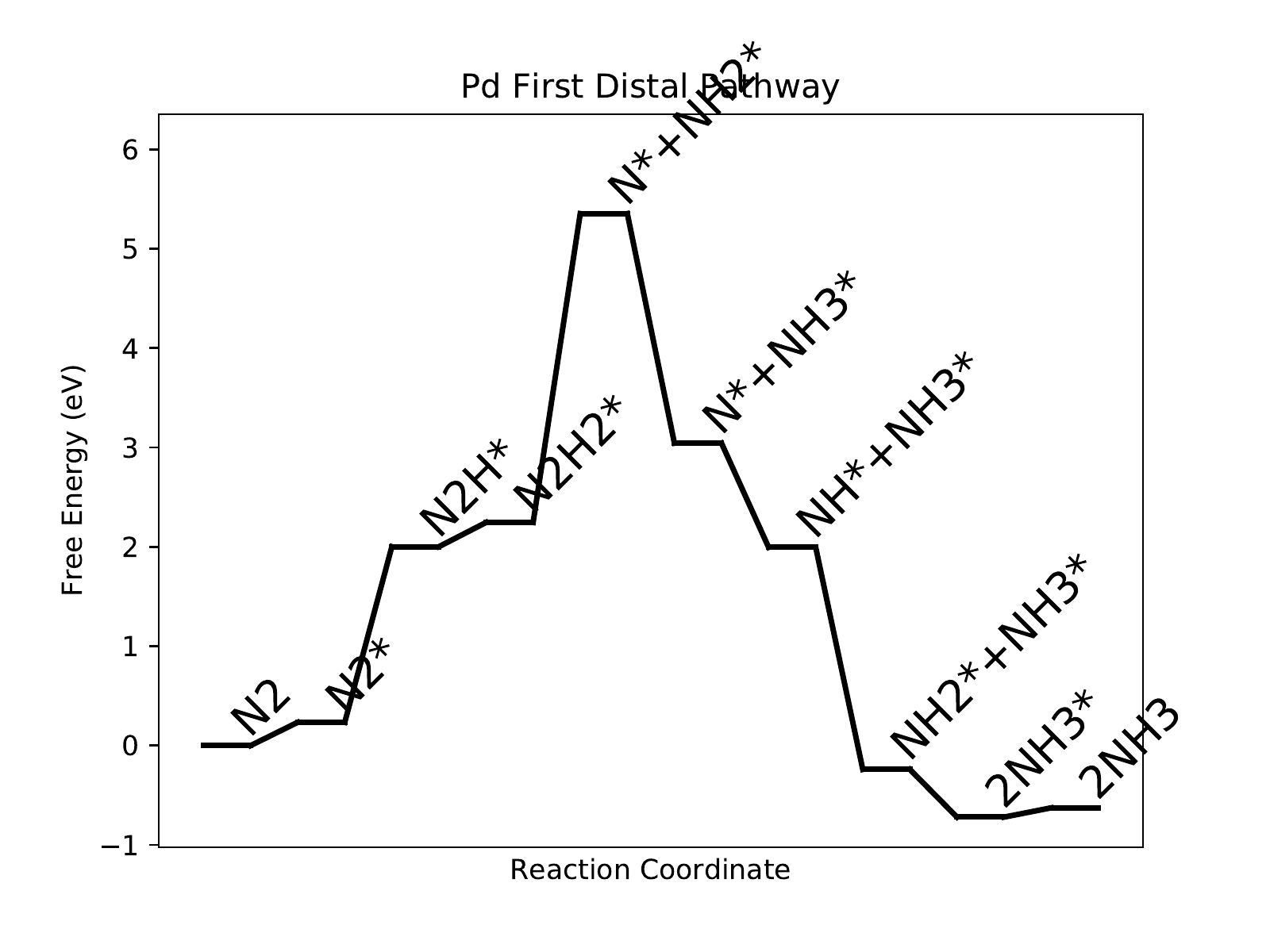}
\label{fig:Pd_distal_1}
\caption{Free energy diagram for Pd}
\end{figure}

\newpage
\begin{figure}
\includegraphics[width=1\linewidth]{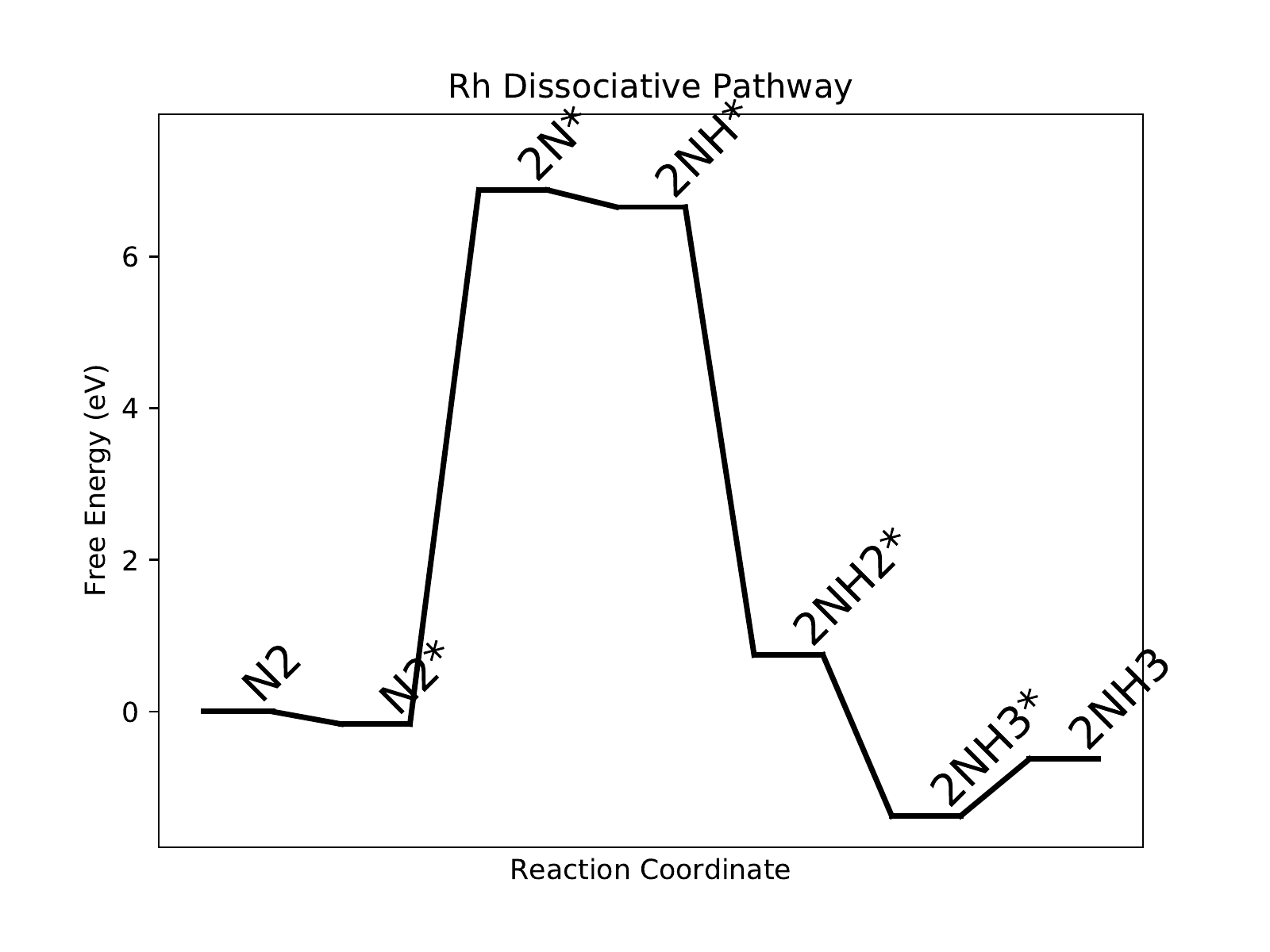}
\label{fig:Rh_dissociative}
\caption{Free energy diagram for Rh}
\end{figure}

\begin{figure}
\includegraphics[width=1\linewidth]{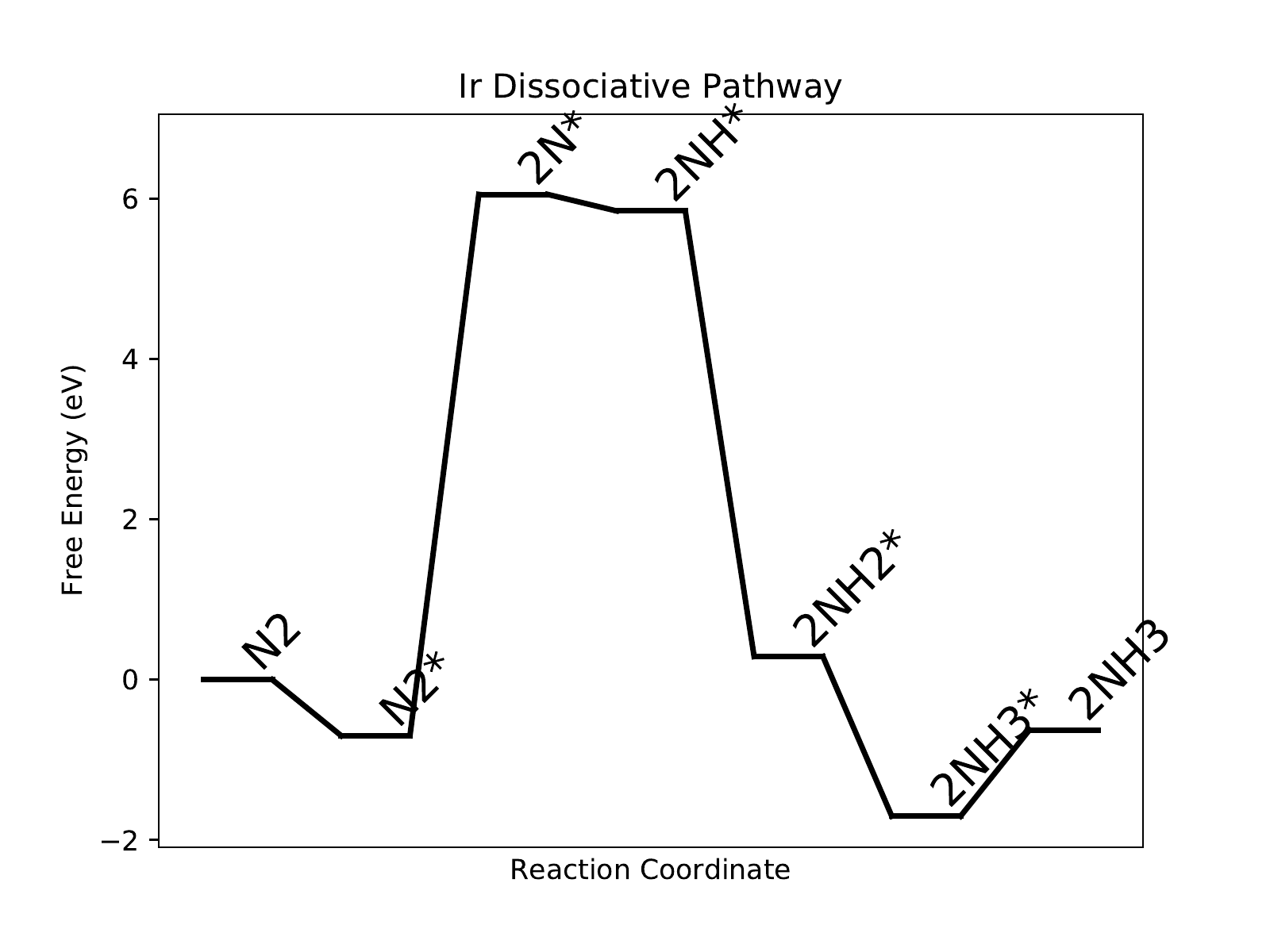}
\label{fig:Ir_dissociative}
\caption{Free energy diagram for Ir}
\end{figure}

\newpage
\begin{figure}
\includegraphics[width=1\linewidth]{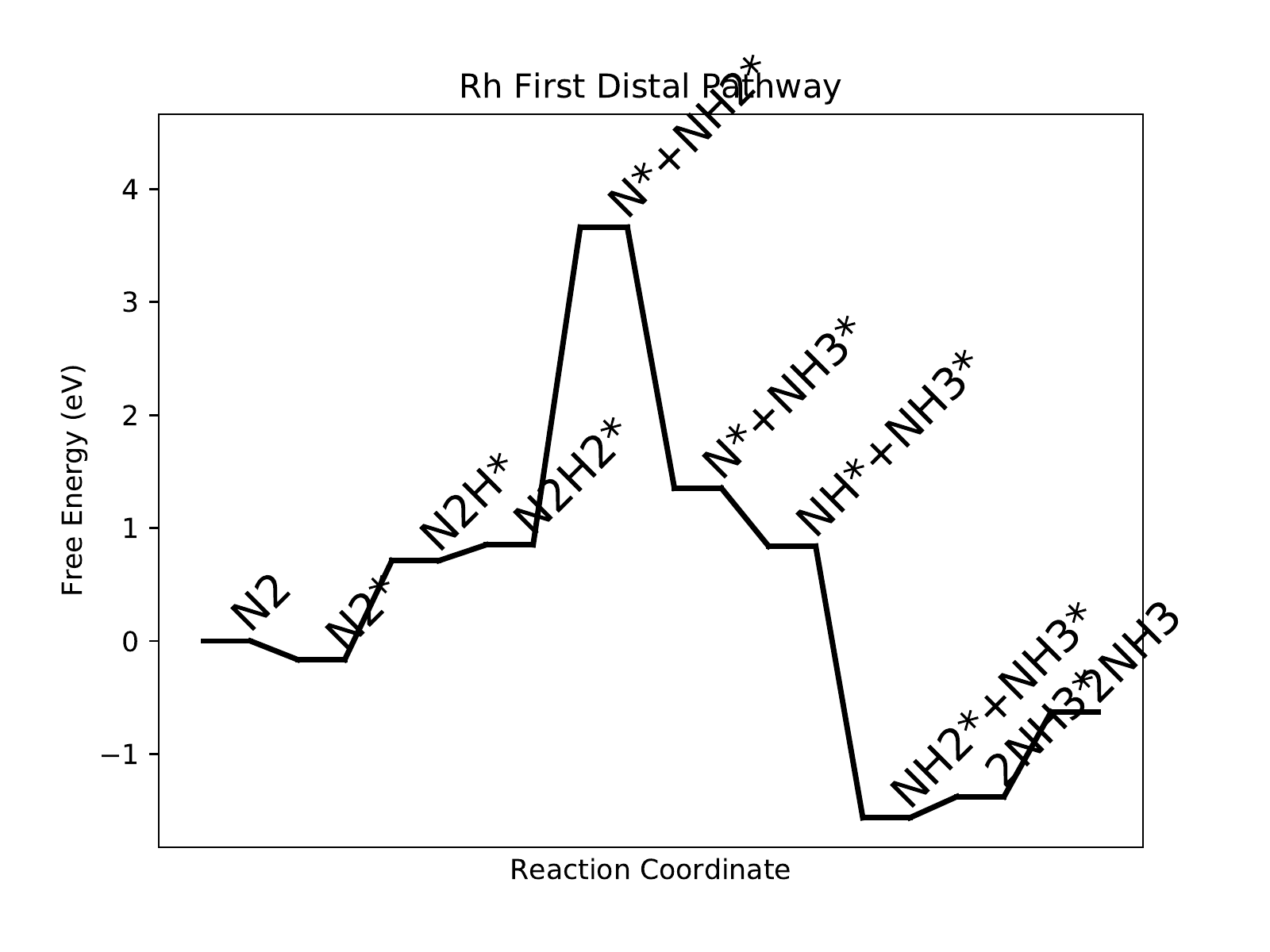}
\label{fig:Rh_distal_1}
\caption{Free energy diagram for Rh}
\end{figure}

\begin{figure}
\includegraphics[width=1\linewidth]{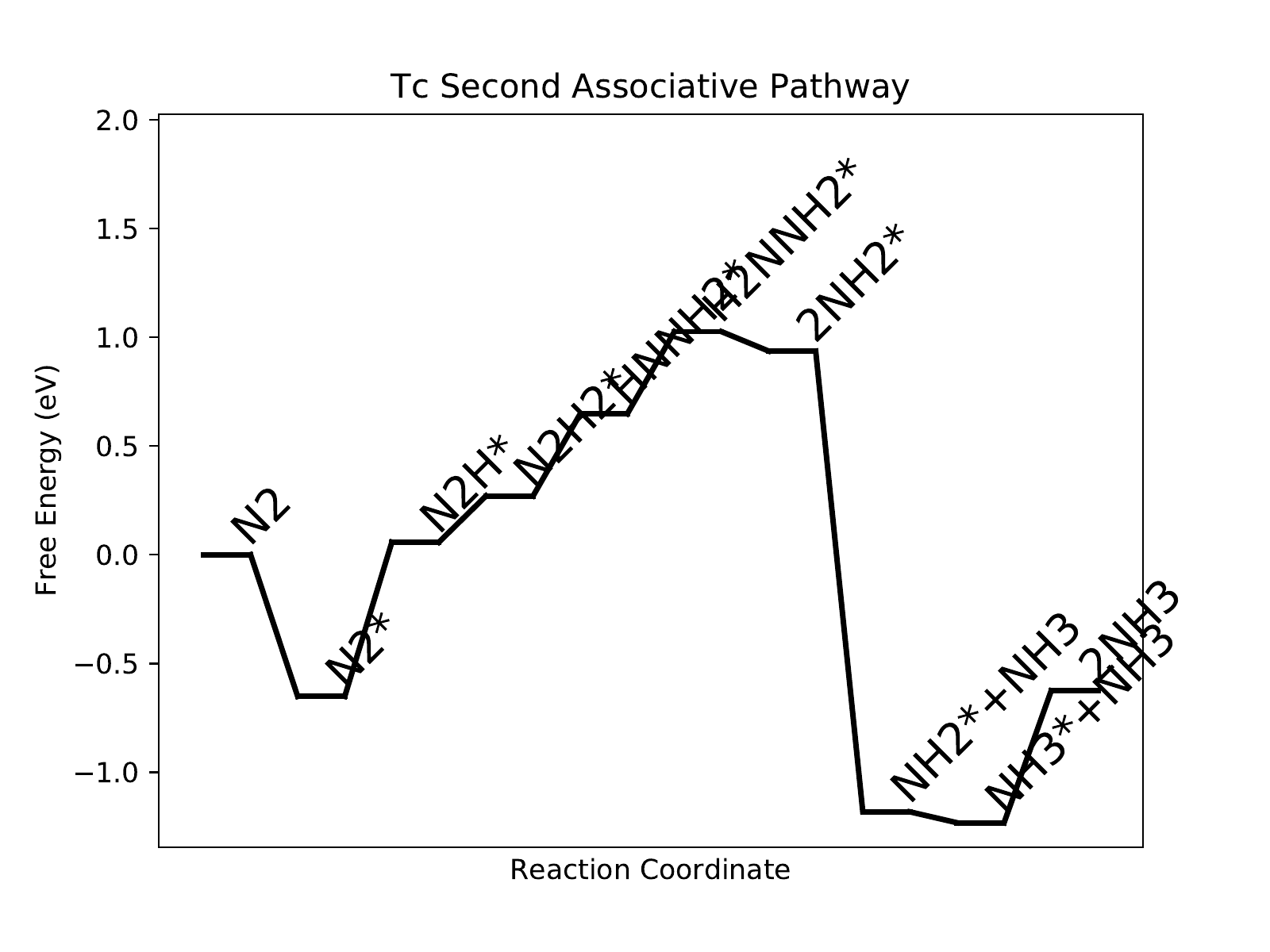}
\label{fig:Tc_associative_2}
\caption{Free energy diagram for Tc}
\end{figure}

\newpage
\begin{figure}
\includegraphics[width=1\linewidth]{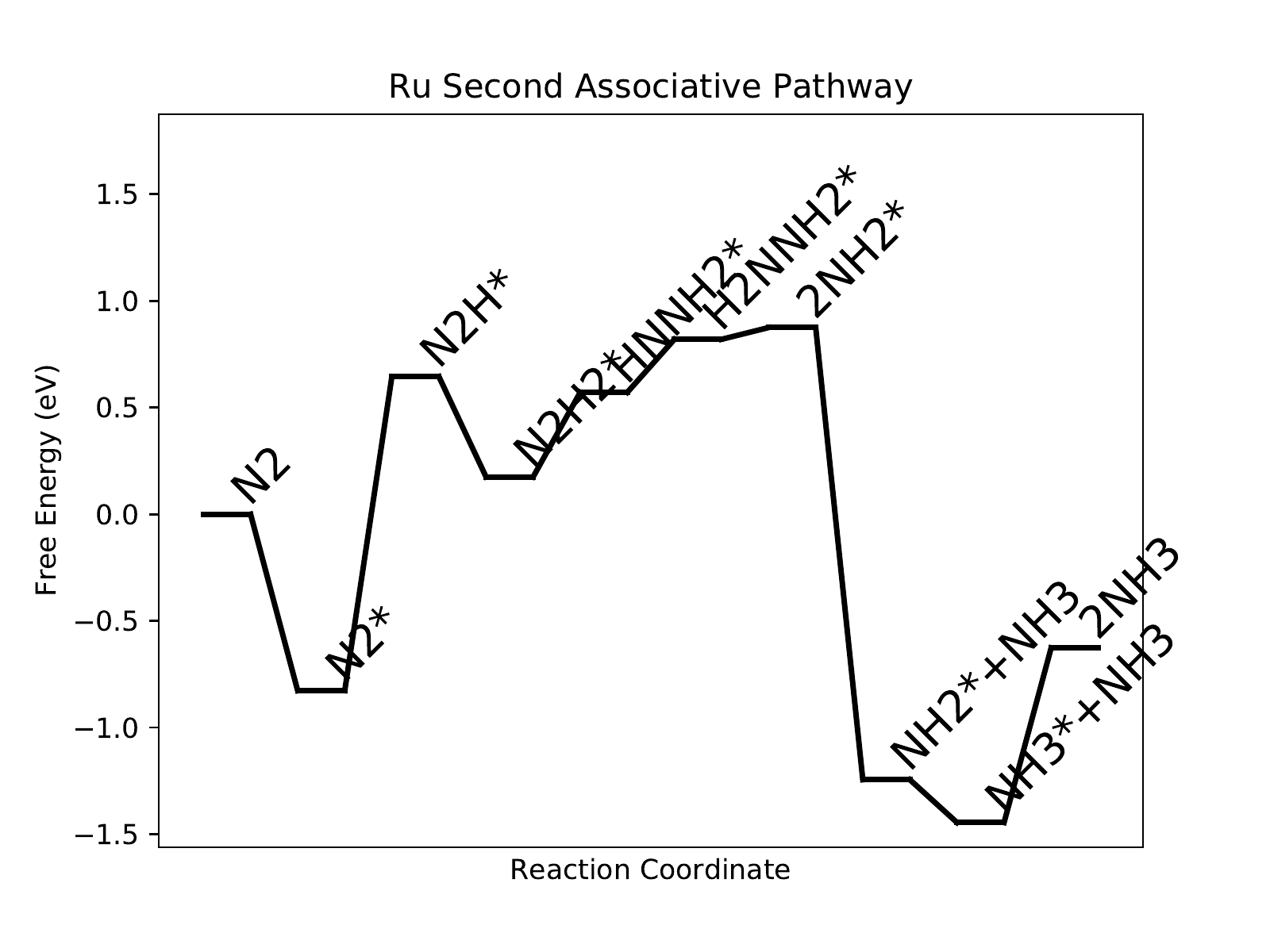}
\label{fig:Ru_associative_2}
\caption{Free energy diagram for Ru}
\end{figure}

\begin{figure}
\includegraphics[width=1\linewidth]{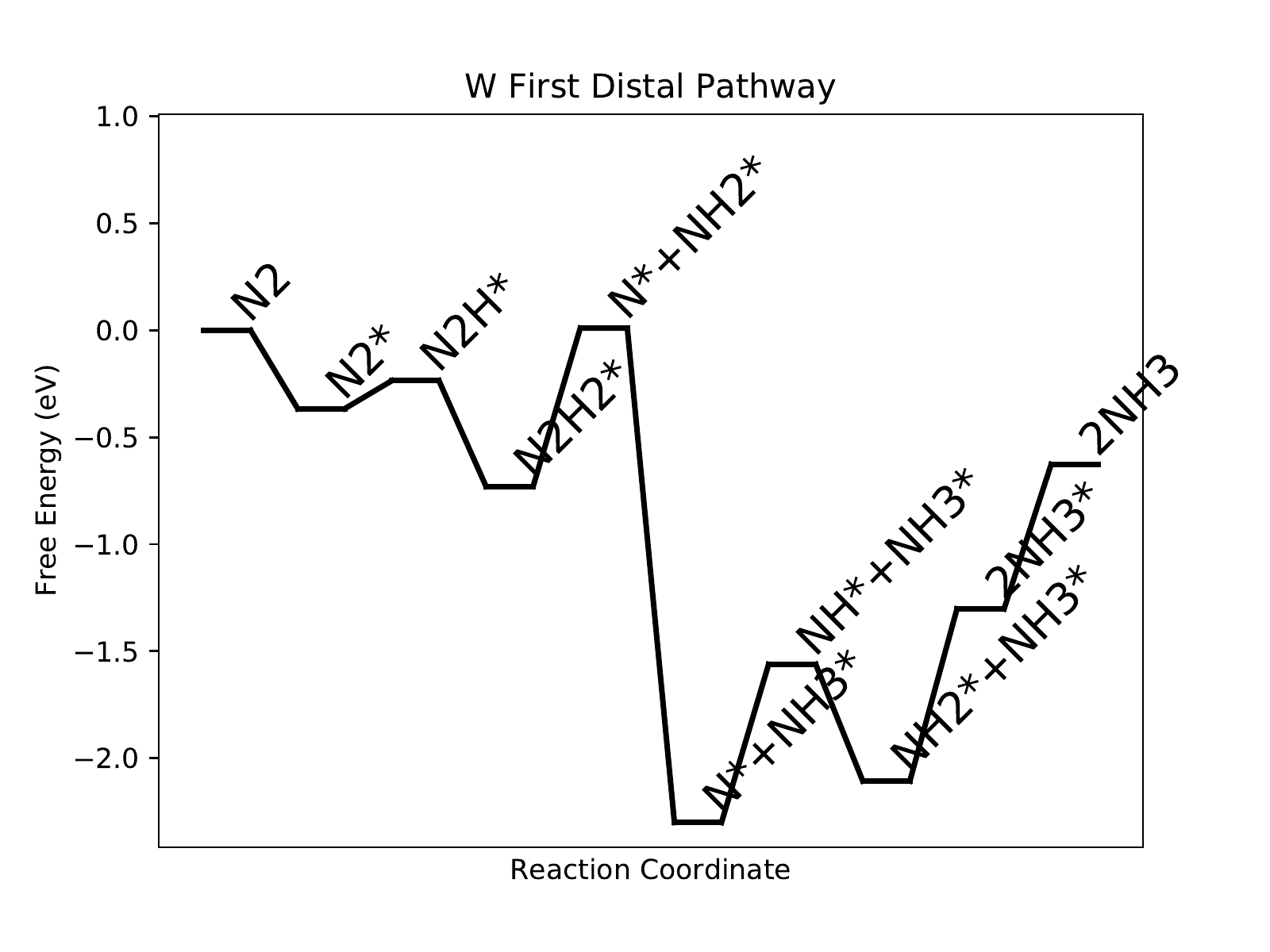}
\label{fig:W_distal_1}
\caption{Free energy diagram for W}
\end{figure}

\newpage
\begin{figure}
\includegraphics[width=1\linewidth]{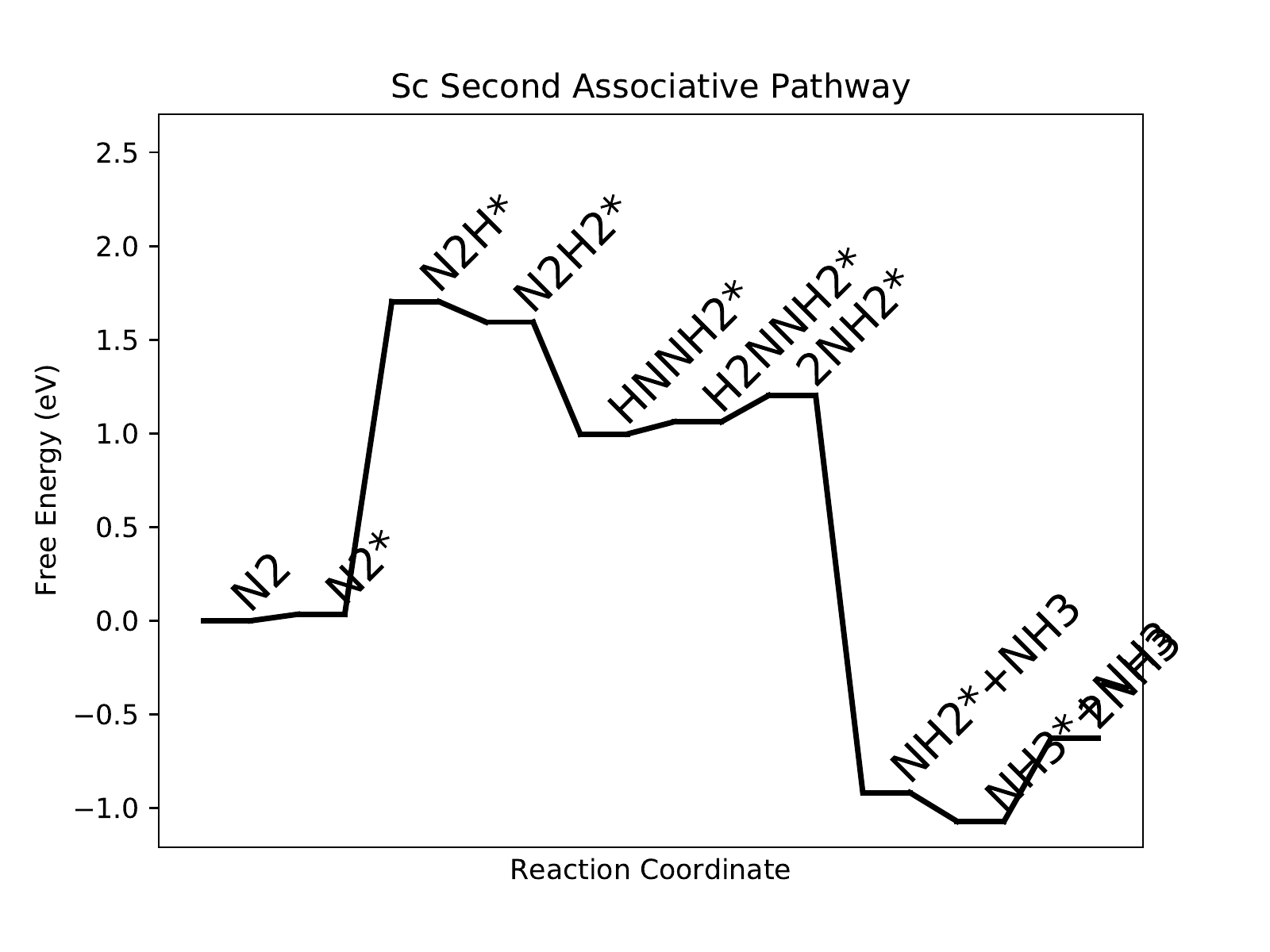}
\label{fig:Sc_associative_2}
\caption{Free energy diagram for Sc}
\end{figure}

\begin{figure}
\includegraphics[width=1\linewidth]{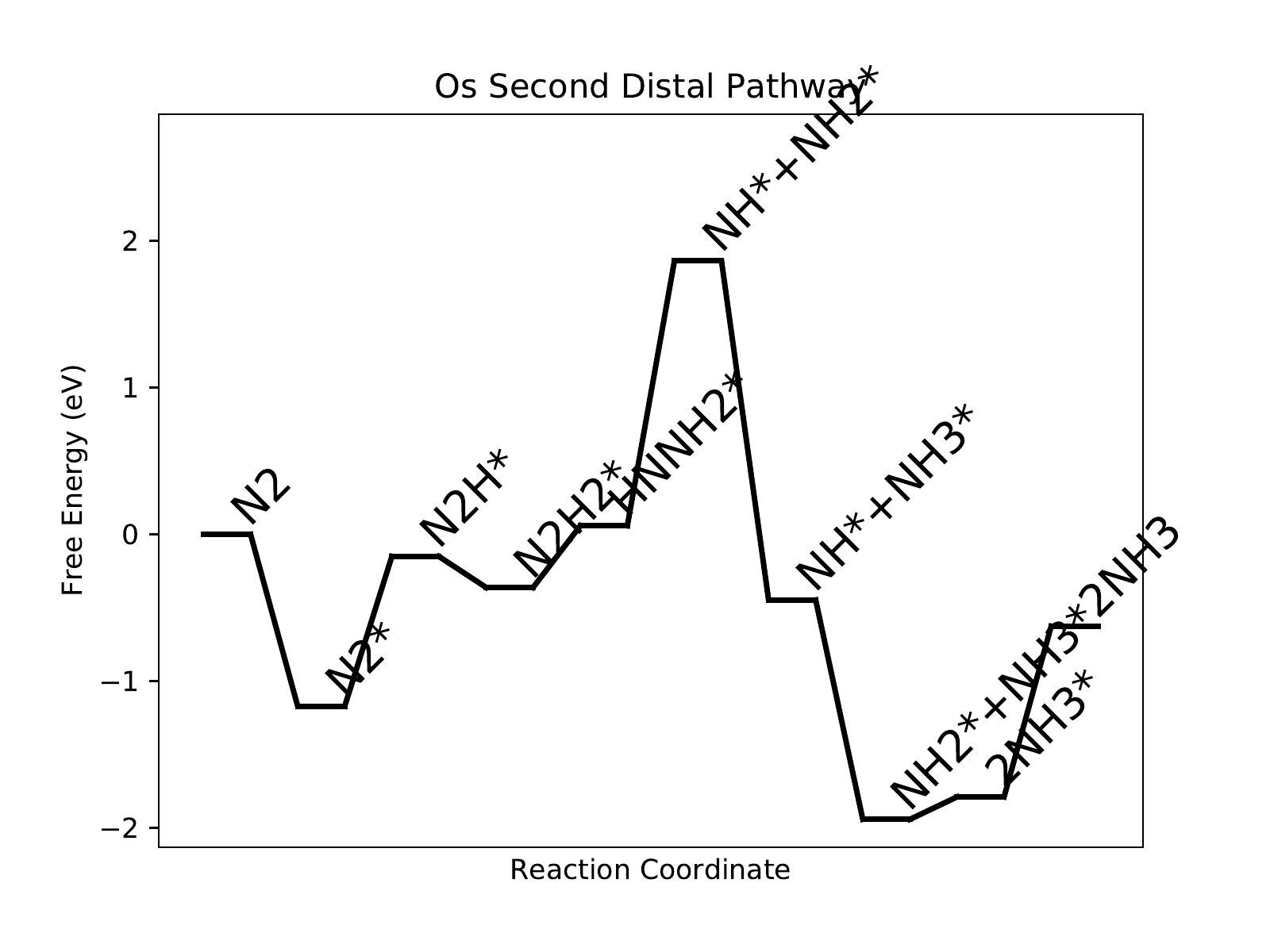}
\label{fig:Os_distal_2}
\caption{Free energy diagram for Os}
\end{figure}

\newpage
\begin{figure}
\includegraphics[width=1\linewidth]{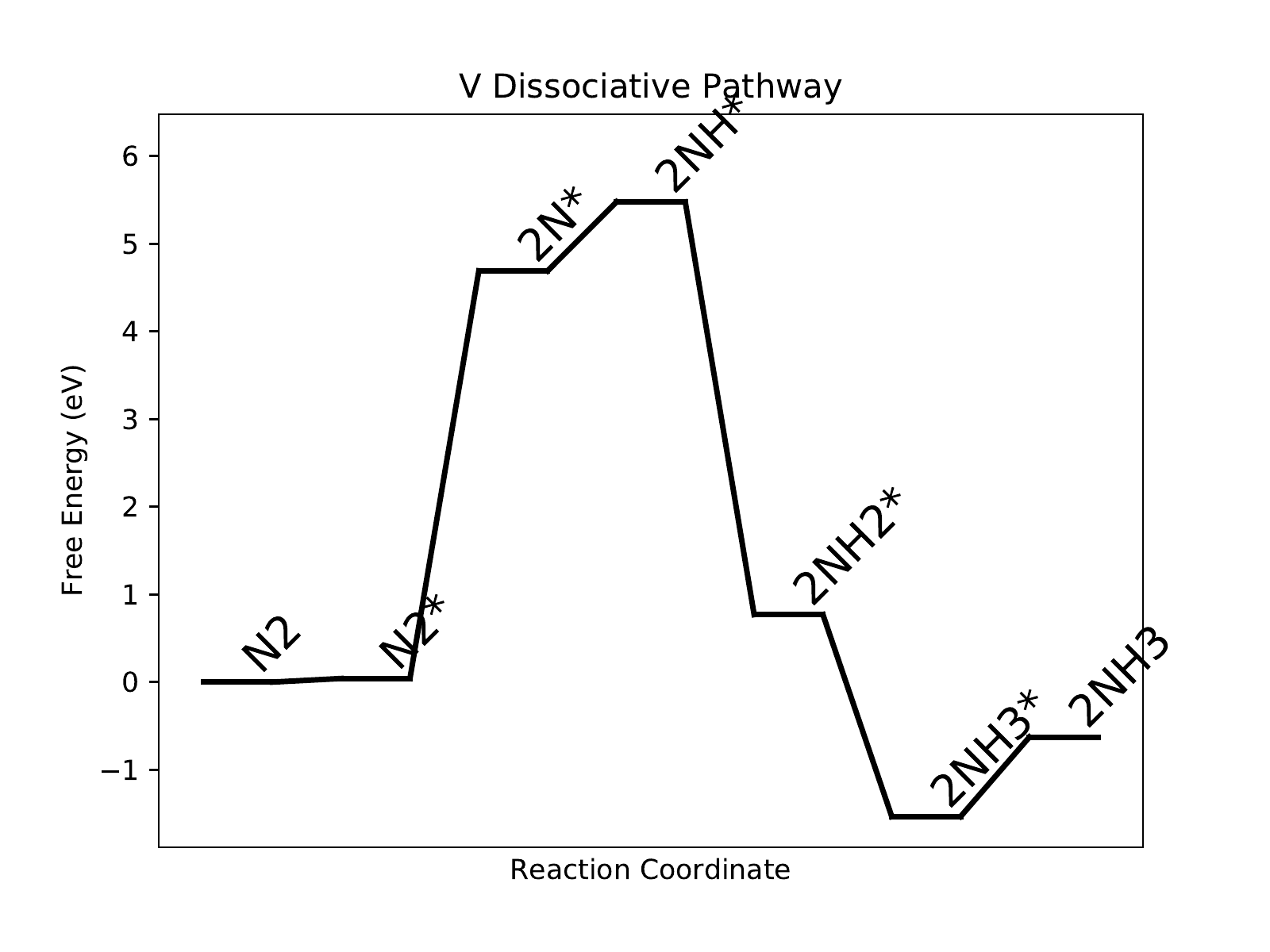}
\label{fig:V_dissociative}
\caption{Free energy diagram for V}
\end{figure}

\begin{figure}
\includegraphics[width=1\linewidth]{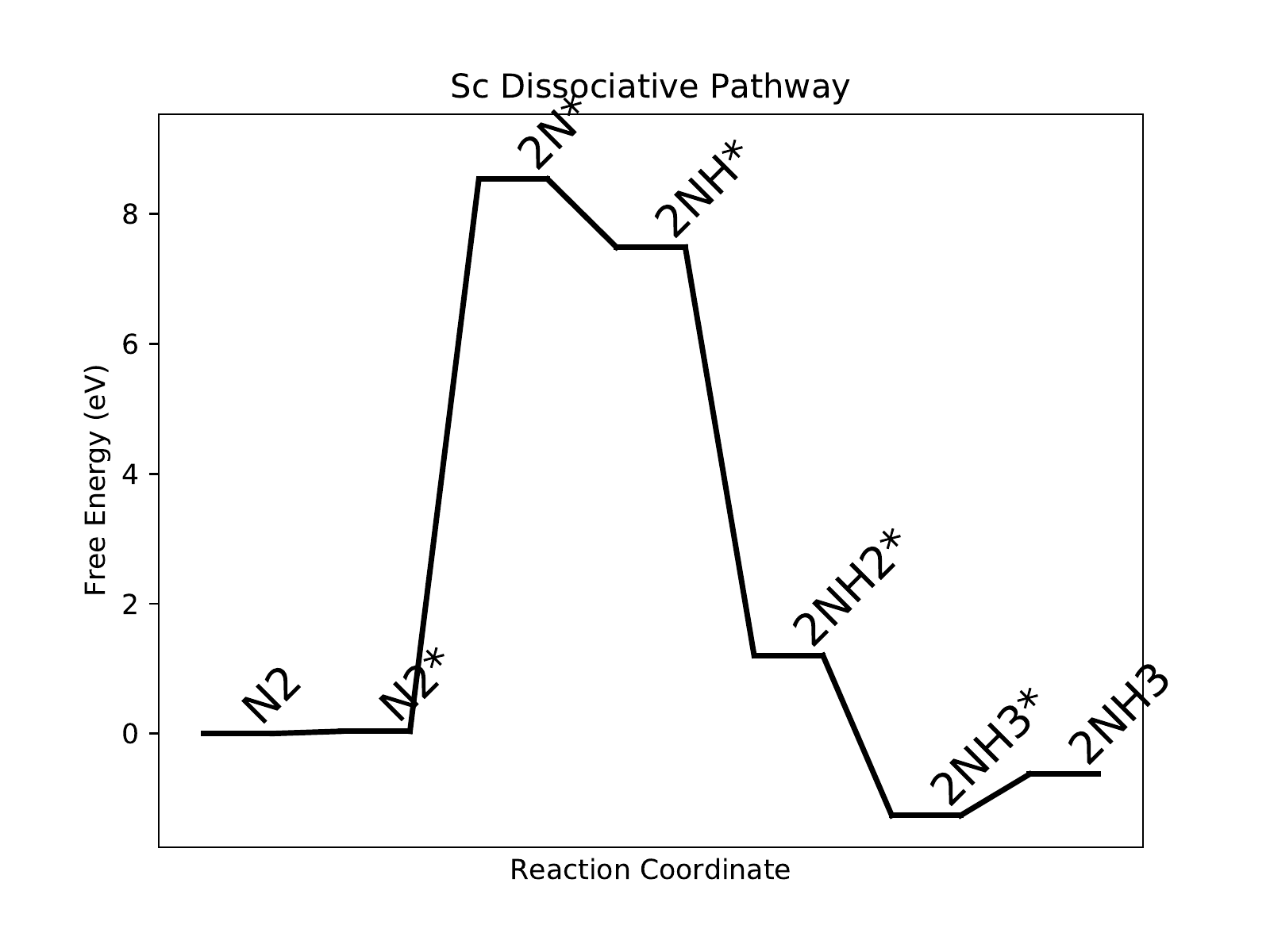}
\label{fig:Sc_dissociative}
\caption{Free energy diagram for Sc}
\end{figure}

\newpage
\begin{figure}
\includegraphics[width=1\linewidth]{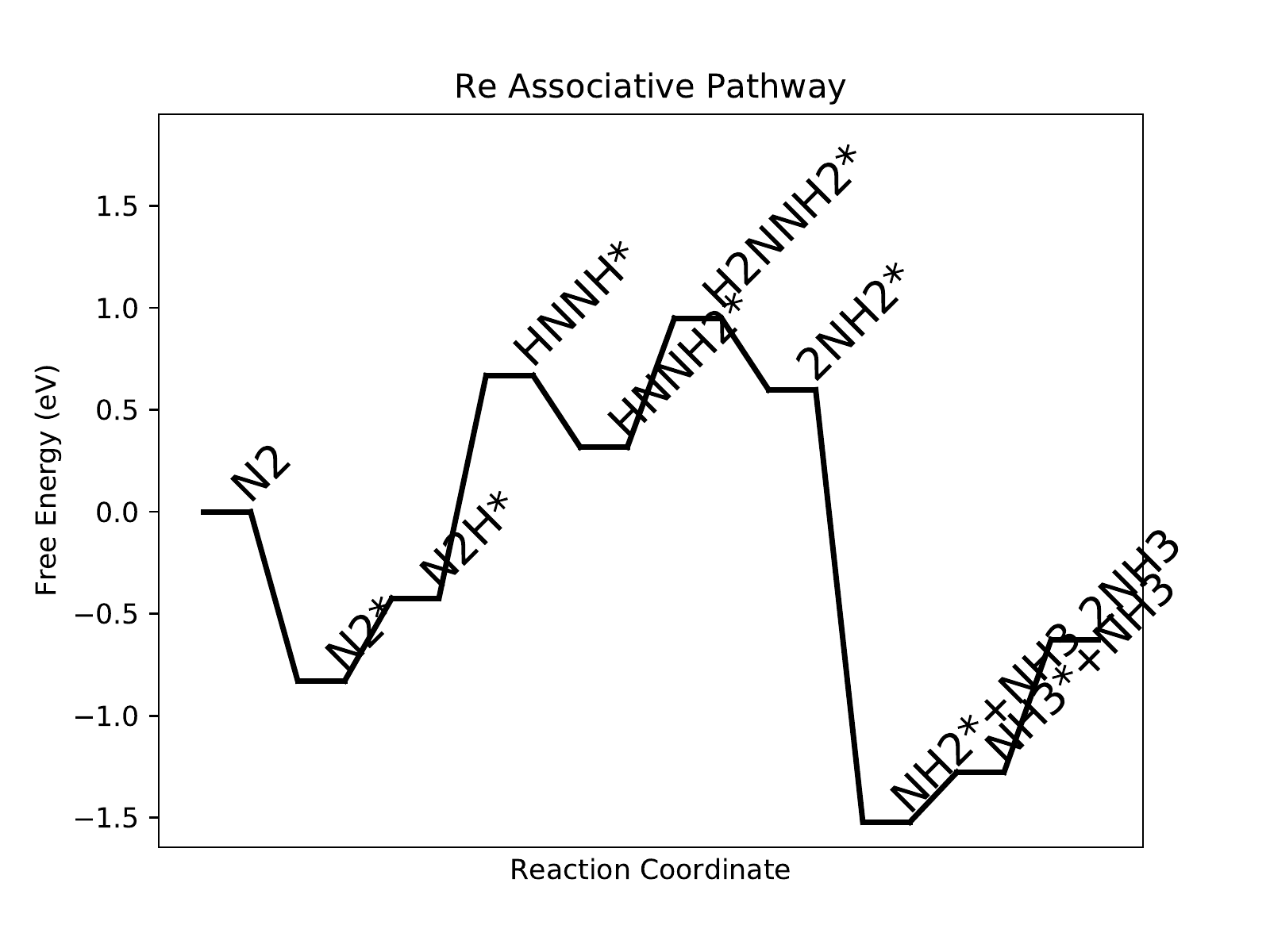}
\label{fig:Re_associative}
\caption{Free energy diagram for Re}
\end{figure}

\begin{figure}
\includegraphics[width=1\linewidth]{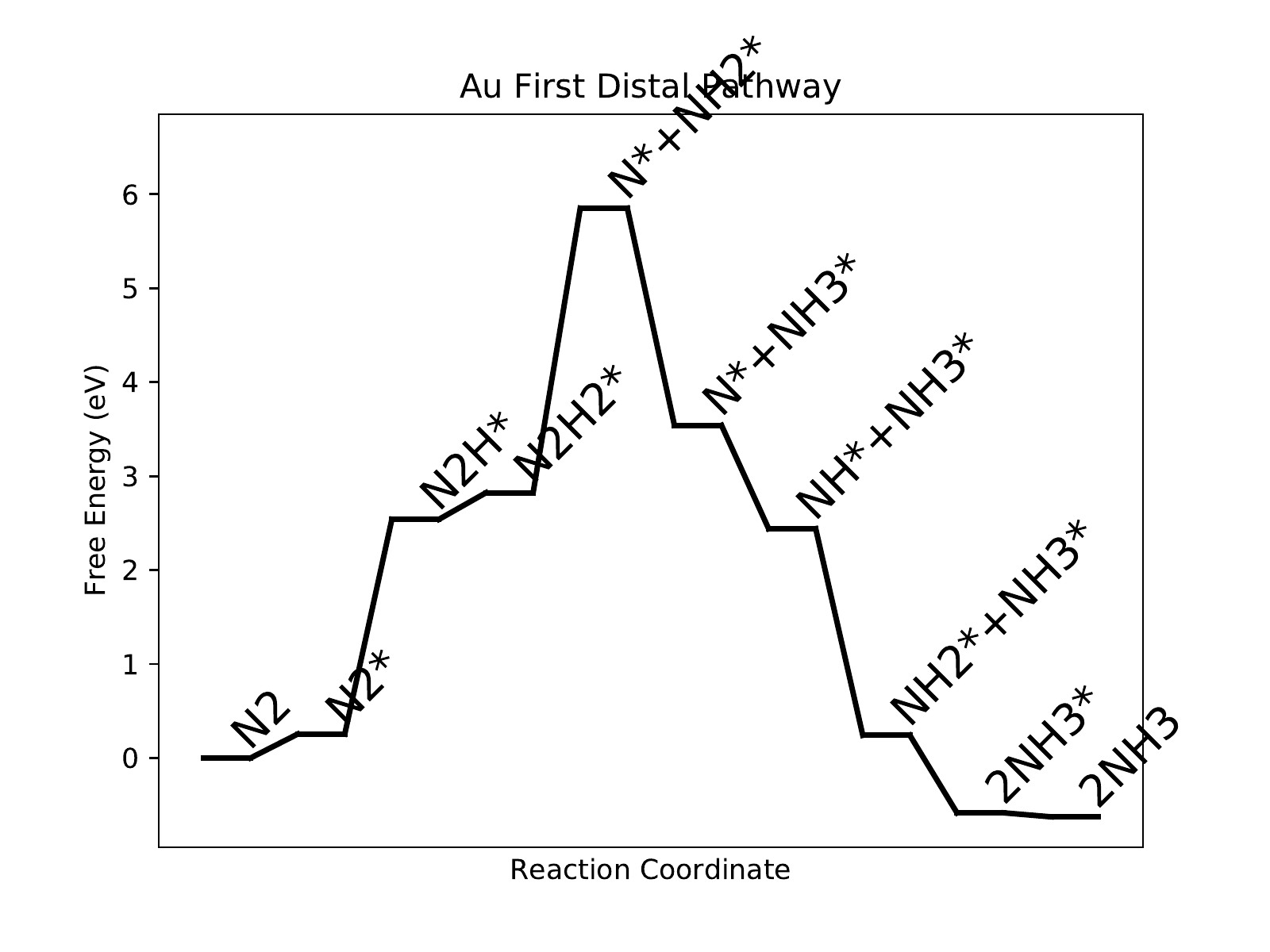}
\label{fig:Au_distal_1}
\caption{Free energy diagram for Au}
\end{figure}

\newpage
\begin{figure}
\includegraphics[width=1\linewidth]{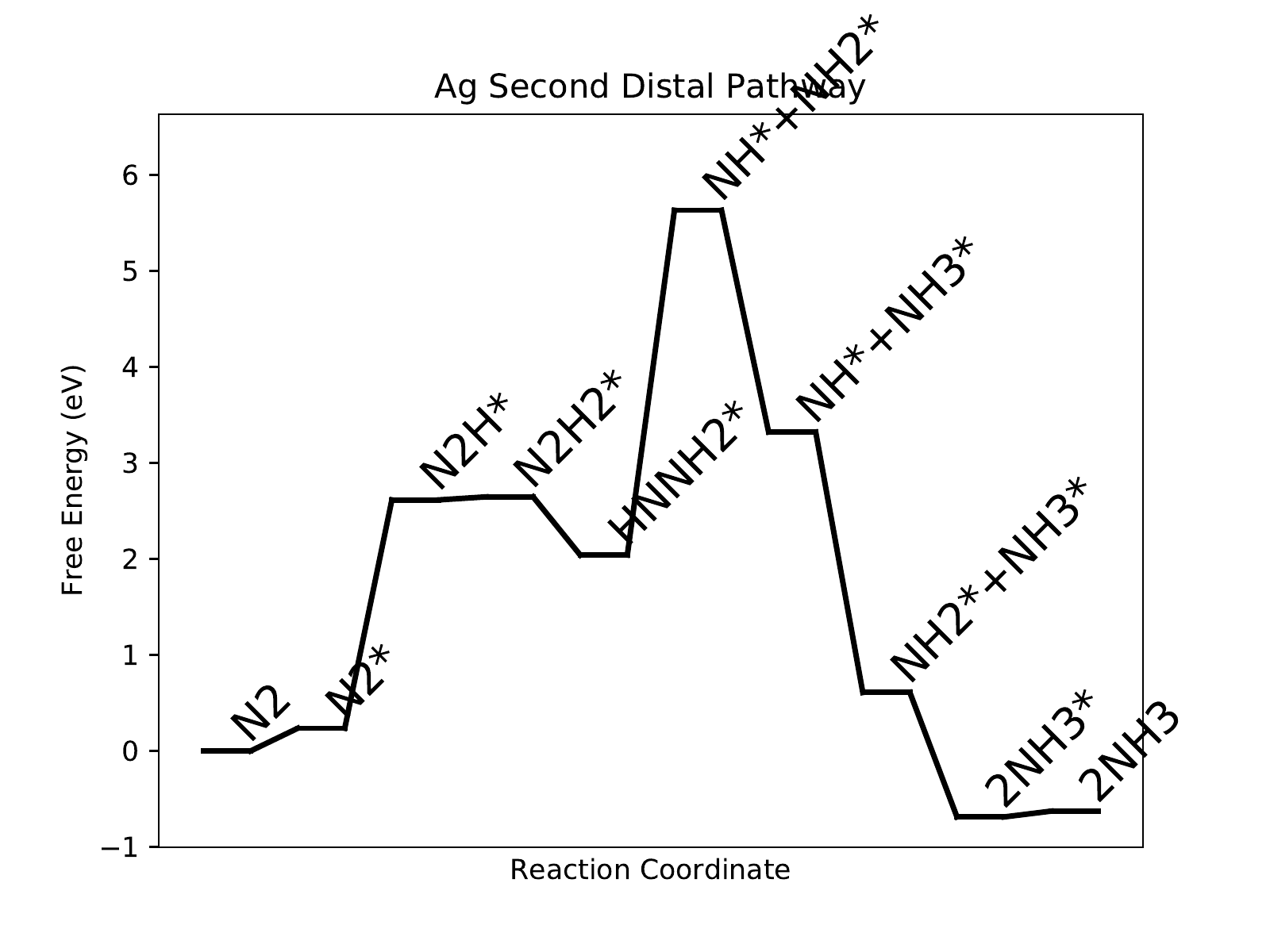}
\label{fig:Ag_distal_2}
\caption{Free energy diagram for Ag}
\end{figure}

\begin{figure}
\includegraphics[width=1\linewidth]{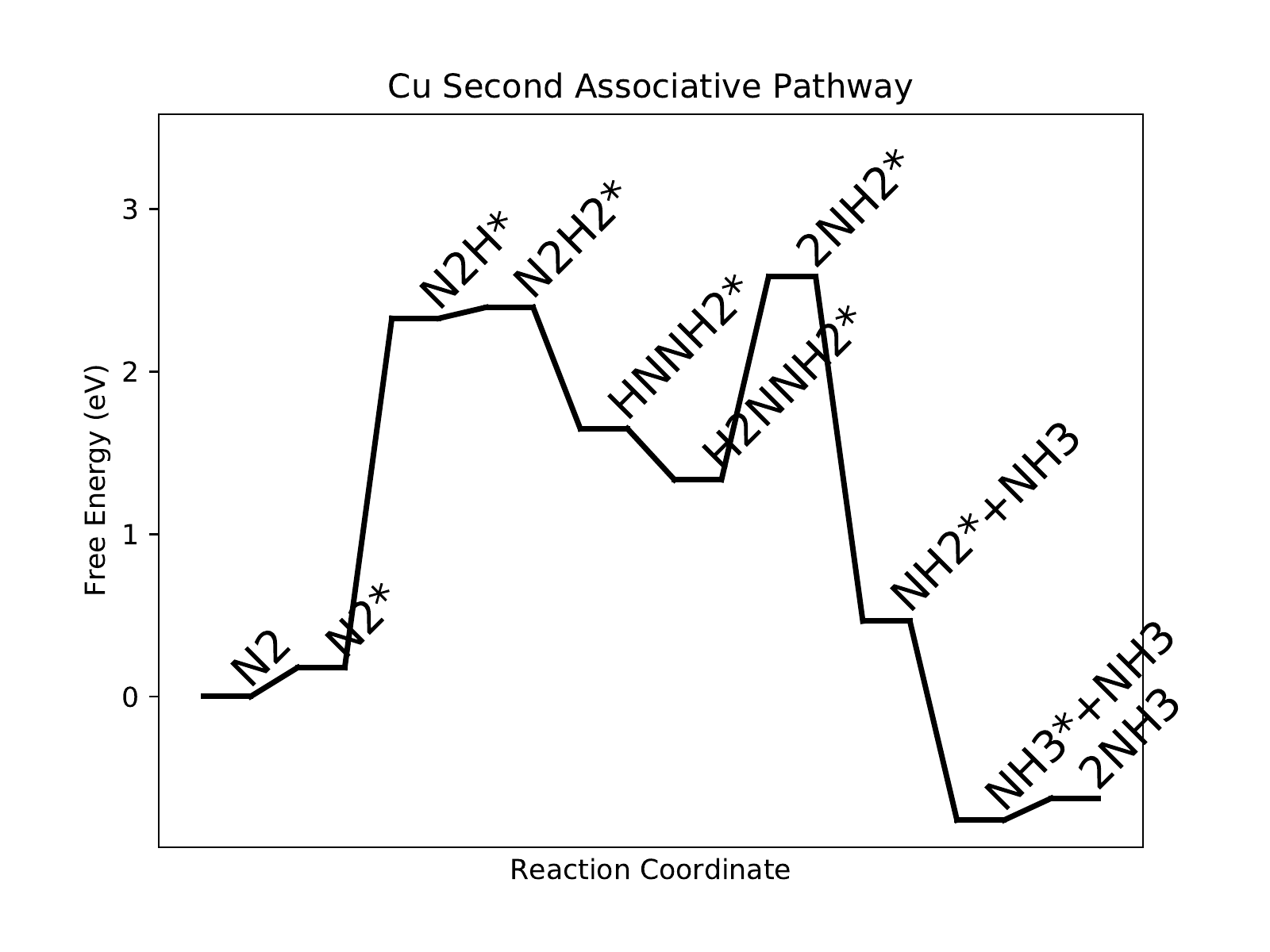}
\label{fig:Cu_associative_2}
\caption{Free energy diagram for Cu}
\end{figure}

\newpage
\begin{figure}
\includegraphics[width=1\linewidth]{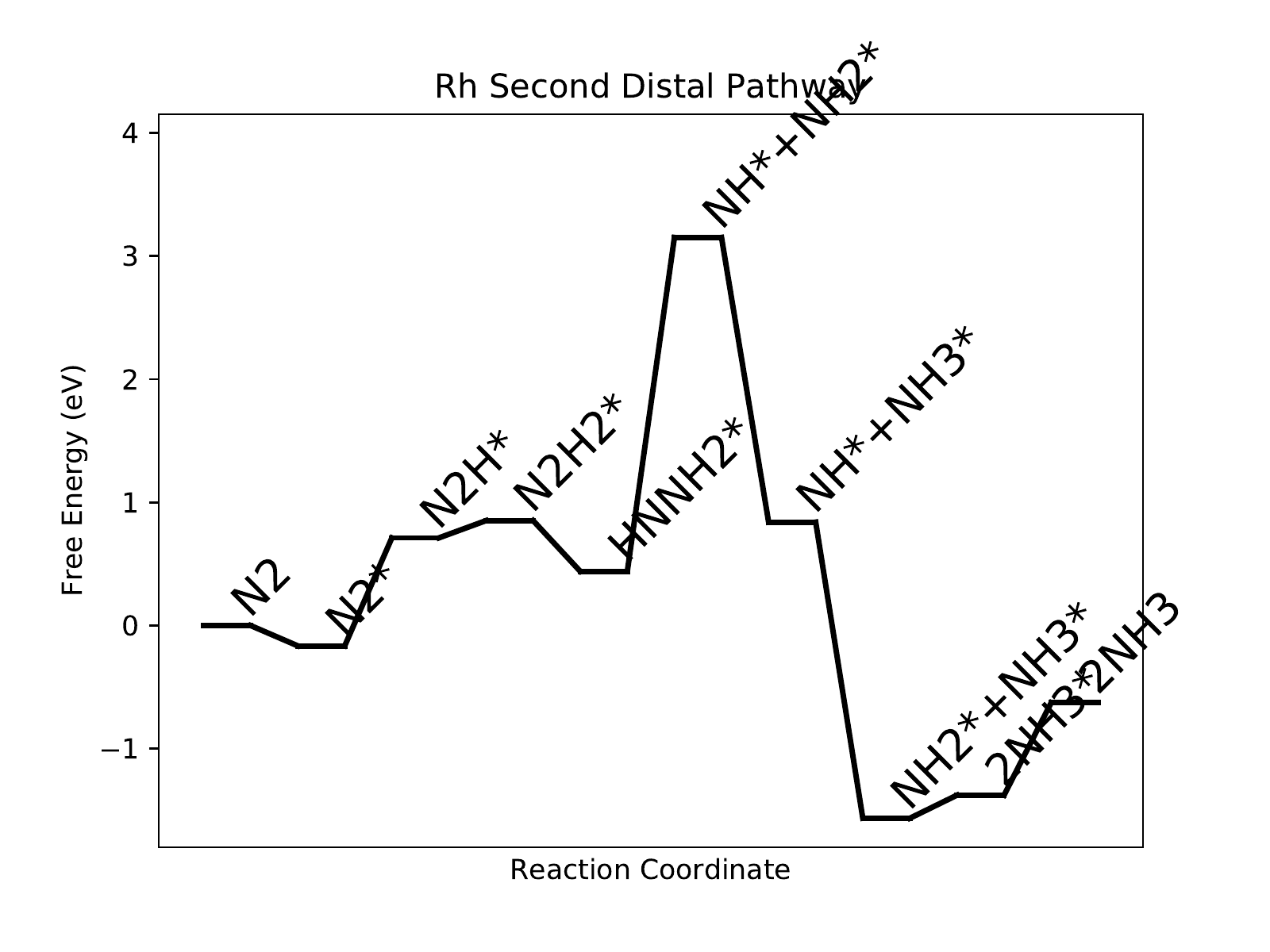}
\label{fig:Rh_distal_2}
\caption{Free energy diagram for Rh}
\end{figure}

\begin{figure}
\includegraphics[width=1\linewidth]{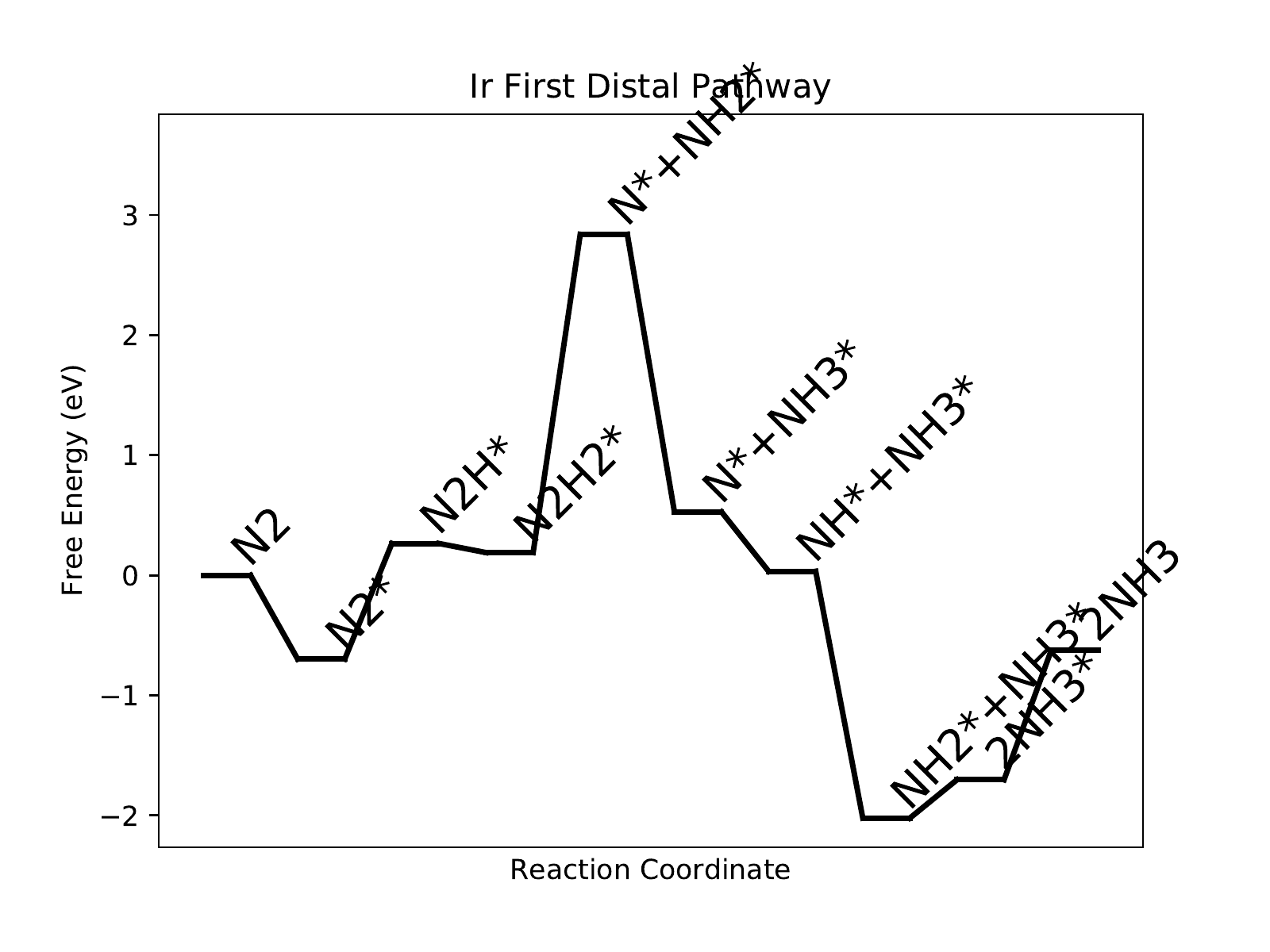}
\label{fig:Ir_distal_1}
\caption{Free energy diagram for Ir}
\end{figure}

\newpage
\begin{figure}
\includegraphics[width=1\linewidth]{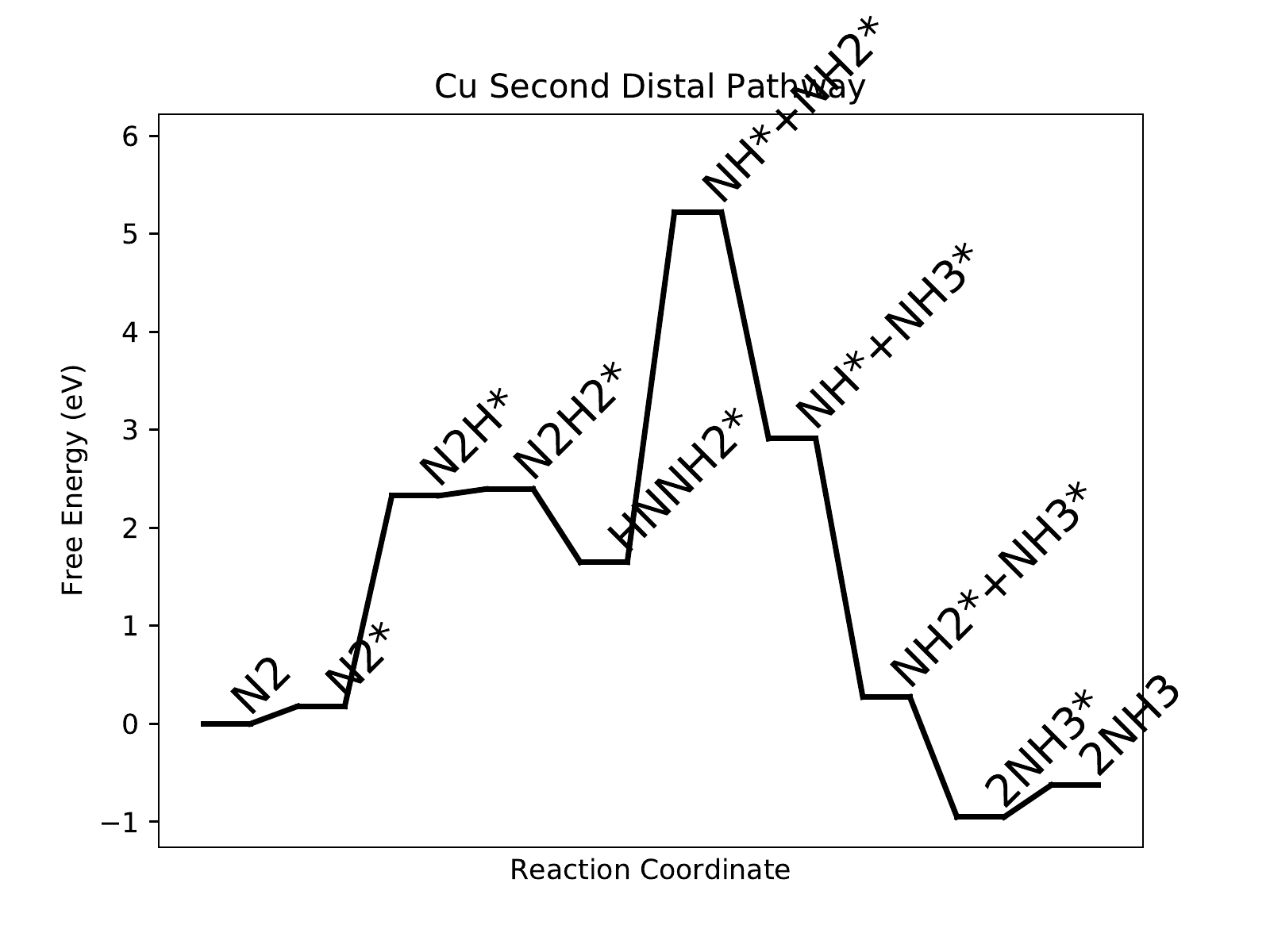}
\label{fig:Cu_distal_2}
\caption{Free energy diagram for Cu}
\end{figure}

\begin{figure}
\includegraphics[width=1\linewidth]{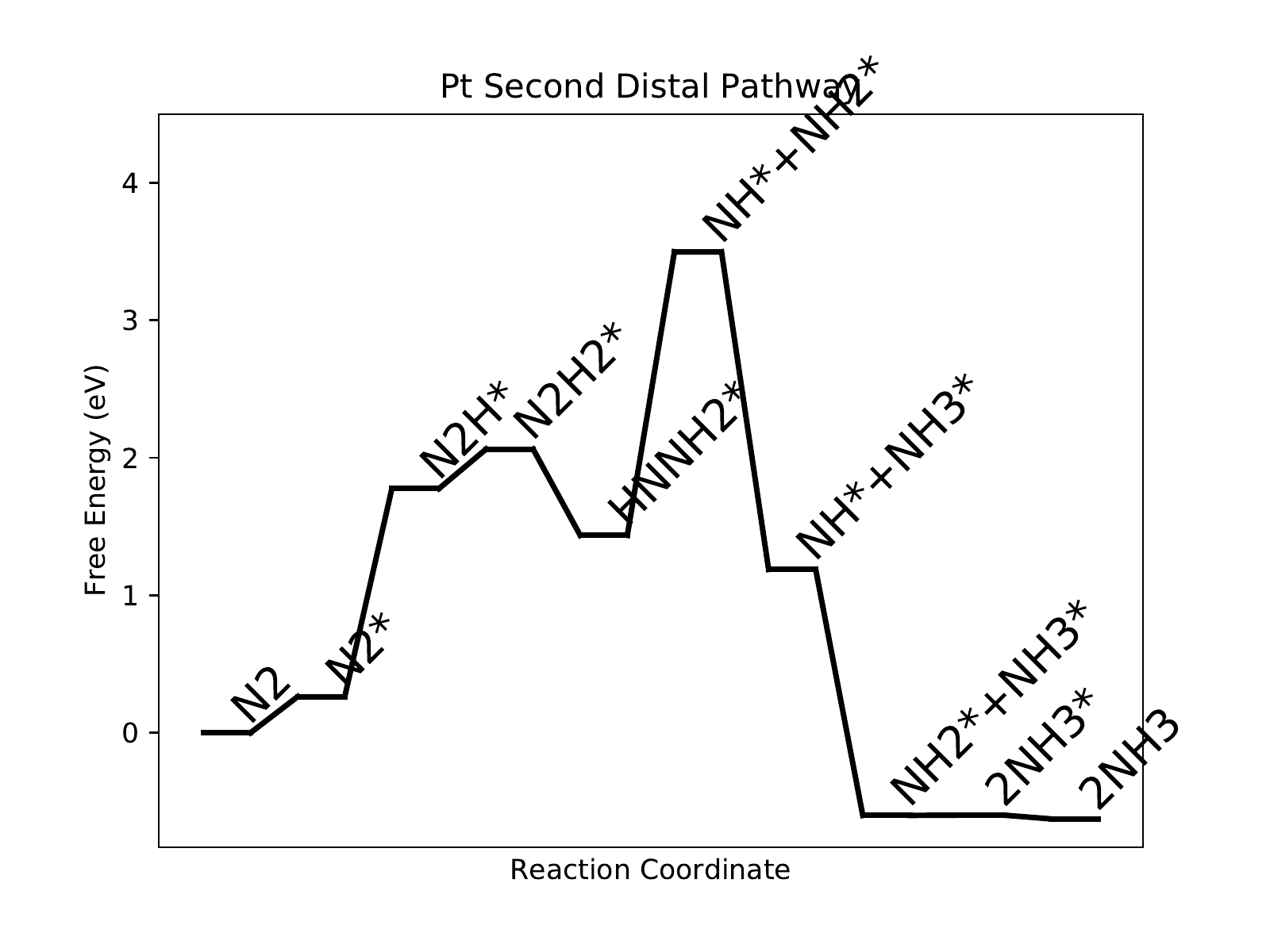}
\label{fig:Pt_distal_2}
\caption{Free energy diagram for Pt}
\end{figure}

\newpage
\begin{figure}
\includegraphics[width=1\linewidth]{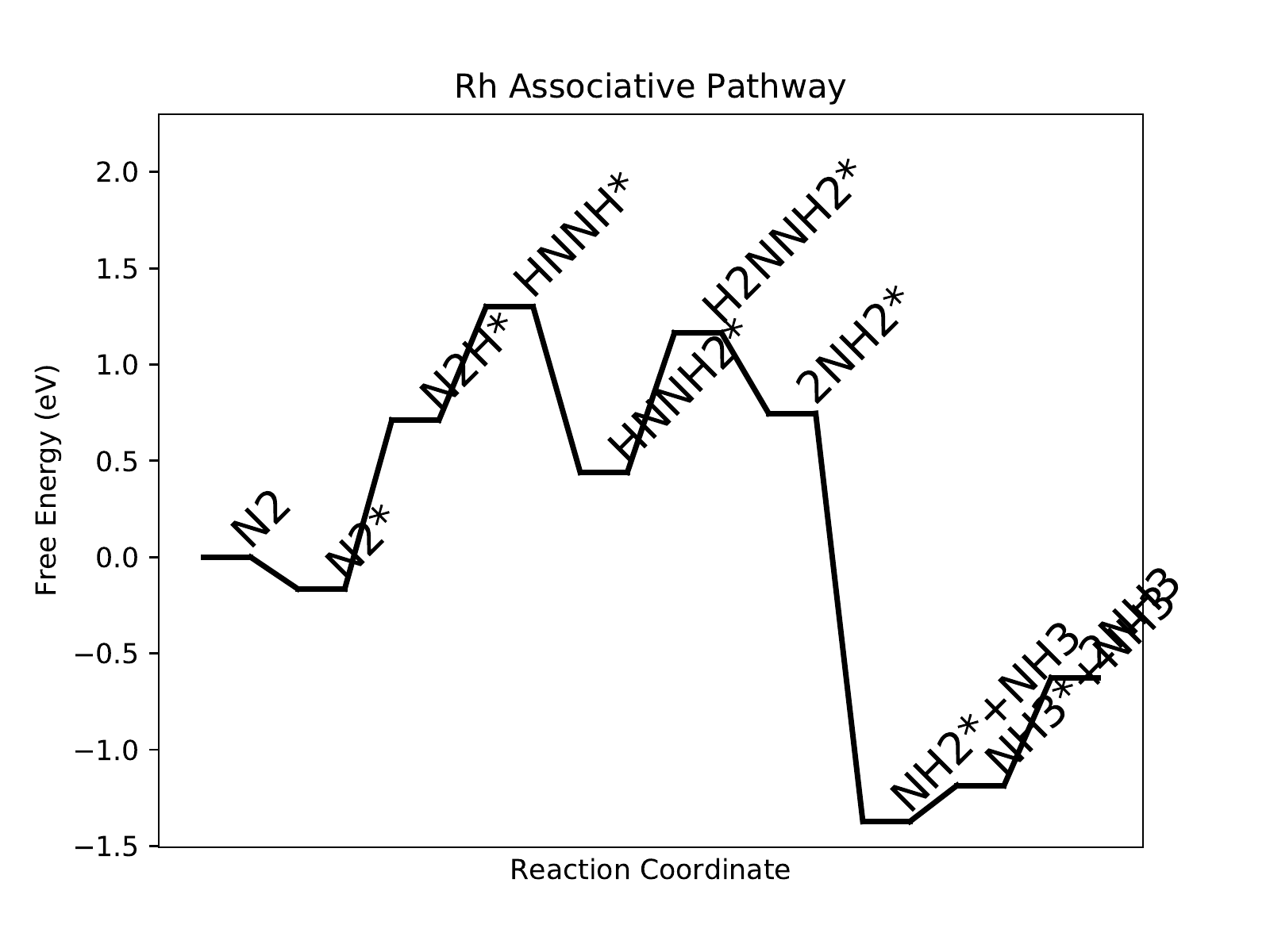}
\label{fig:Rh_associative}
\caption{Free energy diagram for Rh}
\end{figure}

\begin{figure}
\includegraphics[width=1\linewidth]{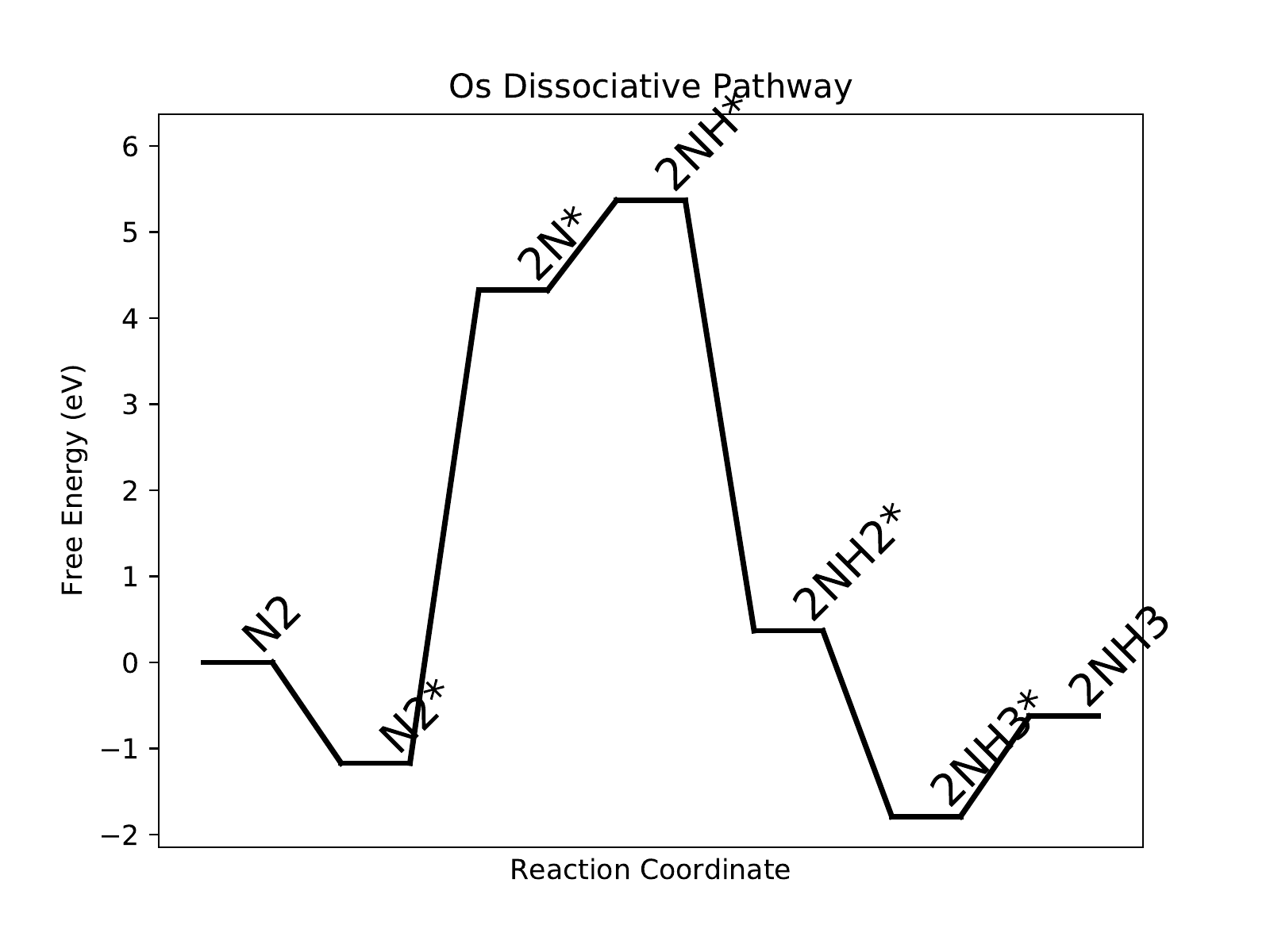}
\label{fig:Os_dissociative}
\caption{Free energy diagram for Os}
\end{figure}

\newpage
\begin{figure}
\includegraphics[width=1\linewidth]{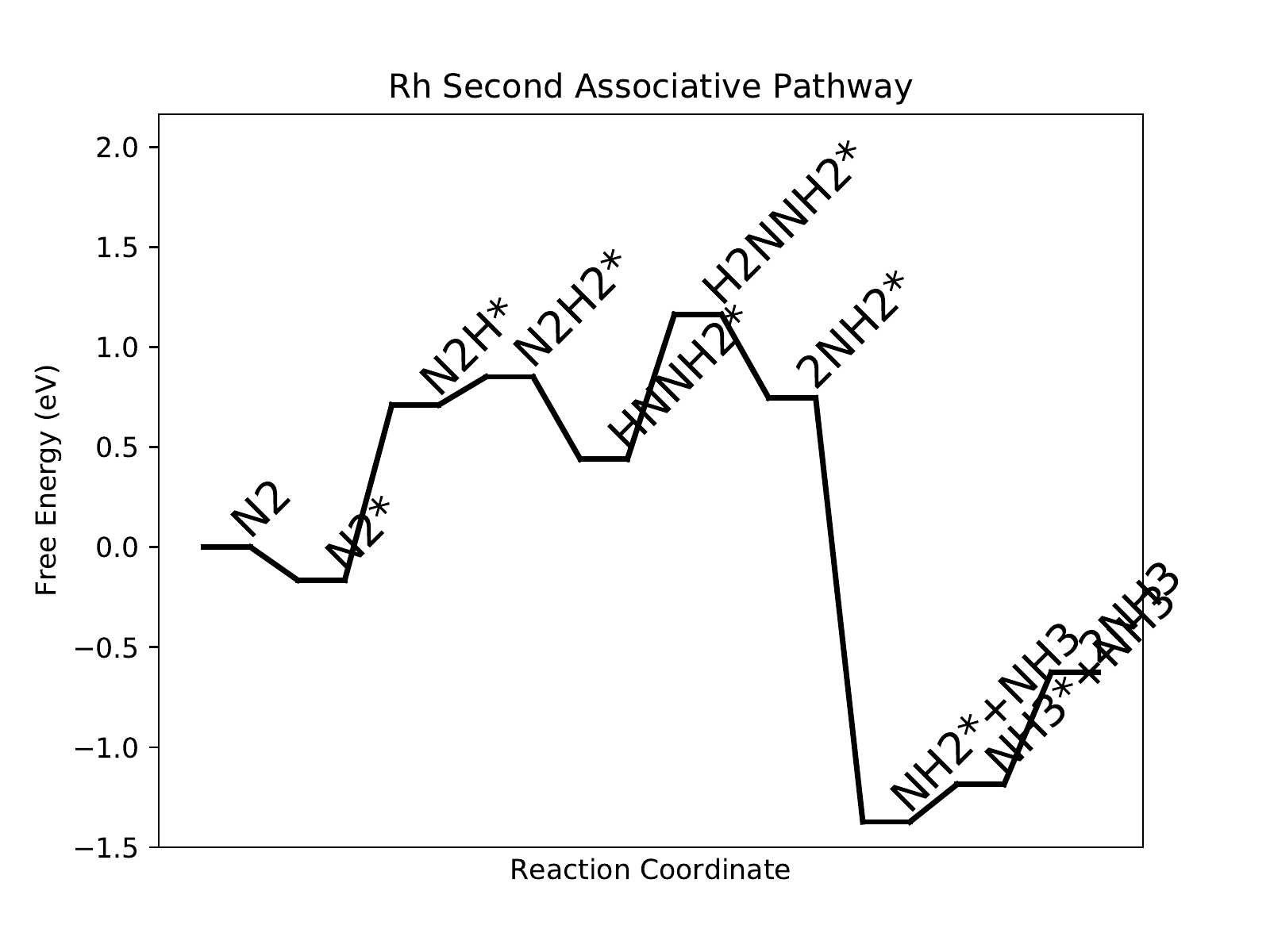}
\label{fig:Rh_associative_2}
\caption{Free energy diagram for Rh}
\end{figure}

\begin{figure}
\includegraphics[width=1\linewidth]{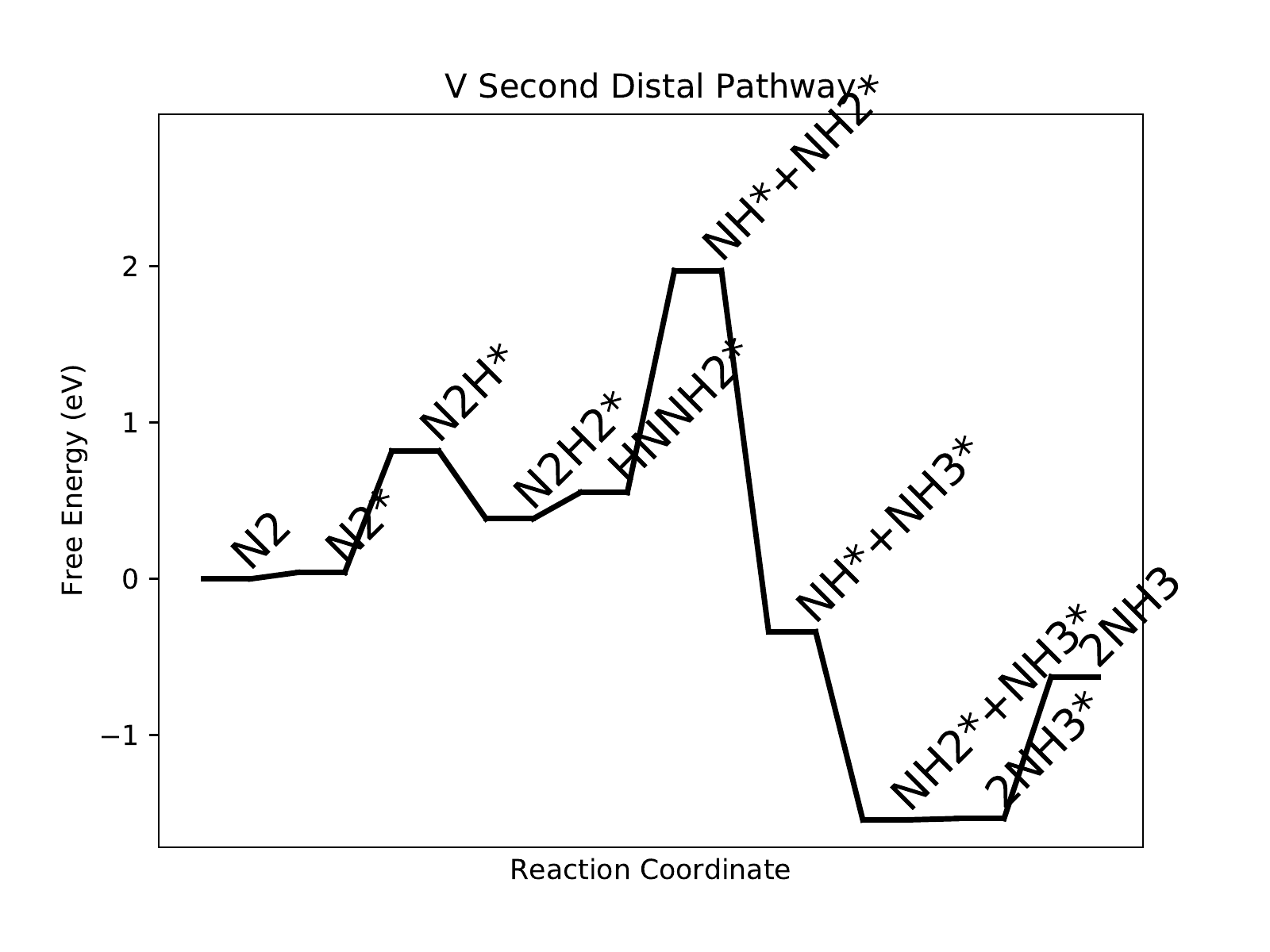}
\label{fig:V_distal_2}
\caption{Free energy diagram for V}
\end{figure}

\newpage
\begin{figure}
\includegraphics[width=1\linewidth]{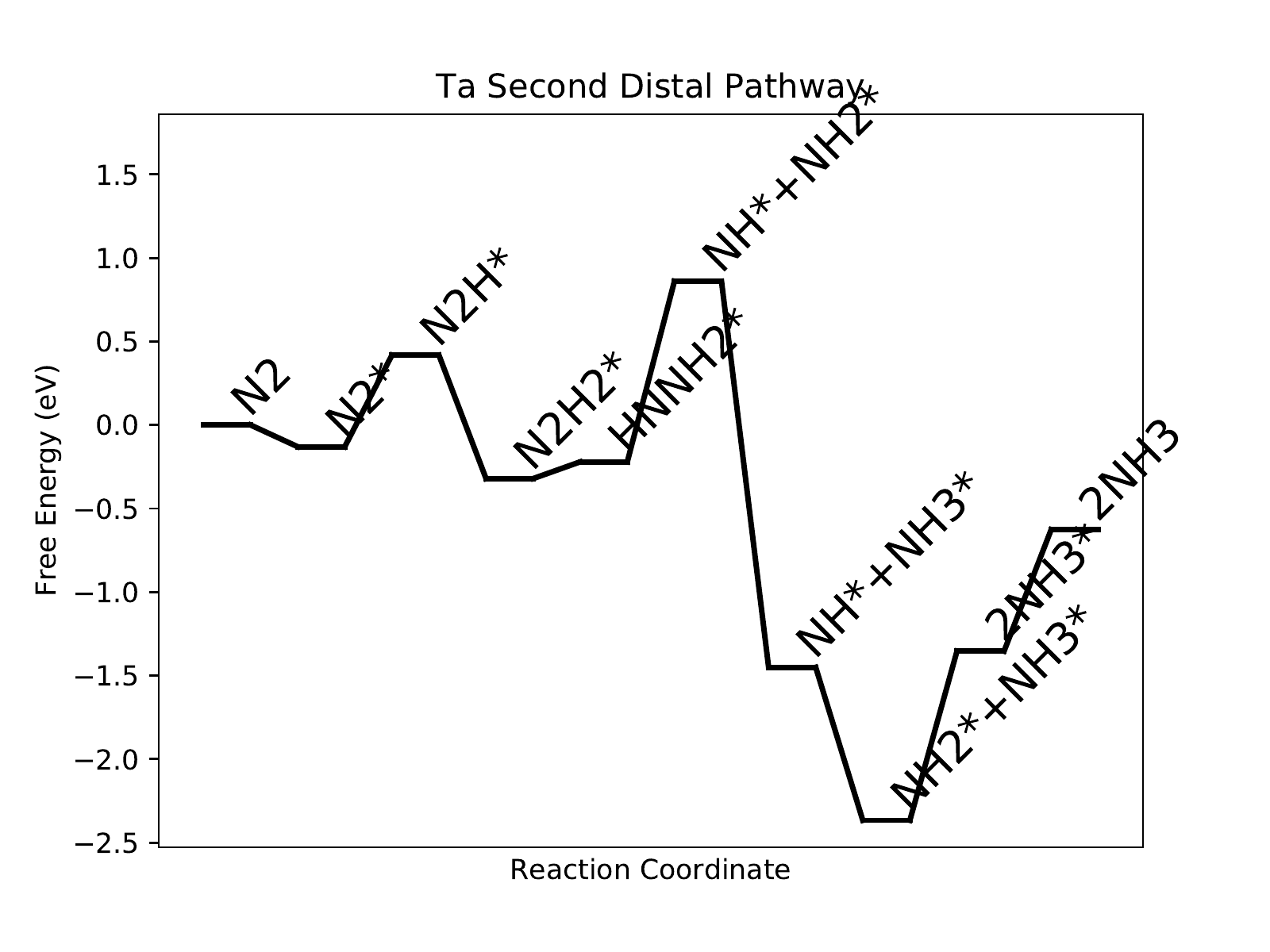}
\label{fig:Ta_distal_2}
\caption{Free energy diagram for Ta}
\end{figure}

\begin{figure}
\includegraphics[width=1\linewidth]{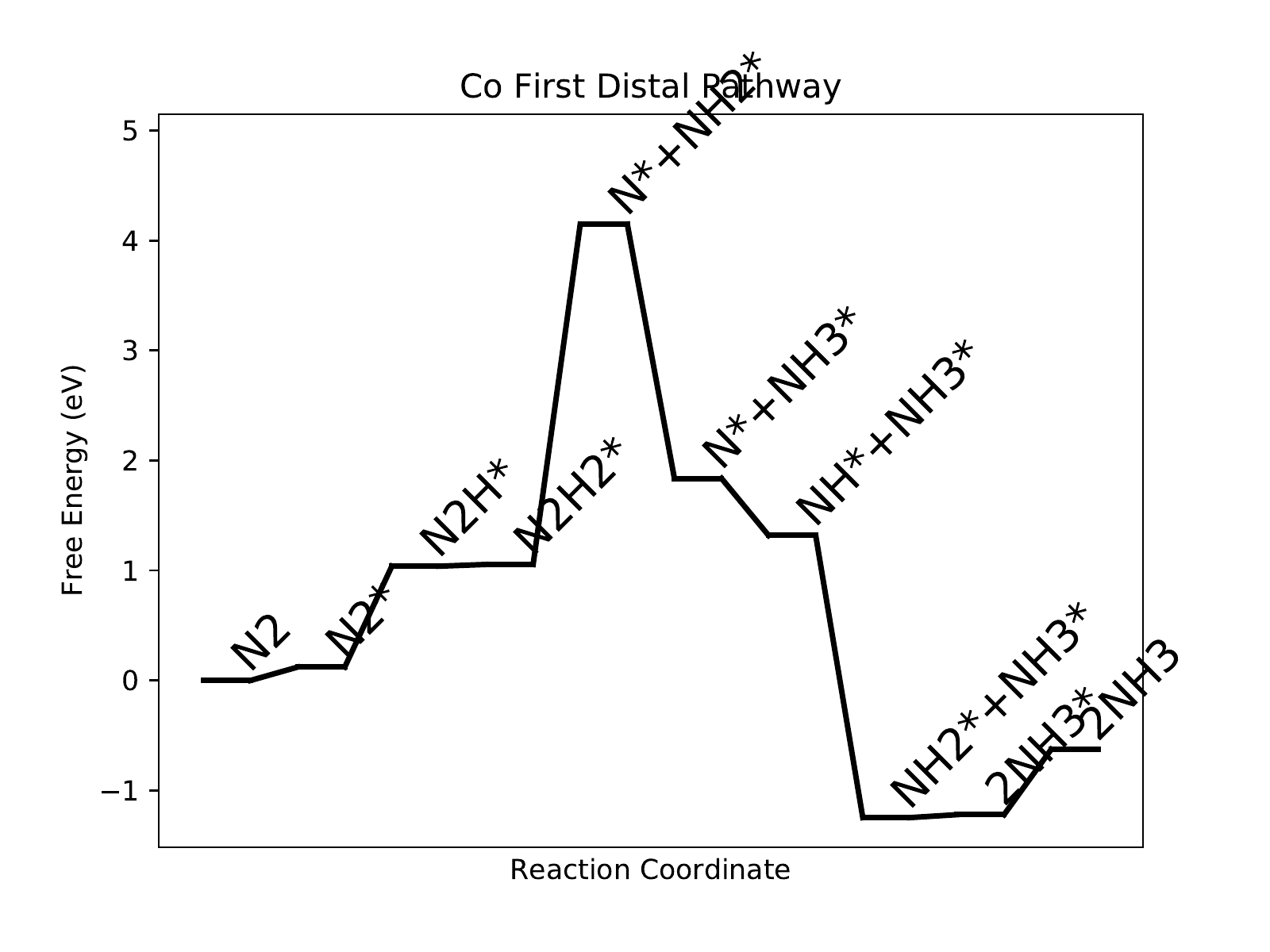}
\label{fig:Co_distal_1}
\caption{Free energy diagram for Co}
\end{figure}

\newpage
\begin{figure}
\includegraphics[width=1\linewidth]{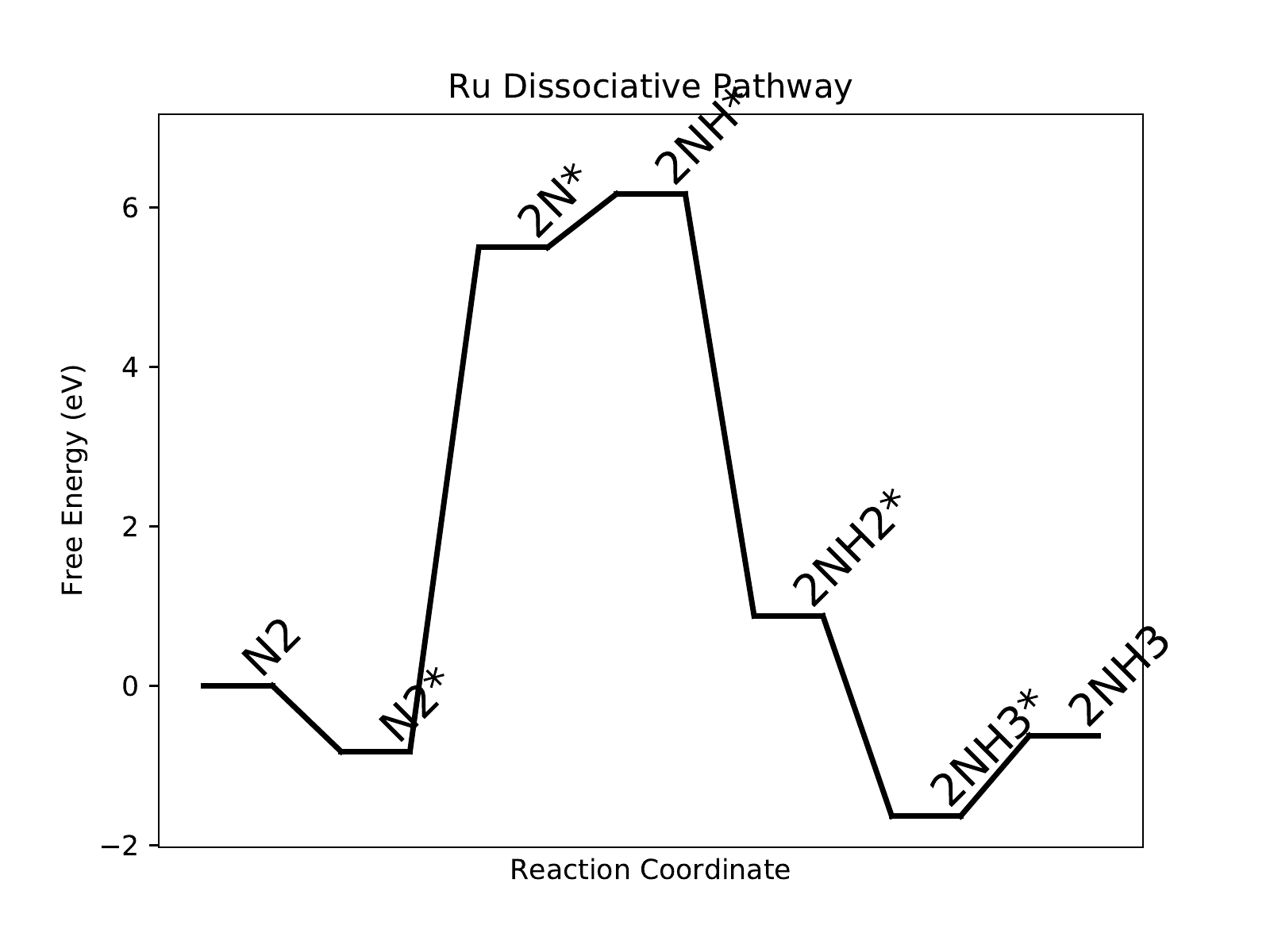}
\label{fig:Ru_dissociative}
\caption{Free energy diagram for Ru}
\end{figure}

\begin{figure}
\includegraphics[width=1\linewidth]{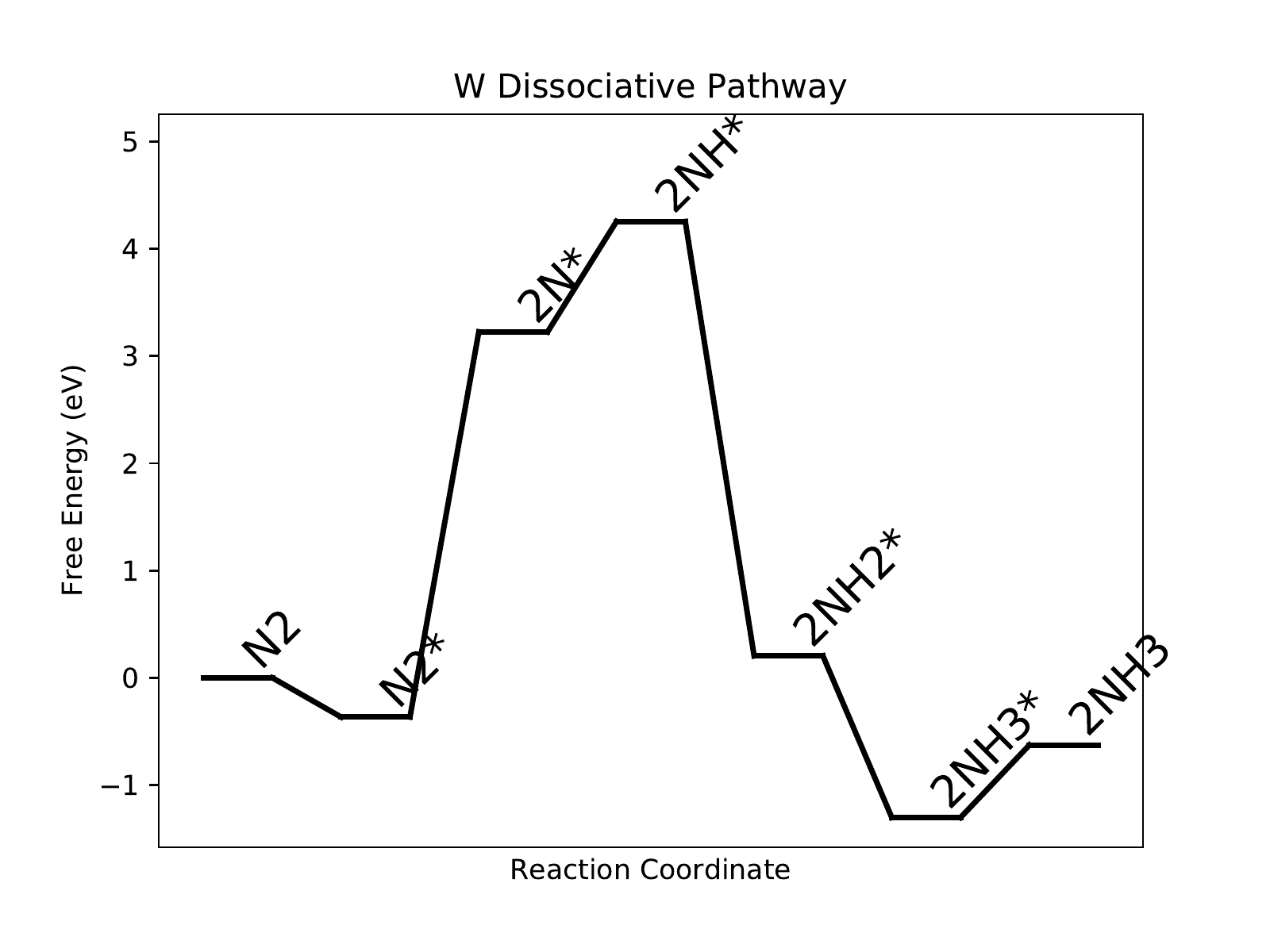}
\label{fig:W_dissociative}
\caption{Free energy diagram for W}
\end{figure}

\newpage
\begin{figure}
\includegraphics[width=1\linewidth]{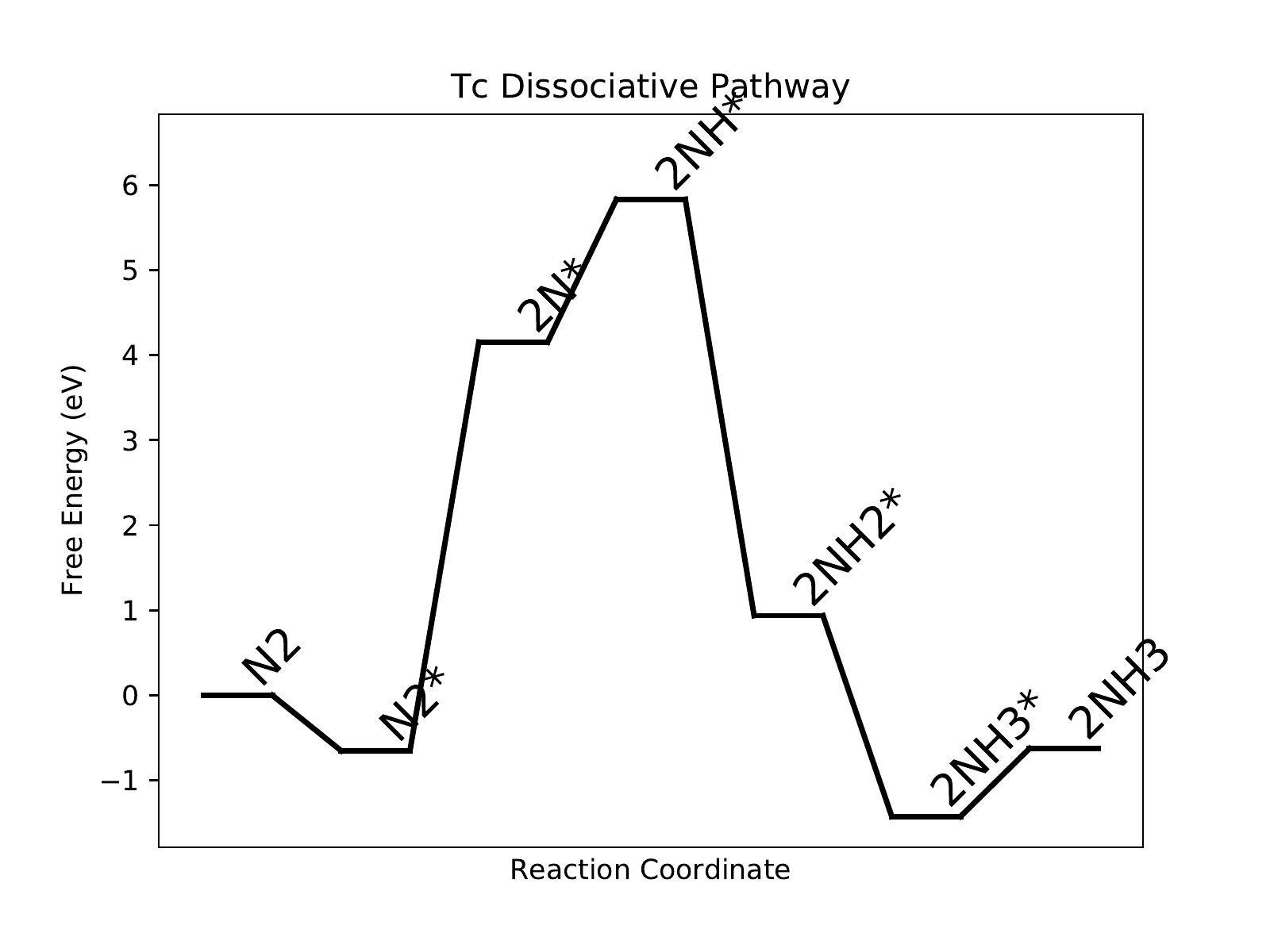}
\label{fig:Tc_dissociative}
\caption{Free energy diagram for Tc}
\end{figure}

\begin{figure}
\includegraphics[width=1\linewidth]{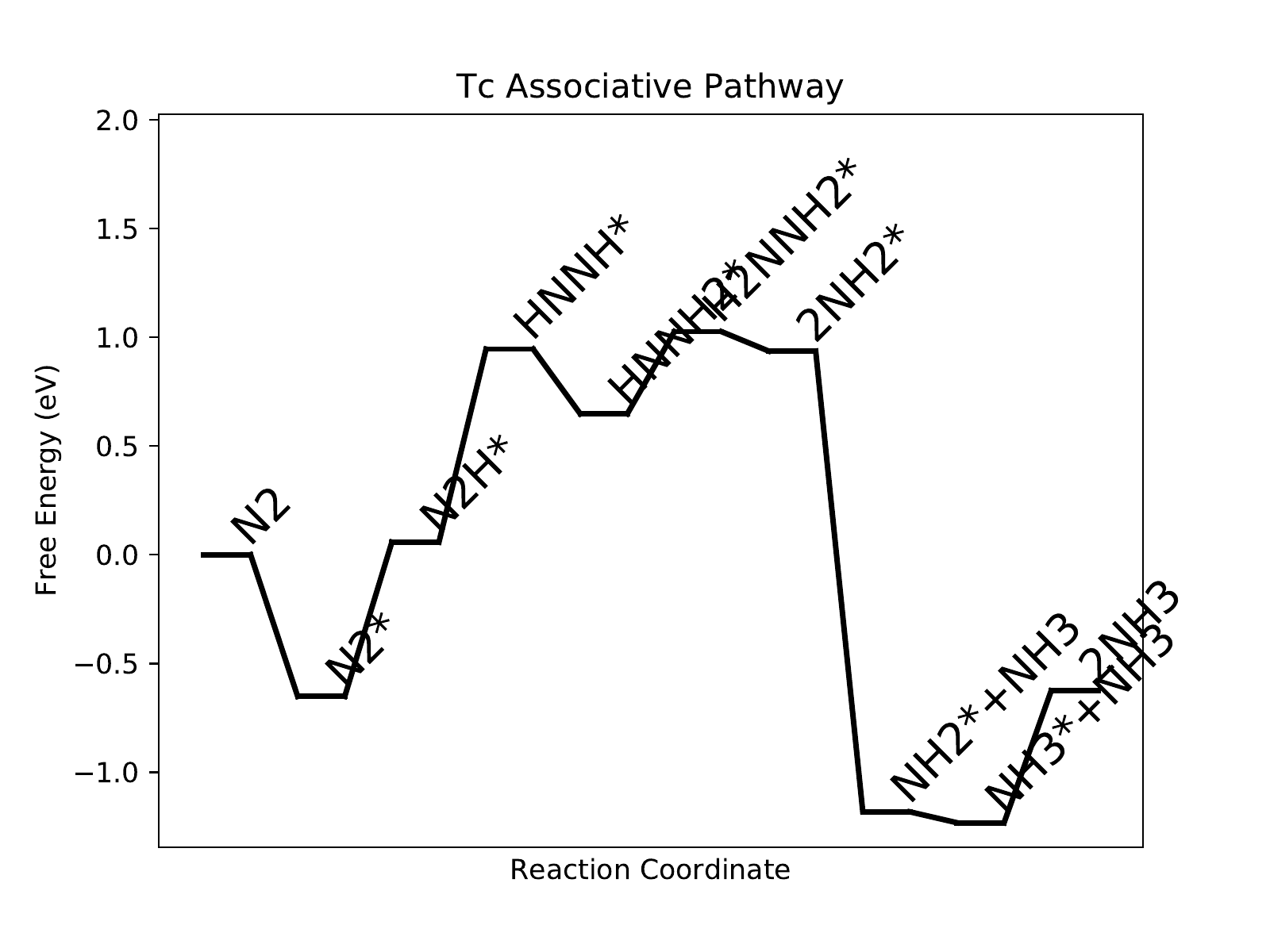}
\label{fig:Tc_associative}
\caption{Free energy diagram for Tc}
\end{figure}

\newpage
\begin{figure}
\includegraphics[width=1\linewidth]{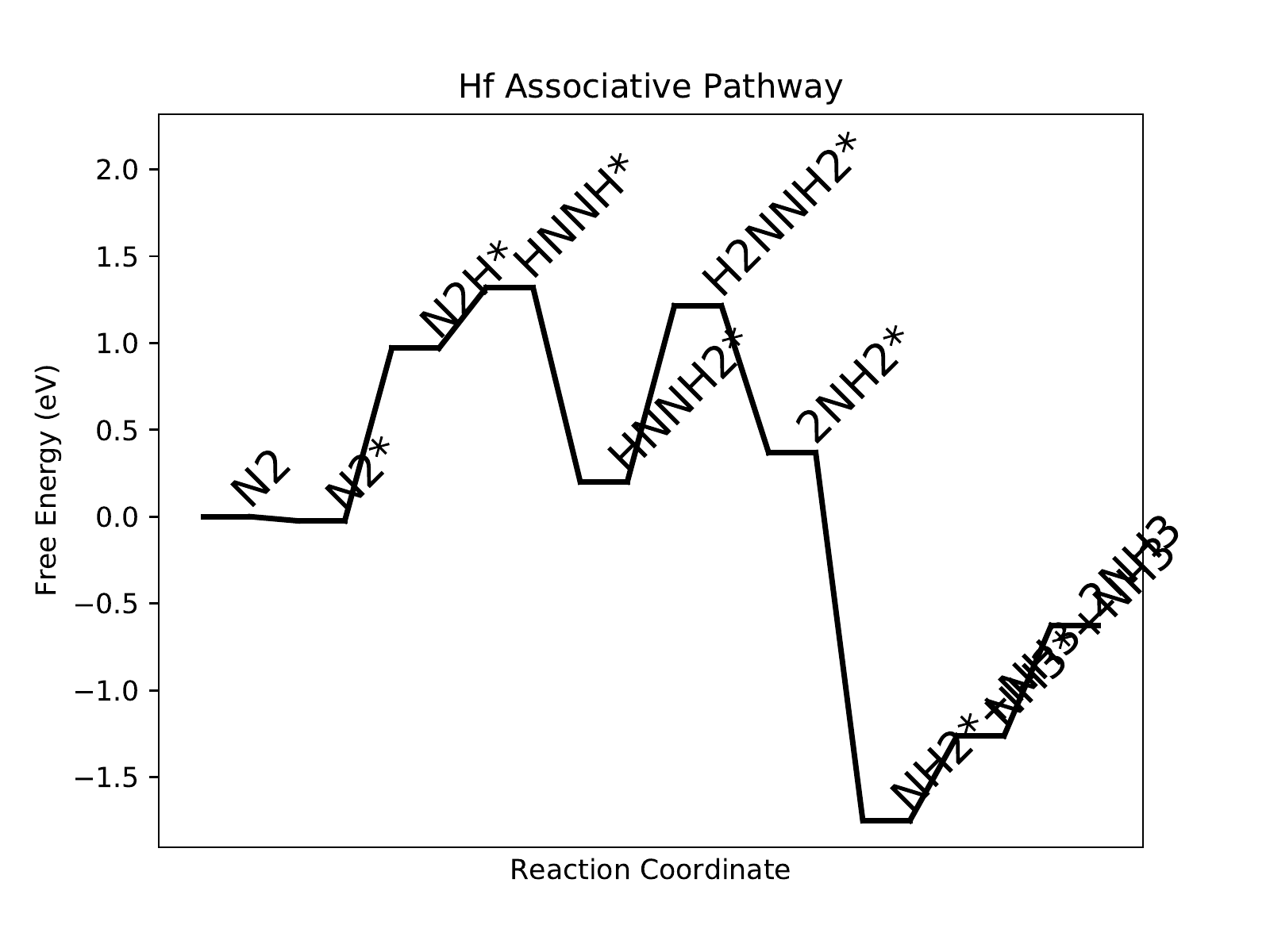}
\label{fig:Hf_associative}
\caption{Free energy diagram for Hf}
\end{figure}

\begin{figure}
\includegraphics[width=1\linewidth]{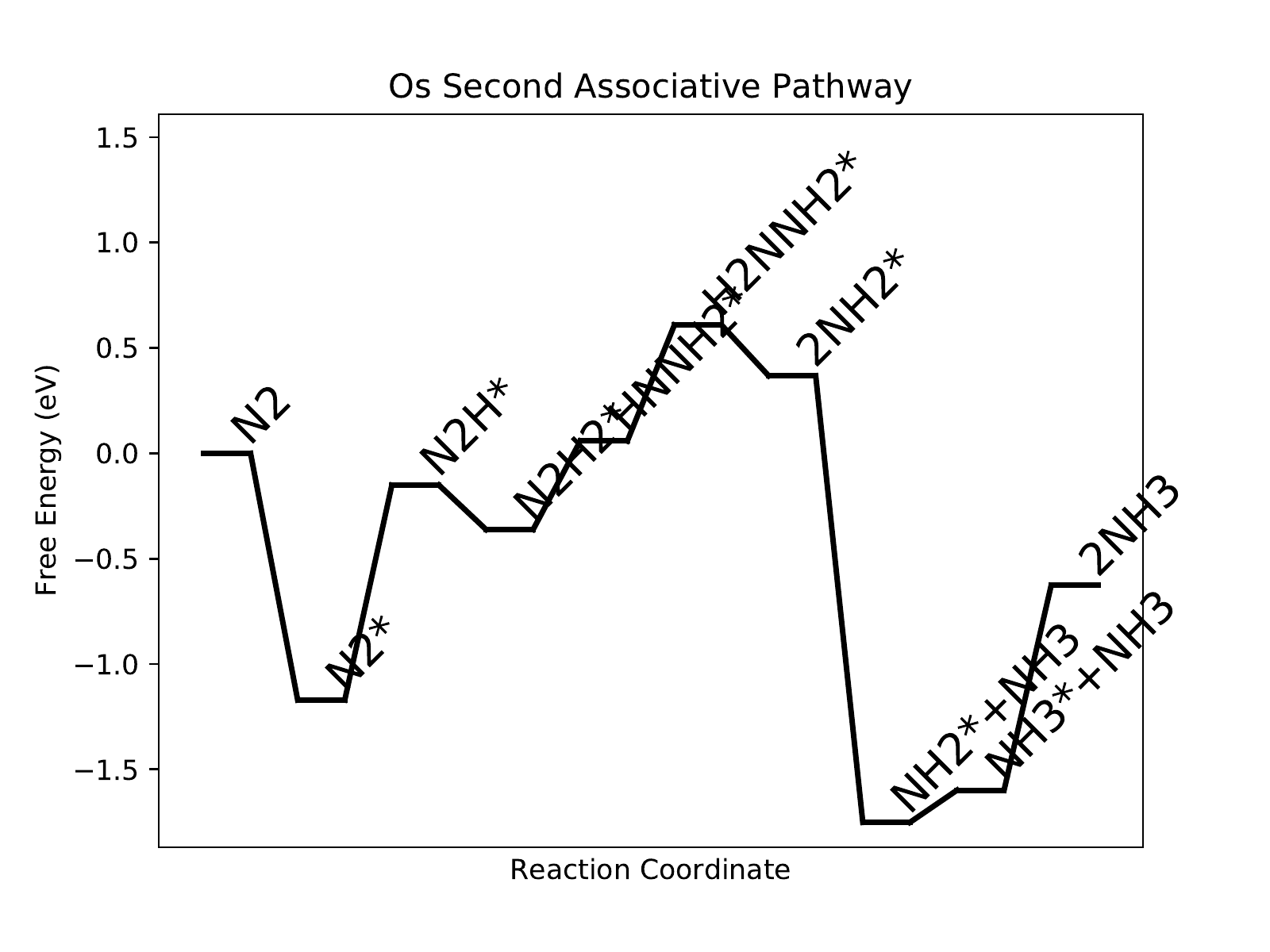}
\label{fig:Os_associative_2}
\caption{Free energy diagram for Os}
\end{figure}

\newpage
\begin{figure}
\includegraphics[width=1\linewidth]{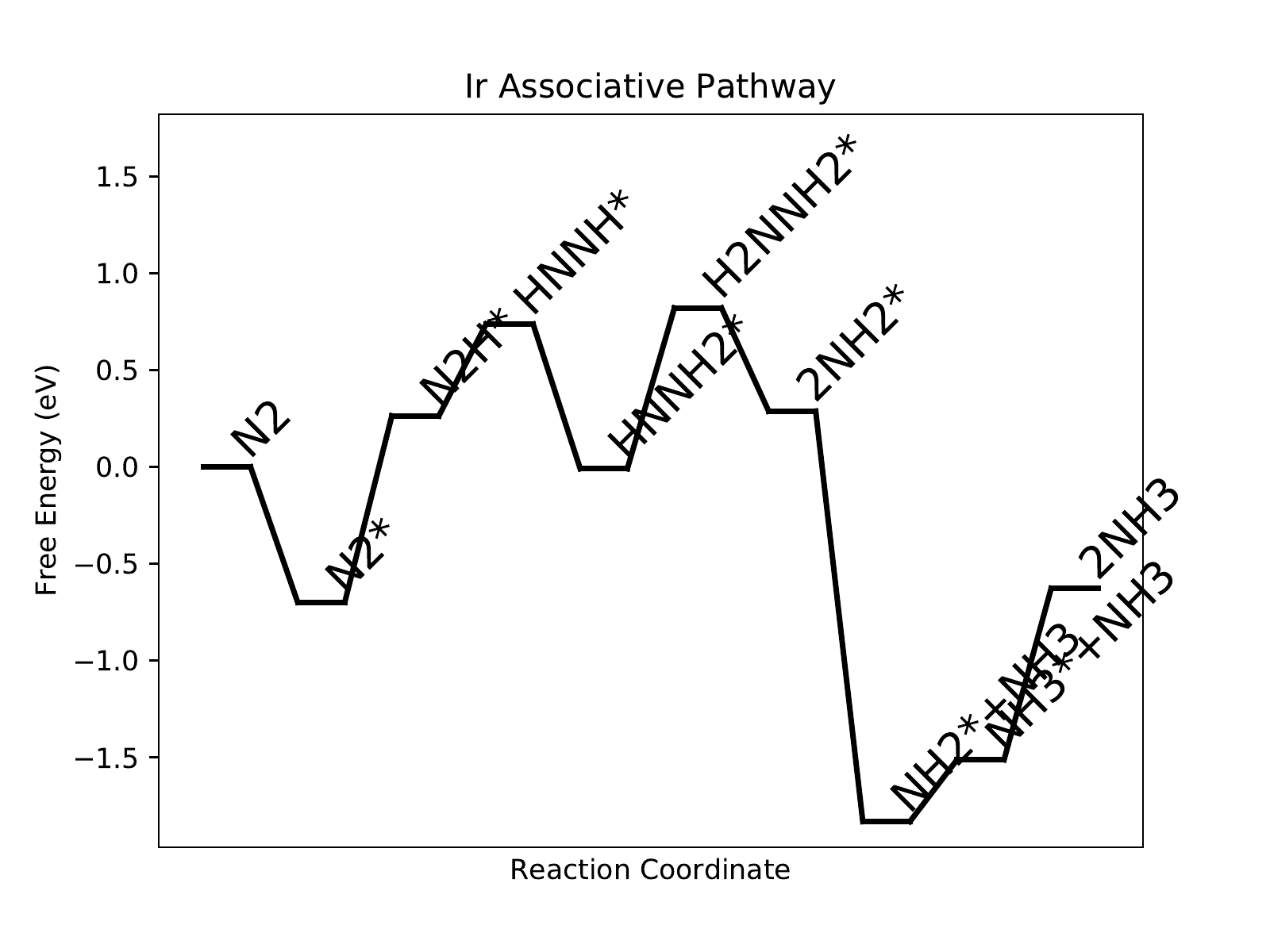}
\label{fig:Ir_associative}
\caption{Free energy diagram for Ir}
\end{figure}

\begin{figure}
\includegraphics[width=1\linewidth]{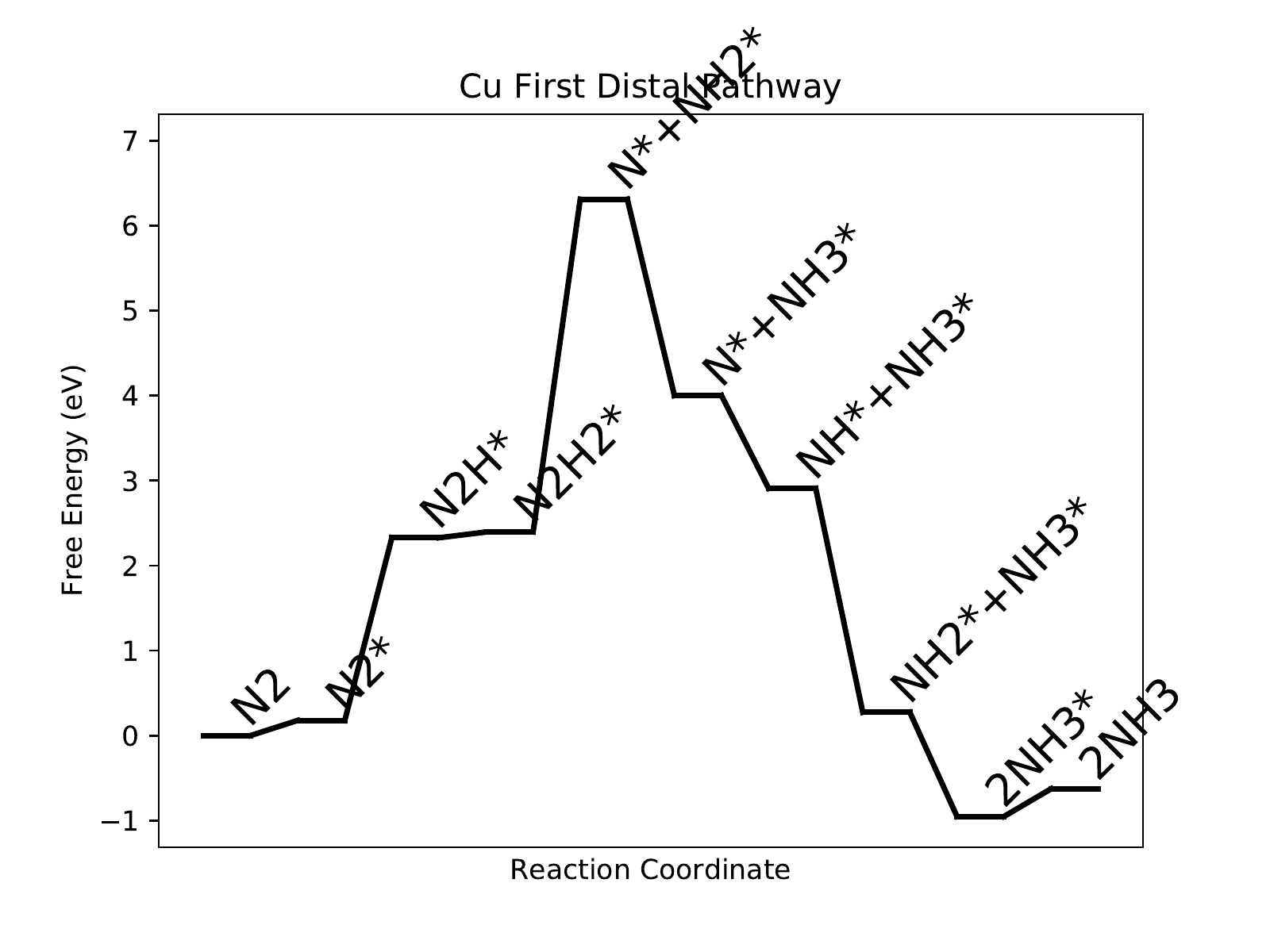}
\label{fig:Cu_distal_1}
\caption{Free energy diagram for Cu}
\end{figure}

\newpage
\begin{figure}
\includegraphics[width=1\linewidth]{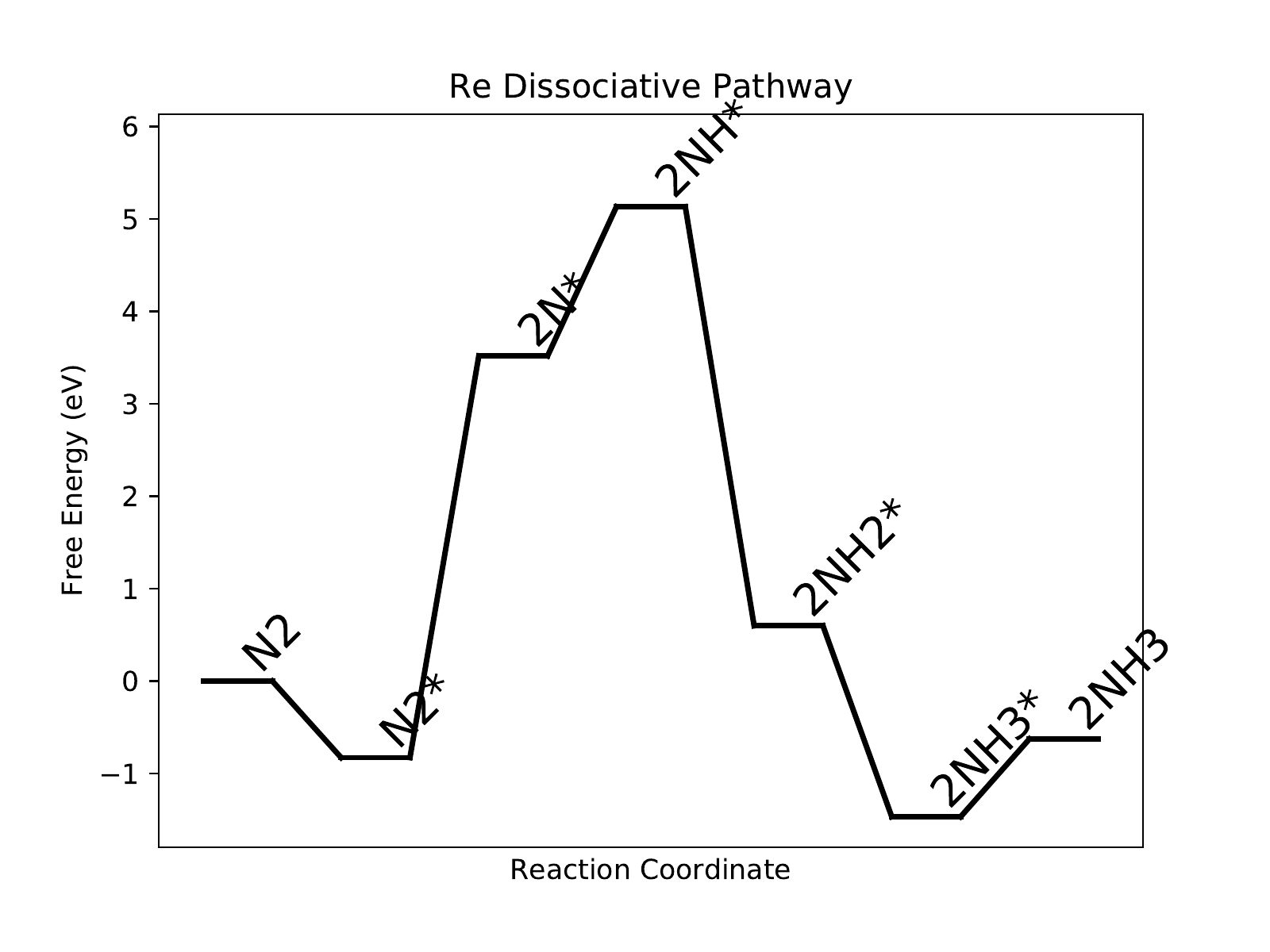}
\label{fig:Re_dissociative}
\caption{Free energy diagram for Re}
\end{figure}

\begin{figure}
\includegraphics[width=1\linewidth]{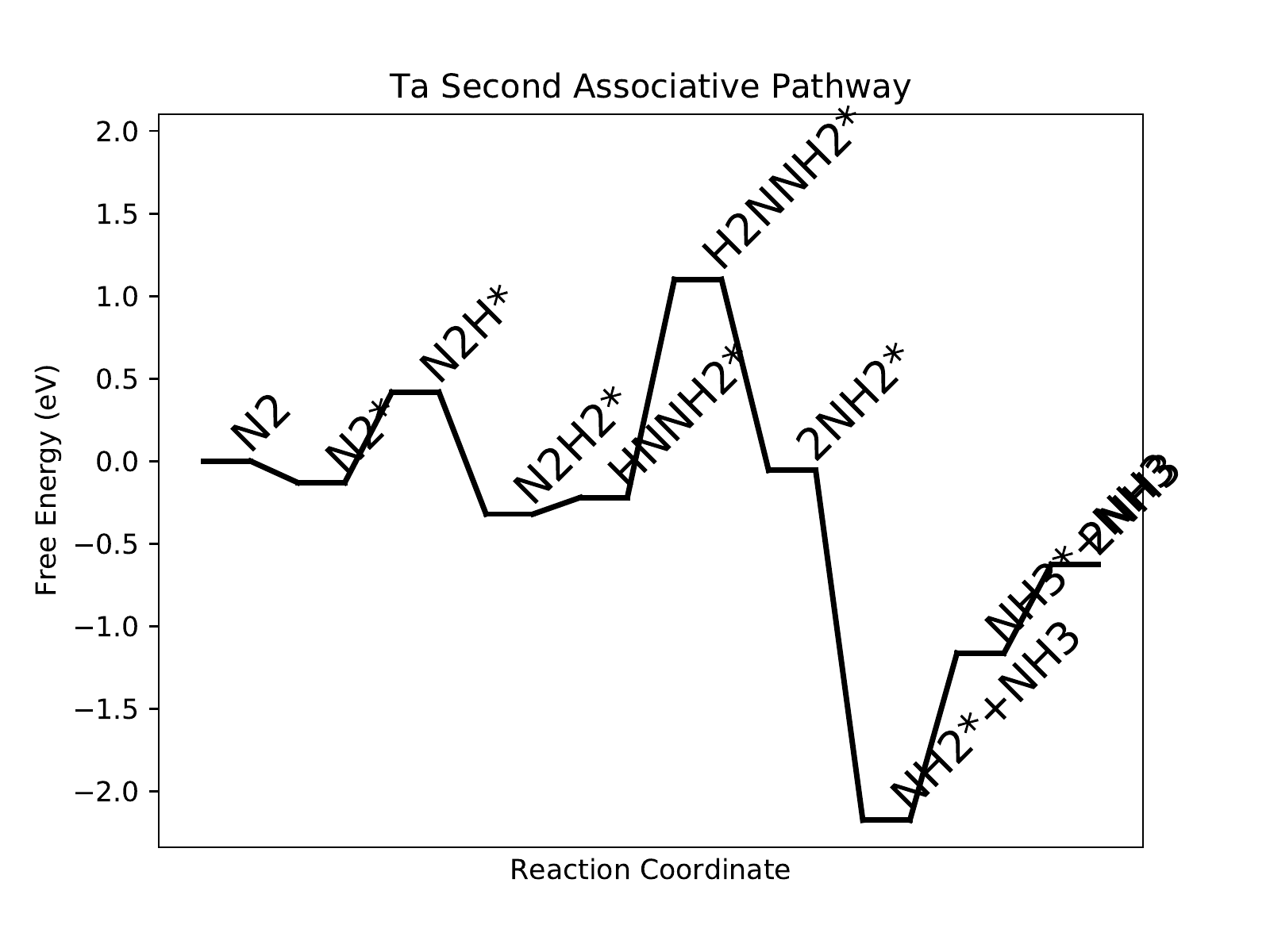}
\label{fig:Ta_associative_2}
\caption{Free energy diagram for Ta}
\end{figure}

\newpage
\begin{figure}
\includegraphics[width=1\linewidth]{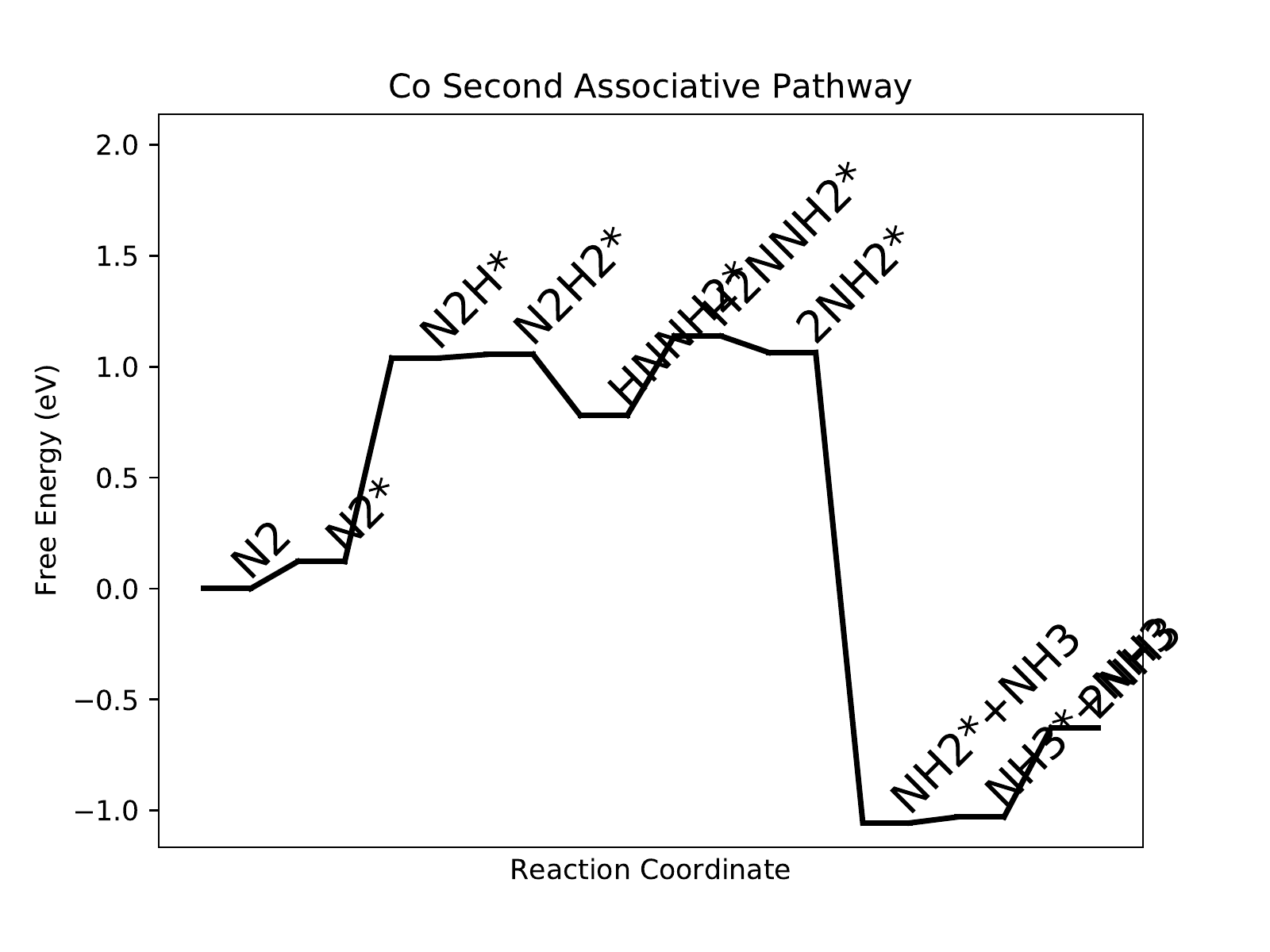}
\label{fig:Co_associative_2}
\caption{Free energy diagram for Co}
\end{figure}

\begin{figure}
\includegraphics[width=1\linewidth]{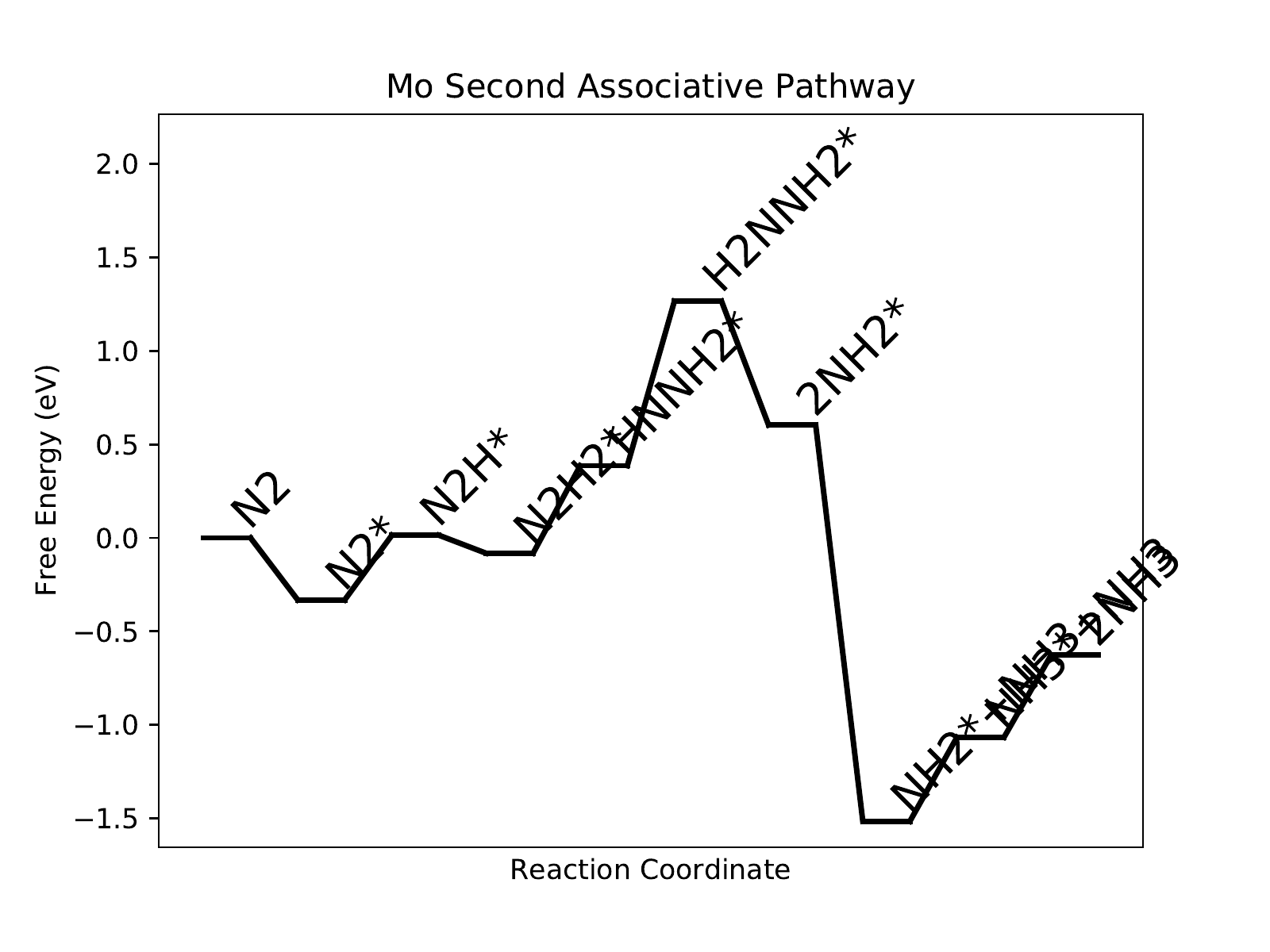}
\label{fig:Mo_associative_2}
\caption{Free energy diagram for Mo}
\end{figure}

\newpage
\begin{figure}
\includegraphics[width=1\linewidth]{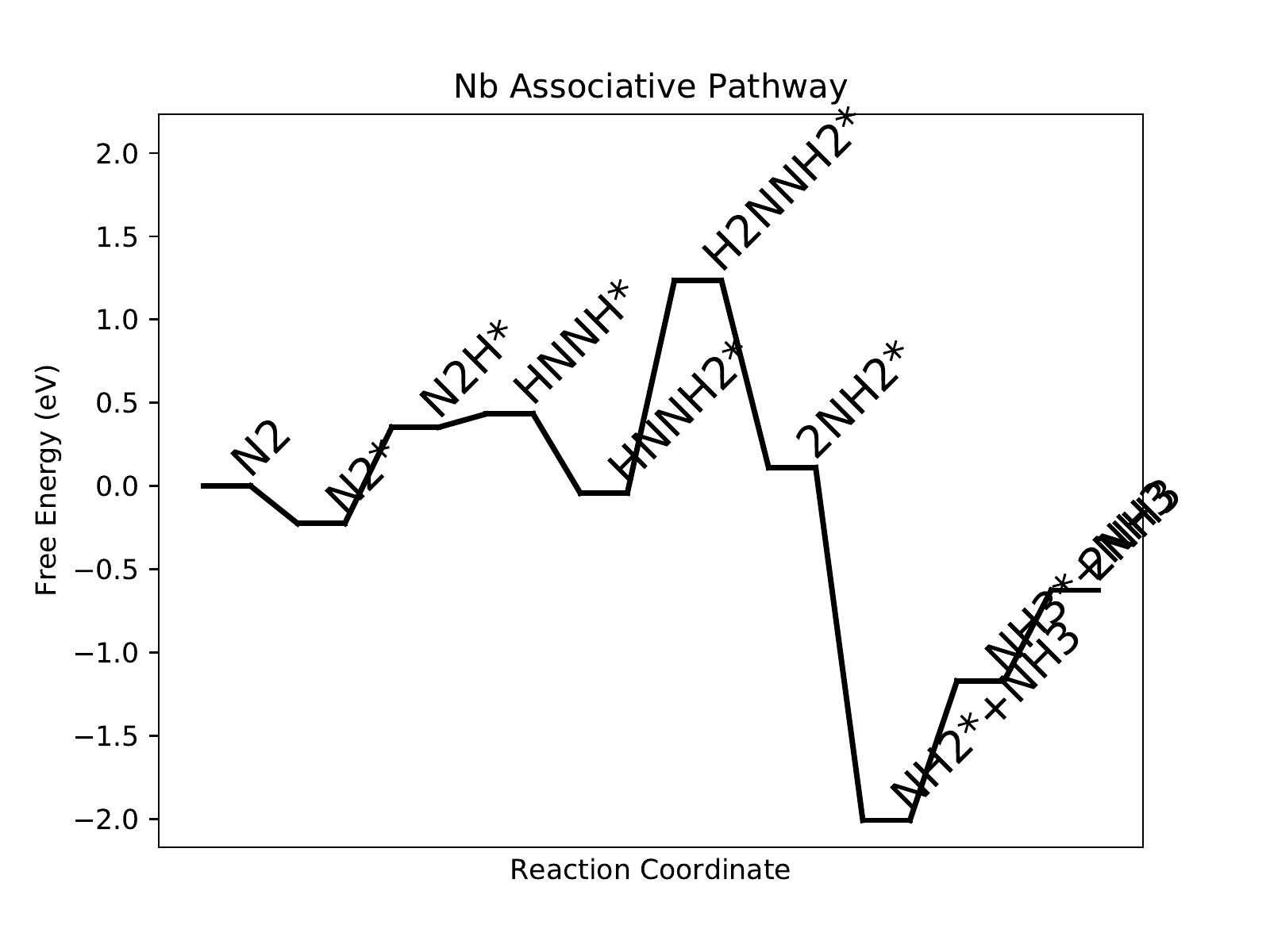}
\label{fig:Nb_associative}
\caption{Free energy diagram for Nb}
\end{figure}

\begin{figure}
\includegraphics[width=1\linewidth]{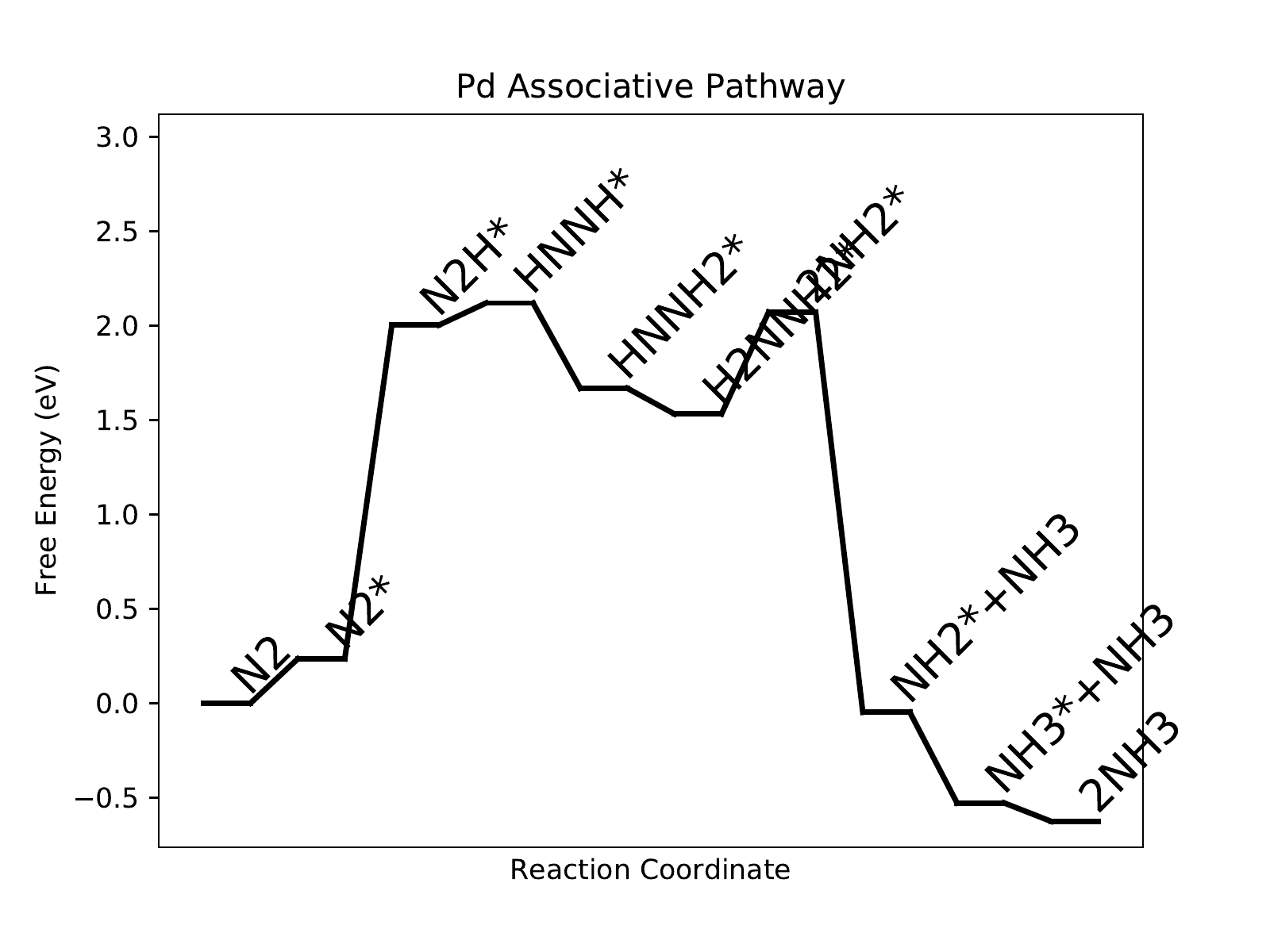}
\label{fig:Pd_associative}
\caption{Free energy diagram for Pd}
\end{figure}

\newpage
\begin{figure}
\includegraphics[width=1\linewidth]{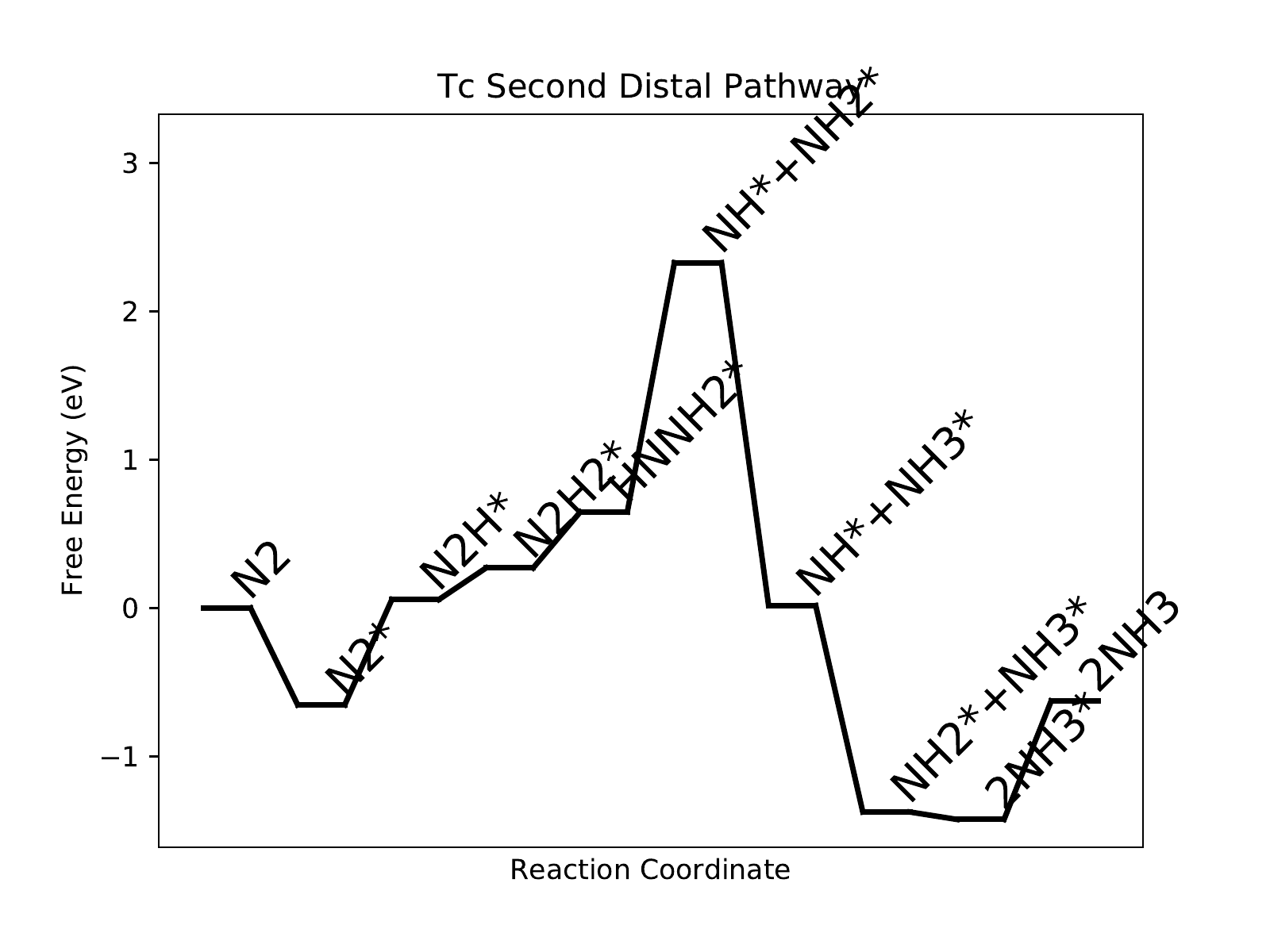}
\label{fig:Tc_distal_2}
\caption{Free energy diagram for Tc}
\end{figure}

\begin{figure}
\includegraphics[width=1\linewidth]{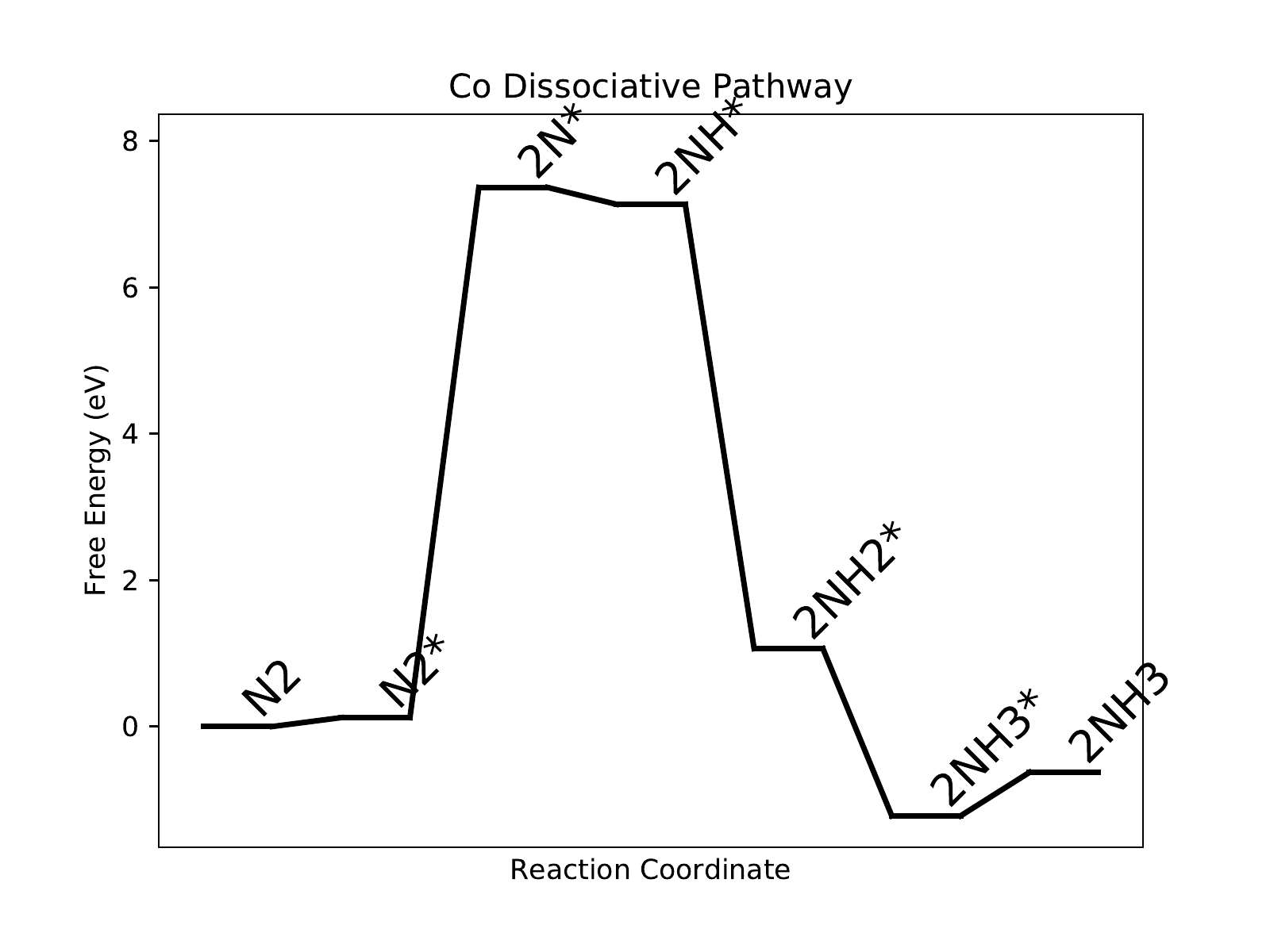}
\label{fig:Co_dissociative}
\caption{Free energy diagram for Co}
\end{figure}

\newpage
\begin{figure}
\includegraphics[width=1\linewidth]{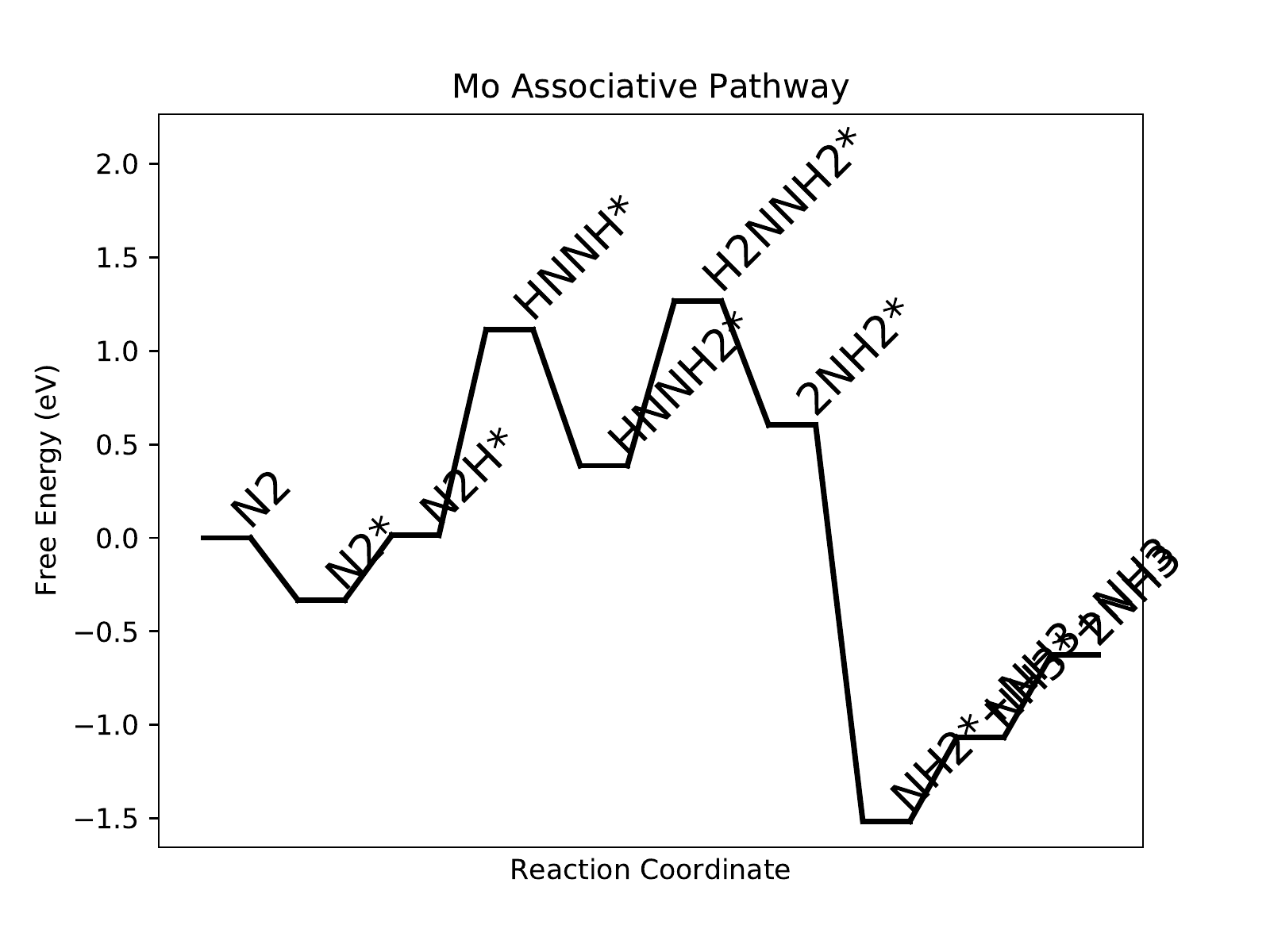}
\label{fig:Mo_associative}
\caption{Free energy diagram for Mo}
\end{figure}

\begin{figure}
\includegraphics[width=1\linewidth]{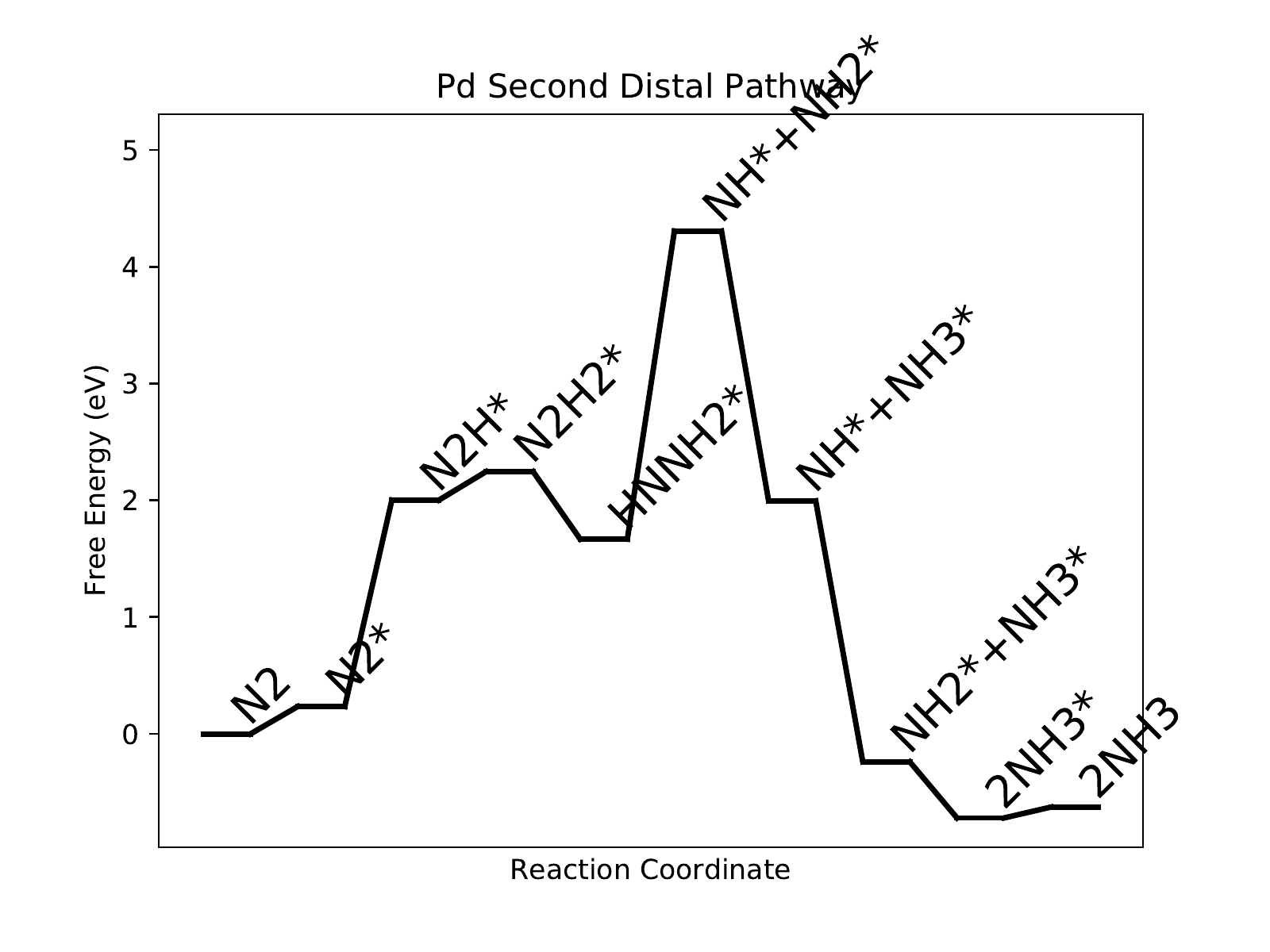}
\label{fig:Pd_distal_2}
\caption{Free energy diagram for Pd}
\end{figure}

\newpage
\begin{figure}
\includegraphics[width=1\linewidth]{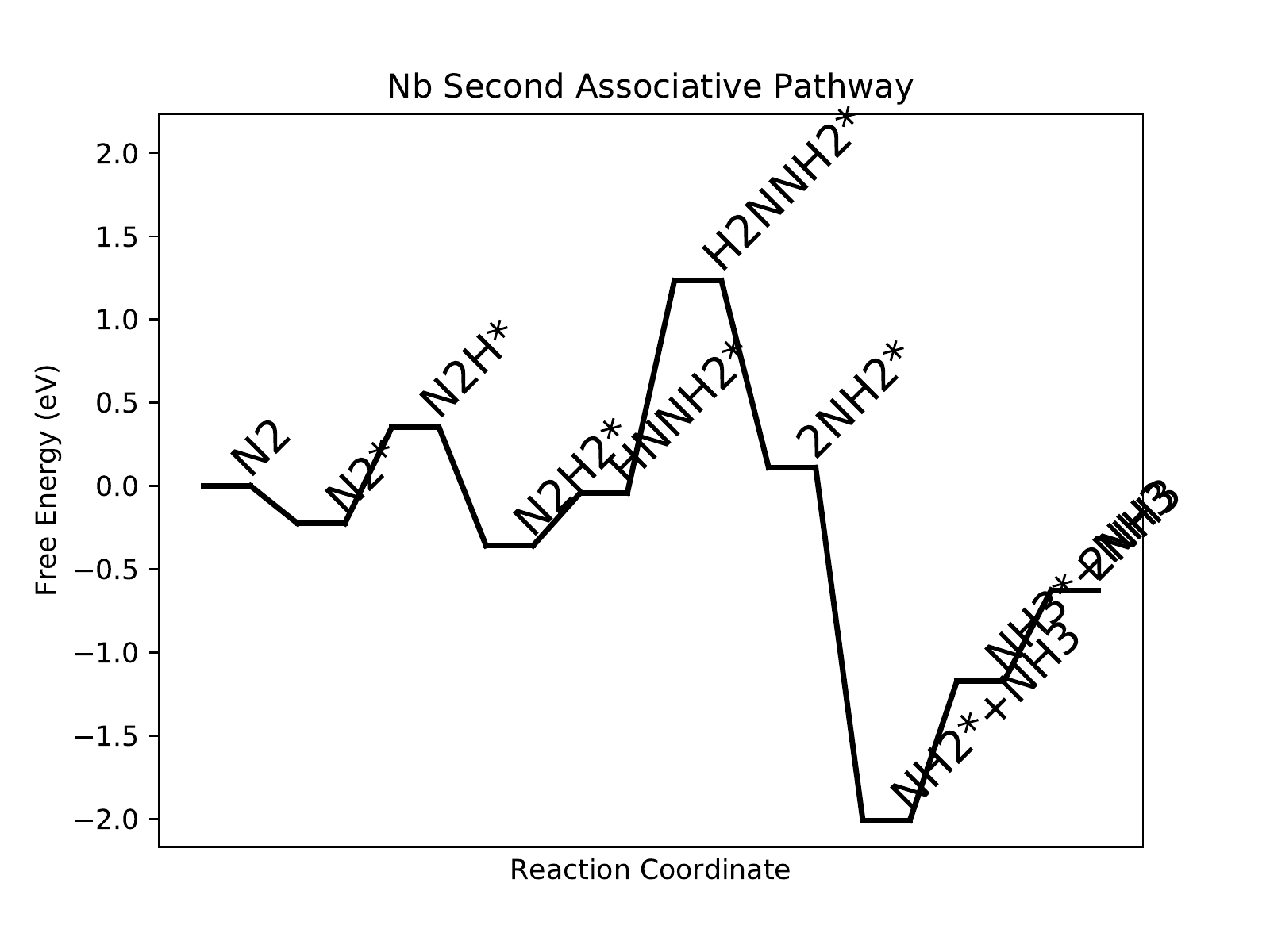}
\label{fig:Nb_associative_2}
\caption{Free energy diagram for Nb}
\end{figure}

\begin{figure}
\includegraphics[width=1\linewidth]{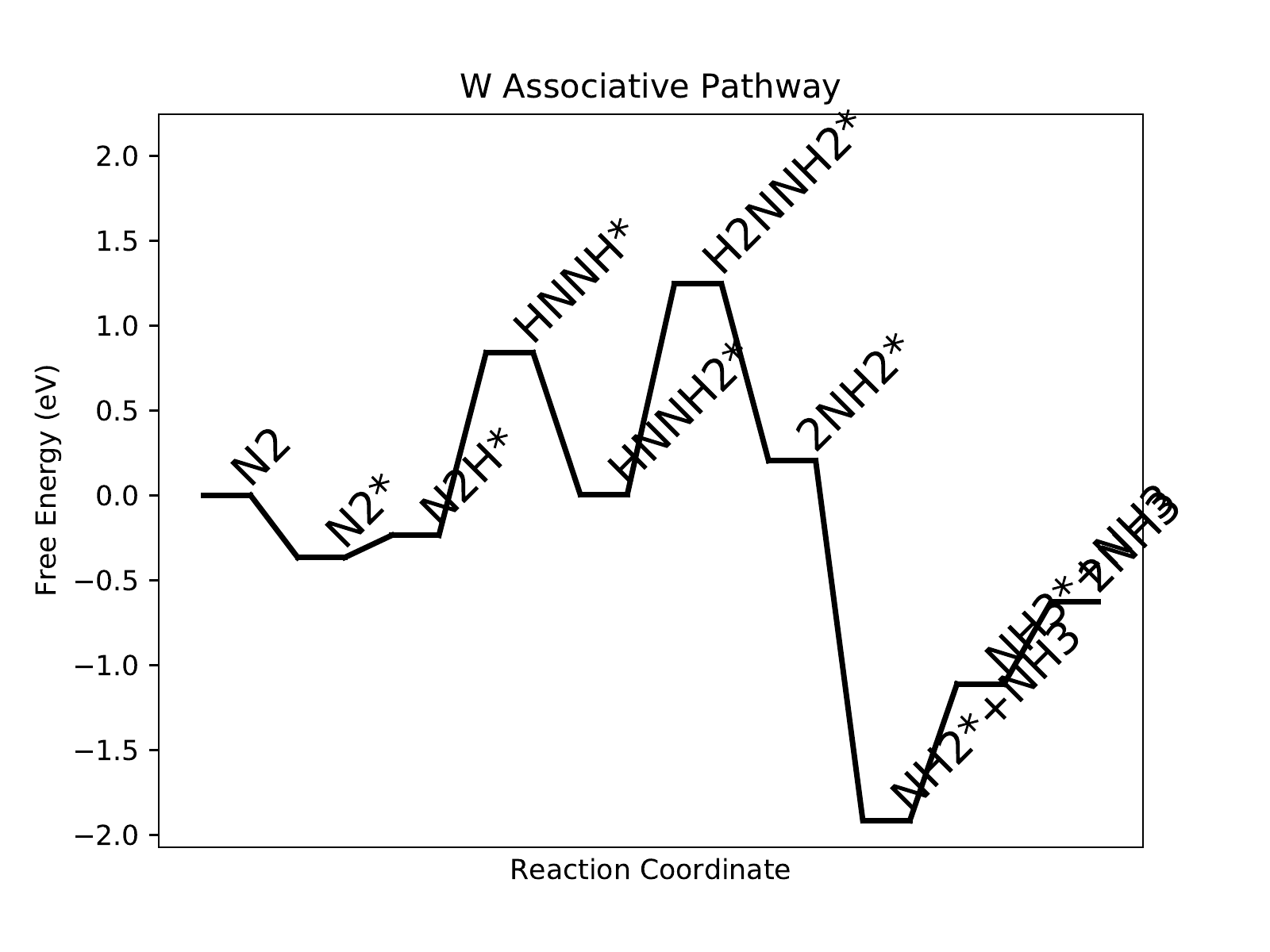}
\label{fig:W_associative}
\caption{Free energy diagram for W}
\end{figure}

\newpage
\begin{figure}
\includegraphics[width=1\linewidth]{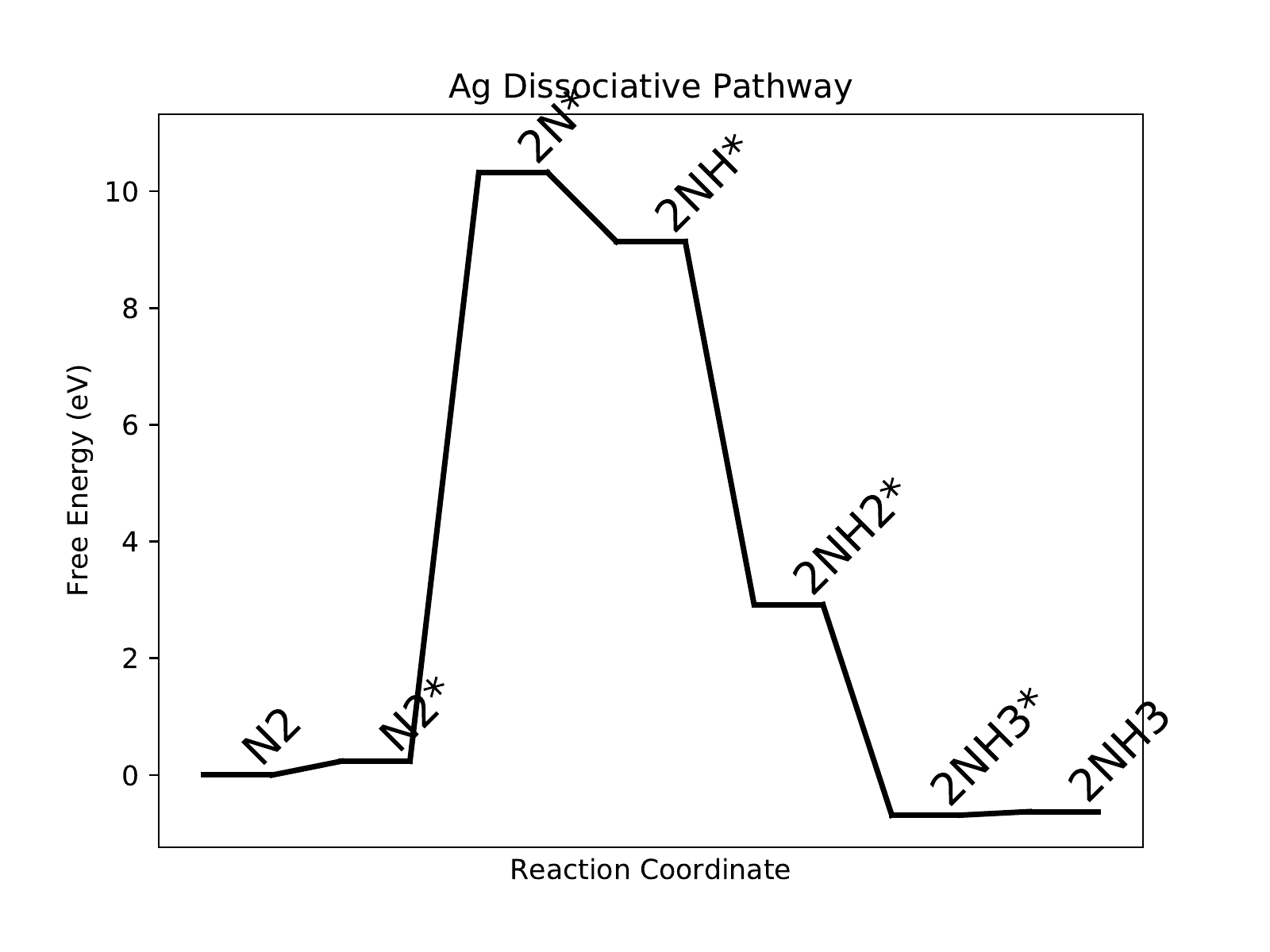}
\label{fig:Ag_dissociative}
\caption{Free energy diagram for Ag}
\end{figure}

\begin{figure}
\includegraphics[width=1\linewidth]{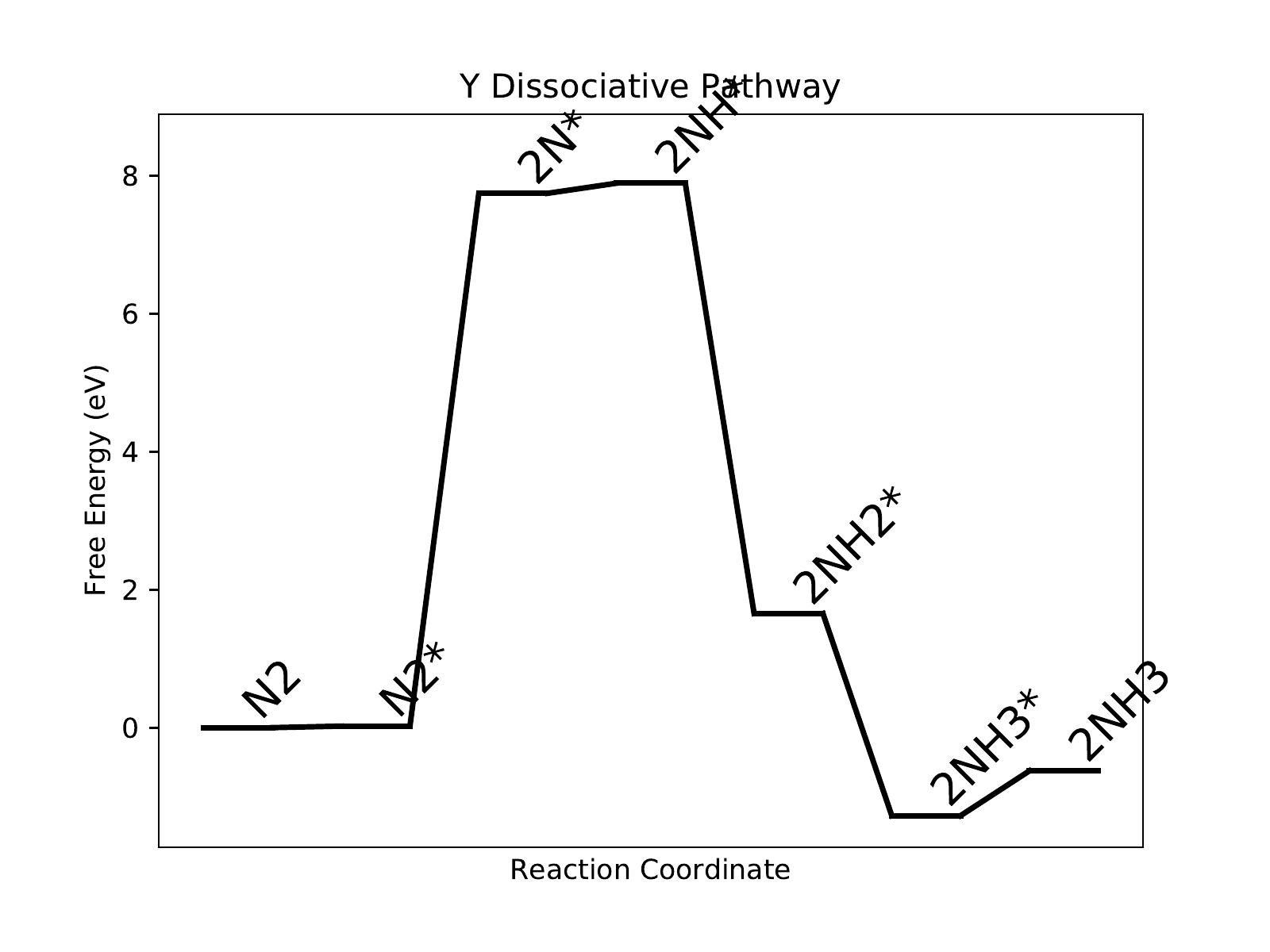}
\label{fig:Y_dissociative}
\caption{Free energy diagram for Y}
\end{figure}

\newpage
\begin{figure}
\includegraphics[width=1\linewidth]{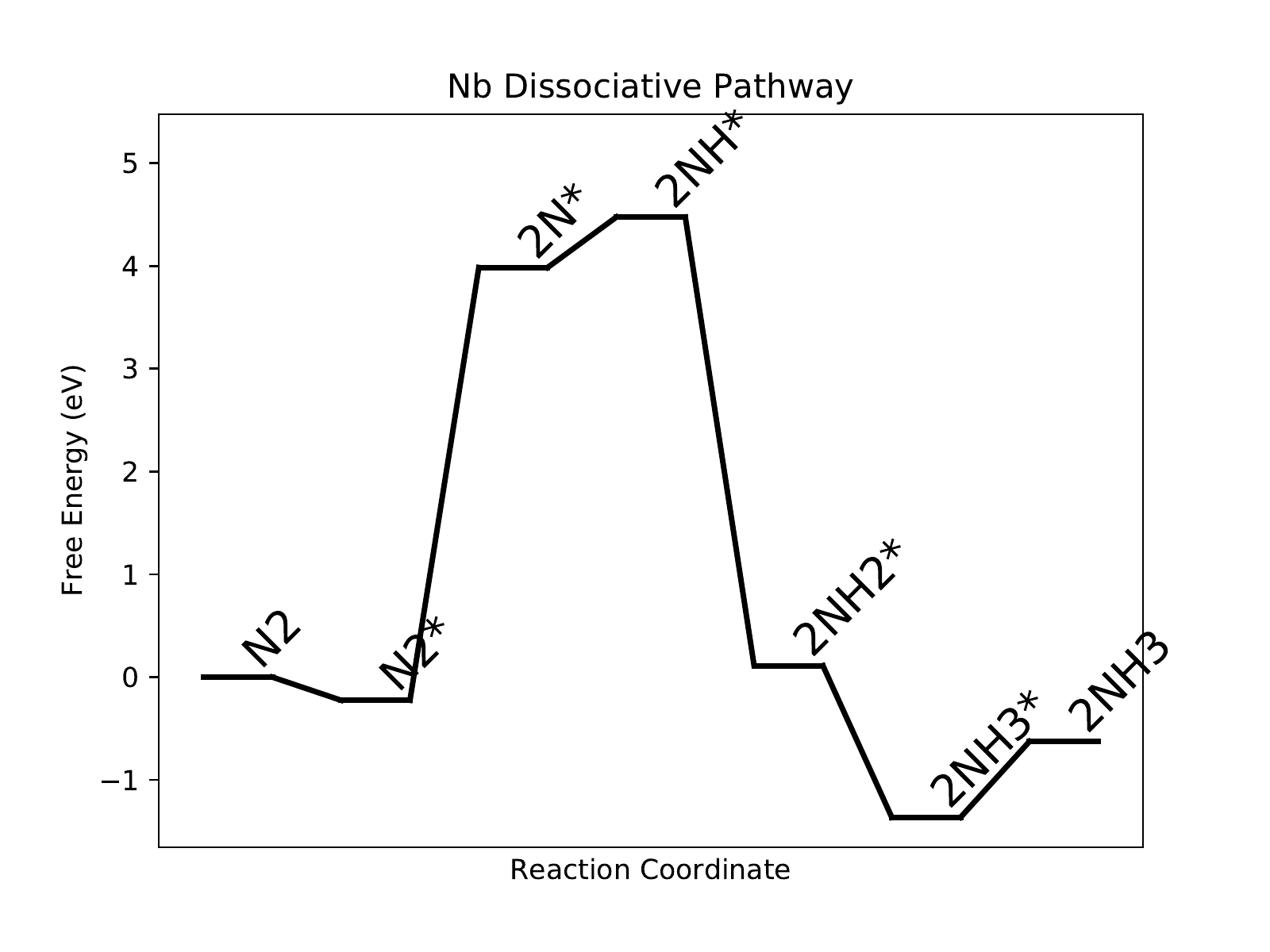}
\label{fig:Nb_dissociative}
\caption{Free energy diagram for Nb}
\end{figure}

\begin{figure}
\includegraphics[width=1\linewidth]{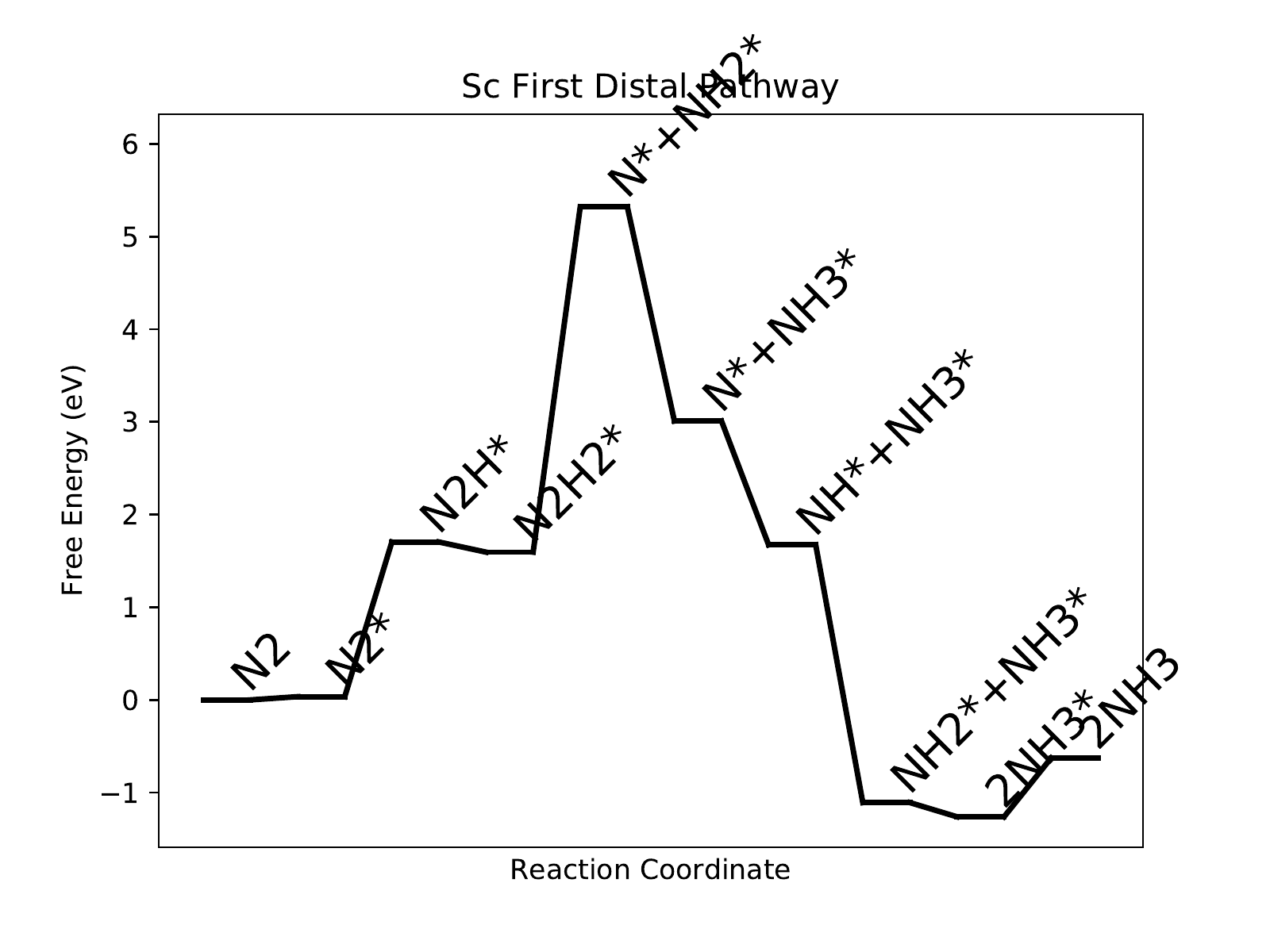}
\label{fig:Sc_distal_1}
\caption{Free energy diagram for Sc}
\end{figure}

\newpage
\begin{figure}
\includegraphics[width=1\linewidth]{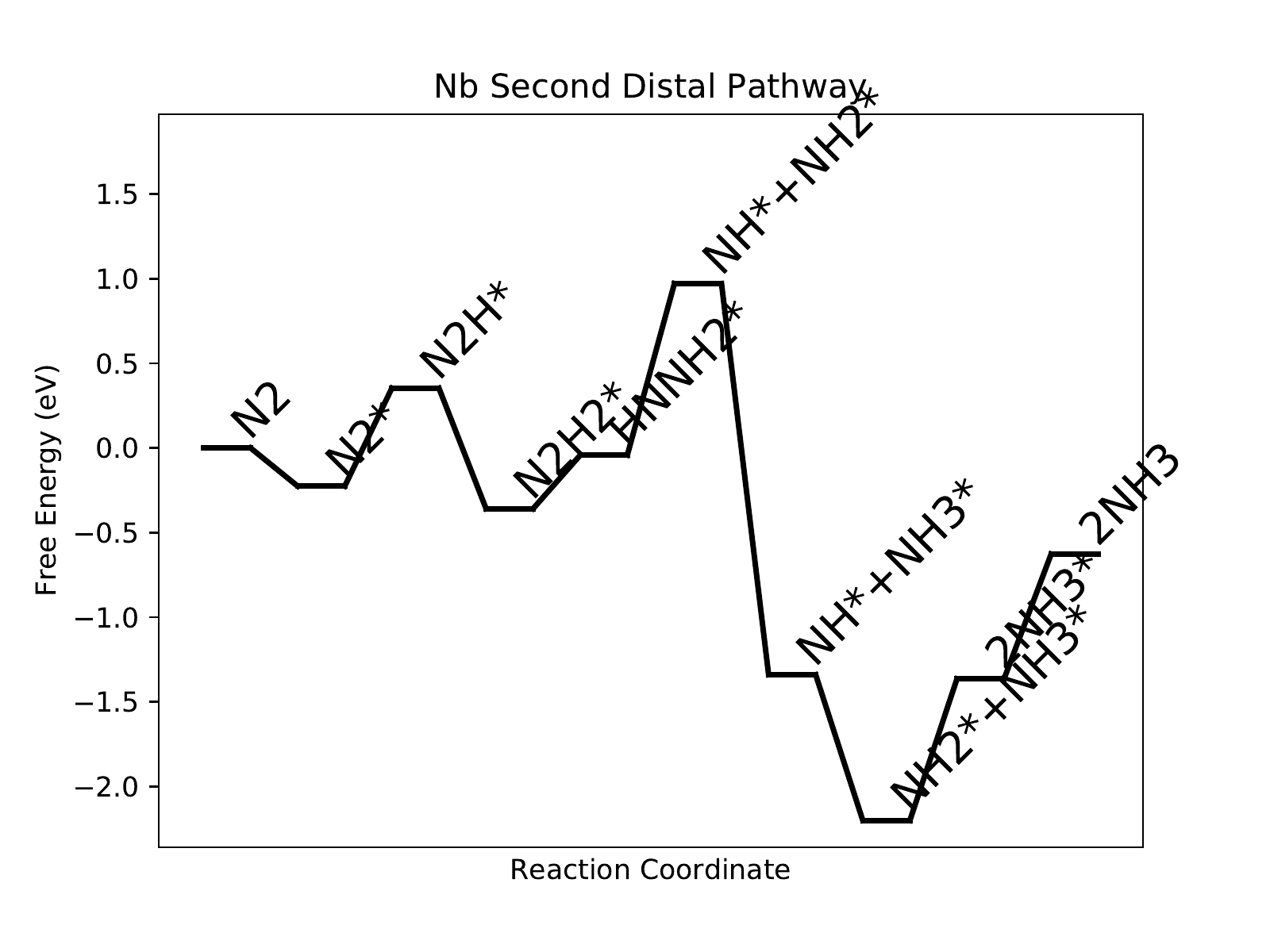}
\label{fig:Nb_distal_2}
\caption{Free energy diagram for Nb}
\end{figure}

\begin{figure}
\includegraphics[width=1\linewidth]{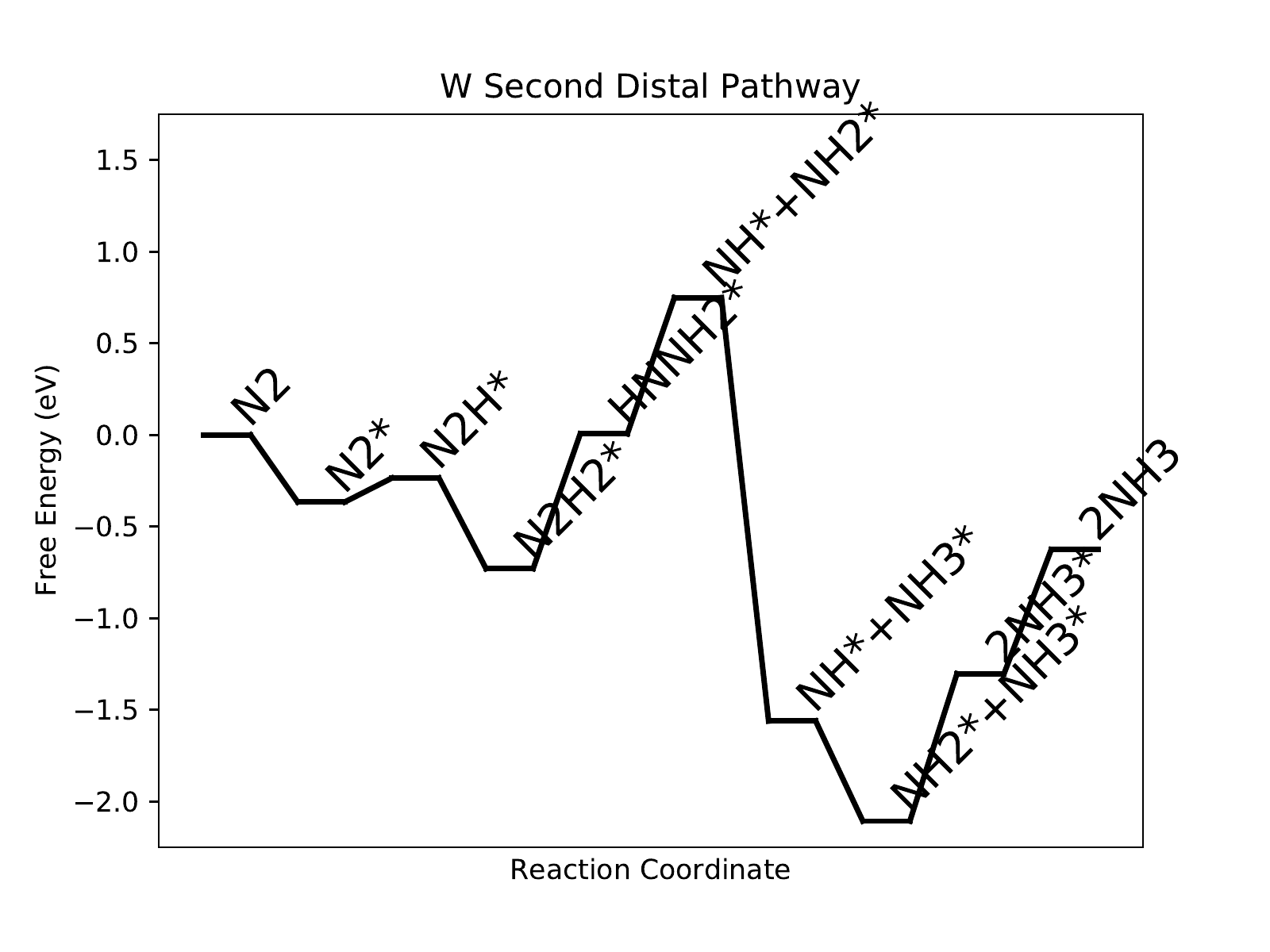}
\label{fig:W_distal_2}
\caption{Free energy diagram for W}
\end{figure}

\newpage
\begin{figure}
\includegraphics[width=1\linewidth]{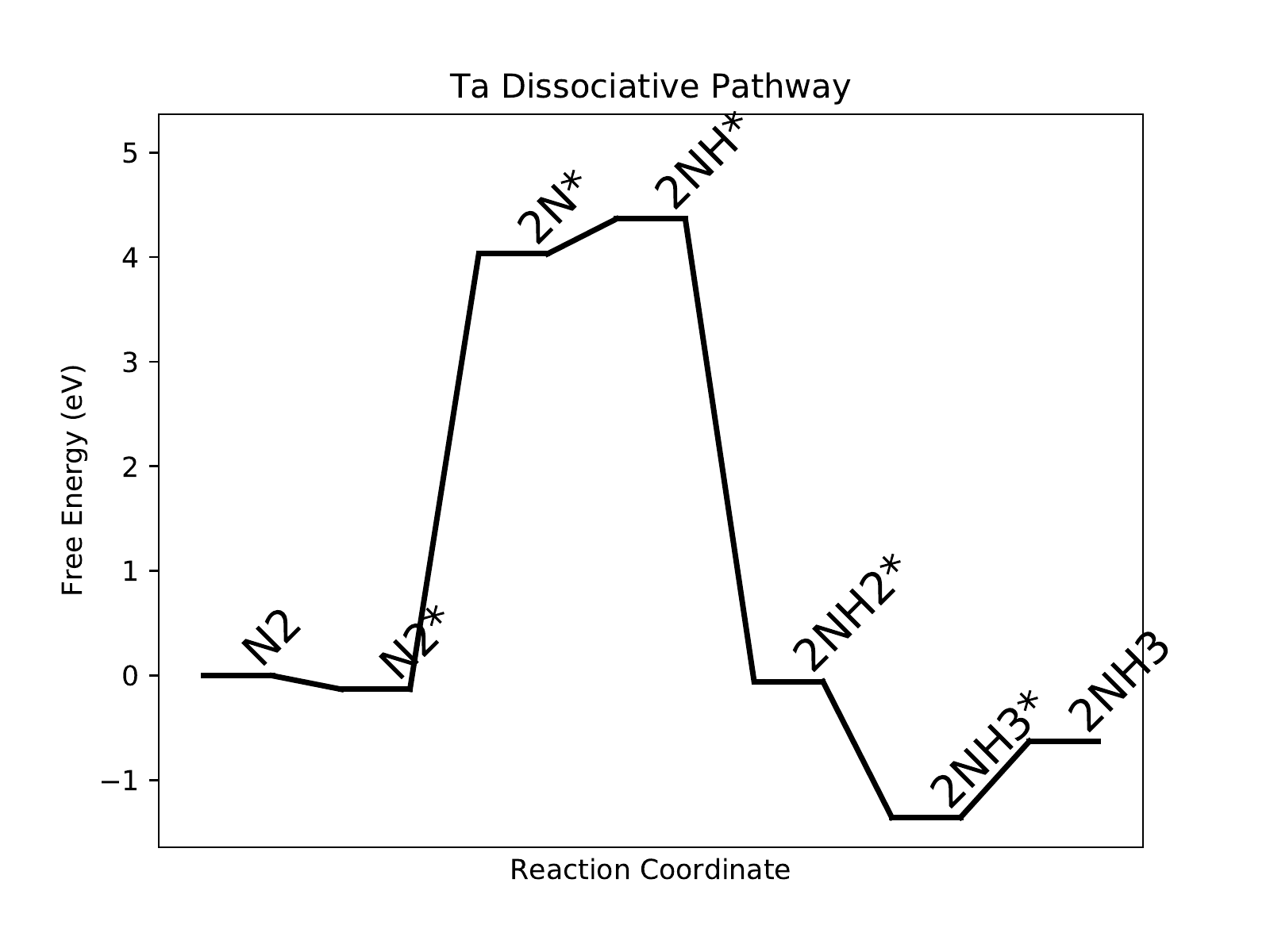}
\label{fig:Ta_dissociative}
\caption{Free energy diagram for Ta}
\end{figure}

\begin{figure}
\includegraphics[width=1\linewidth]{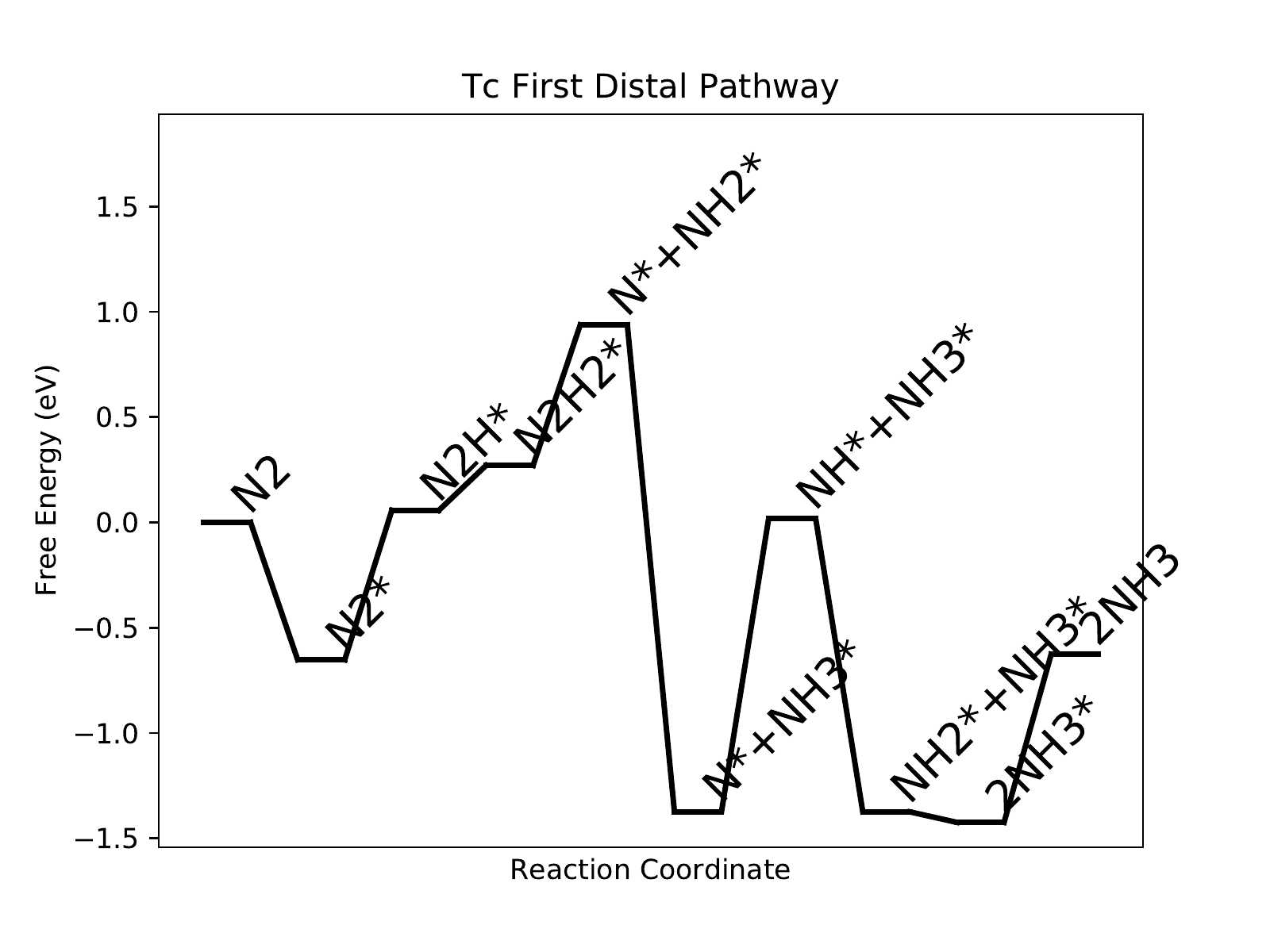}
\label{fig:Tc_distal_1}
\caption{Free energy diagram for Tc}
\end{figure}

\newpage
\begin{figure}
\includegraphics[width=1\linewidth]{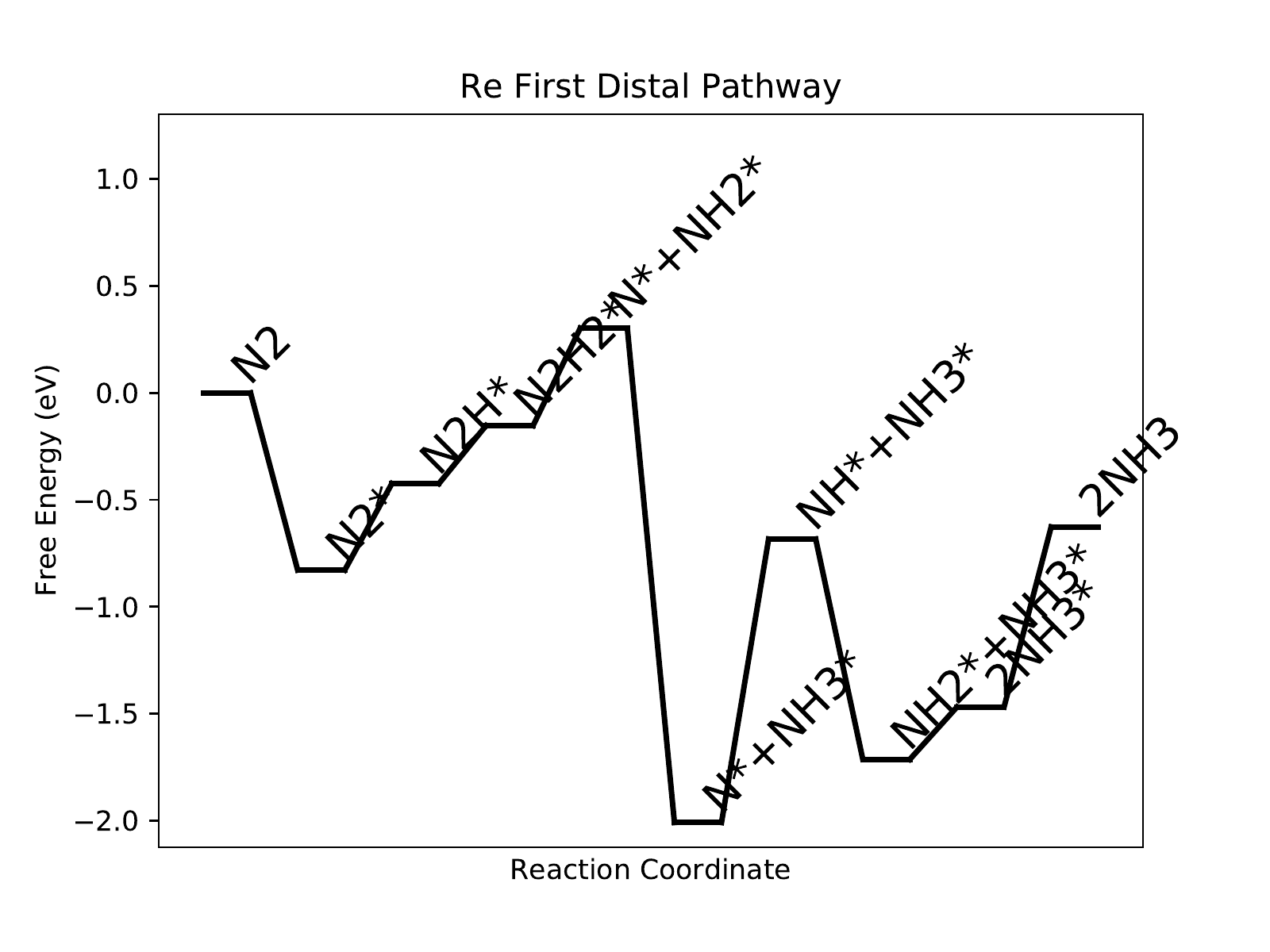}
\label{fig:Re_distal_1}
\caption{Free energy diagram for Re}
\end{figure}

\begin{figure}
\includegraphics[width=1\linewidth]{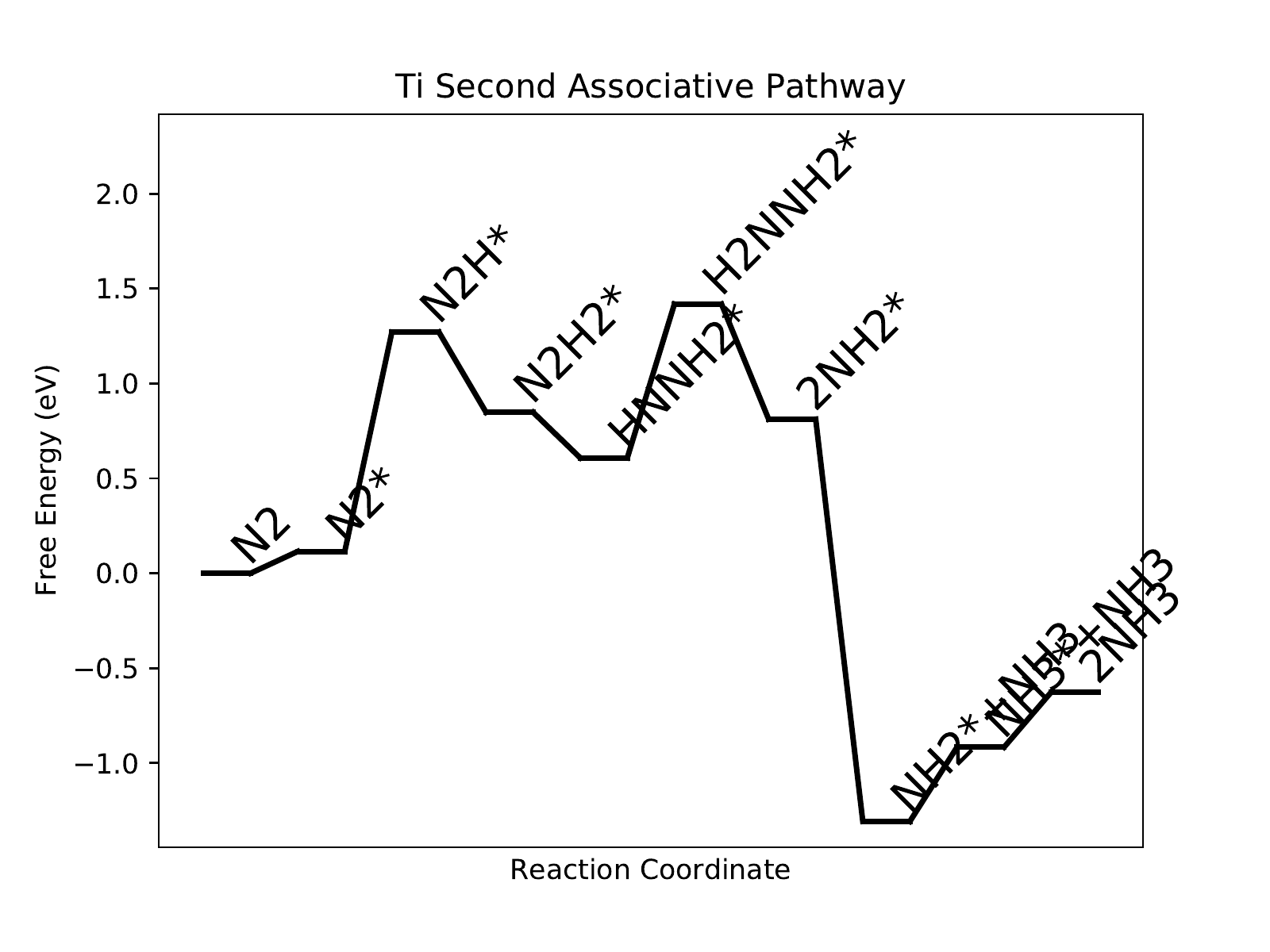}
\label{fig:Ti_associative_2}
\caption{Free energy diagram for Ti}
\end{figure}

\newpage
\begin{figure}
\includegraphics[width=1\linewidth]{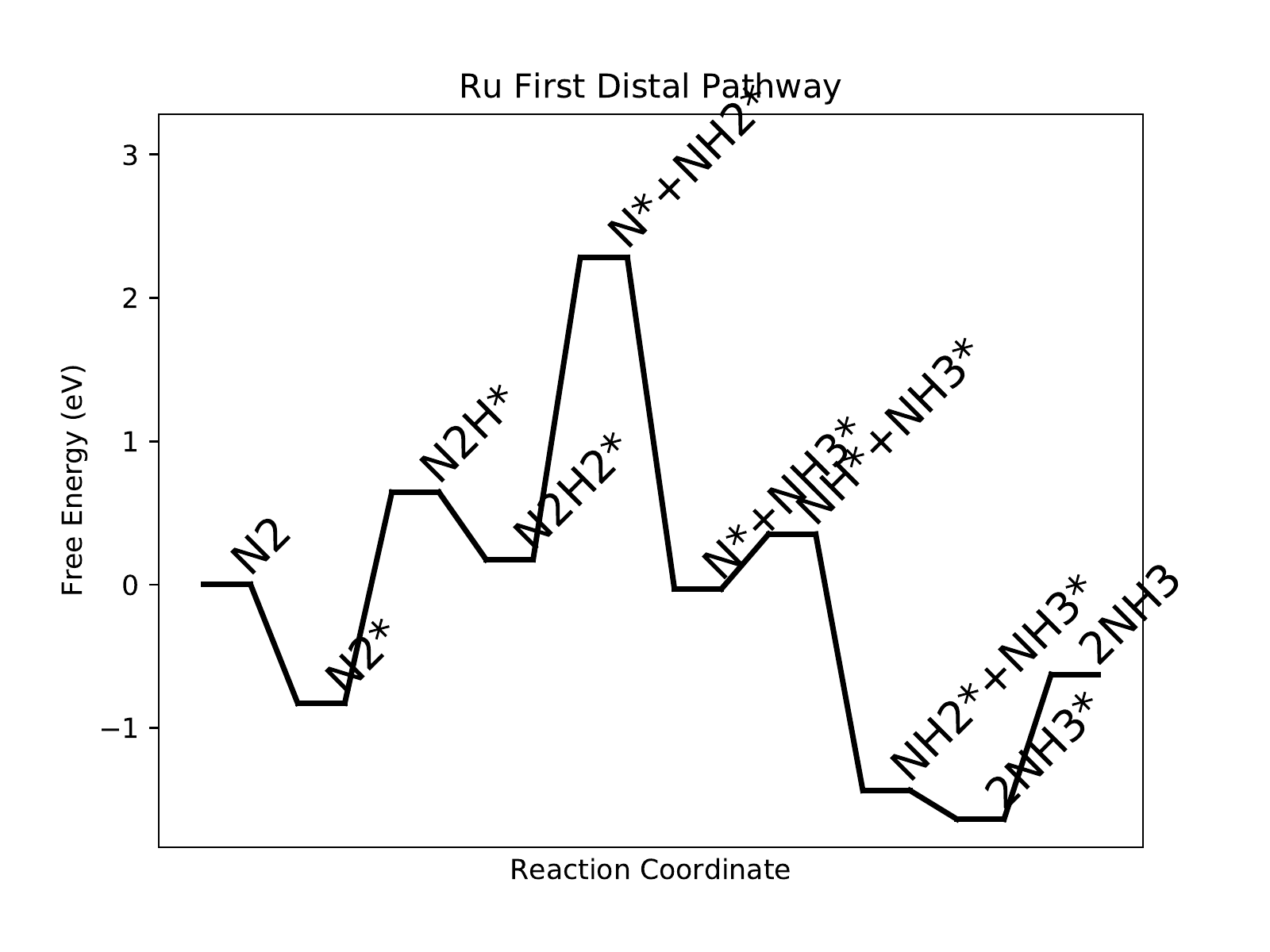}
\label{fig:Ru_distal_1}
\caption{Free energy diagram for Ru}
\end{figure}

\begin{figure}
\includegraphics[width=1\linewidth]{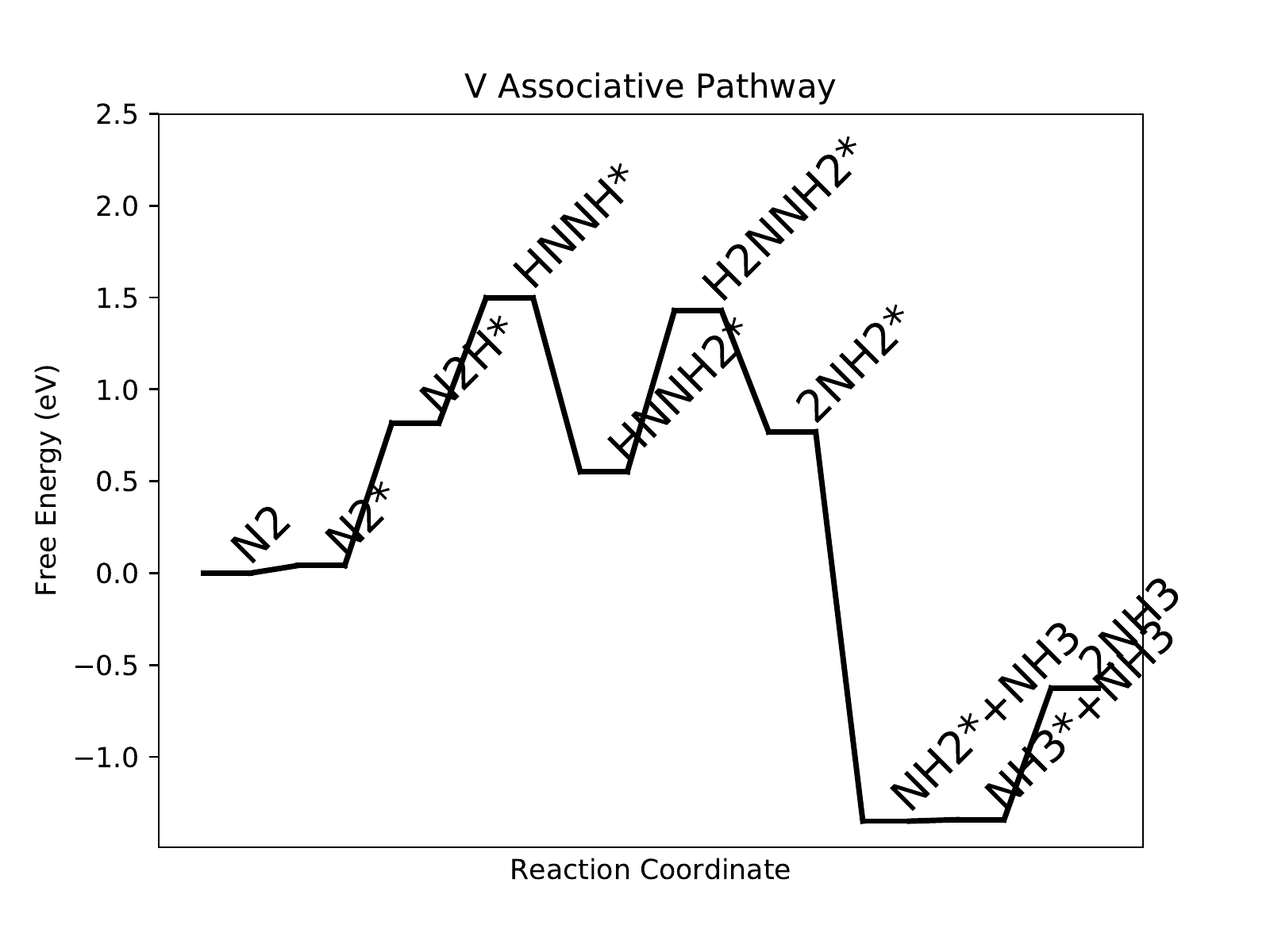}
\label{fig:V_associative}
\caption{Free energy diagram for V}
\end{figure}

\newpage
\begin{figure}
\includegraphics[width=1\linewidth]{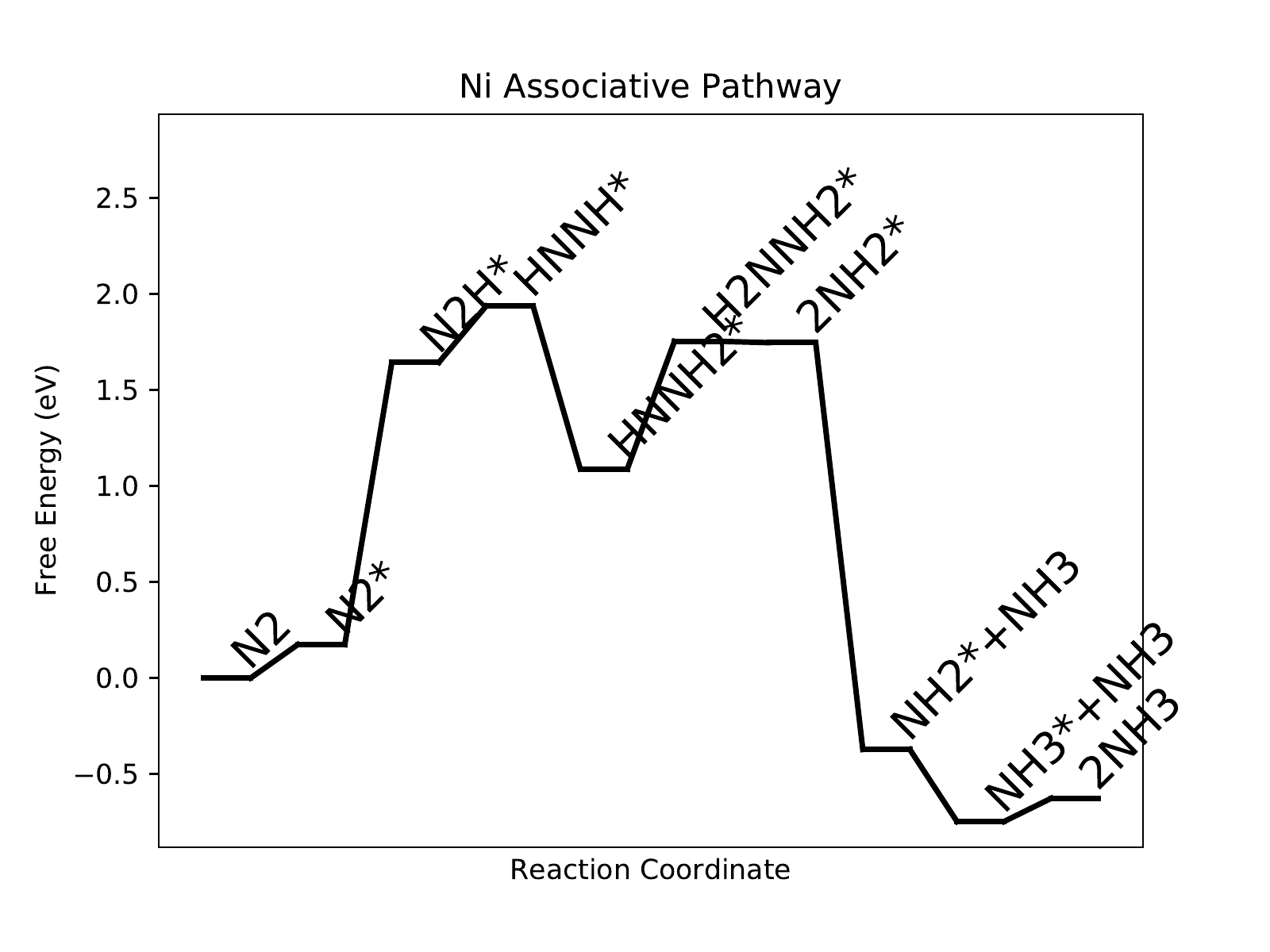}
\label{fig:Ni_associative}
\caption{Free energy diagram for Ni}
\end{figure}

\begin{figure}
\includegraphics[width=1\linewidth]{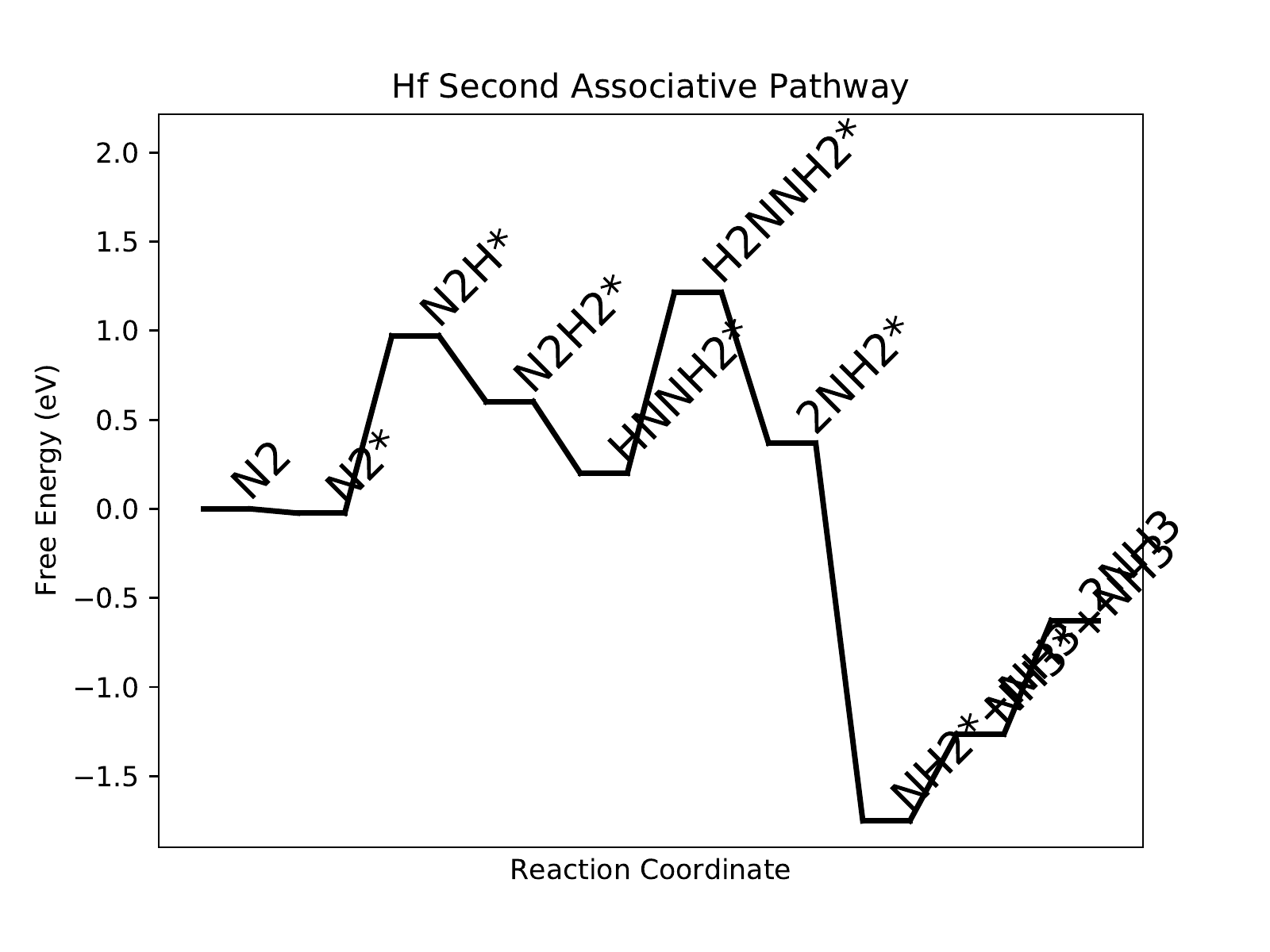}
\label{fig:Hf_associative_2}
\caption{Free energy diagram for Hf}
\end{figure}

\newpage
\begin{figure}
\includegraphics[width=1\linewidth]{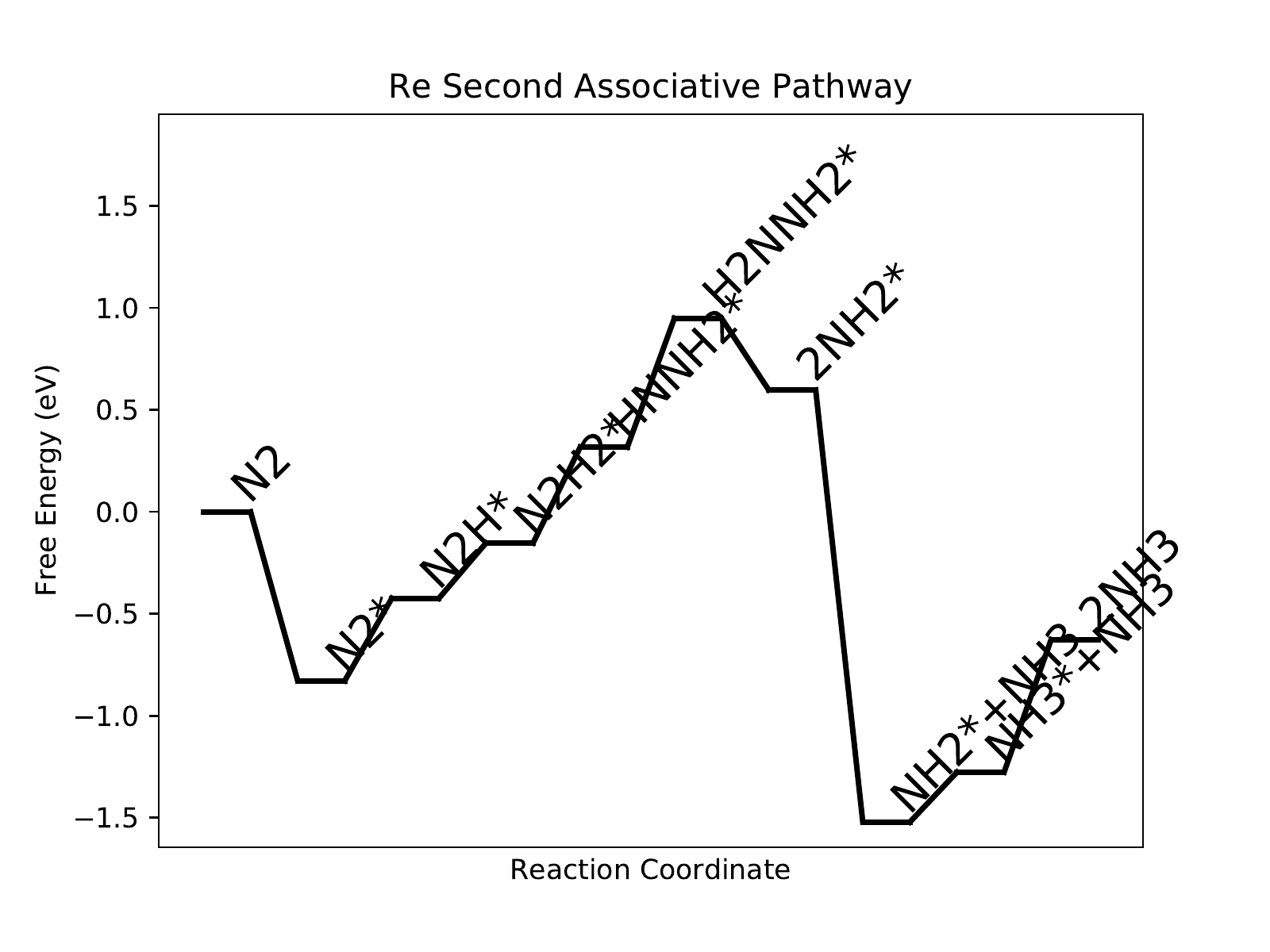}
\label{fig:Re_associative_2}
\caption{Free energy diagram for Re}
\end{figure}

\begin{figure}
\includegraphics[width=1\linewidth]{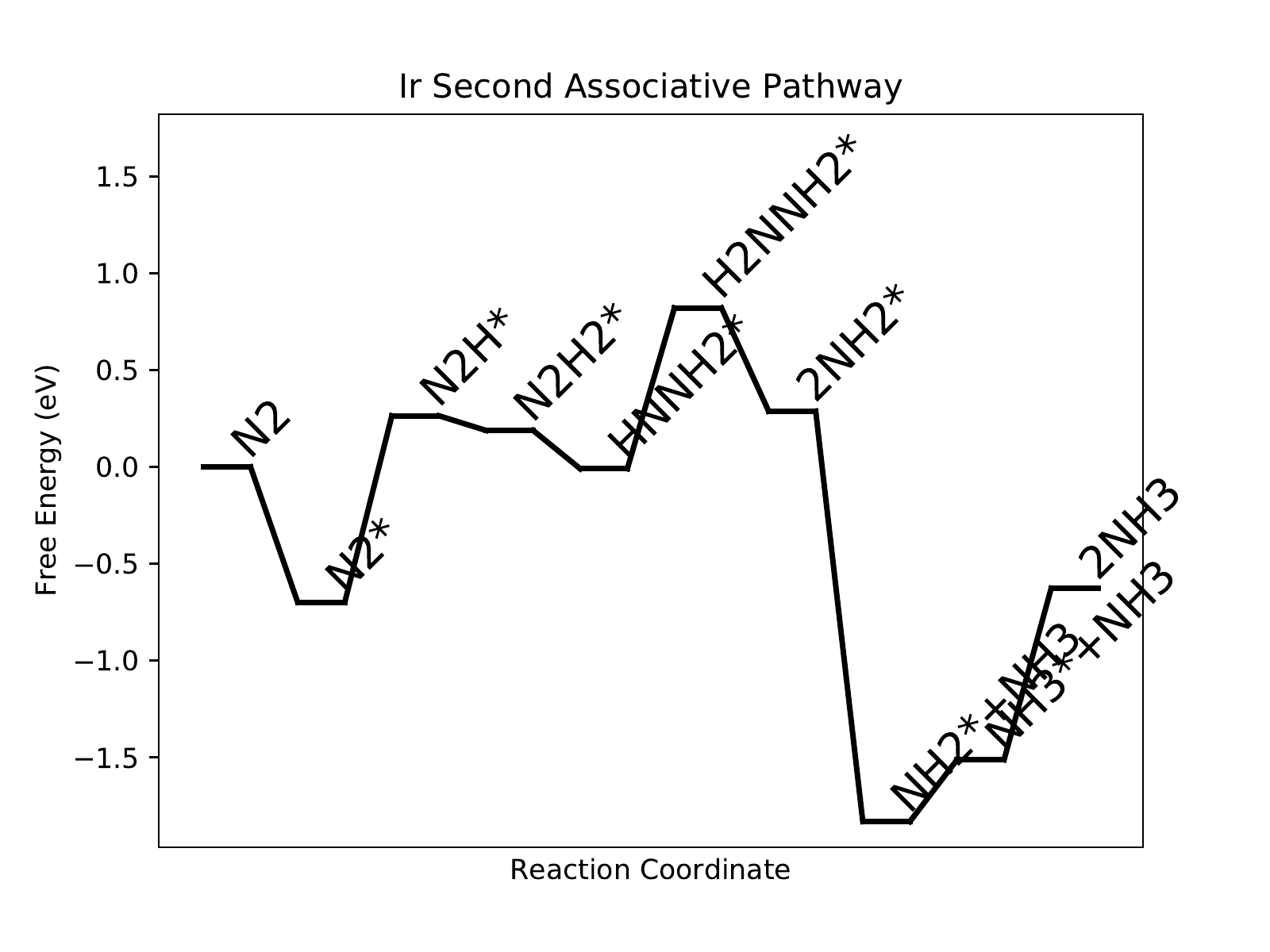}
\label{fig:Ir_associative_2}
\caption{Free energy diagram for Ir}
\end{figure}

\newpage
\begin{figure}
\includegraphics[width=1\linewidth]{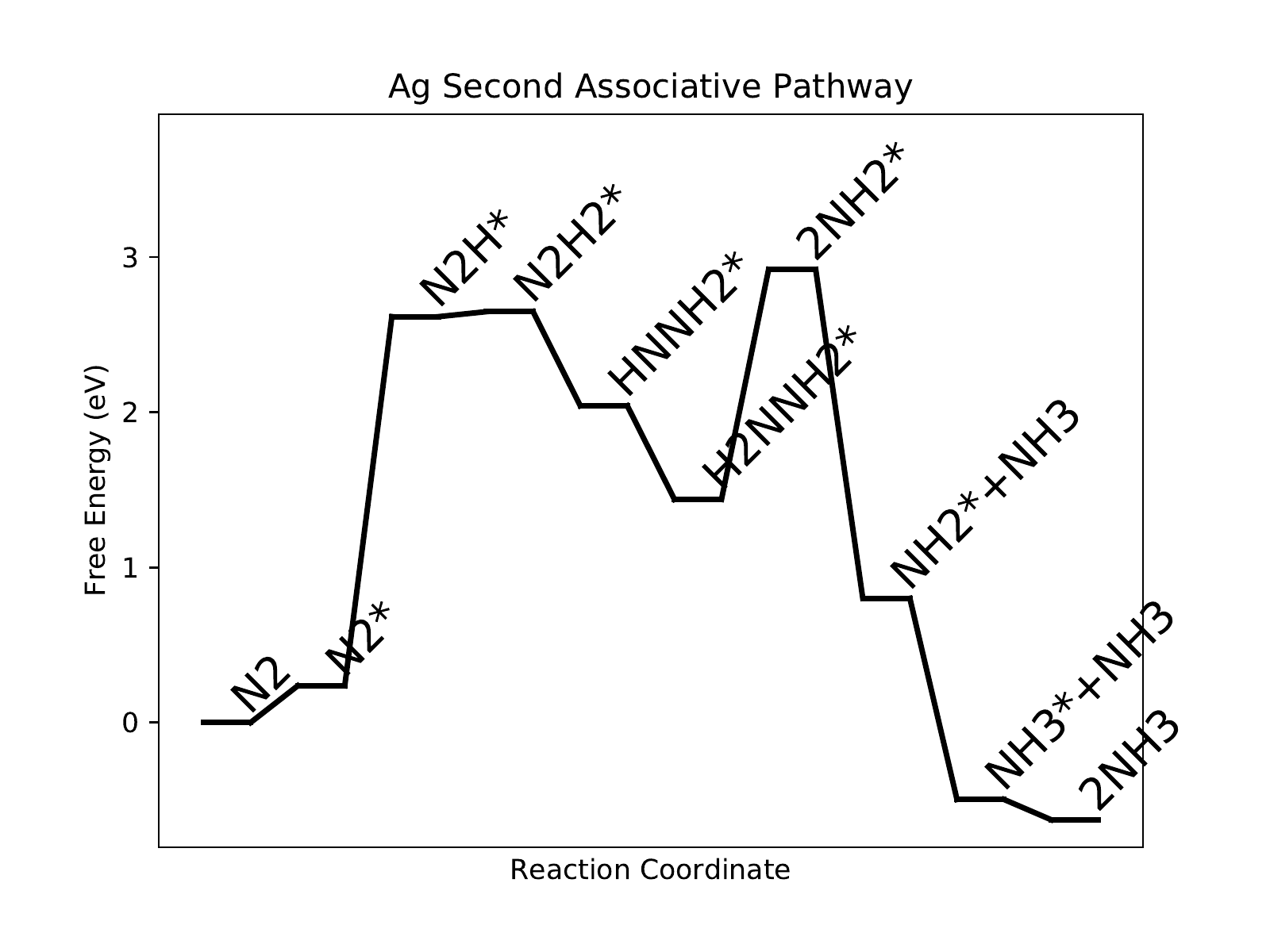}
\label{fig:Ag_associative_2}
\caption{Free energy diagram for Ag}
\end{figure}

\begin{figure}
\includegraphics[width=1\linewidth]{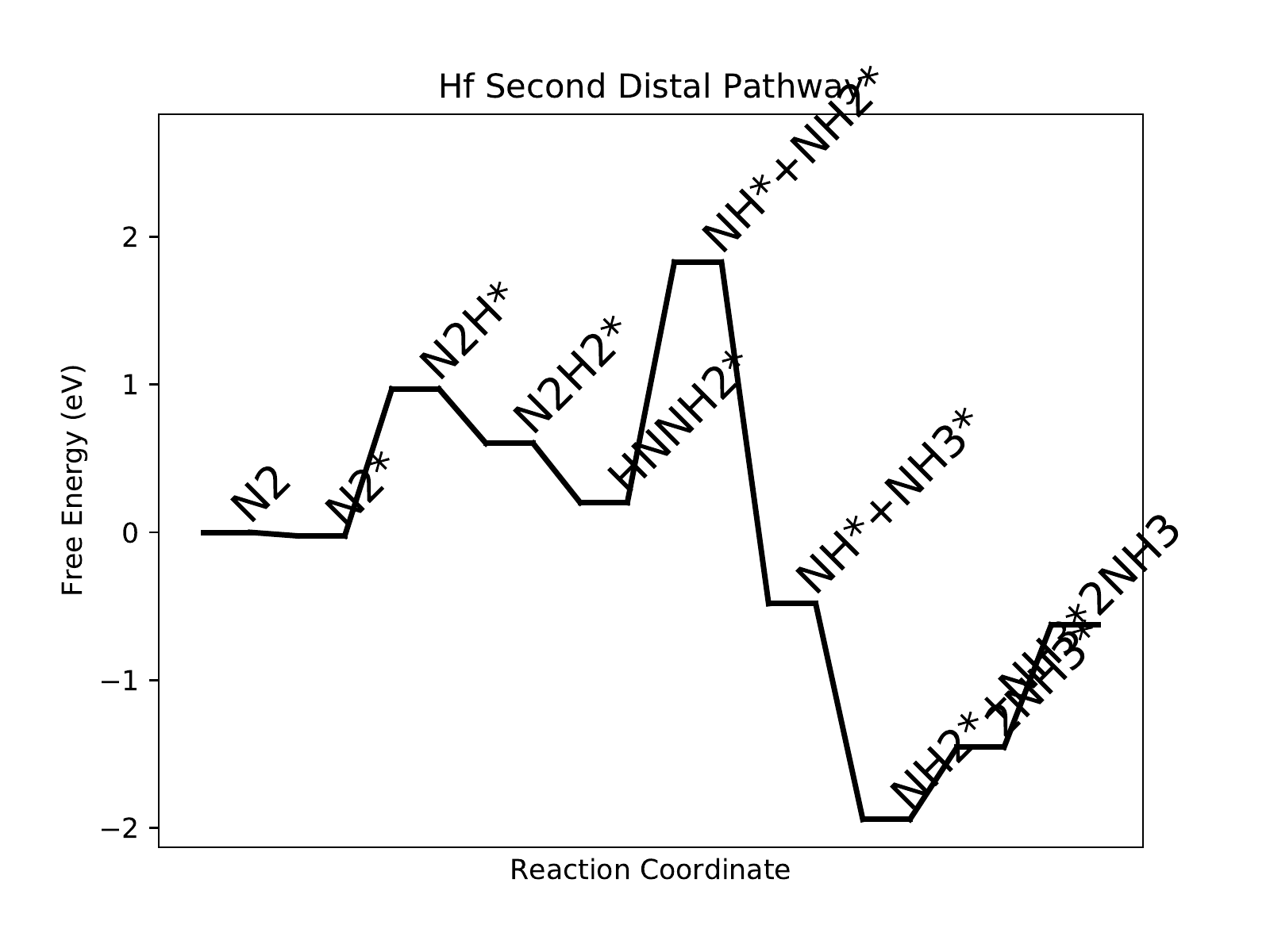}
\label{fig:Hf_distal_2}
\caption{Free energy diagram for Hf}
\end{figure}

\newpage
\begin{figure}
\includegraphics[width=1\linewidth]{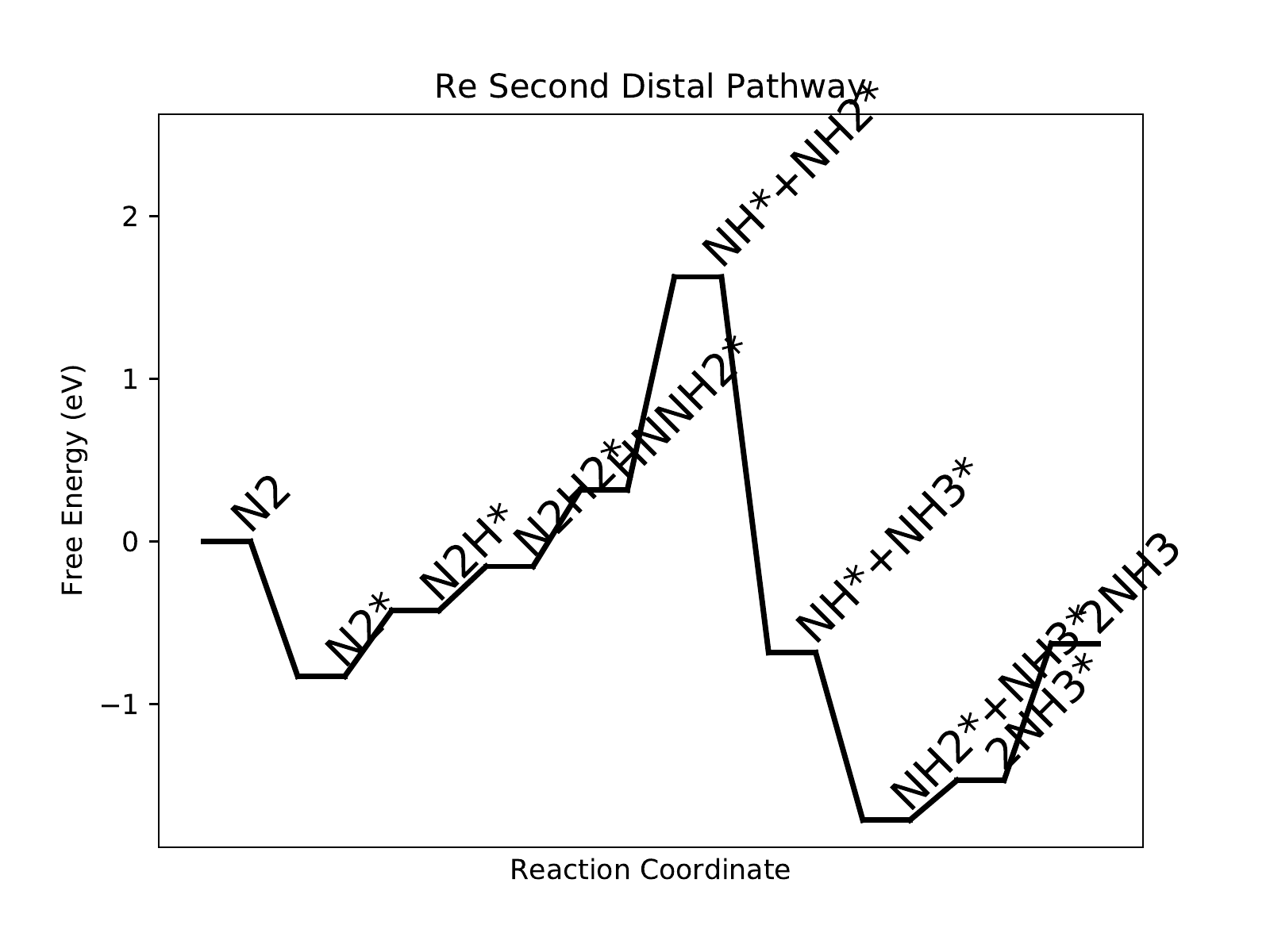}
\label{fig:Re_distal_2}
\caption{Free energy diagram for Re}
\end{figure}

\begin{figure}
\includegraphics[width=1\linewidth]{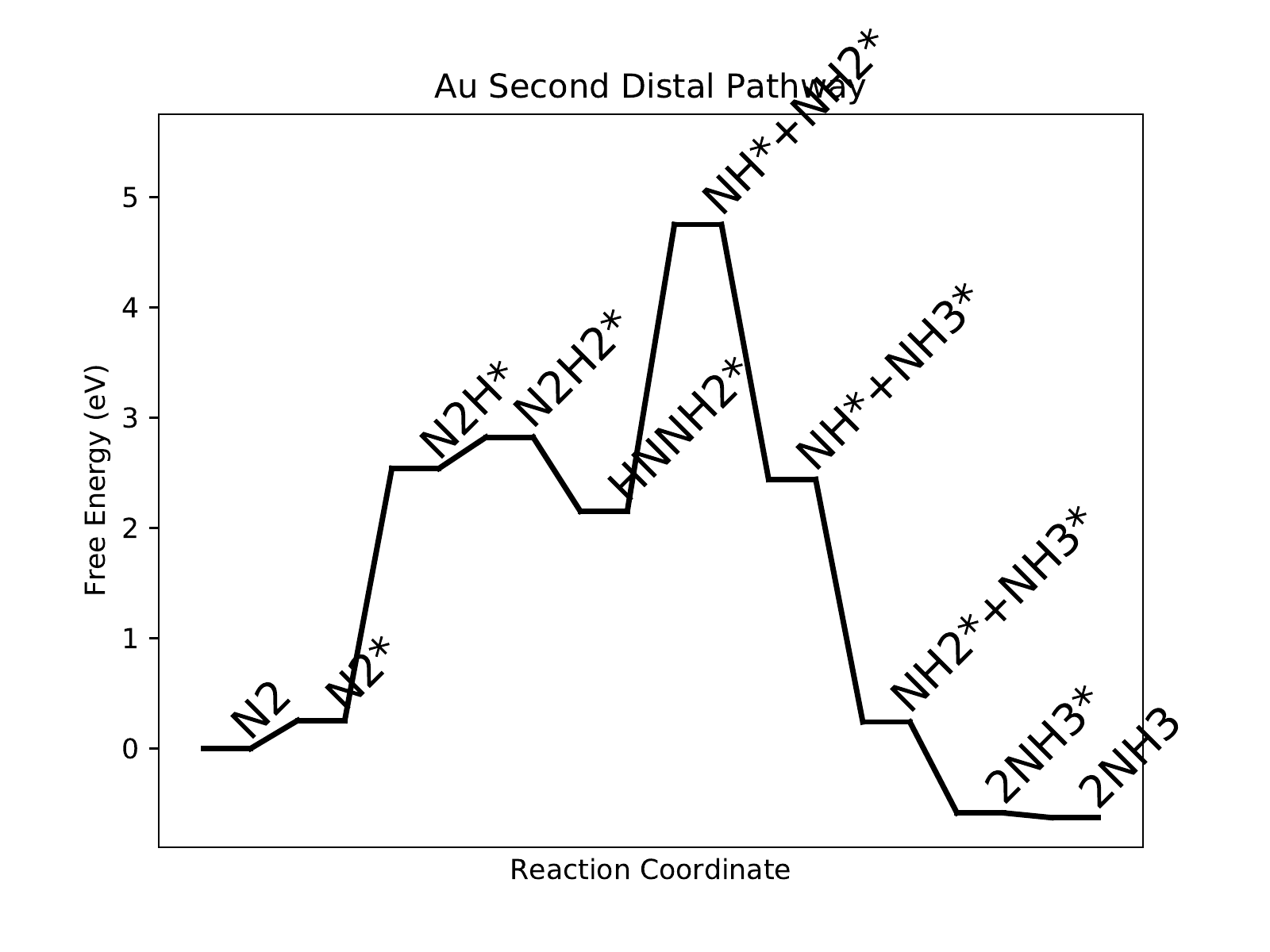}
\label{fig:Au_distal_2}
\caption{Free energy diagram for Au}
\end{figure}

\newpage
\begin{figure}
\includegraphics[width=1\linewidth]{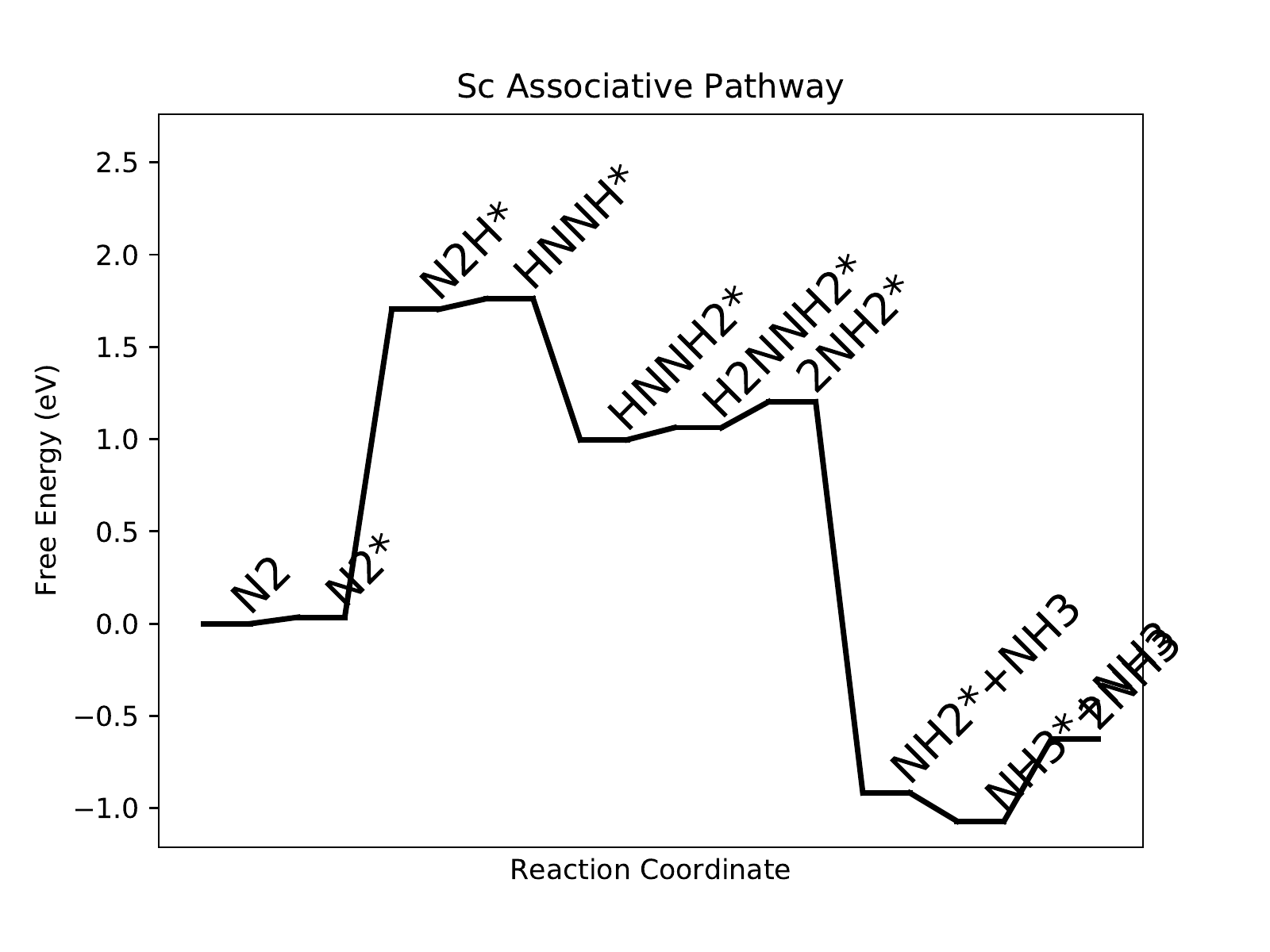}
\label{fig:Sc_associative}
\caption{Free energy diagram for Sc}
\end{figure}

\begin{figure}
\includegraphics[width=1\linewidth]{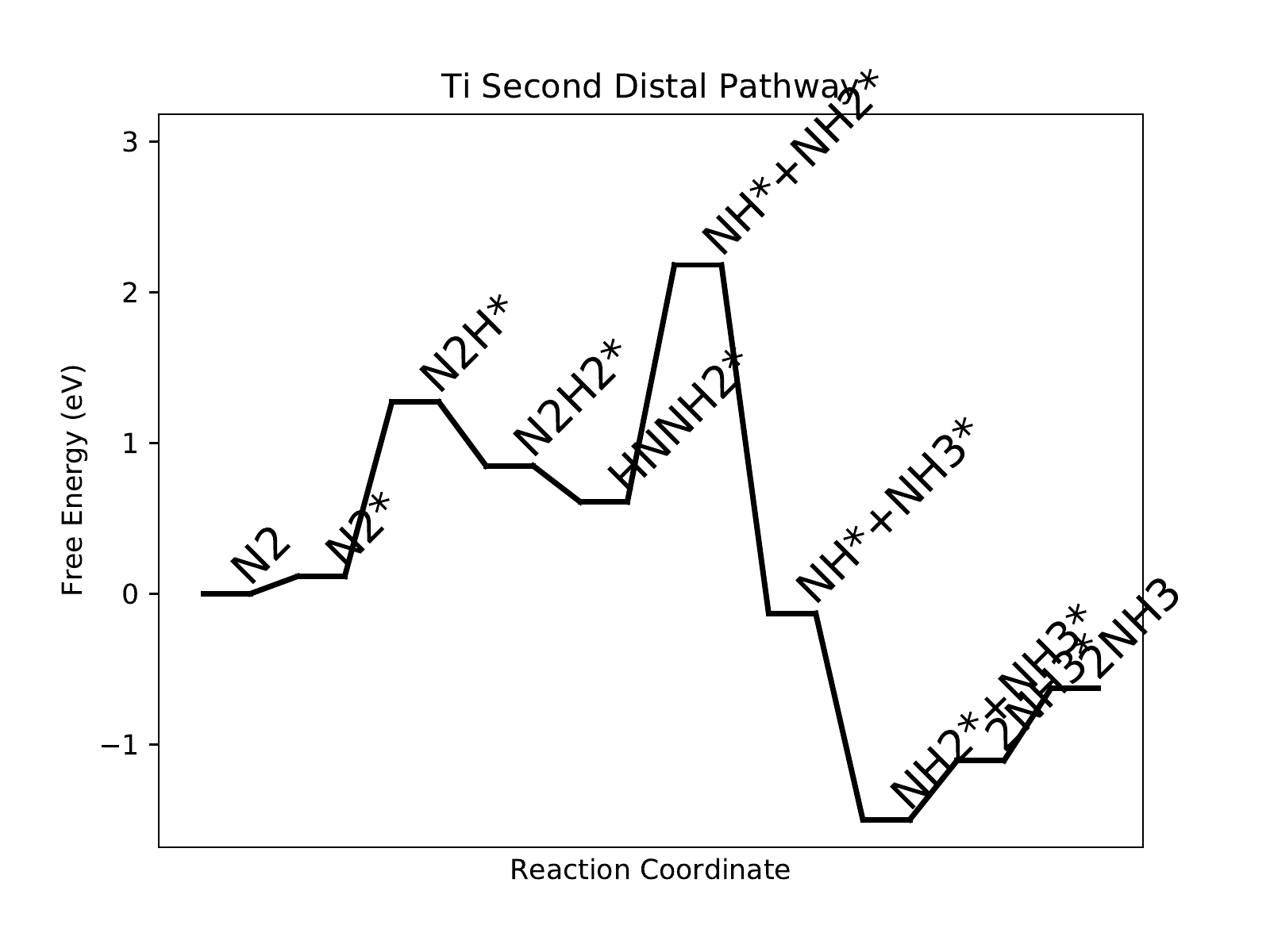}
\label{fig:Ti_distal_2}
\caption{Free energy diagram for Ti}
\end{figure}

\newpage
\begin{figure}
\includegraphics[width=1\linewidth]{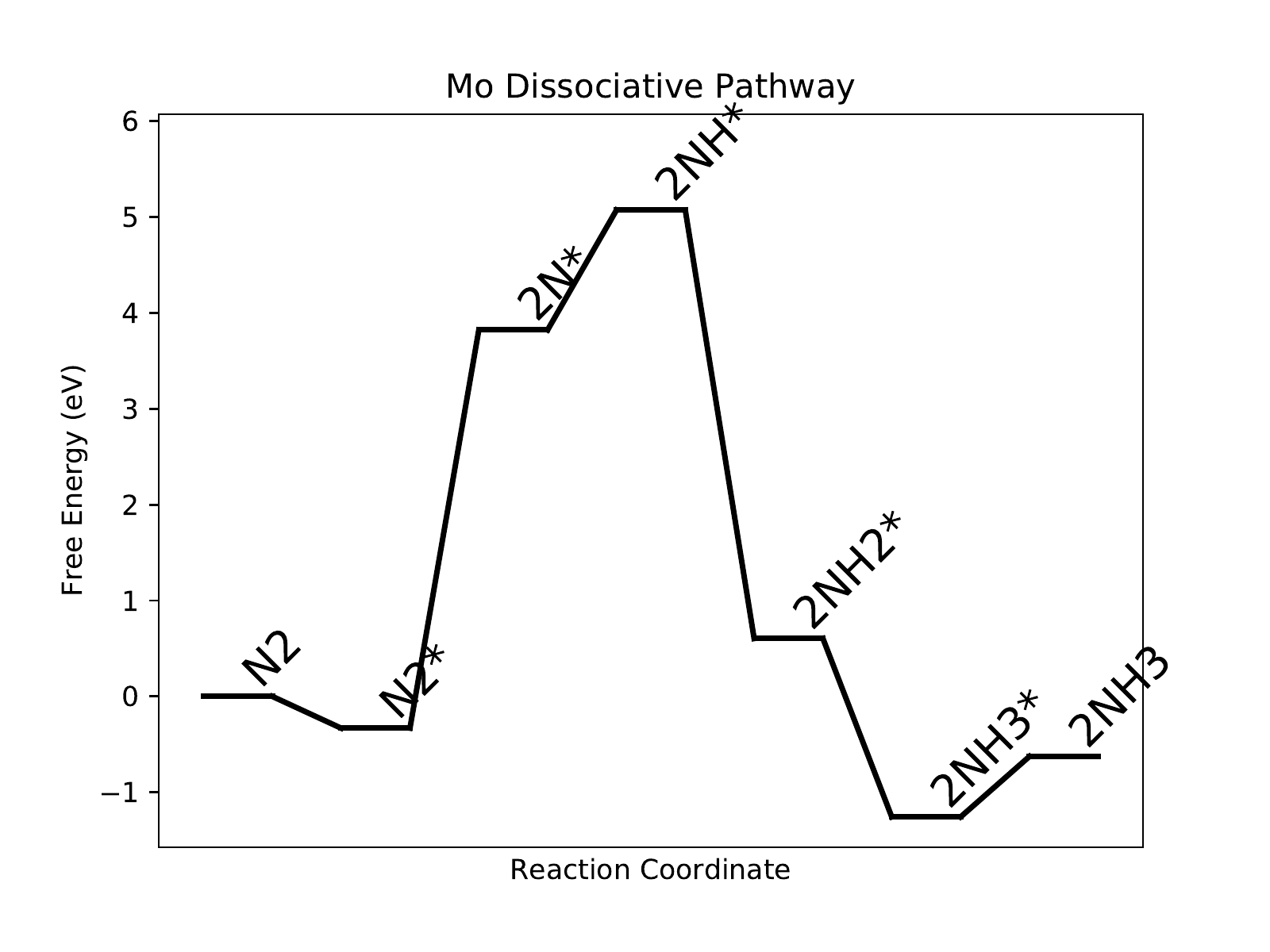}
\label{fig:Mo_dissociative}
\caption{Free energy diagram for Mo}
\end{figure}

\begin{figure}
\includegraphics[width=1\linewidth]{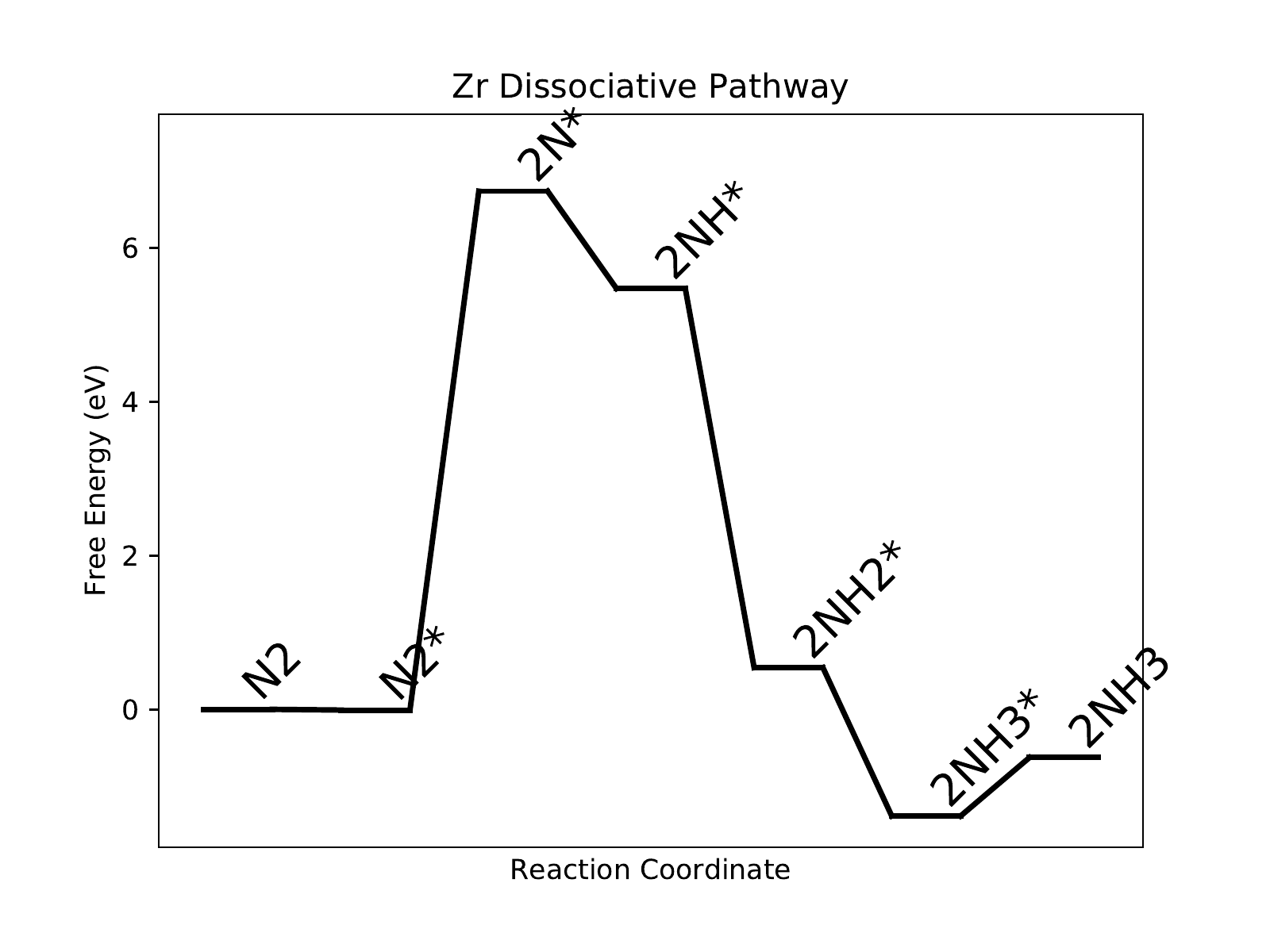}
\label{fig:Zr_dissociative}
\caption{Free energy diagram for Zr}
\end{figure}

\newpage
\begin{figure}
\includegraphics[width=1\linewidth]{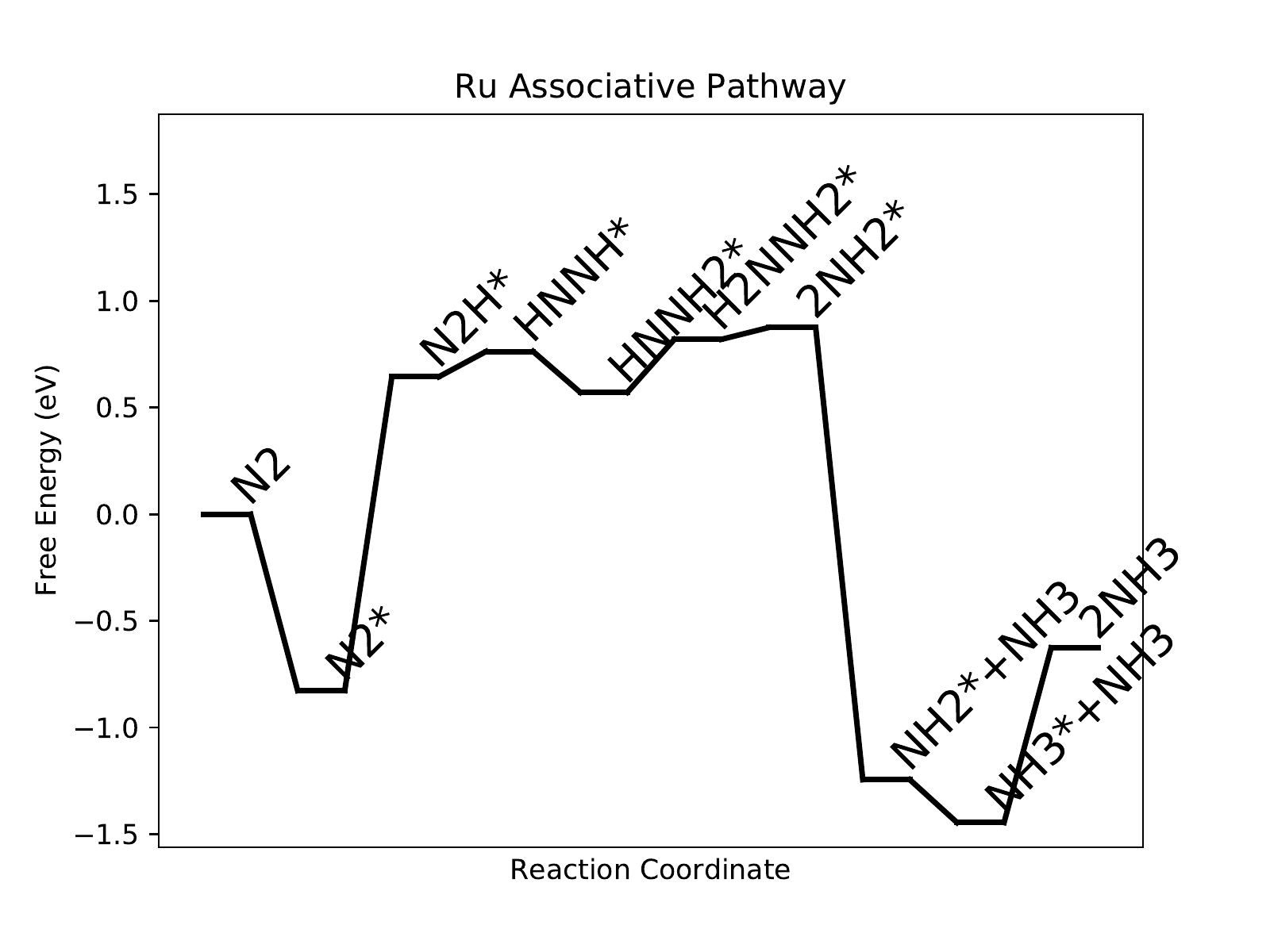}
\label{fig:Ru_associative}
\caption{Free energy diagram for Ru}
\end{figure}

\begin{figure}
\includegraphics[width=1\linewidth]{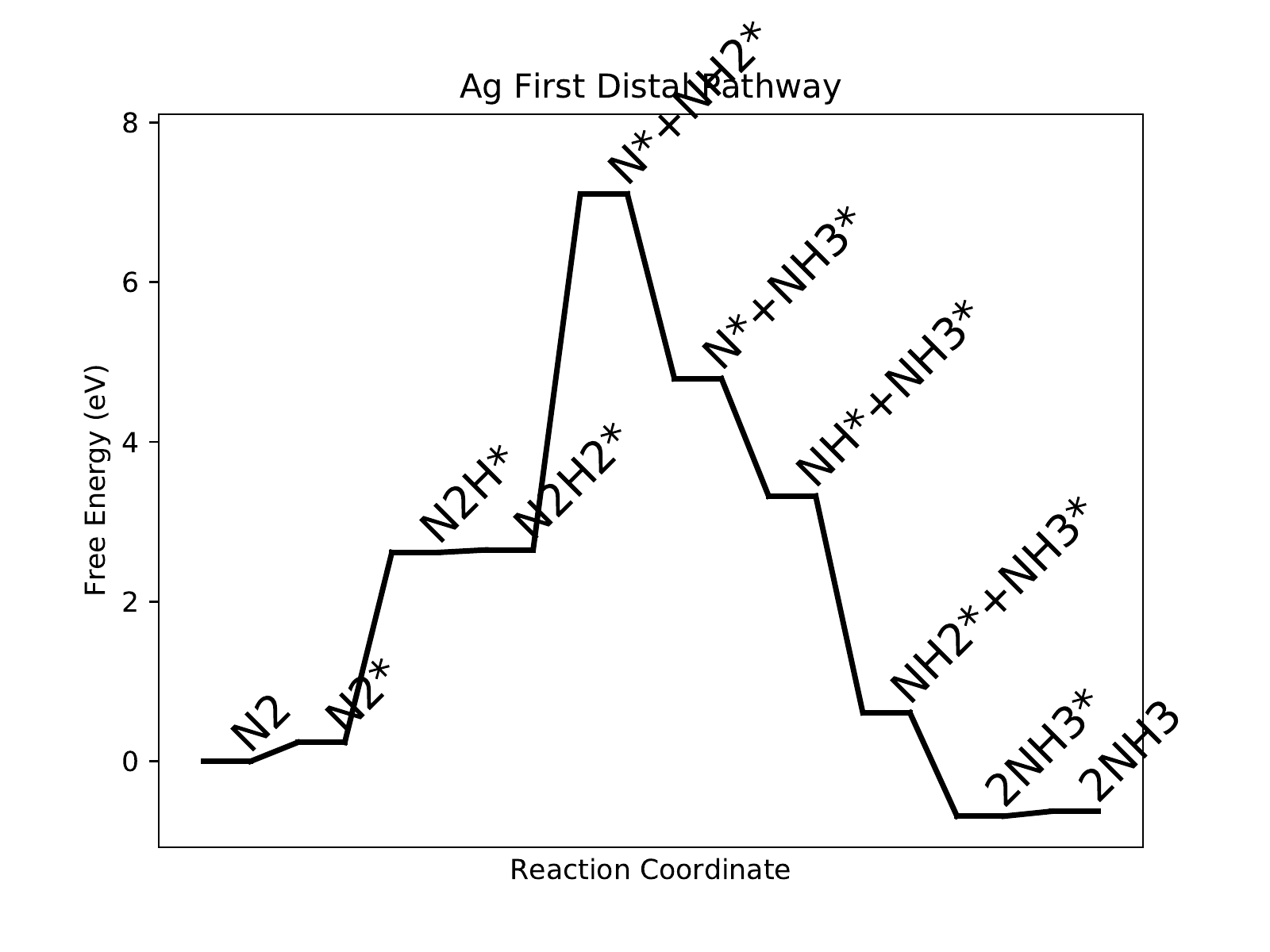}
\label{fig:Ag_distal_1}
\caption{Free energy diagram for Ag}
\end{figure}

\newpage
\begin{figure}
\includegraphics[width=1\linewidth]{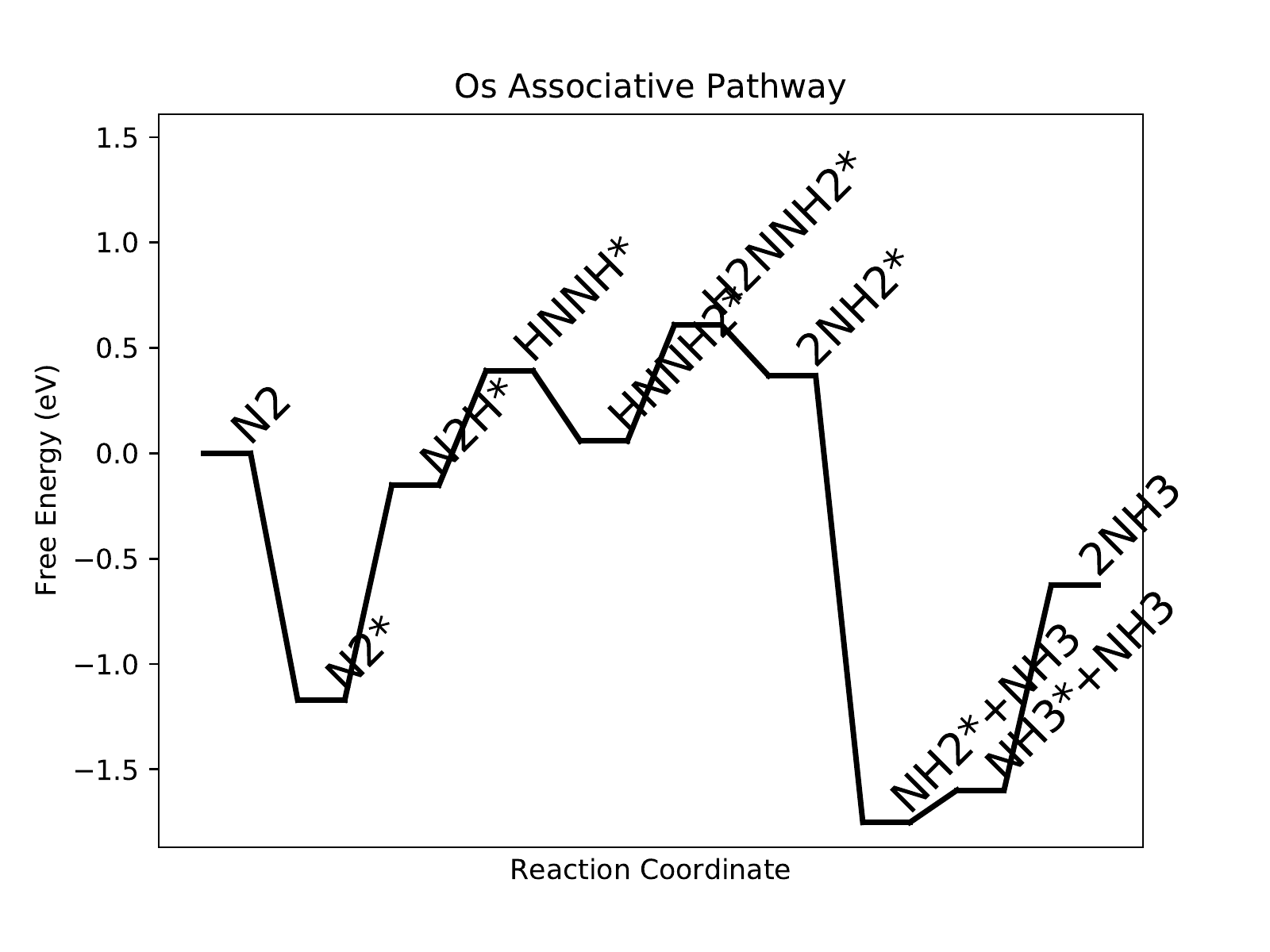}
\label{fig:Os_associative}
\caption{Free energy diagram for Os}
\end{figure}

\begin{figure}
\includegraphics[width=1\linewidth]{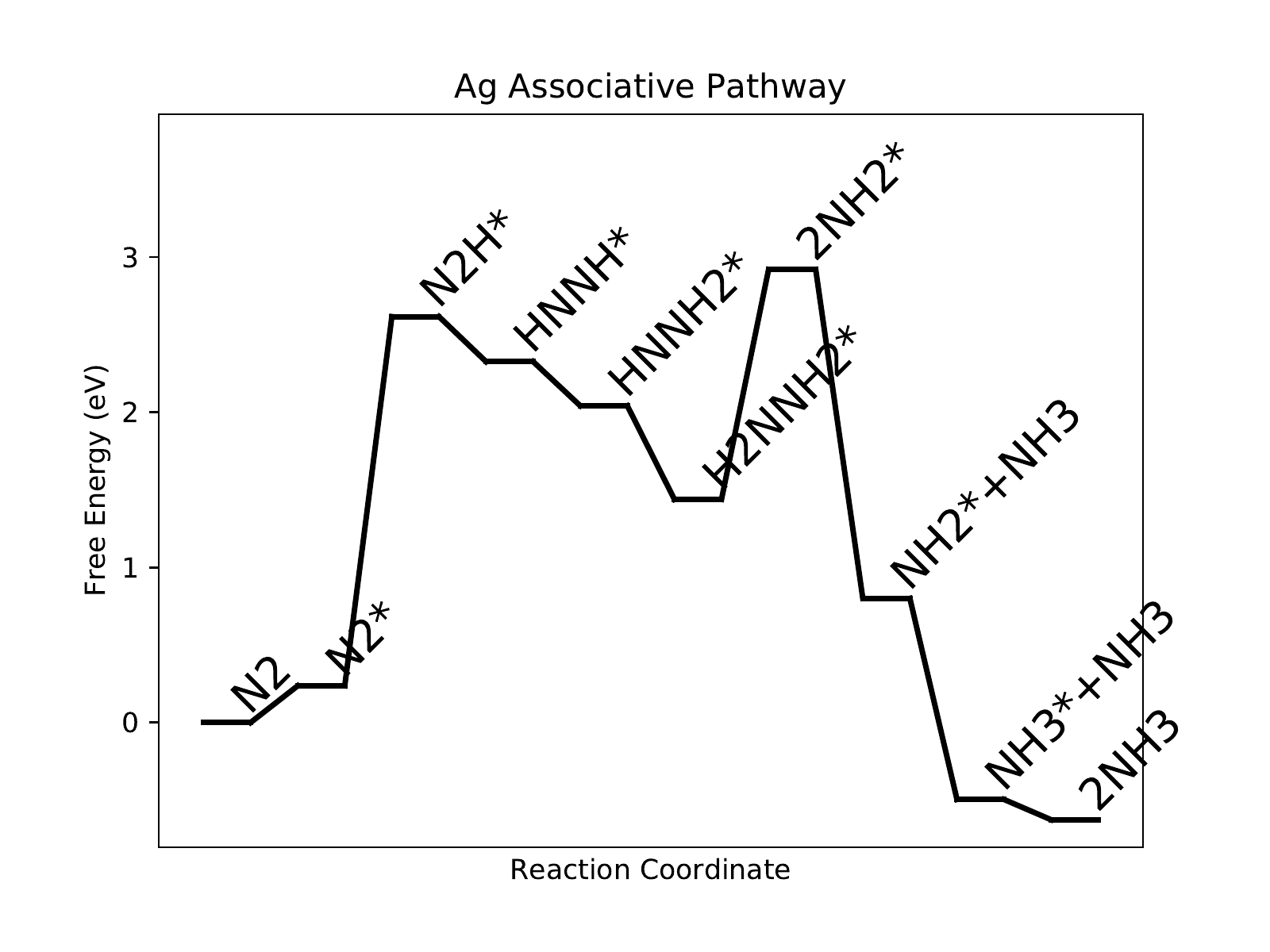}
\label{fig:Ag_associative}
\caption{Free energy diagram for Ag}
\end{figure}

\newpage
\begin{figure}
\includegraphics[width=1\linewidth]{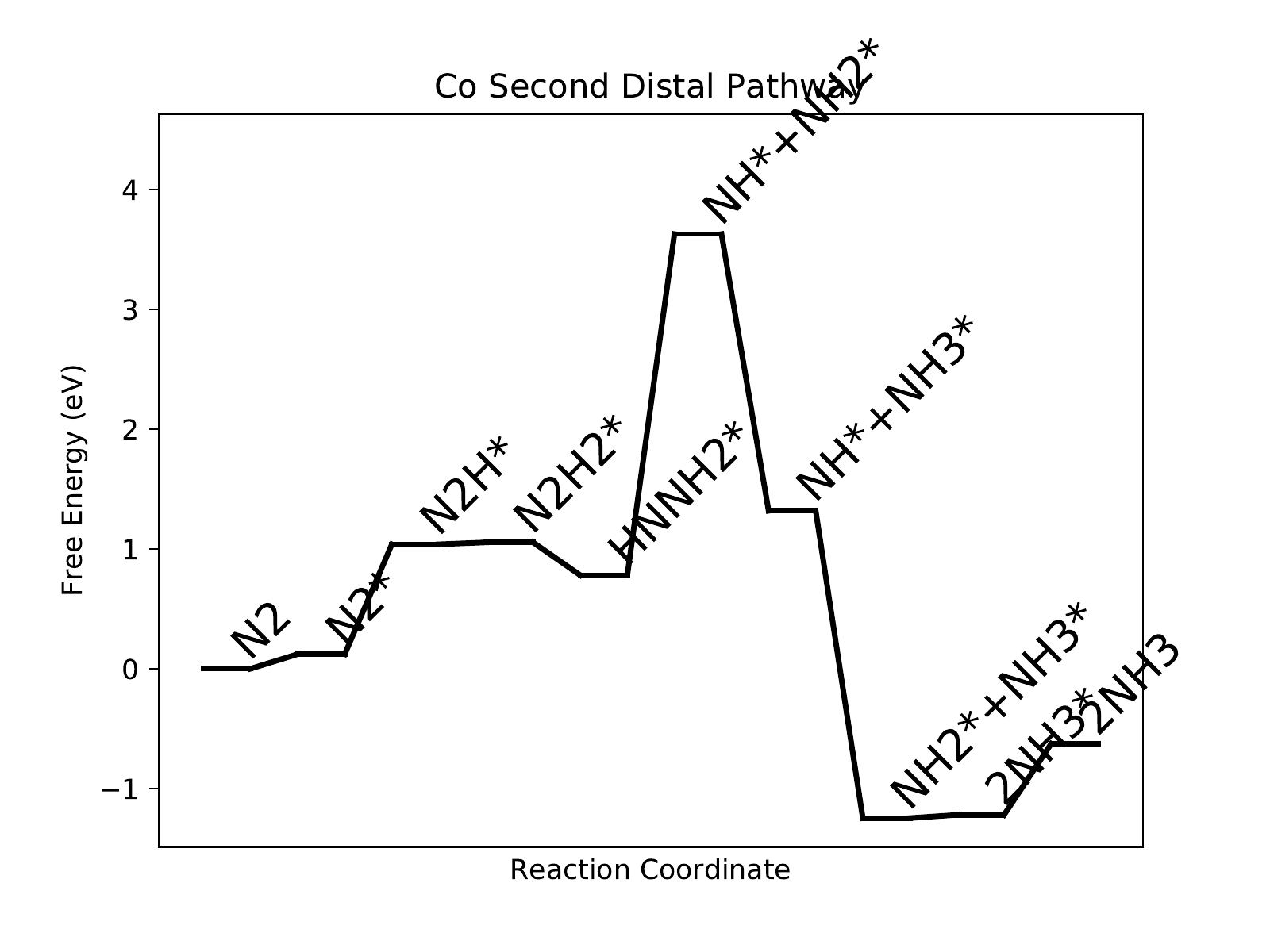}
\label{fig:Co_distal_2}
\caption{Free energy diagram for Co}
\end{figure}

\begin{figure}
\includegraphics[width=1\linewidth]{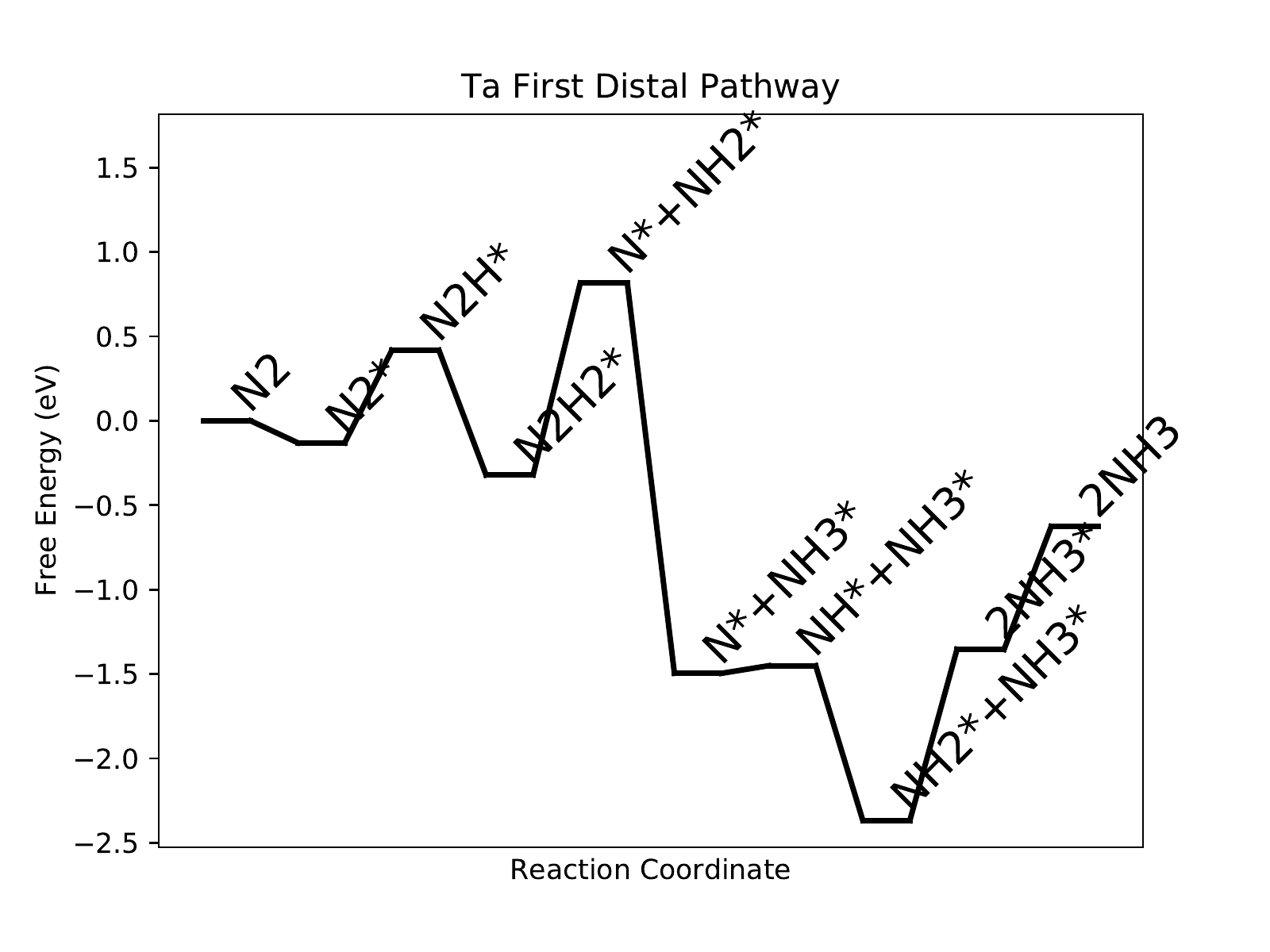}
\label{fig:Ta_distal_1}
\caption{Free energy diagram for Ta}
\end{figure}

\newpage
\begin{figure}
\includegraphics[width=1\linewidth]{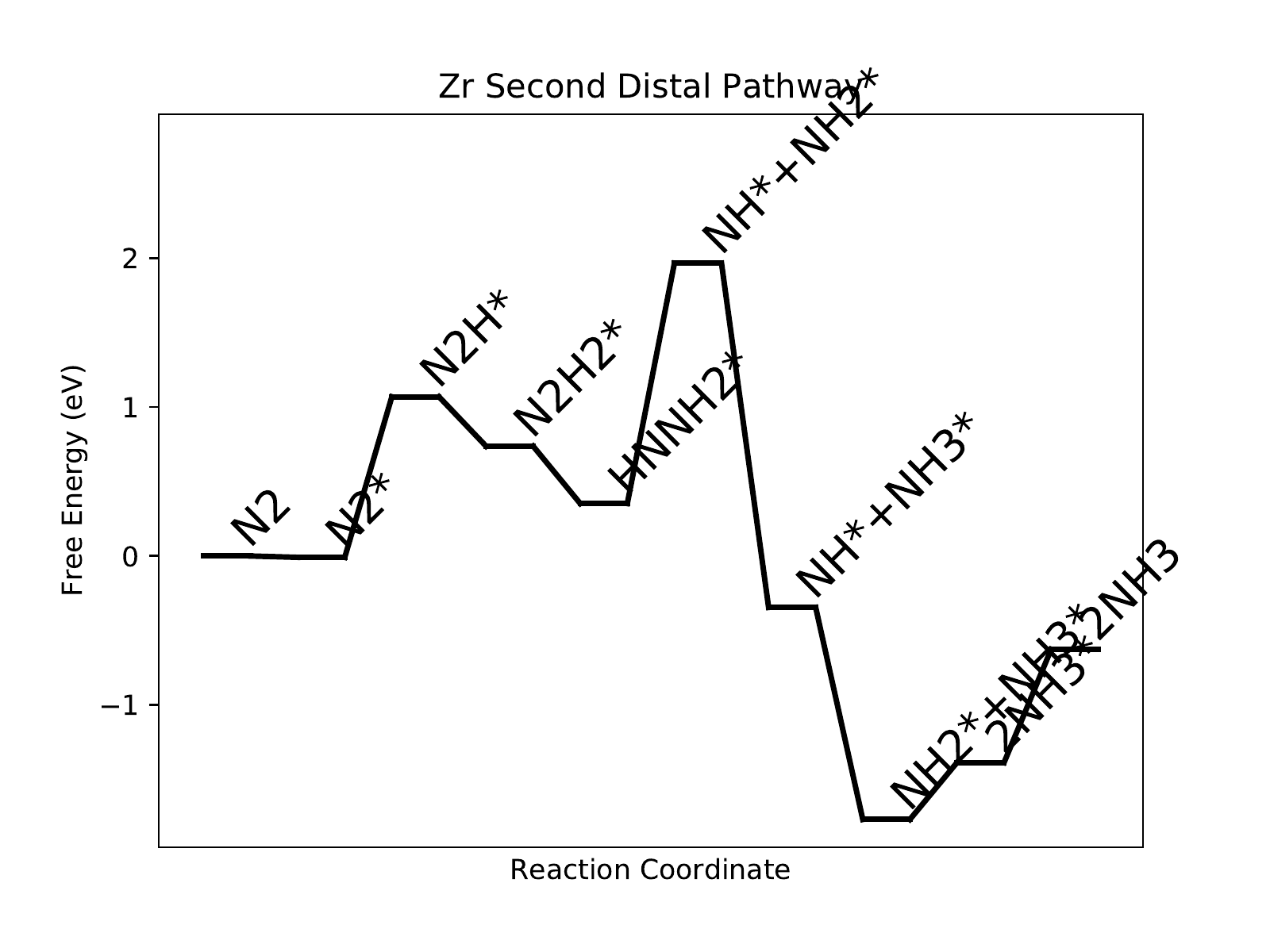}
\label{fig:Zr_distal_2}
\caption{Free energy diagram for Zr}
\end{figure}

\begin{figure}
\includegraphics[width=1\linewidth]{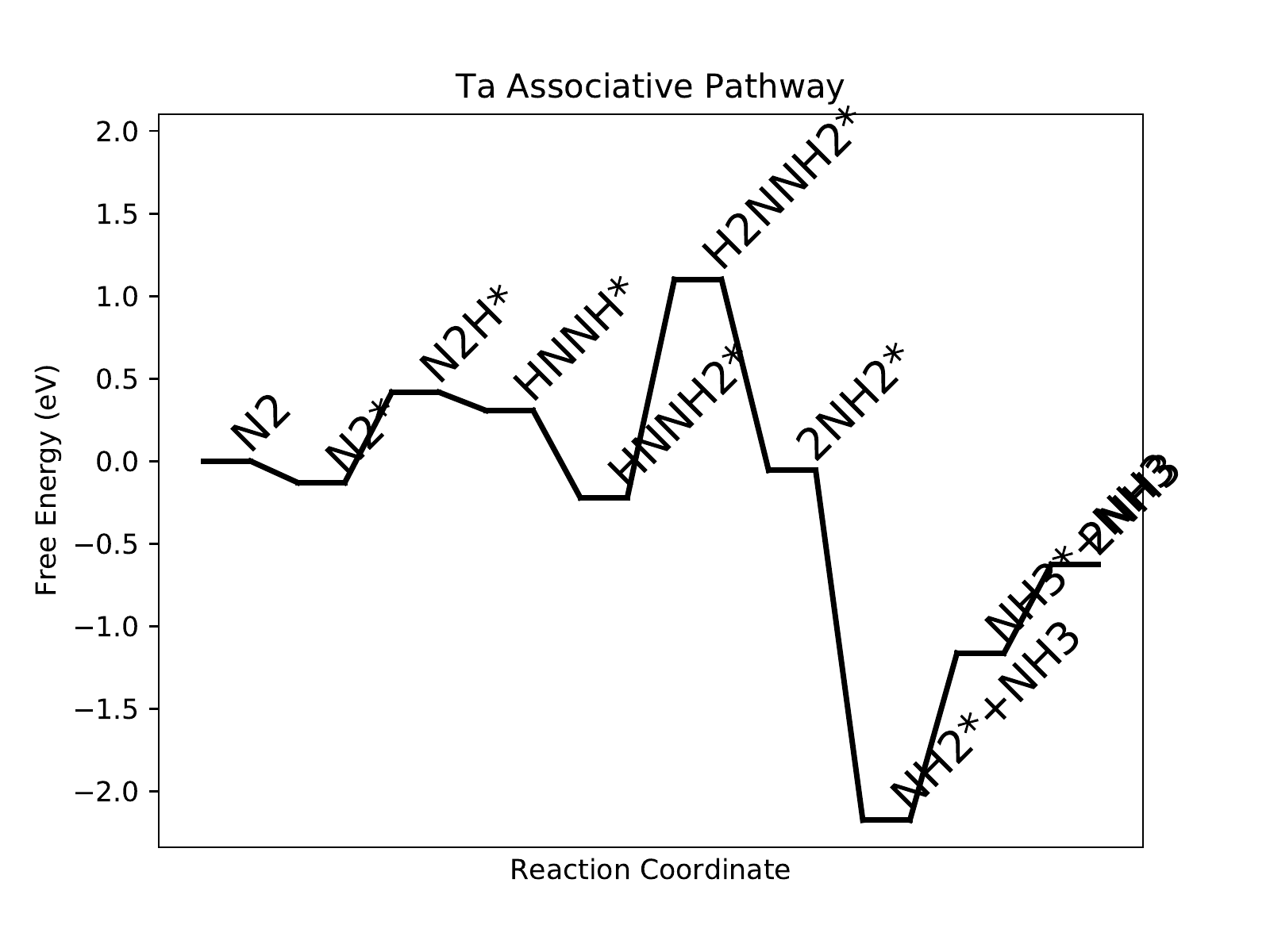}
\label{fig:Ta_associative}
\caption{Free energy diagram for Ta}
\end{figure}

\newpage
\begin{figure}
\includegraphics[width=1\linewidth]{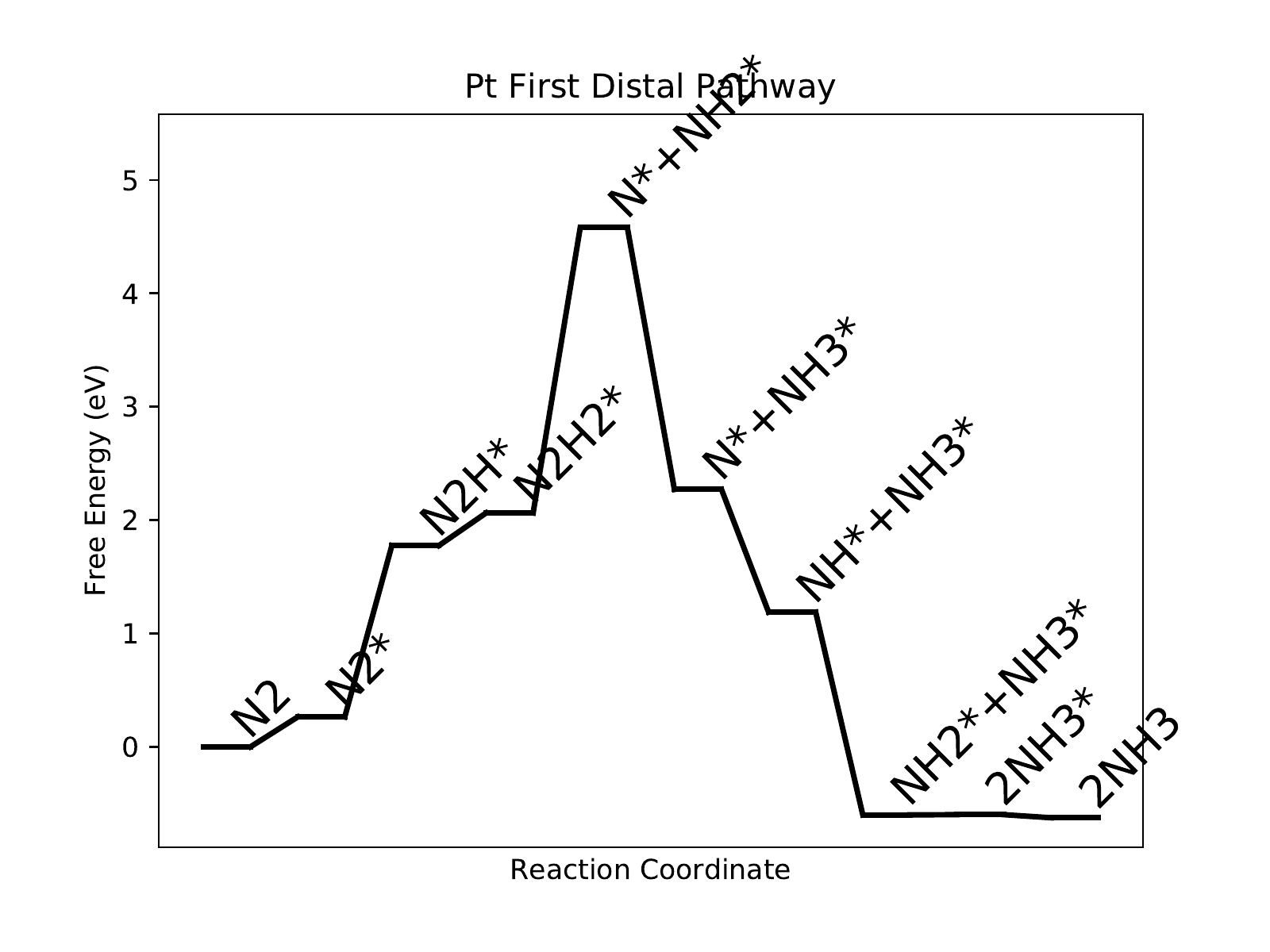}
\label{fig:Pt_distal_1}
\caption{Free energy diagram for Pt}
\end{figure}

\begin{figure}
\includegraphics[width=1\linewidth]{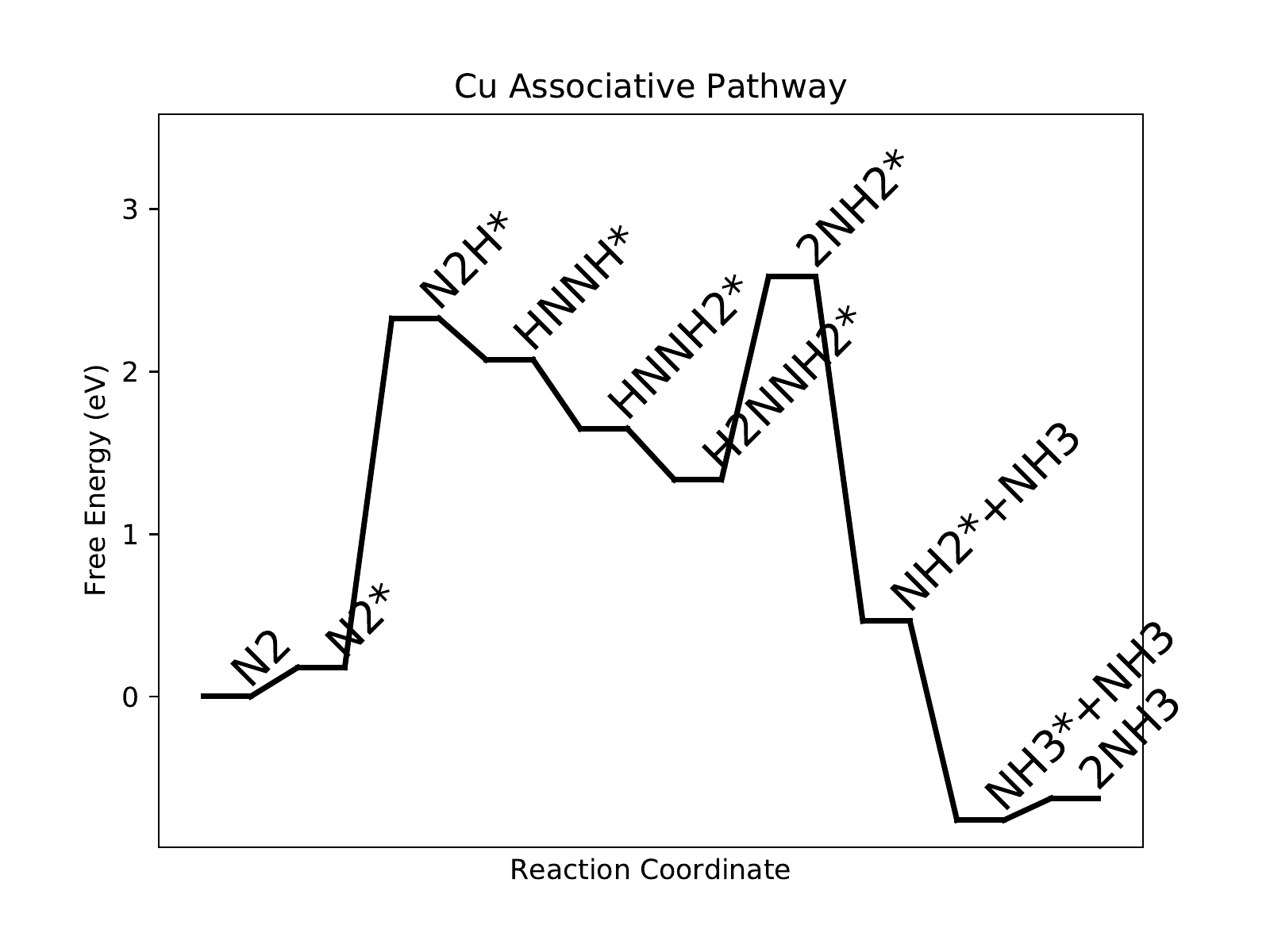}
\label{fig:Cu_associative}
\caption{Free energy diagram for Cu}
\end{figure}

\newpage
\begin{figure}
\includegraphics[width=1\linewidth]{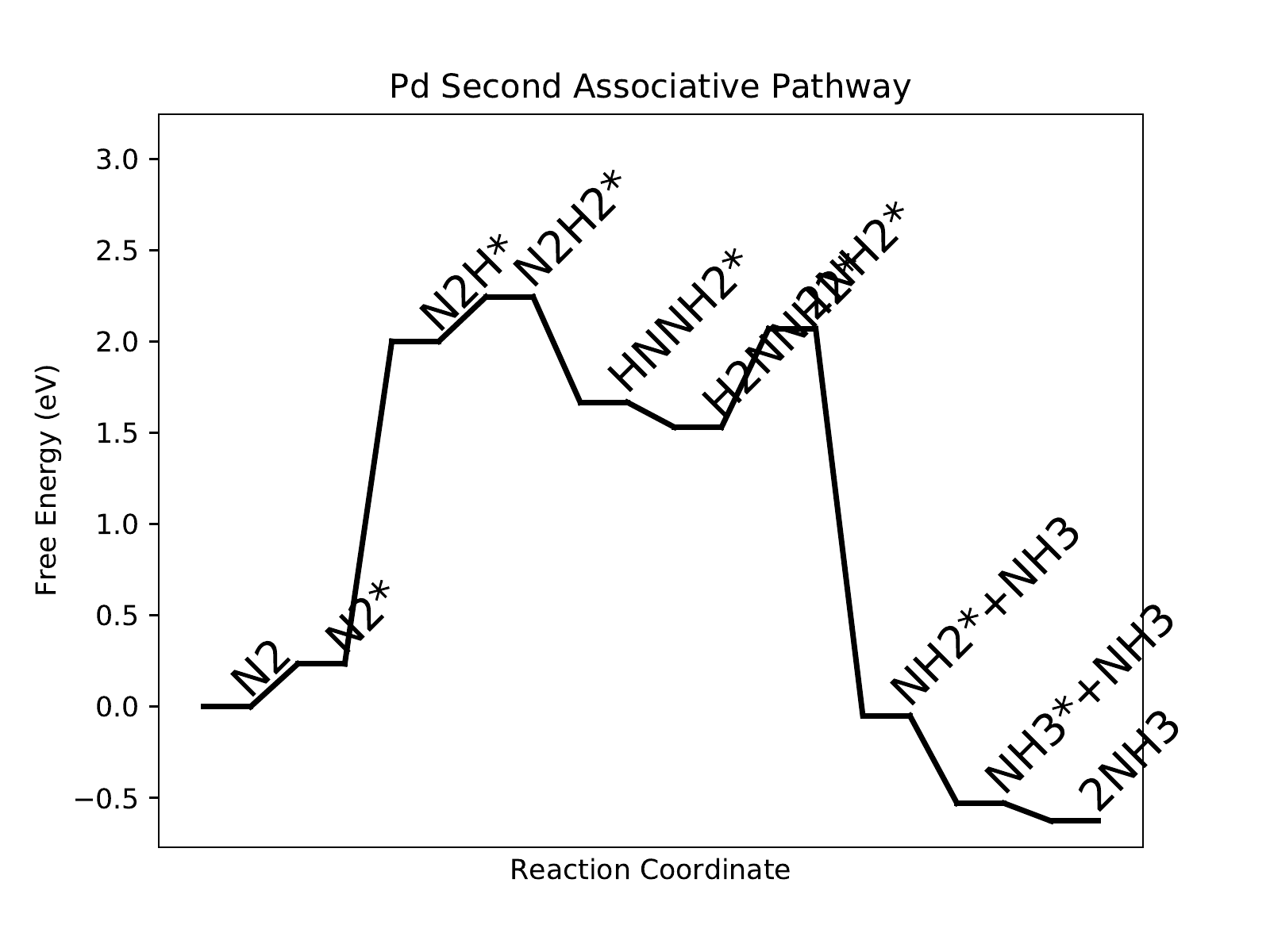}
\label{fig:Pd_associative_2}
\caption{Free energy diagram for Pd}
\end{figure}

\begin{figure}
\includegraphics[width=1\linewidth]{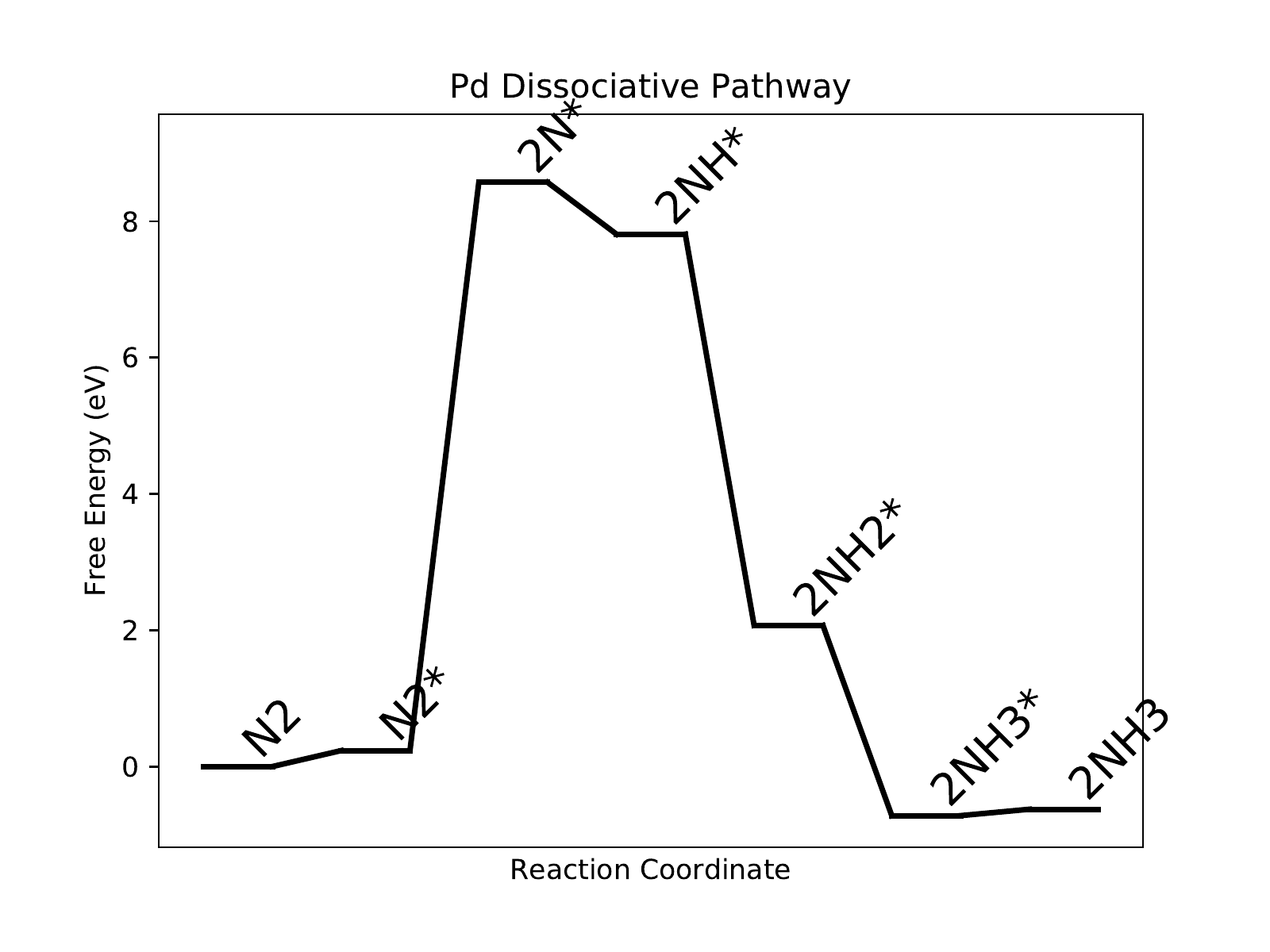}
\label{fig:Pd_dissociative}
\caption{Free energy diagram for Pd}
\end{figure}

\newpage
\begin{figure}
\includegraphics[width=1\linewidth]{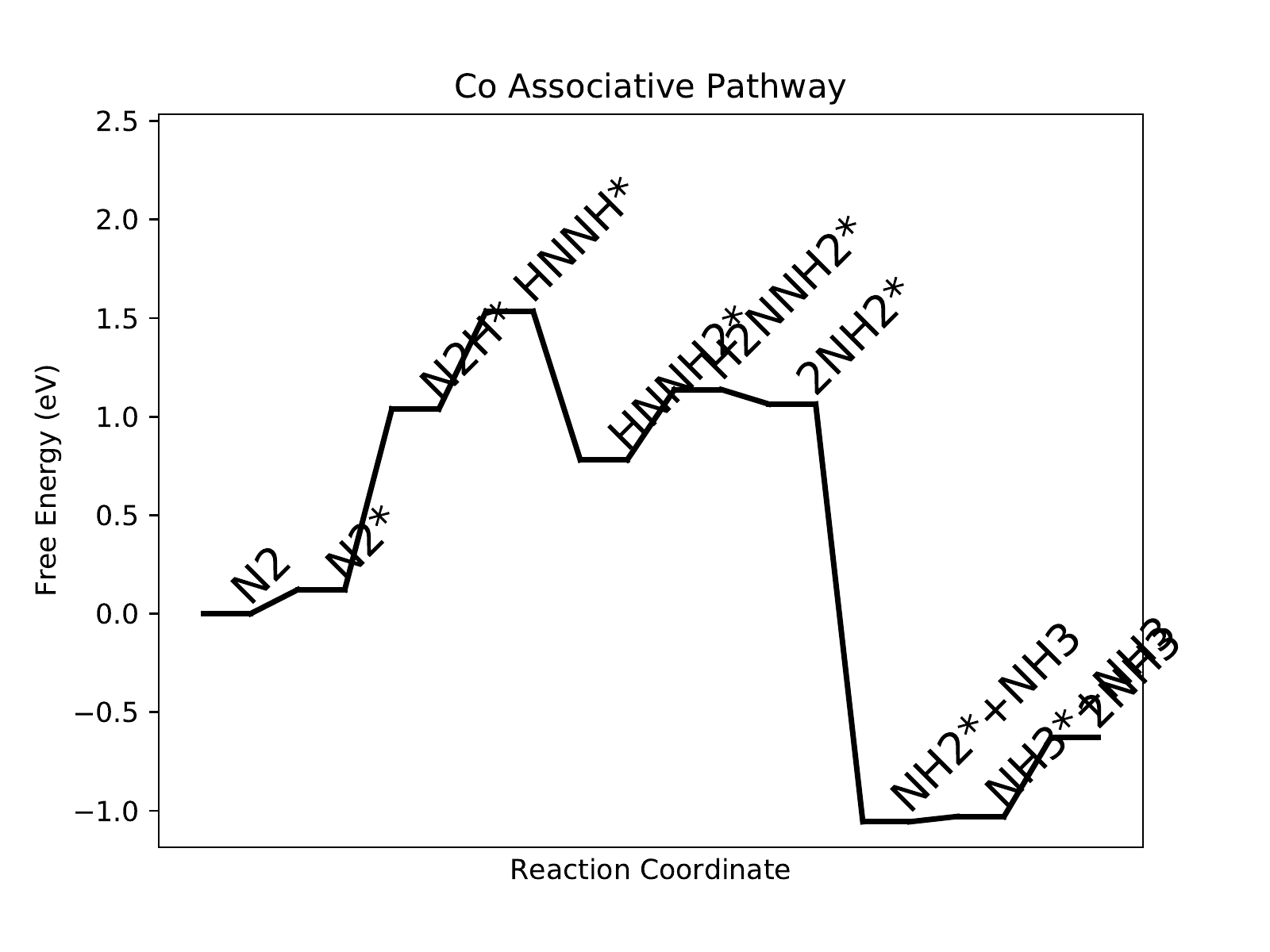}
\label{fig:Co_associative}
\caption{Free energy diagram for Co}
\end{figure}

\begin{figure}
\includegraphics[width=1\linewidth]{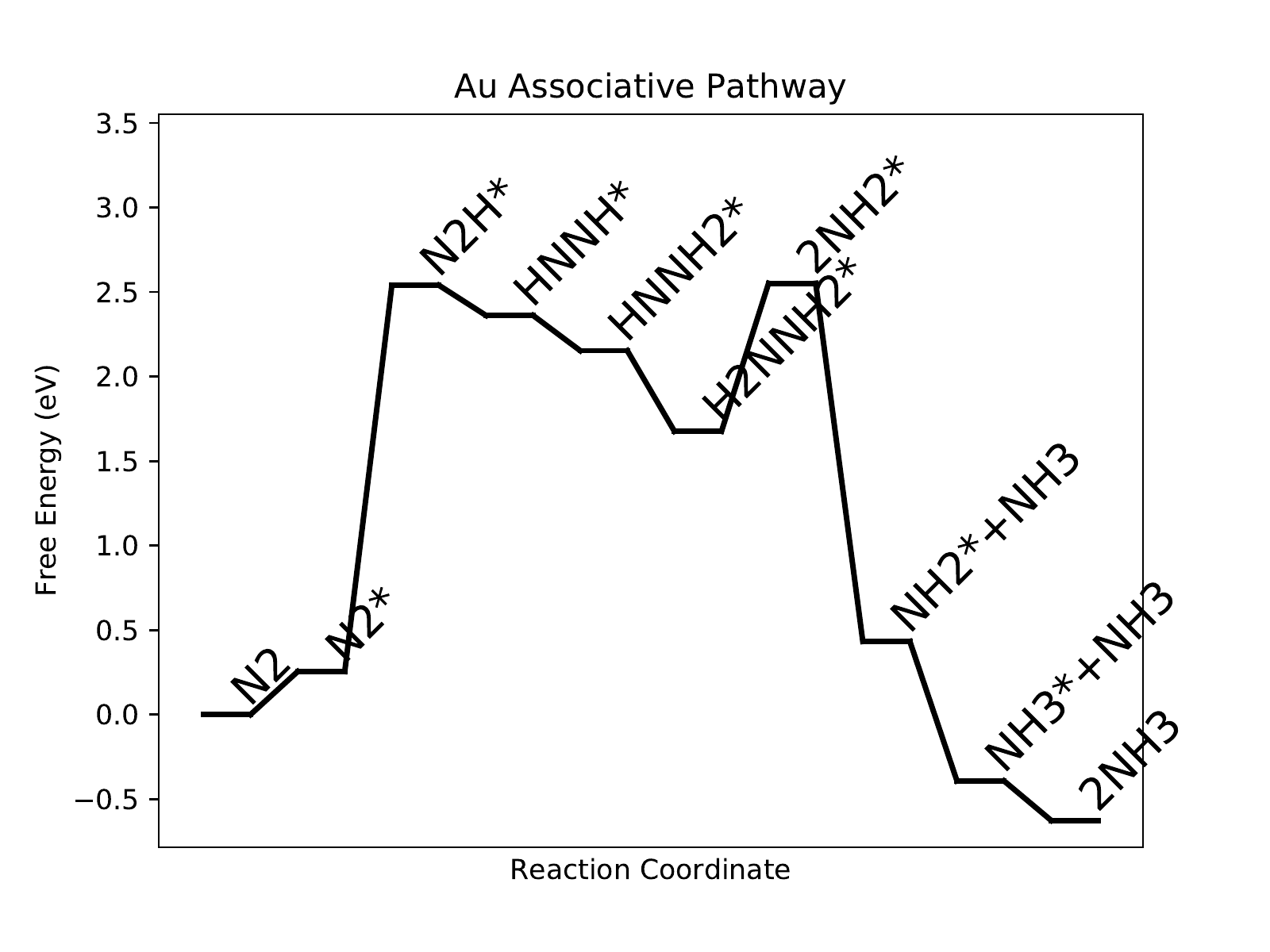}
\label{fig:Au_associative}
\caption{Free energy diagram for Au}
\end{figure}

\newpage
\begin{figure}
\includegraphics[width=1\linewidth]{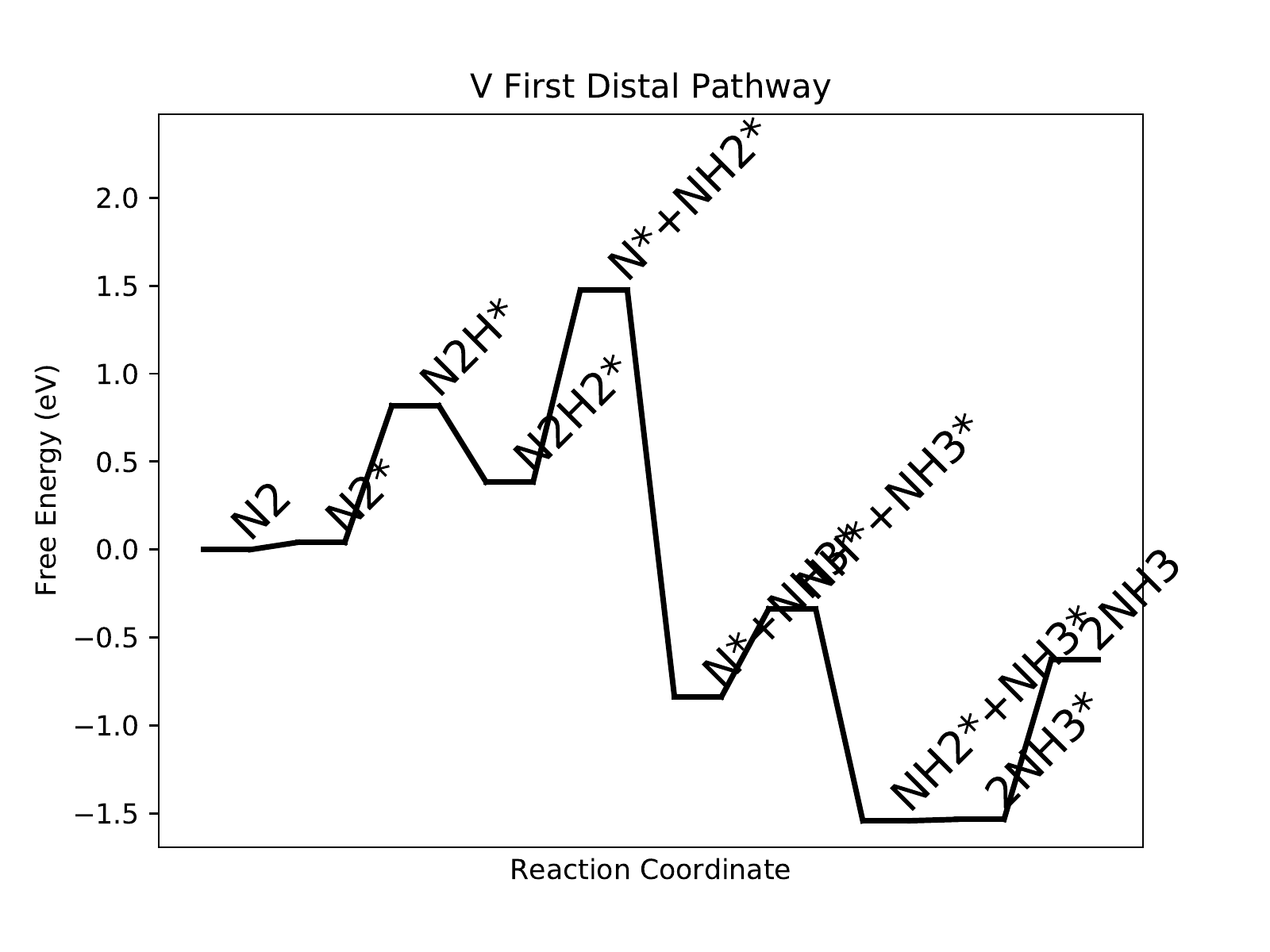}
\label{fig:V_distal_1}
\caption{Free energy diagram for V}
\end{figure}

\begin{figure}
\includegraphics[width=1\linewidth]{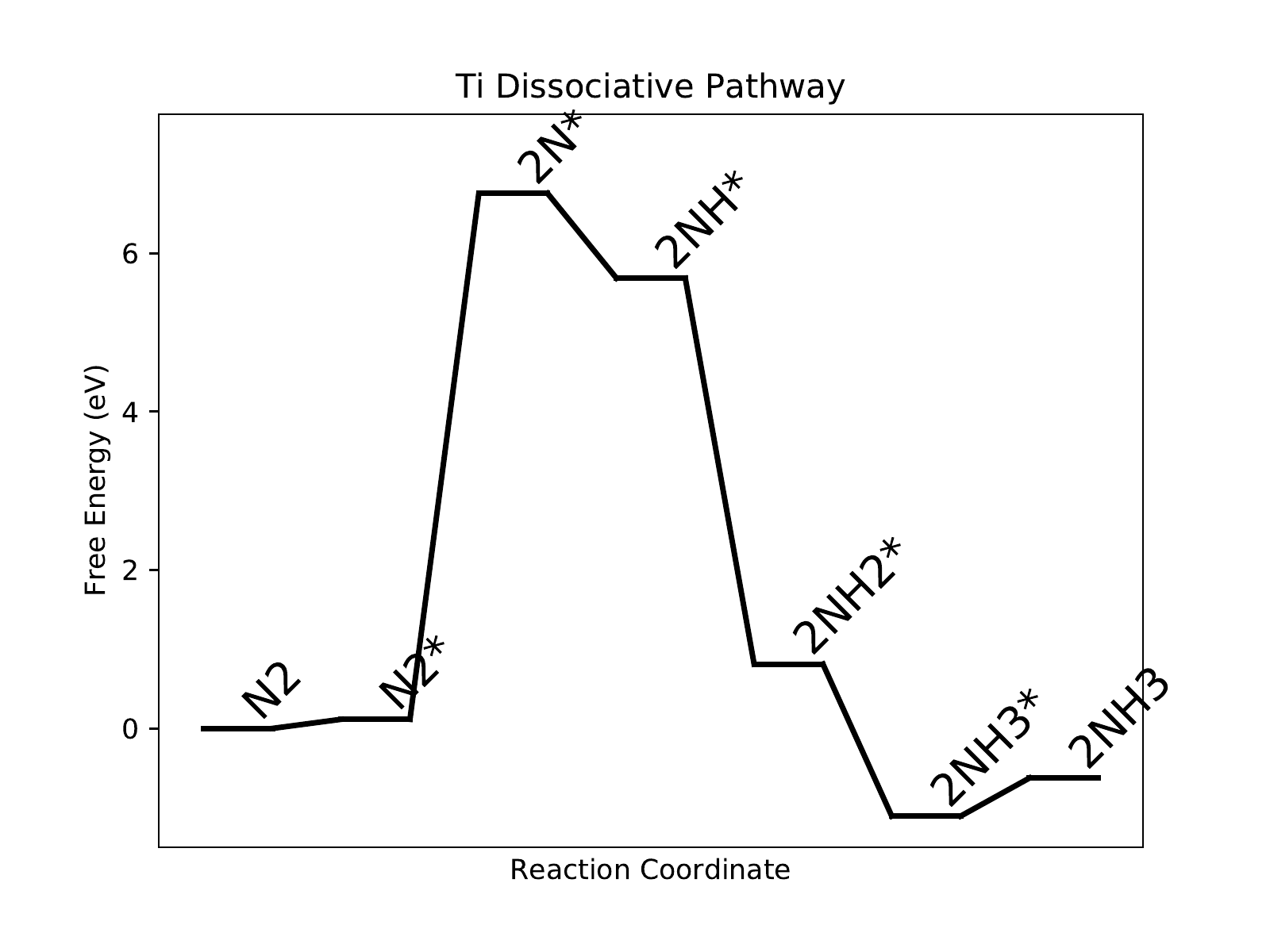}
\label{fig:Ti_dissociative}
\caption{Free energy diagram for Ti}
\end{figure}

\newpage
\begin{figure}
\includegraphics[width=1\linewidth]{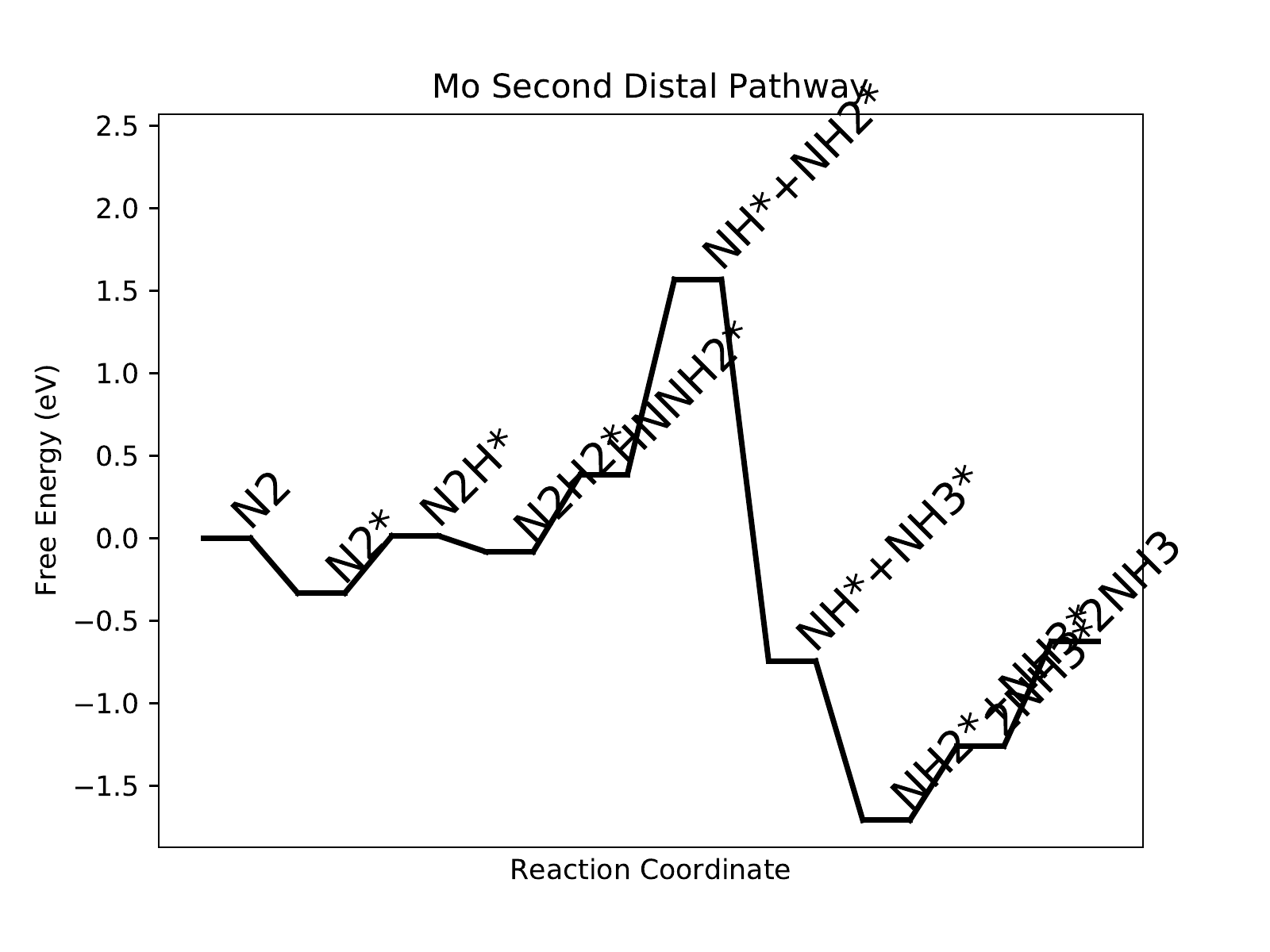}
\label{fig:Mo_distal_2}
\caption{Free energy diagram for Mo}
\end{figure}

\begin{figure}
\includegraphics[width=1\linewidth]{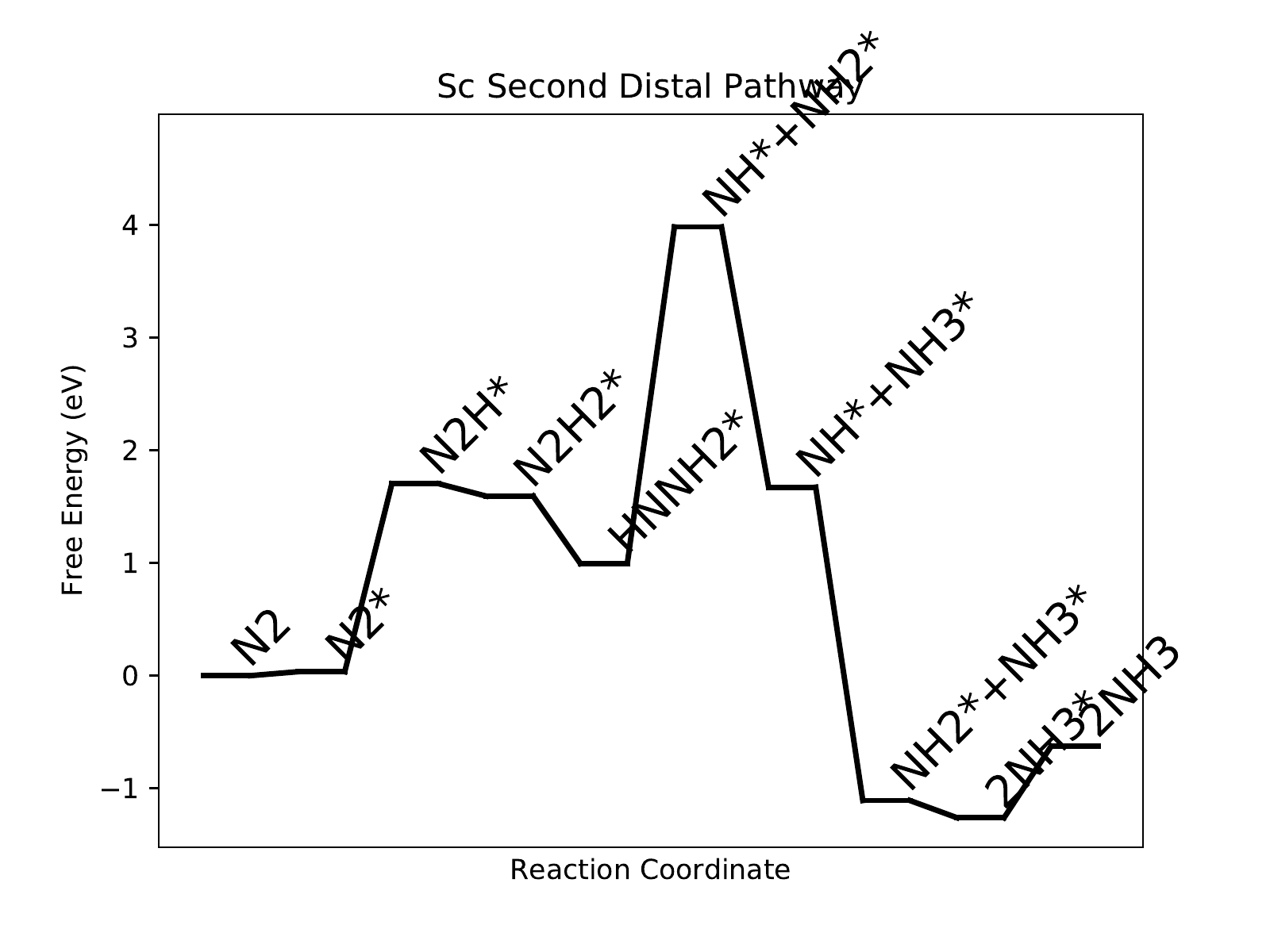}
\label{fig:Sc_distal_2}
\caption{Free energy diagram for Sc}
\end{figure}

\newpage
\begin{figure}
\includegraphics[width=1\linewidth]{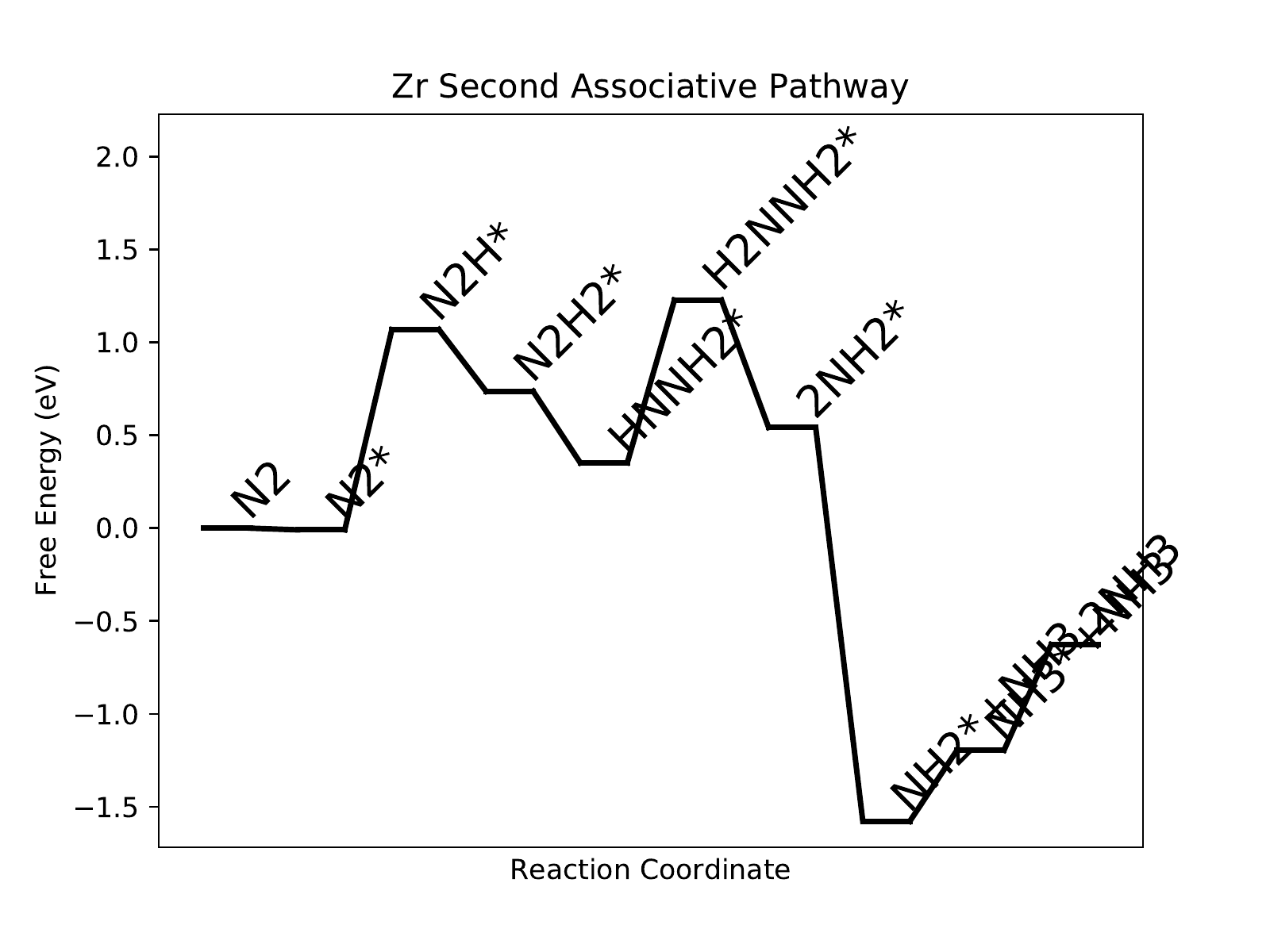}
\label{fig:Zr_associative_2}
\caption{Free energy diagram for Zr}
\end{figure}

\begin{figure}
\includegraphics[width=1\linewidth]{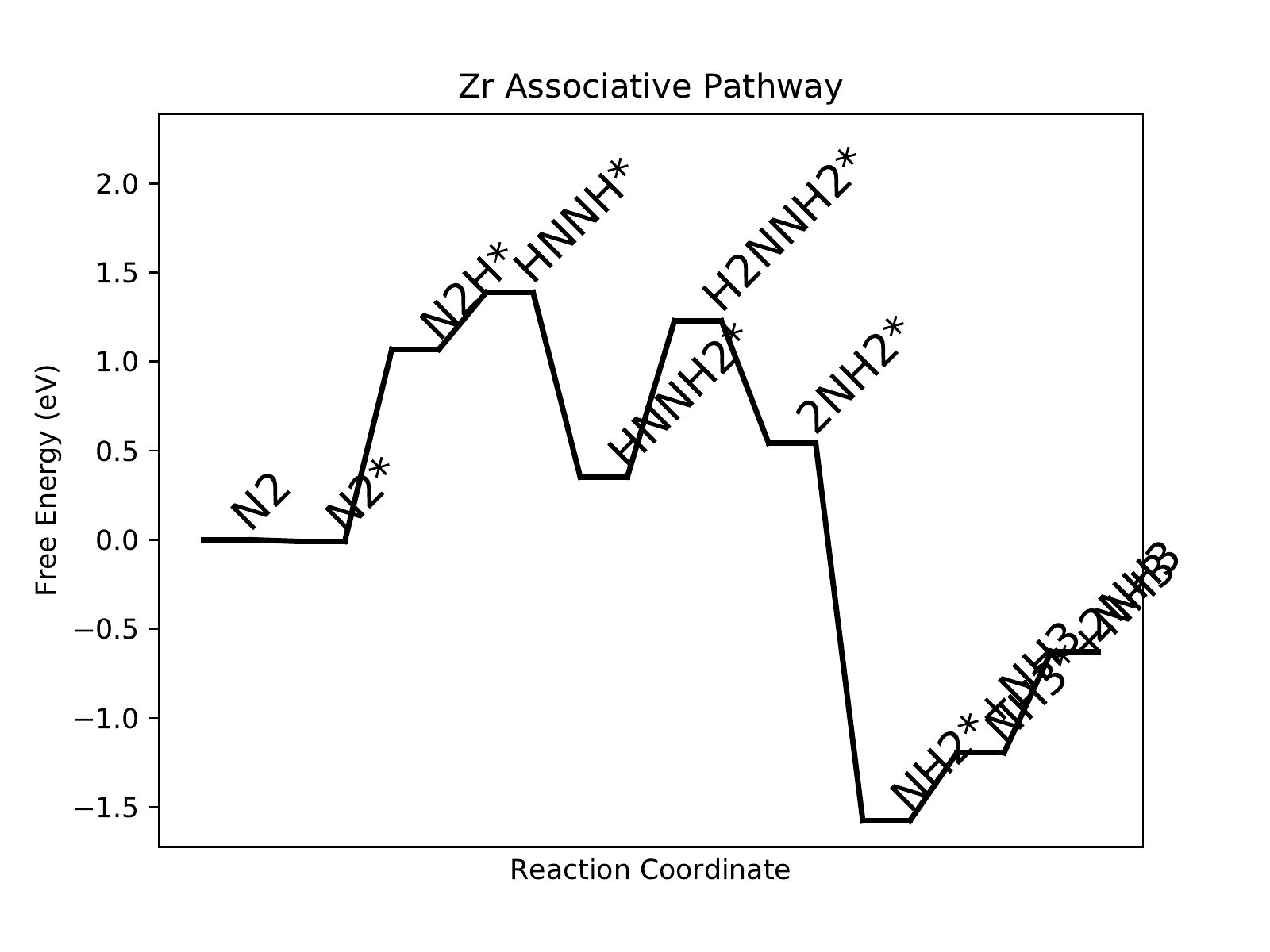}
\label{fig:Zr_associative}
\caption{Free energy diagram for Zr}
\end{figure}

\newpage
\begin{figure}
\includegraphics[width=1\linewidth]{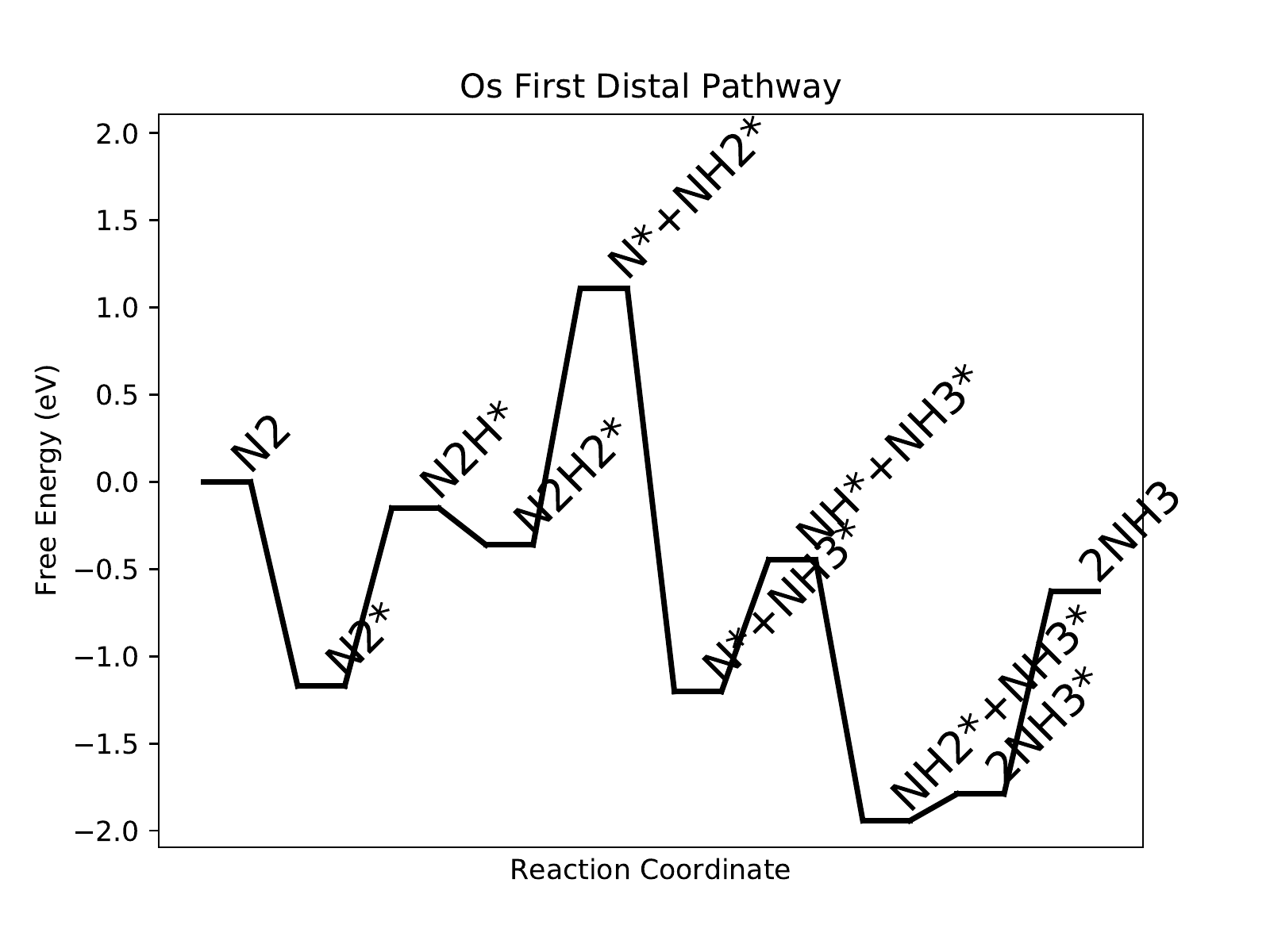}
\label{fig:Os_distal_1}
\caption{Free energy diagram for Os}
\end{figure}

\begin{figure}
\includegraphics[width=1\linewidth]{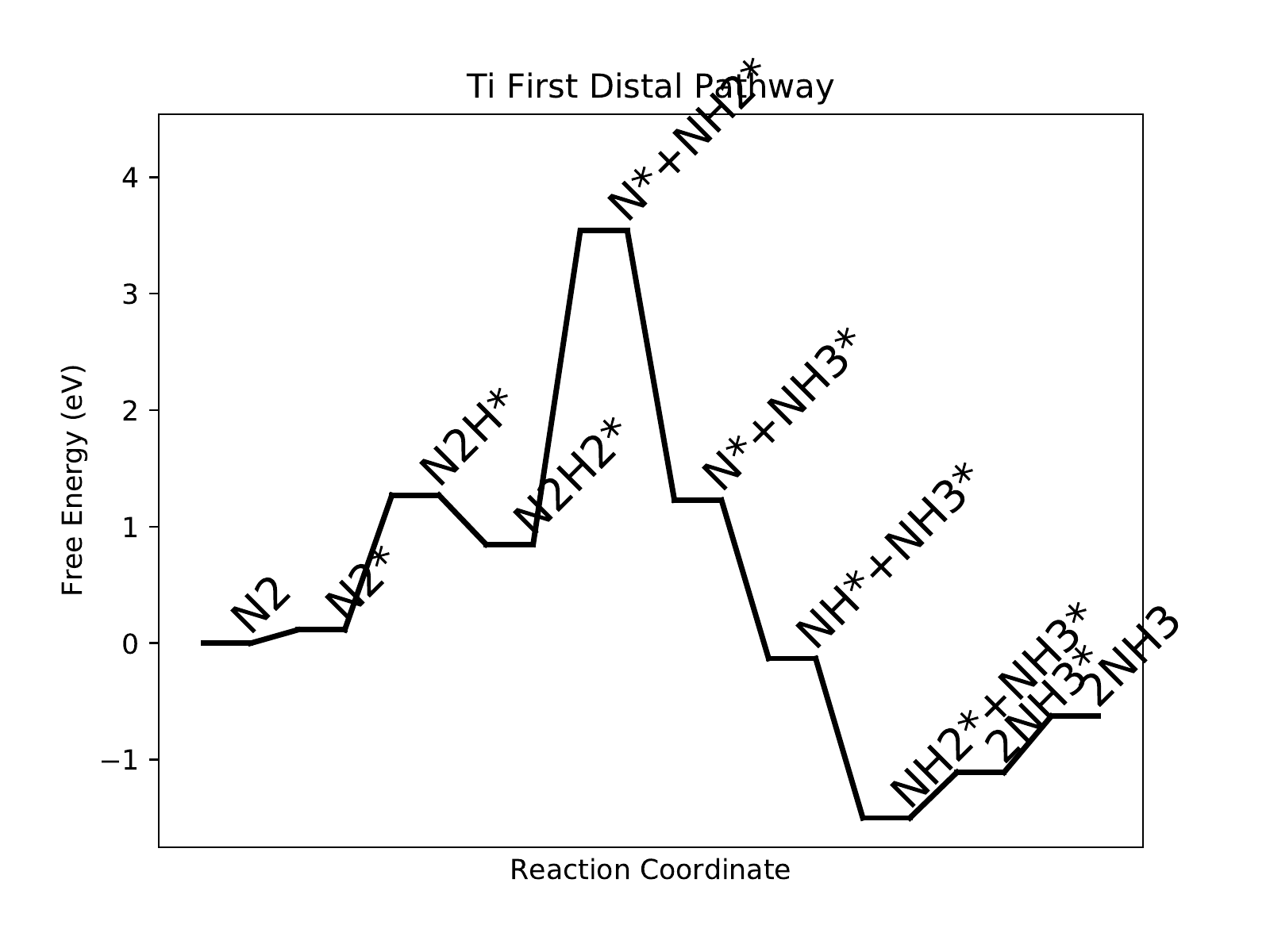}
\label{fig:Ti_distal_1}
\caption{Free energy diagram for Ti}
\end{figure}

\newpage
\begin{figure}
\includegraphics[width=1\linewidth]{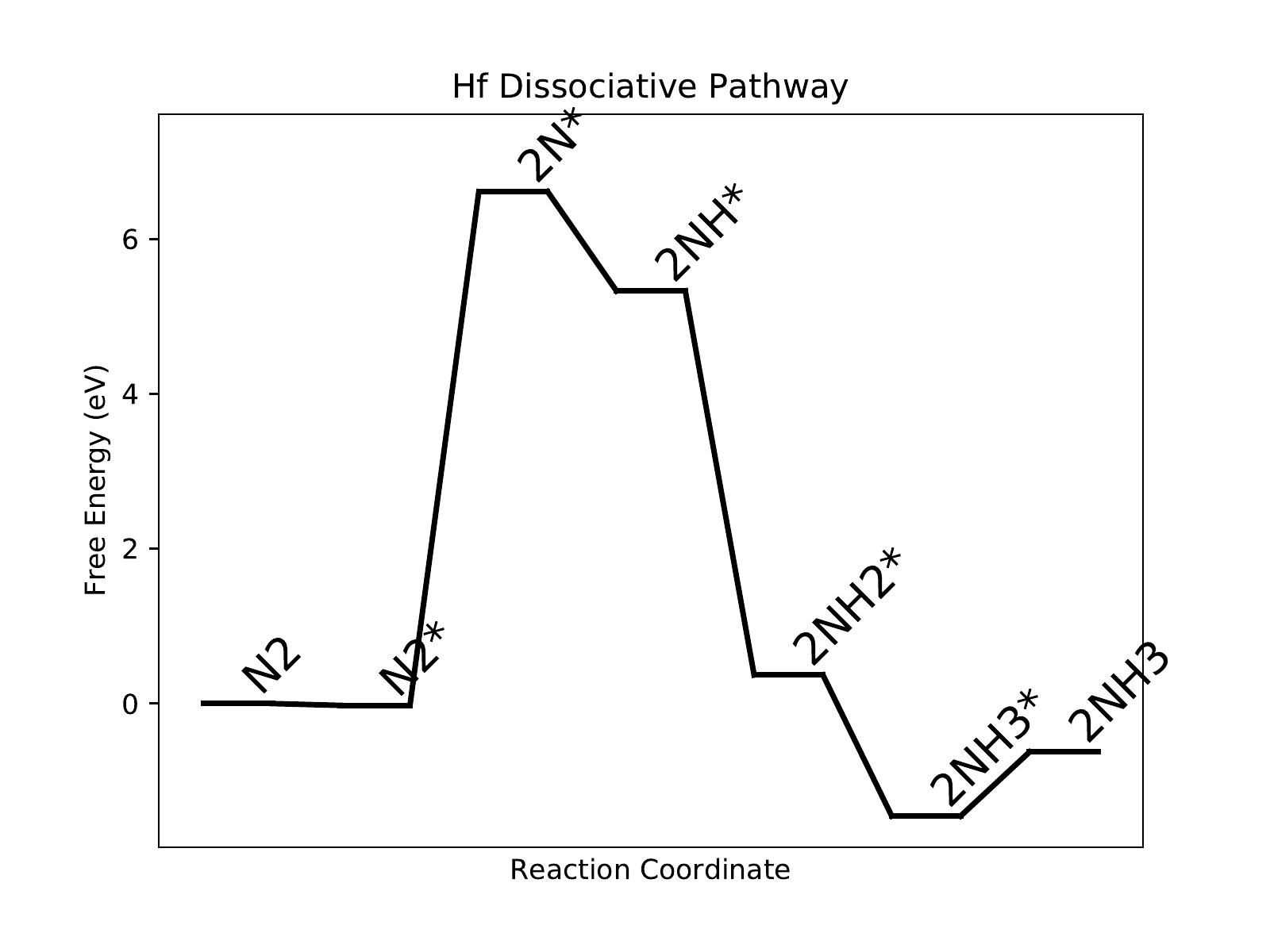}
\label{fig:Hf_dissociative}
\caption{Free energy diagram for Hf}
\end{figure}

\begin{figure}
\includegraphics[width=1\linewidth]{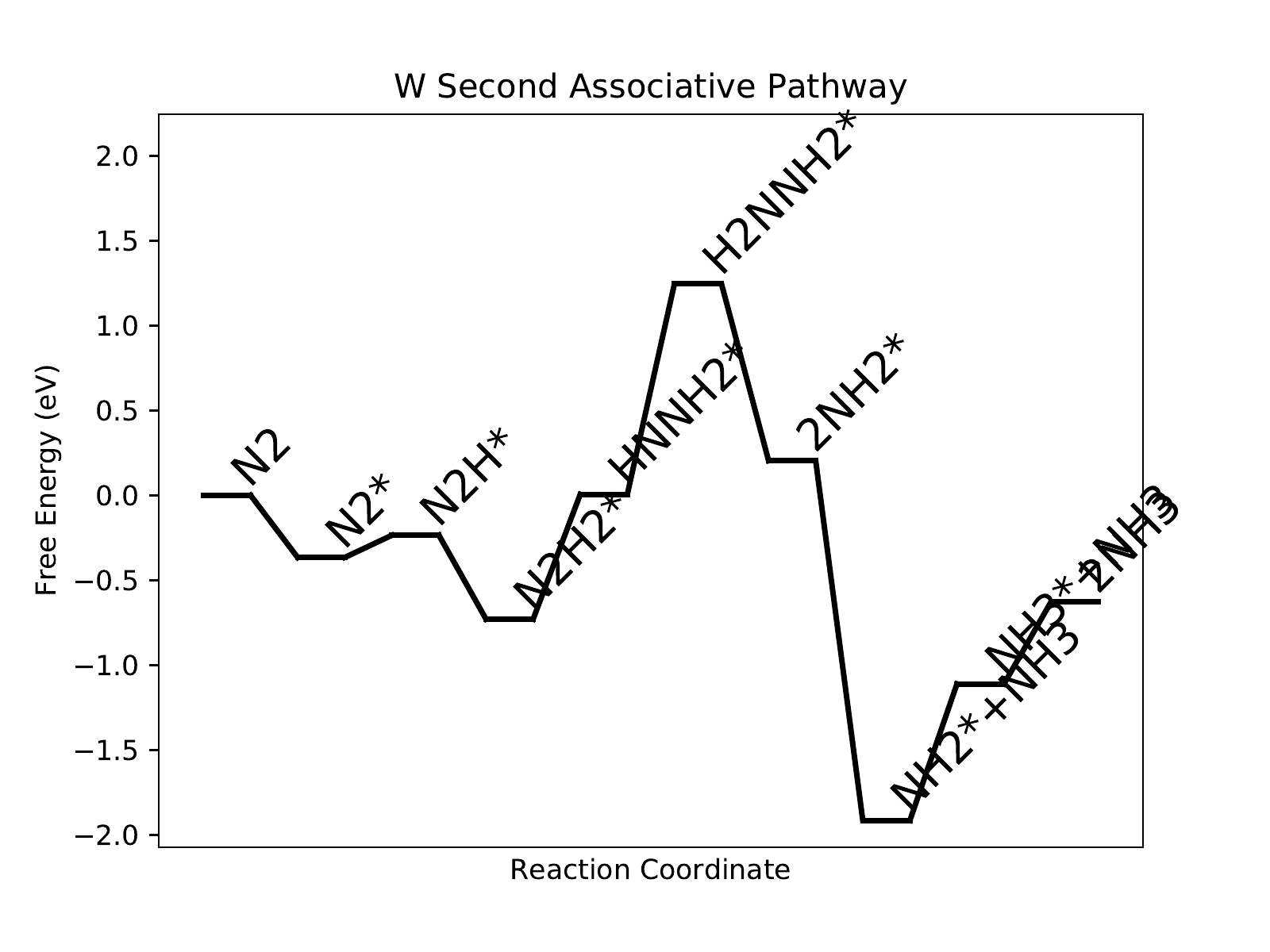}
\label{fig:W_associative_2}
\caption{Free energy diagram for W}
\end{figure}

\newpage
\begin{figure}
\includegraphics[width=1\linewidth]{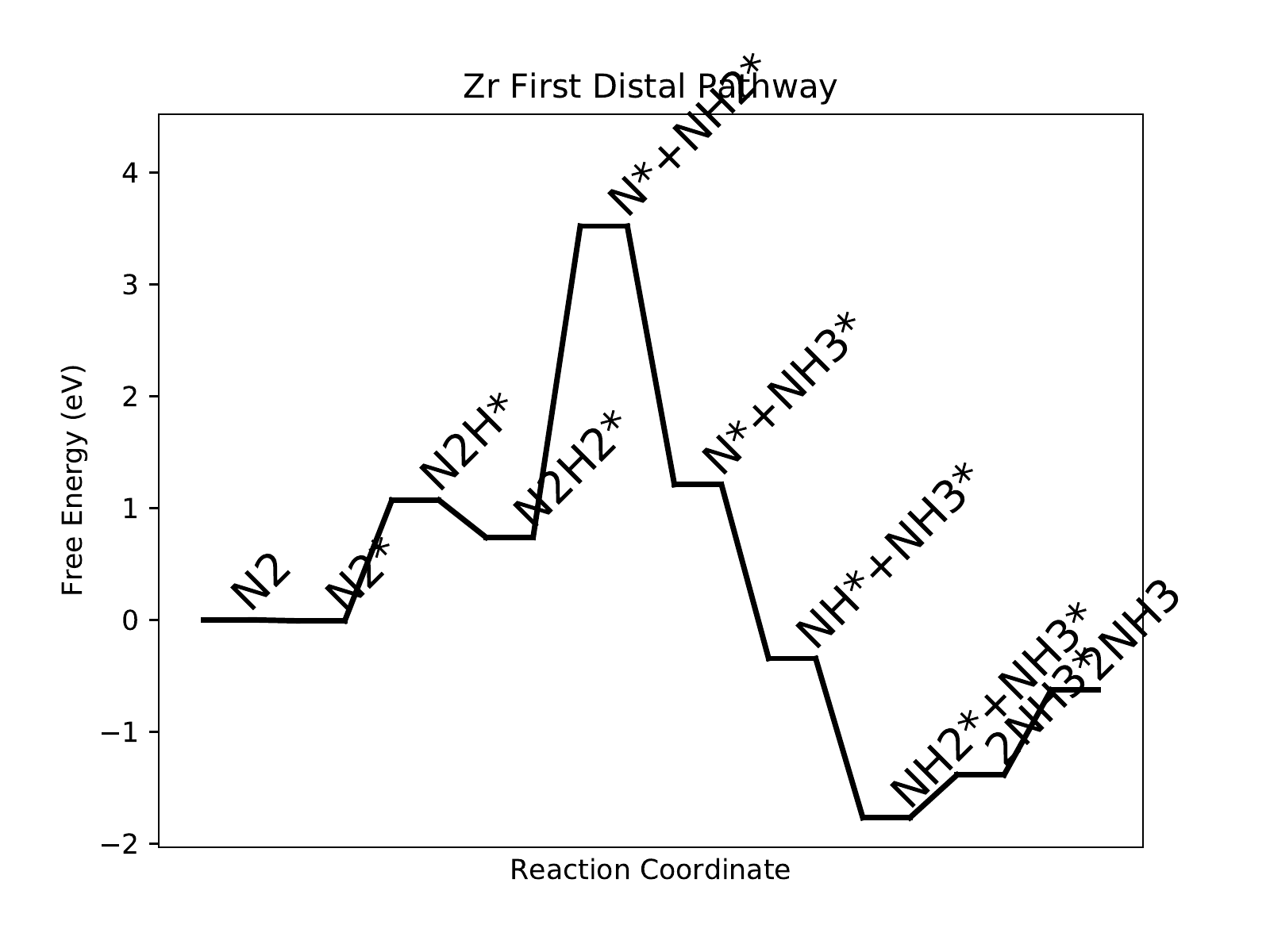}
\label{fig:Zr_distal_1}
\caption{Free energy diagram for Zr}
\end{figure}

\begin{figure}
\includegraphics[width=1\linewidth]{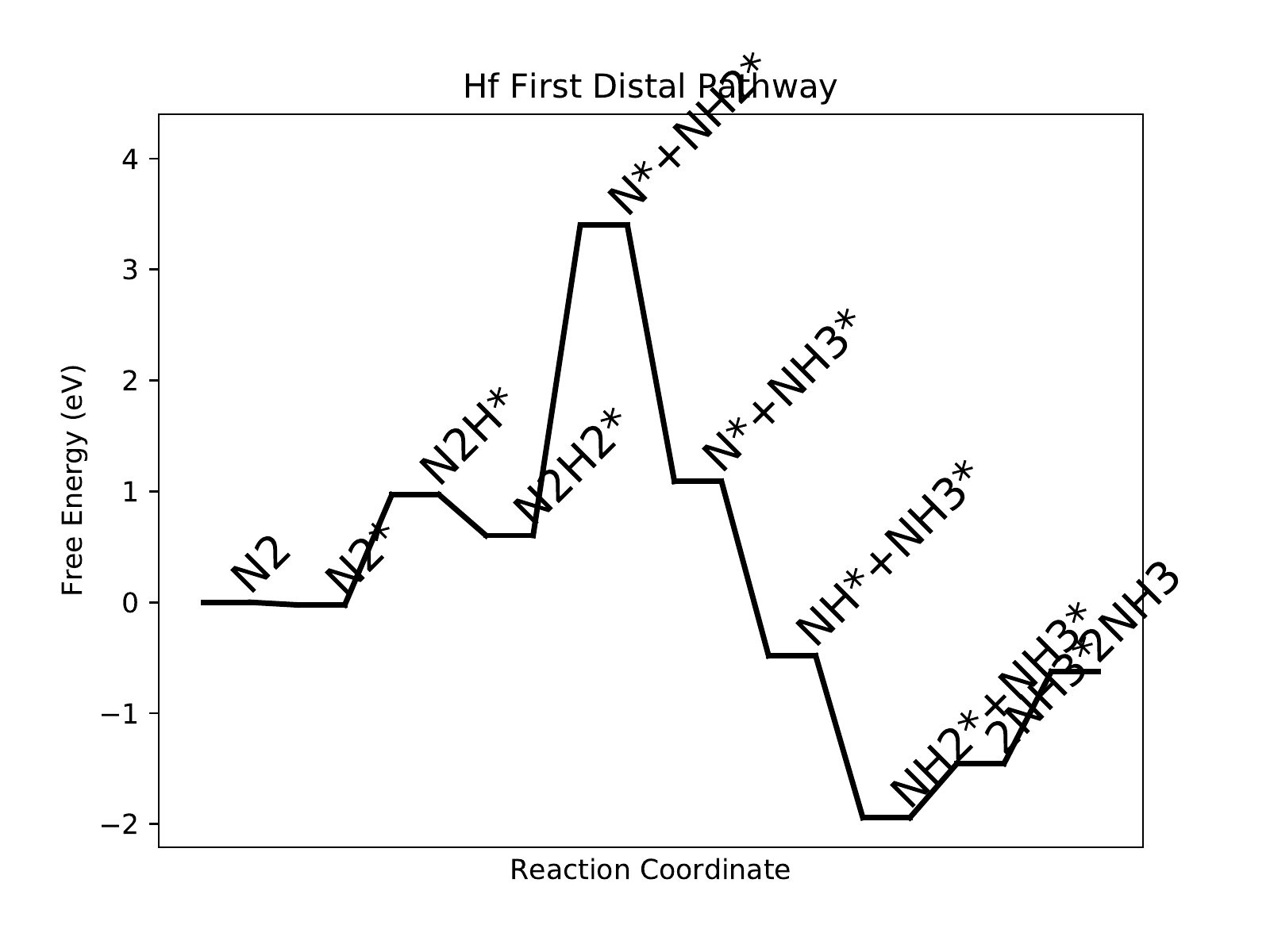}
\label{fig:Hf_distal_1}
\caption{Free energy diagram for Hf}
\end{figure}